\documentclass{elsart}
\usepackage{graphicx}
\usepackage{graphicx,graphics,color,epsfig}
\usepackage{amsmath}
\usepackage{amssymb}
\usepackage{type1cm}

\def\Tr{\mathop{\rm Tr}\nolimits}

\def\real{\mathbb{R}}
\def\complex{\mathbb{C}}
\def\cS{\mathcal{S}}
\def\cD{\mathcal{D}}
\def\cH{\mathcal{H}}

\def\Cov{\mathop{\rm Cov}\nolimits}
\begin{document}

\begin{frontmatter}

\title{Quantum information with Gaussian states}
\author{Xiang-Bin Wang$^1$}\thanks{ Email:
xbwang@mail.tsinghua.edu.cn}\address{ Department of Physics,
Tsingghua University, Beijing 100084, China;}\address{and Imai
Quantum Computation and Information Project, ERATO-SORST, JST, Daini
Hongo White Bldg. 201, 5-28-3, Hongo, Bunkyo-ku, Tokyo 113-0033,
Japan}
\author{Tohya Hiroshima$^2$},
\thanks{ Email: tohya@qci.jst.go.jp}~
\author{Akihisa Tomita},~
\author{Masahito Hayashi}
\address{ Imai Quantum Computation and Information Project, ERATO-SORST, JST,
Daini Hongo White Bldg. 201, 5-28-3, Hongo, Bunkyo-ku, Tokyo
113-0033, Japan}
\begin{abstract}
Quantum optical Gaussian states are a type of important robust
quantum states which are manipulatable by the existing technologies.
So far, most of the important quantum information experiments are
done with such states, including bright Gaussian light and weak
Gaussian light.
 Extending the existing results of quantum information with
discrete quantum states to the case of continuous variable quantum
states is an interesting theoretical job. The quantum Gaussian
states play a central role in such a case. We review the properties
and applications of Gaussian states in quantum information with
emphasis on the fundamental concepts, the calculation techniques and
the effects of imperfections of the real-life experimental setups.
 Topics here include the elementary properties  of
 Gaussian states and relevant quantum information device,
 entanglement-based quantum tasks such as quantum teleportation,
quantum cryptography with weak and strong Gaussian states and the
quantum channel capacity, mathematical theory of quantum
entanglement and state estimation for Gaussian states.
\end{abstract}

\end{frontmatter}

\maketitle
\tableofcontents
\section{Introduction}\label{secint}
Quantum information processing (QIP) is a subject on information
processing with quantum states\cite{nielsen}. In the recent years,
 the subject has
attracted much attention of scientists from various areas. It has
been found that in some important cases, quantum information
processing can have great advantage to any known method in classical
information processing.  A quantum computer can factorize a large
number exponentially more efficiently than the existing classical
methods do\cite{shor}. This means, given a quantum computer, the
widely used RSA system in classical communication is insecure
because one can factorize a huge number very effectively by Shor's
alhorithm. Interestingly, quantum key distribution (QKD) can help
two remote parties share a random binary string which is {\it in
principle} unknown to any third party\cite{bene}. Private
communication based on QKD is proven secure under whatever
eavesdropping including quantum computing. Quantum teleportation can
transfer unknown quantum state to a remote party without moving the
physical system itself\cite{telepor1}.

In classical information processing, all information are carried by
classical bits, which are binary digits of either 0 or 1. The
physical carrier of a classical bit can be any physical quantity
that has two different values, e.g., the electrical potential
(positive or negative voltages).  These are macroscopic quantities
which can be manipulated robustly by our existing technology. In
quantum information processing, we use quantum states to carry
either  quantum information or classical information. Also, we use
quantum entangled states as the resource to assist the effective
processing of quantum information. In principle, lots of different
physical systems can be used to generate the requested
 quantum states and quantum entanglement.
In those tasks related to communication, light seems to be the best
candidate for the physical system to carry the quantum information
and/or the quantum entanglement due to its obvious advantage that it
can be transmitted over a long distance efficiently. Therefore one
may naturally consider to use a single-photon state as a {\em
quantum bit} (qubit) and a two-photon entangled state as the
entanglement resource to assist the  processing.

Towards the final goal of real-life application of quantum
information processing, a very important question in concern is how
robustly we can manipulate the quantum states involved. In
practice, preparing the single-photon states or two-photon entangled
states deterministically are technically difficult. What can be
prepared and manipulated easily is a Gaussian state, e.g., a
coherent state\cite{glauber,KMC65} (the state of light pulses from a
traditional Laser) or a squeezed vacuum
state\cite{KMC65,yuenwf,knight,schu,zhangwm,leonhardt} of one mode
or two modes. Mathematically,  a state is Gaussian if its
distribution function in phase space or its density operator in Fock
space is in the Gaussian form.  (We shall go into details of the
phase space and Fock space later in this chapter and also other
chapters.) Gaussian states have proven to be a type of important
robust states which have been extensively applied for various QIP
tasks in labs. Actually, so far almost all those important
experiments of quantum information are done with Gaussian light.

There are two types of application of Gaussian states in practical
QIP. One is to use weak Gaussian light as approximate qubits or
two-photon entangled states. The other is to use strong Gaussian
light for continuous variable QIP. In applying the weak Gaussian
light, one can
 regard weak coherent light\cite{glauber}
as approximate single-photon source and regard 2-mode entangled weak
Gaussian light as probabilistic 2-photon entanglement resource in
polarization space. Weak Gaussian states have been used in many
experiments such as the quantum teleportation with spontaneous
parametric down conversion (SPDC)\cite{inns} and quantum
cryptography with weak coherent states or SPDC\cite{GRTZ02}.
However, sometimes the higher order terms, i.e., the multi-photon states in the single-mode
Gaussian light or the multi-pair states in the two-mode Gaussian light 
take a very important role even though the Gaussian light in
application is weak. For example, in QKD with weak coherent states,
the final key can actually be totally insecure if we ignore the role
of multi-photon pulses. For another example, in quantum-entanglement
based experiments such as quantum teleportation with SPDC, it is
possible that actually there is no entanglement and the result is a
post-selection result if we don't consider the detailed properties
of the weak Gaussian states. We shall review these types of effects
and possible ways to overcome the drawbacks. One can also use strong Gaussian light to obtain the analog
QIP results of discrete states for continuous variable states.
One important advantage here is that strong Gaussian light can be
used as a resource of deterministic quantum entanglement therefore
the results are deterministic and non-post-selection. This is quite
different from QIP using weak Gaussian states where the results are
often probabilistic and post-selection. We shall review the main
theoretical results of continuous variable QIP with strong Gaussian
light  including quantum teleportation, cloning, error correction
and QKD using a strong Gaussing light. We shall also review the
quantum entanglement properties of multi-mode Gaussian states,
quantum channel capacity and quantum state estimation of Gaussian
states.

This review is arranged as follows: In the remaining part of this
section, we present the elements of QIP and Gaussian states,
including an overview of quantum information  with qubits  and
two-photon maximal entangled states, mathematic foundation
and definition of Gaussian states and properties of a beam-splitter
as an elementary QIP device.
In section \ref{secebt}, we review the entanglement-based quantum
tasks with Gaussian states, which seem to be a very hot topic in
the recent years. The section includes quantum teleportation,
quantum error correction, quantum cloning and non-post-selection
quantum tasks with weak Gaussian states. We then go into the most
important application of QIP, practical quantum key distribution
in section \ref{sectomita} and section \ref{secqkd}. Section
\ref{sectomita} introduces the elements and technology background
of QKD with weak coherent light, section \ref{secqkd} reviews the
protocols and
 security proofs of QKD with weak and strong Gaussian light. We then change
 our direction to the mathematical theory
of QIP with Gaussian states in section \ref{secmet} and
\ref{seccc}. Section \ref{secmet} is on quantum entanglement of
Gaussian states, which have been extensively studied these years.
Section \ref{seccc} is on the properties of quantum Gaussian
channel, which is assumed to be a fundamental subject in quantum
communication. The theory of quantum state estimation of Gaussian
states is reviewed in section \ref{sechayashi}.  This review does
not include quantum computation.
\subsection{Elements of quantum information with 2-level states}
A {\em qubit} is simply a  physical system that carries a two-level quantum state.
For example, a photon can be regarded as a qubit in polarization space or any other two dimensional space.
In general, we have the following mathematical form for
 the state of a qubit
\begin{equation}
|\psi\rangle =\alpha |0\rangle + \beta |1\rangle \label{qubit}
\end{equation}
and $|\alpha|^2+|\beta|^2=1$.
Here $|0\rangle$ and $|1\rangle$ are orthonormal states of any two-level system.
 Mathematically, one can use the
following representation
\begin{equation}
|0\rangle=\left(\begin{array}{c}1\\0\end{array}\right);|1\rangle=\left(\begin{array}{c}0\\1\end{array}\right).
\end{equation}
Consequently, we can use matrices as  representations of
operations to a qubit. In particular, the unity matrix $I$
represents for doing nothing, matrix
$\sigma_x=\left(\begin{array}{cc}0 & 1\\ 1& 0\end{array}\right)$
for a bit-flip operation, $\sigma_z=\left(\begin{array}{cc}1 & 0\\
0& -1\end{array}\right)$ for a phase-flip operation and
$\sigma_y=\left(\begin{array}{cc}0 & i\\ -i& 0\end{array}\right)$
for both bit-flip and phase-flip.

If we use the photon polarization, notations $|0\rangle,|1\rangle$
 represent the horizontal polarization and the vertical polarization (polarization of angle $\pi/2$)
 respectively.
One can replace them with more vivid notations of
$|H\rangle,|V\rangle$. For example, state $\cos\theta|H\rangle +
\sin\theta |V\rangle$ is the  state of polarization angle $\theta$.
In the above equation, $|\psi\rangle$ is linearly superposed by
$|0\rangle$ and $|1\rangle$. The quantum linear superposition is
{\em different} from the classical probabilistic mixture. For
example, consider the polarizations of $\pi/4$ and $3\pi/4$. They
are the linear superposition states of $|\pm\rangle=\frac{1}{\sqrt
2}(|H\rangle\pm |V\rangle)$. Given state $|+\rangle$, i.e.,
polarization of $\pi/4$, if we measure it in the
 $|H\rangle,|V\rangle$ basis, we have equal probability to obtain
an outcome of either $|H\rangle$ or $|V\rangle$. This is due to
the fact
\begin{equation}\label{qubiteq2}
{\left|\langle H|+\rangle\right|}^2={\left|\langle V|+\rangle\right|}^2=1/2.
\end{equation}
However, the state $|+\rangle$, i.e., the polarization of $\pi/4$ is
different from a $mixed$ state which is in a classical mixture of
horizontal and vertical polarizations, with equal probability. Such
a $classical$ mixture can be realized in this way: Source A only
emits photons of horizontal polarization and source B only emits
photons of vertical polarization. In a remote place we receive
photons from both sources. But  we don't know which photon is from
which  source. Our only knowledge is that any photon has equal
probability from A or B. In such a case, any individual photon is in
a classical mixture of state $|H\rangle$ and $|V\rangle$. If we
measure such a mixed state in $\{|H\rangle,~|V\rangle\}$ basis, the
measurement outcome is identical to that of a $pure$ state of
$\pi/4$ polarization. However, if we use a $\pi/4$ polarizer, the
outcome will be different:  the $pure$ state of $\pi/4$ polarization
will always transmit the polarizer while the mixed state only has
half a probability to transmit the polarizer. These can be
interpreted mathematically by Eq.(\ref{qubiteq2}) and the following
equation :
\begin{equation}
\left|\langle +|+\rangle\right|^2=1.
\end{equation}
 Eq.(\ref{qubiteq2})
 shows that both $|H\rangle$ and $|V\rangle$ have half a probability
to transmit the $\pi/4-$polarizer. This means, any classical mixture
of $|H\rangle$ and $|V\rangle$ will always only have half a
probability to pass through the $\pi/4-$polarizers. To have a
universal mathematical picture for both pure states and mixed
states, we can use {\it density operator}  which reflects the
classical probability distribution over different quantum states.
Suppose a certain source consists of $n$ sub-sources. Any sub-source
$i$ will only produce the pure state $|\psi_i\rangle$. Whenever the
source emits a photon, the probability that the photon being emitted
from sub-source $i$ is $p_i$ ($p_1+p_2+\cdots p_n =1$). The state of
any photon from such a source can be described by the density
operator $\sum_ip_i|\psi_i\rangle\langle \psi_i|$.  A density
operator can be represented by a matrix which is called as {\it
density matrix}. The density matrix for pure state $|\psi\rangle=\alpha |0\rangle+\beta|1\rangle$ is
\begin{equation}
|\psi\rangle\langle\psi| = \left(\begin{array}{c}\alpha\\ \beta\end{array}\right) (\alpha^*,\beta^*)
=\left(\begin{array}{cc}|\alpha|^2 & \alpha\beta^*\\ \alpha^*\beta & |\beta|^2\end{array}\right).
\end{equation}
Given any pure state $|\psi\rangle$, there is always a $unitary$ matrix $U$ satisfying
\begin{equation}
|\psi\rangle =U|0\rangle.
\end{equation}
This is to say, we can use the following criterion for a pure
state $|\psi\rangle\langle\psi|=U|0\rangle\langle 0| U^{\dagger}$:
Given a density matrix $M$, if it is for a $pure$ state, there
exists a unitary matrix $U$ so that
\begin{equation}
UMU^\dagger = \left(\begin{array}{cc}1 & 0\\0 & 0\end{array}\right).
\end{equation}
Besides qubits, quantum entanglement is often needed in non-trivial
QIP tasks, such as quantum teleportation\cite{telepor1}. We shall
consider the most well-known case of two-photon maximally entangled
state in the polarization space. Consider the state for spatially
separated two photons, A and B. We have 4 orthogonal states to span
the polarization space of the two-photon system:
\begin{equation}
|\phi^{\pm}\rangle = \frac{1}{\sqrt 2}(|H\rangle_A|H\rangle_B\pm|V\rangle_A|V\rangle_B);\nonumber\\
|\psi^{\pm}\rangle = \frac{1}{\sqrt 
2}(|H\rangle_A|V\rangle_B\pm|V\rangle_A|H\rangle_B).
\end{equation}
All these 4 states are maximally entangled and we shall call any of
them as an EPR (pair) state named after
Eistein-Podolsky-Rosen\cite{EPRname}, or a Bell state. We shall also
call the measurement basis of these 4 EPR states Bell basis.
We use $|\phi^+\rangle$ to demonstrate the non-trivial properties of
an EPR pair. For the pair state $|\phi^+\rangle$, the polarization
of photon A and photon B are correlated: if we measure each of them
in $\{|H\rangle,|V\rangle\}$ basis, we always obtain the same
polarization outcome for two photons. However, this alone does not
show any non-trivial property of quantum entanglement. Because a
certain classical correlation can also produce the same result.
Consider a mixed state $\rho_c=\frac{1}{2}[(|H\rangle\langle
H|)_A\otimes (|H\rangle\langle H|)_B+ (|V\rangle\langle V|)_A\otimes
(|V\rangle\langle V|)_B]$. This state also always produces the same
polarization for two photons, if each photon is measured in
$\{|H\rangle,|V\rangle\}$ basis. One can easily imagine a physical
realization of such a state: a source always produces two
horizontally polarized photons flying in different directions
(photon pairs). However, whenever there is an emission, with half
probability we do nothing to the two photons and with half
probability we rotate $both$ of them by $\pi/2$. In a remote place,
it is unknown which pair is rotated which pair is not. Then any pair
is just in a state with classical correlation as  stated above.
However, an EPR pair has something more than this. Given state
$|\phi^+\rangle$, if each photon is measured in $\{|\pm\rangle\}$
basis, we shall still always find the same polarization correlation
for two photons in the measurement outcome: either both of them are
$|+\rangle$ or both of them are $|-\rangle$, for
\begin{equation}
|\phi^+\rangle= \frac{1}{\sqrt
2}(|+\rangle_A|+\rangle_B+|-\rangle_A|-\rangle_B).
\end{equation}
However, given the state with classical correlation, $\rho_c$, there is no correlation
between two photons if we measure each of them in   $\{|\pm\rangle\}$ basis. As one may easily
check it mathematically:
\begin{equation}
\langle ++|\rho_c|++\rangle =\langle --|\rho_c|--\rangle=\langle +-|\rho_c|+-\rangle=
\langle -+|\rho_c|-+\rangle =1/4.
\end{equation}
Here we have omitted the subscripts for photon A and B.

In the entanglement-assisted QIP tasks, we often need a  Bell
measurement, i.e., a collective measurement in the basis of 4 EPR
states. As shall be shown, the Bell measurement to a photon pair can
be partially done by the currently existing technology through a
beam-splitter.

We often need two elementary ingredients in QIP with two-level
states: a qubit state and an EPR state. Give these resources, we can
carry out a number of non-trivial tasks.
As noted earlier, in practice, a perfect single photon source for
qubits and a perfect entangled-pair source are difficult techniques.
Actually, in almost all existing QIP experiments, these difficult
sources are replaced by weak or strong Gaussian states. The goal of
this review is to present the main results of QIP with weak and
strong Gaussian states.

\subsection{Phase space representation and definition of Gaussian states}\label{fcs}
To clearly define the Gaussian states, we shall use the phase
space\cite{milburn,zoller,barnett,orszag}.  We first take a look at
some elementary properties of the Fock
space\cite{milburn,zoller,barnett,orszag}. Consider the single-mode
field. We have photon-number operator, $\hat N$. Any Fock state
$|l\rangle$ is an eigen-state of photon-number operator, i.e., $\hat
N|l\rangle=l|l\rangle$. When we say that a light pulse is in state
$|l\rangle$, we mean there are $l$ photons in that pulse. Also, we
have creation operator $a^\dagger$ and annihilation operator $a$ and
$\hat N=a^\dagger a$, i.e.
\begin{equation}\label{qed}
a^\dagger  a|l\rangle =l|l\rangle.
\end{equation}
Moreover, $a^\dagger,a$ satisfy the following equations,
\begin{equation}\label{genda}
a^\dagger|l\rangle =\sqrt{l+1}|l+1\rangle
\end{equation}
and
\begin{equation}
a |l\rangle=\sqrt l |l-1\rangle.
\label{gen}\end{equation}
with $ a|0\rangle=0 $.
The operator $a^\dagger$ and $a$ don't commute and
\begin{equation}
[a,a^\dagger]=aa^\dagger-a^\dagger a=1.
\end{equation}
A detailed derivation of the mathematical structure of this Fock space
  from classical electrodynamics can be found in, e.g.\cite{milburn,zoller,barnett,orszag}.
With  eq.(\ref{genda}), it's easy to see that
 mathematically
any photon number state $|l\rangle$ can be generated from the vacuum
by
\begin{equation}\label{exite}
|l\rangle =\frac{{a^\dagger}^l}{\sqrt {l!}}|0\rangle.
\end{equation}
Moreover, an arbitrary pure state in Fock space can be written in the form
\begin{equation}
|\chi\rangle = f(a^\dagger)|0\rangle,
\label{mgen}\end{equation}
where $f(a^\dagger)$ is a functional of $a^\dagger$.
For example, for state $ \frac{1}{\sqrt 3}(|0\rangle+|1\rangle+|2\rangle )$,
we just set $f(a^\dagger)= \frac{1}{\sqrt 3}
(1+a^\dagger+\frac {{a^{\dagger}}^2}{\sqrt 2})$ to rewrite it in the form of
eq.(\ref{mgen}).
%

Straightforwardly, one can extend the above one-mode Fock space to
the multi-mode Fock space. Radiation field can be decomposed into
different radiation modes characterized by the wave number vector
and polarization\cite{milburn,zoller,barnett,orszag}.  Let $\hat
x_{k}$ and $\hat p_{k}$ denote the `position' and `momentum'
operators associated with the $k$th mode, respectively
$(k=1,2,\cdots ,n)$. These operators or canonical variables are
written in terms of the creation and annihilation operators of the
mode;
\[
\hat x_{k}=\sqrt{\frac{1}{2\omega _{k}}}\left(
a_{k}+a_{k}^{\dagger }\right)
\]
and
\[
\hat p_{k}=-i\sqrt{\frac{\omega _{k}}{2}}\left(
a_{k}-a_{k}^{\dagger }\right),
\]
where $\omega _{k}$ denotes the energy of  quanta of the $k$th mode ($\hbar
=1$). Since $[a_{j},a_{k}]=[a_{j}^{\dagger },a_{k}^{\dagger }]=0$
and $[a_{j},a_{k}^{\dagger }]=\delta _{jk}$, we have $[\hat
x_{j},\hat x_{k}]=[\hat p_{j},\hat p_{k}]=0$ and $[\hat x_{j},\hat
p_{k}]=i\delta _{jk}$. Defining $R=(R_{1,}R_{2},\cdots
,R_{2n})^{T}= (\omega _{1}^{1/2}\hat x_{1},\omega _{1}^{-1/2}\hat
p_{1},\cdots ,\omega _{n}^{1/2}\hat x_{n},\omega _{n}^{-1/2}\hat
p_{n})^{T}$, these canonical commutation relations (CCRs) can be
written compactly as $[R_{j},R_{k}]=iJ_{jk}$. Here, $J=\oplus
_{j=1}^{n}J_{1}$ with
\[
J_{1}=\left(
\begin{array}{cc}
0 & 1 \\
-1 & 0
\end{array}
\right) .
\]
For convenience, we shall assume $\omega_k=1$ unless specifically noticed. 
Most generally, the density operator of a state in Fock space can
be written in a functional form of creation and annihilation
operators. But we don't have to always use this operator format
because a density operator can be represented by a (quasi)
distribution function in phase space. The phase space
representation can often simplify the calculations. Actually, the
so called Gaussian state can be easily defined as  a state whose
distribution function  in phase space is in the Gaussian form.

 An $n-$mode density operator $\rho$ is defined on the phase
space that is a $2n$-dimensional real vector space.  The
characteristic function
is\cite{milburn,zoller,barnett,orszag,Pet90,tmsq01,weyl,wigner,wignerr,quasi8,moyl,kimqm,gdkthe,GKD01}
\begin{equation}
\chi (\xi )=\mathrm{Tr}[\rho \mathcal{W}(\xi )].
\end{equation}
 Here, $\mathcal{W}(\xi )=\exp
(i\xi ^{T} R)$ is called Weyl operator
\cite{milburn,zoller,Pet90,tmsq01,weyl,wigner,wignerr,quasi8,moyl,kimqm,gdkthe,GKD01}.
The density operator of any quantum state in Fock space can always
be written in terms of its characteristic function and Weyl
operators as follows
\[
\rho =\frac{1}{(2\pi )^{m}}\int d^{2m} \xi \chi (J\xi
)\mathcal{W}(-J\xi )
\]
   {\bf A  Gaussian state is defined as such a state that its
characteristic function is Gaussian:}
\begin{equation} \label{eq:characteristic_function}
\chi (\xi )=\exp \left[ -\frac{1}{4}\xi ^{T}\gamma \xi +id^{T}\xi
\right] .
\end{equation}
Here, $\gamma > 0$ is a real symmetric matrix and $g\in
\mathbb{R}^{2n}$. As shown below, the quantum vacuum state,
coherent states, squeezed states, and thermal states are typical
Gaussian states and they constitute an important class of states
in  quantum optics. In the picture of distribution function, an
$n$-mode Gaussian state is characterized by the $2n$-dimensional
{\bf covariance matrix} $\gamma$ and the $2n$-dimensional
displacement vector $d$. On the grounds of those, much work has
been reported on Gaussian states, in particular, in quantum
information theory/experiment.

 The first moment is
just the displacement given by $d_{j}=\mathrm{Tr}(\rho R_{j})$ and
the second moment is given by
\begin{equation} \label{eq:covariance}
\gamma _{jk}=2\mathrm{Tr}[\rho
(R_{j}-d_{j})(R_{k}-d_{k})]-iJ_{jk},
\end{equation}
which is called covariance of canonical variables. The real
symmetric matrix $(\gamma _{jk})$ is called the covariance matrix
$\gamma $.
Note that due to our choice of canonical variables,
$R_{2j-1}=\omega _{j}^{1/2}\hat x_{j}$ and $R_{2j}=\omega
_{j}^{-1/2}\hat p_{j}$ ($j=1,2,\cdots ,n$), the trace of the
principal sub-matrix of the $j$th mode of $\gamma $,
\[
\gamma _{[j]}=\left(
\begin{array}{cc}
\gamma _{2j-1,2j-1} & \gamma _{2j-1,2j} \\
\gamma _{2j,2j-1} & \gamma _{2j,2j}
\end{array}
\right)
\]
gives the energy of the $j$th mode if $d_{2j-1}=d_{2j}=0$;
\[
\frac{1}{4}\omega _{k}\left( \gamma _{2j-1,2j-1}+\gamma
_{2j,2j}\right)
=\omega _{k}\left( \left\langle a_{j}^{\dagger }a_{j}\right\rangle +\frac{1}{%
2}\right).
\]

A density operator $\rho $ is a positive (semi-)definite operator
$(\rho \geq 0)$ with $\mathrm{Tr}\rho =1$. If $\rho \ngeq 0$, it
does not describe a physical state. The condition for a physical
Gaussian state is given in terms of the covariance matrix as
follows \cite{Pet90,MV68}.

\smallskip

\noindent{\em Theorem 1.--} Matrix $\gamma $ is the covariance
matrix of a physical state if and only if $\gamma +iJ\geq 0$. A
Gaussian state is pure if and only if $\det \gamma =1$.

\smallskip
From this theorem, it can be shown that $\dim [\mathrm{Ker}(\gamma
+iJ)]=\frac{1}{2}\dim X$. Here, $X$ denotes the phase space.


\noindent{\em Theorem 2.--} If the characteristic function of a
bipartite state density operator $\rho_{AB}$ is
$\chi_{AB}(\xi)=\chi_{AB}(\xi_A,\xi_B)$ with
$\xi^T=(\xi_A^T,\xi_B^T)$, the characteristic function for the
density operator in subspace $B$ is
$\chi_B(\xi_B)=\chi_{AB}(\xi_A=0,\xi_B)$.
%

\smallskip
The proof is very simple: Since $\rho_B=\Tr_A \rho_{AB}$, the
characteristic function in subspace $B$ is $$\chi_B=\Tr_B (\rho_B
e^{i\xi_B^TR_B})=\Tr (\rho_{AB} e^{i\xi_B^TR_B})=\Tr(\rho_{AB}
e^{i\xi_B^TR_B+i\xi_A^TR_A})|_{\xi_A=0}$$ which is just
$\chi_{AB}(0,\xi_B)$.

\smallskip

  Linear transformation is rather useful in calculating the Gaussian
  characteristic functions. A linear transformation on canonical
variables is written as $R\rightarrow R^{\prime }=MR$. Since the new
variables $R ^{\prime } $ also must conserve the CCR $[R_{j}^{\prime
},R_{k}^{\prime }]=iJ_{jk}$, $MJM^{T}=J$ must hold. Such a linear
transformation preserving the CCR is called canonical transformation
and a $2n \times 2n$ real matrix satisfying $MJM^{T}=J$ is called
symplectic matrix, $M\in Sp(2n,\mathbb{R})$. The inverse of $M$ is
given by $M^{-1}=JM^{T}J^{-1}$. If $M$ is Symplectic, $M^{-1}$ and
$M^{T}$ are also symplectic and $\det M=1$ \cite{SSM87}. One of the
most important properties of symplectic matrices is the following
Williamson's theorem \cite{Arn89}.

\smallskip

 For a real symmetric positive-definite
$2n\times 2n$ matrix $A=A^{T}>0$, there exists a symplectic matrix
$M\in Sp(2n,\mathbb{R})$ such that $MAM^{T}=\mathrm{diag}(\kappa
_{1},\kappa _{1},\kappa _{2},\kappa _{2},\cdots ,\kappa _{n}\kappa
_{n})$ with $\kappa _{i}>0$ $(i=1,2,\cdots ,n)$.

A linear canonical transformation corresponds to a unitary
transformation in the Hilbert space. Such a unitary transformation
is defined by
$$
S_{M,d'}^{\dagger} R S_{M,d'} =MR + d'
$$
and $d'$ is a $2m$-dimensional real vector. Consequently,
$S_{M,d'}^{\dagger}\mathcal{W}(\xi )S_{M,d'}=\mathcal{W}(M\xi
)\exp(i\xi^Td')$ and the density operator is changed to
$\widetilde{\rho }=S_{M,d'}\rho S_{M,d'}^{\dagger }$.  Suppose
$\rho$ is a Gaussian state with covariance matrix $\gamma$ and
displacement $d$. Omitting the subscripts of $S_{M,d'}$, we can
formula the characteristic function of the new state
$\widetilde{\rho }=S\rho S^{\dagger}$ as follows
\begin{eqnarray}
&&\mathrm{Tr}[S\rho S^{\dagger }\mathcal{W}(\xi )] =\mathrm{Tr}[\rho
S^{\dagger }\mathcal{W}(\xi )S]=\mathrm{Tr}[\rho
\exp(i\xi^T S^\dagger R S )] \nonumber\\
&=&\exp \left[ -\frac{1}{4}\xi^{T}M\gamma M^T \xi +i\xi^T(d'+Md)
\right]
=\exp \left[ -\frac{1}{4}\xi ^{T}\widetilde{\gamma }\xi +i\xi^T \widetilde{d}%
 \right] .\label{lqtch}
\end{eqnarray}
and $\widetilde{\gamma }=M\gamma M^{T}$ and $\widetilde{d}=d'+Md$.
Therefore we conclude

\noindent {\em Theorem 3:} Suppose the covariance matrix is
$\gamma$ and the displacement vector is $d$ for density operator
$\rho$.  The characteristic function for density operator
$\tilde\rho=S\rho S^{\dagger}$ must be
$\exp\left(-\frac{1}{4}\xi^T \tilde \gamma \xi+ i\xi^T \tilde
d\right)$ with $\widetilde{\gamma }=M\gamma M^{T}$ and
$\widetilde{d}=d'+Md$, if the unitary operator $S$ satisfies
$S^\dagger R S= MR +d'$.

 The linear transforms can be implemented by
optical elements such as beam splitters, phase shifters, and
squeezers. For example, a (phase-free) beam splitter transforms
the field operators $a_{1}$ and $a_{2}$ as follows.
\[
\left(
\begin{array}{c}
a_{1} \\
a_{2}
\end{array}
\right) \rightarrow \left(
\begin{array}{cc}
\cos \theta  & \sin \theta  \\
-\sin \theta  & \cos \theta
\end{array}
\right) \left(
\begin{array}{c}
a_{1} \\
a_{2}
\end{array}
\right) .
\]
Thus the corresponding symplectic matrix takes the form
\begin{equation} \label{eq:beam_splitter}
M_{bs}=\left(
\begin{array}{cccc}
\cos \theta  & 0 & \sin \theta  & 0 \\
0 & \cos \theta  & 0 & \sin \theta  \\
-\sin \theta  & 0 & \cos \theta  & 0 \\
0 & -\sin \theta  & 0 & \cos \theta
\end{array}
\right),
\end{equation}

Given that the transmission rate  and reflection rate of the beam
splitter are given by $T=\cos ^{2}\theta $ and $R=\sin ^{2}\theta $,
respectively. The (single mode) squeezer with squeezing parameter $r$
transforms the field operator $a$ into $a\cosh r+a^{\dagger }\sinh
r$ and the corresponding symplectic matrix $S_{sq}$ takes the form
\begin{equation} \label{eq:squeezer}
M_{sq}=\mathrm{diag}(e^{-r},e^{r}) .
\end{equation}

Besides the characteristic function, there are other forms of
distribution functions which can also represent the quantum
states. One of them is the famous Winger function. The Wigner
function is defined as the symplectic Fourier transform of
characteristic function,
\begin{equation} \label{eq:Wigner_function_by_cf}
W(\xi )=\frac{1}{(2\pi )^{2n}} \int d^{2n} \eta e^{-i\xi^T \eta
}\chi (\eta )d\eta .
\end{equation}
For a Gaussian state, its Wigner function is also a Gaussian
function. After performing the Gaussian integration, we obtain
\[
W(\xi )=\frac{1}{\pi ^{n}\sqrt{\det \gamma }}\exp \left[ -(\xi
-d)^{T}\gamma ^{-1}(\xi -d)\right] .
\]
Here, $\gamma ^{-1}$ is the Wigner {\bf correlation matrix}.
Therefore the Wigner correlation matrix of $\widetilde{\rho
}=S\rho S^\dagger$ is $M^{-1}\gamma ^{-1}(M^T)^{-1}$, if
$\gamma^{-1}$ is the correlation matrix of $\rho$ and $S^\dagger R
S=MR$. The Wigner function is similar to a classical probability
distribution in position-momentum. For example, consider the
one-mode Wigner function and denote $\xi_1=x,~\xi_2=p$,
\begin{equation}
\int W(x,p){\rm d}x {\rm d}p= 1
\end{equation}
and $\int W(x,p)dx,~ \int W(x,p)dp $ are for the classical
probability distributions over momentum space (variant $p$) and
position space (variant $q$), respectively.

 An $n-$mode density
operator $\rho $ can be formally expressed as the diagonal form
with respect to the $n$ mode coherent state $\left| \alpha
_{1},\alpha _{2},\cdots ,\alpha _{n}\right\rangle =\otimes
_{k=1}^{n}\left| \alpha _{k}\right\rangle $ as
\begin{equation} \label{eq:density_oerator_by_Pf}
\rho =\int \prod_{k=1}^{n}d^{2}\alpha _{k}P(\alpha _{1},\alpha
_{2},\cdots ,\alpha _{n})\left| \alpha _{1},\alpha _{2},\cdots
,\alpha _{n}\right\rangle \left\langle \alpha _{1},\alpha
_{2},\cdots ,\alpha _{n}\right| .
\end{equation}
This expression is called the $P$ representation of the density
operator $\rho $, and $P(\alpha _{1},\alpha _{2},\cdots ,\alpha
_{n})$ is called the $P$ function. Such a representation exists for
any density operator provided that the $P$ function is permitted to
be  singular\cite{KMC65}. In Eq.~(\ref{eq:density_oerator_by_Pf}),
\begin{equation} \label{eq:coherent_state}
\left| \alpha _{k}\right\rangle =\exp \left( -\frac{1}{2}\left|
\alpha
_{k}\right| ^{2}\right) \sum_{j=0}^{\infty }\frac{(\alpha _{k})^{j}}{\sqrt{j!%
}}\left| j\right\rangle
\end{equation}
with $\left| j\right\rangle =(j!)^{-1/2}(a_{k}^{\dagger })^{j}\left|
0\right\rangle $, and $d^{2}\alpha =d \mathrm{Re} \alpha d
\mathrm{Im} \alpha $. The coherent state
[Eq.~(\ref{eq:coherent_state})] is an eigenstate of $a_{k}$;
\begin{equation} \label{eq:eigen_equation_of_coherent_state}
a_{k}\left| \alpha _{k}\right\rangle =\alpha _{k}\left| \alpha
_{k}\right\rangle
\end{equation}
and satisfies the completeness relation,
\begin{equation} \label{eq:overcompleteness}
\int \frac{d^{2}\alpha }{\pi }\left| \alpha \right\rangle
\left\langle \alpha \right| =\mathbf{I}.
\end{equation}
The $P$ function is written as
\begin{equation}
P(\alpha _{1},\alpha _{2},\cdots ,\alpha _{n}) =\frac{1}{\pi
^{2n}}\int \prod_{k=1}^{n}d^{2}\eta _{k}\exp \left(
\sum_{k=1}^{n}\alpha _{k}\eta _{k}^{*}-\alpha _{k}^{*}\eta
_{k}\right) \chi _{N}(\eta _{1},\eta _{2},\cdots ,\eta _{n}) .
\end{equation}
Here,
\[
\chi _{N}(\eta _{1},\eta _{2},\cdots ,\eta _{n})=\mathrm{Tr}\left[
\rho \exp \left( \sum_{k=1}^{n}\eta _{k}a_{k}^{\dagger }\right)
\exp \left( -\sum_{k=1}^{n}\eta _{k}^{*}a_{k}\right) \right]
\]
is the normally ordered characteristic function. The normally
ordered characteristic function is also written as
\begin{equation} \label{eq:nocf}
\chi _{N}(\eta _{1},\eta _{2},\cdots ,\eta _{n})=\mathrm{Tr}\left[
\rho \exp \left( \sum_{k=1}^{n}\eta _{k}a_{k}^{\dagger }-\eta
_{k}^{*}a_{k}\right) \right] \exp \left(
\frac{1}{2}\sum_{j=1}^{n}\left| \eta _{j}\right| ^{2}\right) .
\end{equation}
Here, if we write $\xi _{2k-1}=\sqrt{2} \mathrm{Re} \eta _{k}$ and
$\xi _{2k}=-\sqrt{2} \mathrm{Im} \eta _{k}$ $(k=1,2, \cdots, n)$,
$\sum_{k=1}^{n}(\eta _{k}a_{k}^{\dagger }-\eta
_{k}^{*}a_{k})=\sum_{k=1}^{n}(-i\xi _{2k}\hat x_{k}+i\xi
_{2k-1}\hat p_{k})=i\xi ^{T}JR $ . This shows that
$\mathrm{Tr}[\cdots ]$ in the right-hand side of
Eq.~(\ref{eq:nocf}) is nothing but the characteristic function.
The $P$ function can be thus written in terms of the
characteristic function.

Here, we derive the trace formula of Weyl operator, which is used
in Sec.~\ref{subsec:Distillability}. Using
Eqs.~(\ref{eq:eigen_equation_of_coherent_state}) and
(\ref{eq:overcompleteness}), we have
\begin{eqnarray*}
W(\xi ) &=&\prod_{k=1}^{n}\exp \left( \eta _{k}a_{k}^{\dagger
}-\eta
_{k}^{*}a_{k}\right) \\
&=&\prod_{k=1}^{n}\exp \left( \frac{1}{2}\left| \eta _{k}\right|
^{2}\right) \exp \left( -\eta _{k}^{*}a_{k}\right) \exp \left(
\eta _{k}a_{k}^{\dagger
}\right) \\
&=&\prod_{k=1}^{n}\exp \left( \frac{1}{2}\left| \eta _{k}\right|
^{2}\right) \int \frac{d^{2}\alpha _{k}}{\pi }\sum_{r,s=0}^{\infty
}\frac{(-\eta _{k}^{*})^{r}}{r!}\frac{(\eta
_{k})^{s}}{s!}a_{k}^{r}\left| \alpha
_{k}\right\rangle \left\langle \alpha _{k}\right| a_{k}^{\dagger s} \\
&=&\prod_{k=1}^{n}\exp \left( \frac{1}{2}\left| \eta _{k}\right|
^{2}\right) \int \frac{d^{2}\alpha _{k}}{\pi }\left| \alpha
_{k}\right\rangle \left\langle \alpha _{k}\right| \exp \left( \eta
_{k}\alpha _{k}^{*}-\eta _{k}^{*}\alpha _{k}\right) .
\end{eqnarray*}
Therefore,
\[
\mathrm{Tr}W(\xi )=\prod_{k=1}^{n}\exp \left( \frac{1}{2}\left|
\eta _{k}\right| ^{2}\right) \int \frac{d^{2}\alpha _{k}}{\pi
}\exp \left( \eta _{k}\alpha _{k}^{*}-\eta _{k}^{*}\alpha
_{k}\right) .
\]
Since
\[
\int \frac{d^{2}\alpha }{\pi }\exp \left( \eta \alpha ^{*}-\eta
^{*}\alpha \right) =\pi \delta (\mathrm{Re}\eta )\delta
(\mathrm{Im}\eta ),
\]
we have
\begin{equation} \label{eq:trace_formula_for_Weyl_opeartor}
\mathrm{Tr}W(\xi )=(2\pi )^{n}\delta (\xi ).
\end{equation}
\subsection{Coherent states}
Lets start from the simplest state, the vacuum, i.e., the density
operator $|0\rangle\langle 0|$. According to the definition, the
characteristic function is
\begin{equation}\label{vacuumch}
\chi(\xi_1,\xi_2) = {\rm Tr}e^{i(\xi_1 \hat x+\xi_2\hat p)
}|0\rangle\langle 0|\nonumber =\sum_{0}^\infty \langle n|e^{i(\xi_1
\hat x+\xi_2\hat p) }|0\rangle\langle 0|n|\rangle\nonumber = \langle
0|e^{i(\xi_1 \hat x + \xi_2\hat p) }|0\rangle .
\end{equation}
Given that $\hat x=\frac{1}{\sqrt 2}(a^\dagger + a)$ and $\hat
p=\frac{i}{\sqrt 2}(a^\dagger-a)$, we have
\begin{equation}
e^{i(\xi_1 \hat x+\xi_2\hat p)}=e^{\frac{i\xi_1-\xi_2}{\sqrt
2}a^\dagger+\frac{i\xi_2+\xi_2}{\sqrt 2}a}
=e^{-(\xi_1^2+\xi_2^2)/4}e^{\frac{i\xi_1-\xi_2}{\sqrt
2}a^\dagger}e^{\frac{i\xi_1+\xi_2}{\sqrt 2}a}.
\end{equation}
Here we have used the following well known operator
identity\cite{glauber,KMC65}
\begin{equation}\label{opid01}
\exp(\Omega_1+\Omega_2)=\exp(-\frac{1}{2}[\Omega_1,\Omega_2])\exp
\Omega_1\exp {\Omega_2}
\end{equation}
if
$[\Omega_1,[\Omega_1,\Omega_2]]=[\Omega_2,[\Omega_1,\Omega_2]]=0$\cite{KMC65}.
 Putting this back into
Eq.(\ref{vacuumch}) we immediately obtain the following explicit
formula for the characteristic function of vacuum
\begin{equation}\label{vchf}
\chi(\xi_1,\xi_2)=\exp[-(\xi_1^2+\xi_2^2)/4]=\exp\left[-\frac{1}{4}(\xi_1,\xi_2)\gamma_0\left(\begin{array}{c}
\xi_1\\\xi_2\end{array}\right)\right]
\end{equation}
Therefore, the characteristic function of vacuum state is indeed
in the Gaussian form and the covariance matrix $\gamma_0$ here is
simply a $2\times 2$ unity matrix.
 Based on this fact, through linear
transformations, one can easily write out the characteristic
function of a coherent state\cite{glauber}, the characteristic
function a squeezed (vacuum) state, and many other Gaussin states.
The state of a light beam out of a laser device is coherent
state\cite{glauber}.  (For a systematic study of coherent states and
squeezed states, one may go into more specific reviews, such
as\cite{KMC65,knight,schu,zhangwm} or quantum optics textbooks,
e.g.\cite{knight,milburn,zoller,barnett,orszag}.)  The following
well known Baker-Campbell-Hausdorff (BCH) formula is particularly
useful in studying the linear transformations:
\begin{equation}\label{bch}
e^\mu\nu e^{-\mu}=\nu+[\mu,\nu]+\frac{1}{2!}[\mu,[\mu,\nu]]+
\frac{1}{3!}[\mu,[\mu,[\mu,\nu]]]+\cdots
\end{equation}
Here $\mu$ and $\nu$ are two arbitrary operators and
$[\mu,\nu]=\mu\nu-\nu\mu$.
 A
displacement operator is defined as
\begin{equation}\label{pmdis}
D(d_x,d_p)=\exp( id_p\hat x- id_x\hat p)=\exp [(\alpha a^\dagger
-\alpha^* a)/\sqrt 2]={\mathcal D}(\alpha/\sqrt 2)
\end{equation}
 and $d_x,~ d_p$ are real numbers satisfying $d_x+id_p=\alpha$. Using Eq.(\ref{bch}) we can easily obtain
\begin{equation}
\hat {\mathcal D}^\dagger (\alpha) (a^\dagger,a)\hat {\mathcal
D}(\alpha)=(a^\dagger +\alpha^*,a+\alpha);~~ D^\dagger (d_x,d_p)
(\hat x,\hat p) D(d_x,d_p) =(\hat x+d_x, \hat p+d_p).
\end{equation}
 The coherent
state
\begin{equation}
|\alpha\rangle = e^{-|\alpha|^2/2}\sum_{n=0}^{\infty}\frac{\alpha^{n}}{\sqrt {n!}}|n\rangle.\label{coh00}
\end{equation}
is equivalent to
\begin{equation}\label{cs11}
|\alpha\rangle = e^{\alpha a^\dagger-|\alpha|^2/2}|0\rangle=\hat {\mathcal D}
(\alpha)|0\rangle.
\end{equation}
Here we have used the the operator identity of Eq.(\ref{opid01}) and
fact $|0\rangle=e^{\alpha^*a}|0\rangle $.

The coherent states is a member of Gaussian states. Since the
coherent state $|\alpha\rangle$ can be regarded as the displaced
vacuum of $\hat {\mathcal D} (\alpha ) |0\rangle $, according to theorem 3
its characteristic function is simply
\begin{equation}\chi(\xi_1,\xi_2)=
\exp\left[-\frac{1}{4}(\xi_1^2+\xi_2^2)+i\xi_1(\alpha+\alpha^*)/\sqrt
2 +i\xi_2 (\alpha-\alpha^*)/\sqrt 2\right]
\label{chacoh}\end{equation} based on the characteristic function of
vacuum state in Eq.(\ref{vchf}). This is a Gaussian function.

Coherent states are a type of minimum uncertainty states for
quadrature operator $\hat x,~\hat p$.
 Since $\hat x$ and $\hat p$ don't commute, the
following uncertainty relation must be respected  given any state:
\begin{equation}
(\langle \hat x^2\rangle -\langle \hat x\rangle^2)\cdot (\langle
\hat p^2\rangle -\langle \hat p\rangle^2)\ge \left|[\hat x,\hat
p]/2\right|^2=1/4.
\end{equation}
Here we have used the fact
\begin{equation}
[\hat x,\hat p]=i.
\end{equation}
The  commutation relation of quadrature  operators $\hat x,\hat p$
is the same  with that of position operator and momentum operator
in quantum mechanics, we also call operators $\hat x,\hat p$ as
``position operator'' or ``momentum operator'' for convenience.
According to Eq.(\ref{chacoh}), the covariance matrix of any
coherent state is a unity matrix which indicates $\langle(\hat
x-\langle \hat x\rangle)^2\rangle=\langle(\hat p-\langle \hat
p\rangle)^2\rangle=1/2$. This is equivalent to $(\langle \hat
x^2\rangle -\langle \hat x\rangle^2)\cdot (\langle \hat p^2\rangle
-\langle \hat p\rangle^2)=1/4$ which is the minimum value as
requested by the uncertain principle.

\begin{figure}
\begin{center}
\epsffile{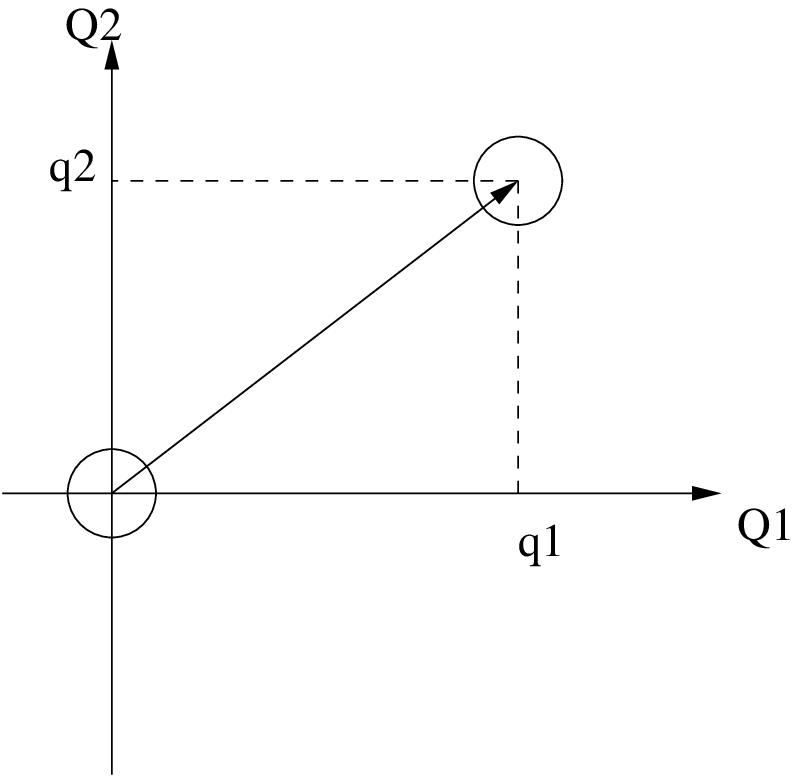}
\end{center}
\caption{Any coherent state $|\alpha\rangle$ can be represented by
an error circle with radius of 1/2 and a displacement $q_1=\sqrt 2
{\rm Re}(\alpha)$ and $q_2=\sqrt 2{\rm Im} \alpha$.
}\label{cohfig}
\end{figure}

Intuitively,  the coherent state $|\alpha\rangle $ can be
represented by a displaced error circle  as shown in
fig.(\ref{cohfig}).

\\
The averaged photon number of coherent state $|\alpha\rangle$ is
\begin{equation}
\mu=\langle\alpha|a^\dagger a|\alpha\rangle =|\alpha|^2.
\end{equation}
This is the intensity of coherent state $|\alpha\rangle$. A coherent
state can also be characterized by its intensity and phase, i.e., a
coherent state with intensity $\mu$ and phase $\theta$ is
\begin{equation}
|\mu,\theta\rangle=e^{\sqrt \mu e^{i\theta}a^\dagger-\mu/2}|0\rangle
\end{equation}
with $\mu=|\alpha|^2$ and $e^{i\theta}=\alpha/|\alpha|$. Doing
Taylor expansion to operator $e^{\alpha a^\dagger}$ and using the
property of $a^\dagger|n\rangle=\sqrt{n+1}|n+1\rangle$ we have the
following linear superposition form for the coherent state in
photon-number space:
\begin{equation}
|\mu,\theta\rangle=e^{- \mu/2}\sum_{n=0}^\infty \frac{\mu^{n/2}e^{in\theta}}{\sqrt {n!}}|n\rangle.
\end{equation}
 The phase $\theta$ of
light pulse from an ordinary laser device is { random}. If the phase
is random, a coherent state with intensity $\mu$ is actually a
classical mixture with Poissonian distribution of photon number
states:
\begin{equation}
\rho_\mu=\frac{1}{2\pi}\int_0^{2\pi}|\mu, \theta\rangle\langle \mu, \theta|
= e^{-\mu}\sum_n \frac{\mu^n}{n!}|n\rangle\langle n|.
\end{equation}

If $\mu$ is small, a light pulse in the state above is almost the
same with a single-photon pulse if it is not vacuum. Therefore the
weak coherent light is often regarded as the approximate
single-photon light.
 All properties of a coherent state can be simulated
by a classical Gaussian light. Therefore a coherent state is
regarded as a {\it classical} state.  Also, any classical
probabilistic mixture of different coherent states is also a
classical state, i.e., state
\begin{equation}
\int P(\alpha)|\alpha \rangle\langle \alpha| {\rm d}^2\alpha
\end{equation}
is  classical  if $P(\alpha)$ is positive-definite.

\subsection{Squeezed states}
Besides coherent states, a more general class of minimum-uncertainty
states are known as the {\it squeezed} states. In general, squeezed
states may have less uncertainty in one quadrature than that of
coherent states. Due to the requirement of the uncertainty
principle, the noise in the other quadrature of a squeezed state
must be larger than that of the coherent states. Most generally, a
state is called squeezed if its covariance matrix has an eigenvalue
smaller than one.

\subsubsection{One-mode squeezed states}
The simplest single mode squeezed state is the squeezed vacuum
state,
\begin{equation}
|\zeta,0\rangle= S(\zeta)|0\rangle
\end{equation}
generated by the squeezing operator
\begin{equation}
 S(\zeta)=\exp\left(-\frac{\zeta}{2}{a^\dagger}^2+\frac{\zeta^*}{2}a^2\right),
\label{jpsq348}\end{equation}
 from vacuum $|0\rangle$
where $\zeta=r\exp(i\phi)$ is an arbitrary complex number with
modulus $r$ and argument $\phi$. The squeezing operator $ S(\zeta)$
is unitary because $ S^\dagger(\zeta)
=\exp\left(\frac{\zeta}{2}{a^\dagger}^2-\frac{\zeta^*}{2}a^2\right)=
S^{-1}(\zeta)$ is the inverse of operator $ S(\zeta)$.

\begin{figure}
\begin{center}
\epsffile{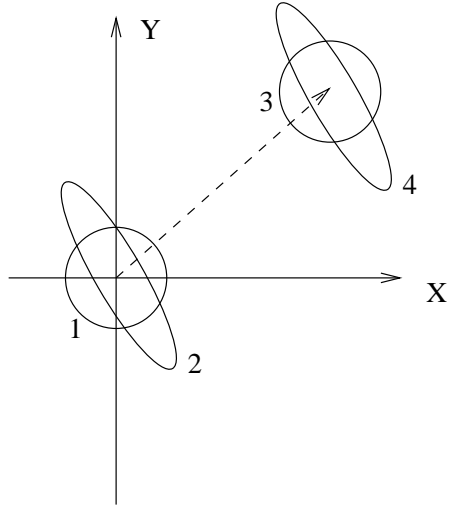}
\end{center}
\caption{ Comparison of the variance shape of a squeezed vacuum state and vacuum, a displaced squeezed vacuum state and a coherent state.
1: Error circle of vacuum; 2: Error ellipse of a squeezed vacuum;
3: error circle of a coherent state (displaced vacuum); 4: error ellipse of a displaced squeezed vacuum state.}\label{ppsq}
\end{figure}

The squeezed vacuum state is a minimum-uncertainty state with the
variance of two quadratures being {\em different}. To show this, we
calculate the variance of quadrature operators $\hat x,~\hat p$ .
The BCH formula of Eq.(\ref{bch}) gives rise to the following
transformation
\begin{equation}\label{setran1}
 S^{-1}(\zeta)(a^\dagger, a)^T S(\zeta)
=M_\zeta(a^\dagger,a)^T\end{equation} and
\begin{equation}M_\zeta=\left(\begin{array}{cc}\cosh r & -e^{-i\phi}\sinh r\\
-e^{i\phi}\sinh r & \cosh r\end{array} \right) .\end{equation} and
it's inverse transformation is
\begin{equation}\label{setraninv}
 S(\zeta)(a^\dagger, a)^T S^{-1}(\zeta)
=M_\zeta^{-1}(a^\dagger,a)^T.\end{equation} 
Here $$(a^\dagger, a)^T=\left(\begin{array}{c}a^\dagger\\a\end{array}\right) $$
The transformation in
$\hat x,~\hat p$ space is
\begin{equation}
S^{-1}(\hat x,~\hat
p)^TS=S^{-1}[(a^\dagger,a)K]^TS=[(a^\dagger,a)M_\zeta
K]^T={\mathcal M}_\zeta (\hat x,~\hat p)^T\end{equation} with
\begin{equation}{\mathcal M}_\zeta=K^{-1}M_\zeta K\label{sqpmtran}\end{equation} and
$K=\frac{1}{\sqrt
2}\left(\begin{array}{cc}1&1\\i&-i\end{array}\right)$.
 In the special case that $\zeta$ is
real,
\begin{eqnarray}
S^{-1}(\zeta) (\hat x,~\hat p)^TS(\zeta)=
\left(\begin{array}{cc}e^{-r} &
0\\0&e^{r}\end{array}\right)\left(\begin{array}{c}\hat x\\\hat
p\end{array}\right).
\end{eqnarray}
This is just Eq.(\ref{eq:squeezer}). According to theorem 3, the
covariance matrix of the squeezed vacuum state
$S(\zeta)|0\rangle\langle 0| S^{-1}$ is simply
\begin{equation}
\gamma_{sq}(r)=\left(\begin{array}{cc}e^{-r}&0\\0&e^{r}\end{array}\right)\gamma_0
\left(\begin{array}{cc}e^{-r}&0\\0&e^{r}\end{array}\right)=
\left(\begin{array}{cc}e^{-2r}&0\\0&e^{2r}\end{array}\right)
\end{equation}
Such a covariance matrix indicates that the variance of position and
momentum are $e^{-2r}$ and $e^{2r}$ respectively, while the product
of these two variance keeps to be the minimum uncertainty. The
position (momentum) noise is ``squeezed" if $r$ is positive
(negative). Most generally, $\zeta$ is a complex number and the
maximum squeezing is for the observable of rotated position. This
minimum-uncertainty relation actually holds for the squeezed vacuum
state with any complex parameter $\zeta$. Moreover, since a
displacement does not change the variance, this minimum-uncertainty
relation and the squeezed variance of $\hat x$ are also true for an
arbitrary displaced squeezed vacuum state $\hat
{\mathcal D}(\alpha)S(\zeta)|0\rangle$. Pictorially, one can draw the
variance ellipse for a displaced squeezed state as shown by
fig.\ref{ppsq}.

 The squeezing operator can be normally ordered as\cite{barnett}:
\begin{eqnarray}
S(\zeta)=\frac{1}{\sqrt{\cosh |\zeta |}}\nonumber\\ \times
\exp\left[-\frac{{a^{\dagger}}^2}{2}e^{i\phi}\tanh |\zeta| \right]
\exp\left[-a^\dagger a(\ln \cosh |\zeta|)\right]
\exp\left[\frac{1}{2}a^2e^{i\phi}\tanh|\zeta|\right]\label{sqnormal1}
\end{eqnarray}
This leads to
\begin{equation}
 S(\zeta)|0\rangle=\frac{1}{\sqrt {\cosh r}}\exp\left[-\frac{1}{2}{a^\dagger}^2e^{i\phi}\tanh r\right]|0\rangle
\end{equation}
given the fact $a|0\rangle=0$. Furthermore, by taking Taylor
expansion to the exponential operator and using eq.(\ref{exite}) we
have
\begin{equation}\label{even}
| S(\zeta)|0\rangle=\sqrt{{\rm sech} r} \sum\limits_{n=0}^{\infty}
\frac{\sqrt{(2n)!}}{n!}\left[-\frac{1}{2}e^{i\phi}\tanh
r\right]^n|2n\rangle.
\end{equation}
From this equation, we see that state $|\zeta,0\rangle$ is actually
a linear superposition of even photon number states.

Different squeezed vacuum states are not orthogonal. By the
transformation property, overlap between two squeezed states can
be easily calculated\cite{knight}.
\begin{equation}
\langle \zeta,0|\zeta',0\rangle =\langle 0| S^\dagger(\zeta)
S(\zeta')|0\rangle .\end{equation} It can be simplified with the
normally ordered form of  Eq.(\ref{sqnormal1}) as
$$
\langle \zeta,0|\zeta',0\rangle$$
\begin{equation}= \sqrt{{\rm sech} r {\rm sech}
r'}\langle 0|\exp\left[-\frac{1}{2}a^2e^{-i\phi}\tanh r\right]\exp\left[
-\frac{1}{2}{a^\dagger}^2e^{i\phi'}\tanh r'\right]|0\rangle
\end{equation}
 One can transform the product
operator between the two vacuum states in the above equation into
the normally ordered form\cite{barnett,bchwang0,bchwang1,wangE} and
obtain the overlap of two squeezed states\cite{schu}
\begin{equation}
\langle \zeta,0|\zeta',0\rangle=\sqrt{\frac{{\rm sech} r{\rm sech}
r'}{1-e^{i(\phi'-\phi)}\tanh r\tanh r'}}
.\end{equation}
\subsubsection{Two-mode squeezed states}
The two-mode squeezed states\cite{tmsq01} are especially useful in
QIP, because they can be used as entanglement resource.
Mathematically, similar to the one-mode squeezed state, a two-mode
squeezed state can be generated from the vacuum by a two-mode
squeezing operator,
\begin{equation} S  =  \exp \left(
-\zeta^\ast a_1 a_2 + \zeta a_1^\dagger a_2^\dagger \right),
\end{equation} where $a_1^\dagger$ and $a_2^\dagger$ ($a_1$ and
$a_2$) are the photon creation (annihilation) operator of each
mode. Operators of different modes commute.

The two mode squeezing operator has the following transformation
property:
\begin{equation}
 S^{-1} (a_1^\dagger, a_1,a_2^\dagger,a_2)^T S=M_\zeta(a_1^\dagger,
 a_1, a_2^\dagger,a_2)^T
\end{equation}
and
\begin{equation}
M_\zeta=\left(
\begin{array}{cccc}
\cosh r  & 0 & 0 &e^{-i\phi}\sinh r
\\ 0 & \cosh r  &  e^{i\phi}\sinh r & 0
\\ 0 & e^{-i\phi}\sinh r & \cosh r & 0
\\ e^{i\phi}\sinh r & 0 & 0 & \cosh r
\end{array}
\right). \label{tsetran1n}\end{equation} Here $r=|\zeta|$ and $r
e^{i\phi}=\zeta$ . If $r$ is real, this transformation in the
position-momentum space is
\begin{equation}\label{eq80}
 S^{-1} (\hat x_1, \hat x_2,\hat p_1,\hat p_2)^T S=
 L(\hat x_1, \hat x_2,\hat p_1,\hat p_2)^T
\end{equation}
and $L=H {\rm diag}(e^{-r},e^r,e^r,e^{-r})H^T$,
$H=\frac{1}{\sqrt
2}\left(\begin{array}{cccc}1&0&-1&0\\1&0&1&0\\0&1&0&-1\\0&1&0&1\end{array}\right)$.
This means that $S^\dagger (\hat x_1-\hat x_2)S=e^{-r}(\hat
x_1-\hat x_2)$ and $S^\dagger (\hat p_1+\hat p_2)S=e^{-r}(\hat
p_1+\hat p_2)$, i.e., both $\hat x_1-\hat x_2$ and $\hat p_1+\hat
p_2$ are squeezed.
Therefore the covariance matrix for the two-mode
squeezed state with squeezing factor $\zeta=r$ is
\begin{equation}\label{2gamma}\gamma_{sq2}=H^T{\rm
diag}(e^{-2r},e^{2r},e^{2r},e^{-2r})H.
\end{equation}
 The formula above gives rise to the following formula for the
characteristic function of two-mode squeezed state
\begin{equation}\label{chsq2}
\chi_{sq2}(\xi_1,\xi_2,\xi_3,\xi_4)=\exp
\left\{-\frac{e^{-2r}}{8}[(\xi_1-\xi_3)^2+(\xi_2+\xi_4)^2]-\frac{e^{2r}}{8}[(\xi_1+\xi_3)^2+(\xi_2-\xi_4)^2]\right\}
.\end{equation}
  The two mode squeezing operator can also be written in the
normally ordered form\cite{barnett,bchwang0}
\begin{eqnarray} \label{ntse}
S(\zeta)=\frac{1}{\cosh |\zeta|} \exp \left[{a_1}^{\dagger}
{a_2}^{\dagger}e^{i\phi}\tanh |\zeta|\right]
\nonumber\\
\times \exp\left[ (a_1^\dagger a_1 +a_2^\dagger a_2)\ln(\cosh
|\zeta |)\right] \cdot\exp\left[-e^{-i\phi}\tanh |\zeta | a_1a_2
\right] ,\end{eqnarray}
  With
this normally ordered form,  the two-mode squeezed vacuum state
can be simplified:
\begin{equation}\label{2sim}
S(\zeta)|00\rangle={\rm sech} r
\sum\limits_{n=0}^{\infty}\left[\exp(i\phi) \tanh
r\right]^n|n\rangle_1|n\rangle_2
\end{equation}
The state of each individual mode in the two-mode squeezed state
is a mixed state. The density operator of each mode can be
calculated by taking the partial trace of the whole state. For
example, the state for mode $1$ is
\begin{equation}
\rho_1={\rm tr}_2 |\zeta,0\rangle \langle \zeta,0|={\rm sech}^2 r
\sum\limits_{n=0}^{\infty}\tanh^{2n}r|n\rangle\langle  n|
.\end{equation} This is just a thermal state of
\begin{equation}
\rho_{th}(\beta)=(1-e^{-\beta}) \exp (-\beta a^\dagger a)
\end{equation}
and $\beta=-\ln \tanh^2 r$. According to our theorem 2, the
characteristic function of the thermal state above can be easily
deduced from Eq. (\ref{chsq2})
\begin{equation}\label{chthm}
\chi_{th,\,\beta}(\xi_1,\xi_2)(\beta) =\exp \left[-\frac{1}{4}\nu
(\xi_1^2+\xi_2^2)\right]
\end{equation}
and $\nu=\cosh 2r $. This means that the characteristic function
in the above form is for a thermal state $\rho_{th}(\beta)$ with
\begin{equation}e^{-\beta}=\frac{\nu-1}{\nu+1}.\label{thtemp}\end{equation}

The wave-function of a two-mode squeezed state in position space is\cite{leonhardt} is
 \begin{equation}
 S(\zeta)|00\rangle
 = \frac{1}{\sqrt \pi}\int {\rm d}x_1 {\rm d}x_2
 \exp\left[-\frac{e^{2r}}{4}(x_1-x_2)^2-\frac{e^{-2r}}{4}(x_1+x_2)^2
 \right].
 \end{equation}
 The same state in the momentum representation is
 \begin{equation}
 \frac{1}{\sqrt \pi}\int {\rm d}p_1 {\rm d}p_2
 \exp\left[-\frac{e^{2r}}{4}(p_1+p_2)^2-\frac{e^{-2r}}{4}(p_1-p_2)^2
 \right].
 \end{equation}
 Therefore, in the limit that
$r\rightarrow\infty$, the two-mode squeezed state has the
following un-normalized form:
\begin{equation}
 \sim \int |x\rangle_1|x\rangle_2 {\rm d}x=\int
|p\rangle_1|-p\rangle_2 {\rm d}p.
\end{equation}
From this we can also see clearly the squeezing properties for a
two-mode squeezed state in the extreme case: It is a simultaneous
eigen-state of both $(\hat x_1-\hat x_2)$ and $ (\hat p_1+\hat
p_2)$ with both eigenvalues being 0. We shall call this type of
entangled state as the EPR state or Bell state in
position-momentum space.  We shall call the simultaneous
measurement of $(\hat x_1-\hat x_2)$ and $(\hat p_1+\hat p_2)$
{\em Bell measurement} in position-momentum space. Most generally,
the simultaneous eigen-state of $\hat x_1-\hat x_2$ and $\hat
p_1+\hat p_2$ with eigenvalues $\Delta$ and $\Sigma$ is a
displaced two-mode squeezed state of
\begin{equation}\label{bellout}
|B(\Delta,\Sigma)\rangle = D(\Delta,\Sigma)\exp [r(a_1^\dagger
a_2^\dagger -a_1 a_2)]|00\rangle =D(\Delta,\Sigma){\rm sech}
r\sum_{k=0}^{\infty}\gamma^k |kk\rangle
\end{equation}
and $r\rightarrow\infty$. Here $D(\Delta,\Sigma)$ is defined in
Eq.(\ref{pmdis}) and $\gamma=\tanh r\rightarrow 1$, and we have used
Eq.(\ref{2sim}).

{\em Remark}: The continuous variable Bell state implies the
infinite squeezing which doesn't exist. However, a two-mode squeezed
state of finite squeezing is also entangled and can be used to carry
out non-trivial entanglement-based tasks.

\subsection{ Beam-splitter}
Beam-splitters \cite{bsreview0,bsreview1,campos,btheorem,bwang} are
widely used in the quantum information processing. A beam-splitter
can work as a type of interferometric device, a type of quantum
entangler and also a device for Bell measurement. Lets first
consider the elementary properties of a beam-splitter: the relation
between the output state and input state of a beam-splitter. For
clarity, we use the Schrodinger
picture\cite{btheorem,bwang,wangsingle,bscm} here. We shall define
the mode by the propagation direction, for both input and output
states.
\begin{figure}
\begin{center}\centerline{\includegraphics[bb=291 589 398 684, width=5.6 cm,
clip]{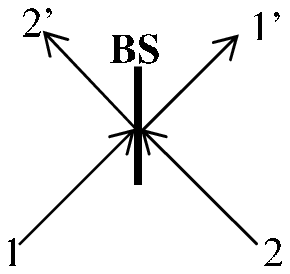}}
\end{center}
\caption{ A schematic diagram of the beam-splitter. Both the input
and the output are two mode states. The different mode is
distinguished by the propagating direction of the field. BS:
beam-splitter.} \label{bs0}
\end{figure}
Consider figure \ref{bs0}. Both the input state and output state are
two-mode states. Even though we have only sent one light beam into
the beam-splitter,
 we shall still regard the total input state
as a two-mode state by setting the state of another input mode to
be vacuum. For example, the input light beam of mode 1 contains
one photon and there is nothing in mode 2, the total input state
is $|1\rangle_1\otimes|0\rangle_2$ which is simplified by notation
$|10\rangle$. The output state is
$|\psi\rangle_{out}=U_B|10\rangle$ and $U_B$ is the time evolution
operation caused by the beam-splitter. The specific parameters in
$U_B$ are determined by the material properties of the
beam-splitter itself. However, in calculation, normally we don't
need the explicit formula of $U_B$ itself, we only need the
transformation properties of $U_B$. Given any  input of
pure state, we can always write it in the two-mode form of
\begin{equation}
|\psi\rangle_{in}=f(a_1^\dagger,a_2^\dagger)|00\rangle.
\end{equation}
After passing the beam-splitter, the state is changed to
\begin{equation}
|\psi\rangle_{out}=U_Bf(a_1^\dagger,a_2^\dagger)|00\rangle
=U_Bf(a_1^\dagger,a_2^\dagger)U_B^{-1}U_B|00\rangle.
\end{equation}
Considering the fact of no input no output for any beam-splitter,
one immediately finds $U_B|00\rangle=|00\rangle$, therefore
\begin{equation}
|\psi\rangle_{out}
=U_Bf(a_1^\dagger,a_2^\dagger)U_B^{-1}|00\rangle
=f(U_B a_1^\dagger U_B^{-1},U_B a_2^\dagger U_B^{-1})|00\rangle.
\label{relation}\end{equation}
Denote $(a_1^\dagger, a_1, a_2^\dagger, a_2)^T
=\left(\begin{array}{c}a_1^\dagger\\ a_1\\ a_2^\dagger\\ a_2
\end{array}\right)$ .Suppose
\begin{equation}
U_B (a_1^\dagger, a_1, a_2^\dagger, a_2)^T
U_B^{-1}=M_B(a_1^\dagger, a_1, a_2^\dagger, a_2)^T \label{bmatrix}
\end{equation}
then a beam-splitter is fully characterized by matrix $M_B$, with
the output state being explicitly given by Eq.(\ref{relation}), given
whatever input state. Since any time evolution operator must be
unitary, matrix $M_B$ of any beam-splitter in $SU(2)$. Therefore,
the most general form of matrix $M_B$ is
\begin{equation}\label{bbmatrix}
M_B =\left(\begin{array}{cccc}\cos\theta e^{i\phi_0}& 0 & -\sin\theta e^{-i\phi_1} & 0\\
 0 & \cos\theta e^{-i\phi_0} & 0& -\sin\theta e^{-i\phi_1}
 \\\sin\theta e^{i\phi_1} & 0& \cos\theta e^{-i\phi_0}& 0
  \\0&\sin\theta e^{i\phi_1}& 0 & \cos\theta e^{i\phi_0}.
\end{array}\right)
\end{equation}
In the special case of a 50:50 beam-splitter, $\theta =\pi/4$ and
$\phi_0=\phi_1=0$, the matrix is
\begin{equation}\label{50:50}
M_B =\frac{1}{\sqrt 2}\left(\begin{array}{cccc}1 & 0 & -1 & 0\\
 0 & 1 & 0& -1
 \\ 1 & 0& 1 & 0
  \\0& 1& 0 & 1.
\end{array}\right)
\end{equation}
 In the position-momentum space,
the transformation formula is
\begin{equation}\label{bbmpm}
U_B (\hat x_1,~\hat p_1,~\hat x_2,~\hat p_2)^TU_B^{-1}={\mathcal
M}_B(\hat x_1,~\hat p_1,~\hat x_2,~\hat p_2)^T
\end{equation}
with \begin{equation}{\mathcal M}_B=(K^{-1}\oplus K^{-1}) M_B(
K\oplus K)\end{equation} and $
K=\left(\begin{array}{cc}1&1\\i&-i\end{array}\right)/\sqrt 2$. In
the special case that $\theta=\pi/4$ and $\phi_0=\phi_1=0$,
${\mathcal M}_B=M_B$, which is given by Eq.(\ref{50:50}).
\\{\em Summary}: To write the output state of a beam-splitter, one
only needs to write the input state in the form of
$f(a_1^\dagger,a_2^\dagger)|00\rangle$ and then replace
$a_1^\dagger,a_2^\dagger$ by $U_B a_1^\dagger U_B^{-1}$ and $U_B
a_2^\dagger U_B^{-1}$, respectively.
  $U_B a_1^\dagger U_B^{-1}$ and
$U_B a_2^\dagger U_B^{-1}$ are determined by eq.(\ref{bmatrix}).
Matrix $M_B$ is determined by the material properties of the
beam-splitter
itself.\\
\subsubsection {Collective measurement in polarization space}
So far we have assumed that the input light beams of each mode
have the same polarization and frequency. If the two input beams
have different polarization or frequency distribution, the
 result will be different. Beams with perpendicular
 polarizations or different frequencies  will have no interference.
For simplicity,
we consider the effects caused by different polarizations here only.
Since perpendicularly polarized photons don't have quantum interference, the transformation
matrix $M_B$ $only$ applies to horizontal photons and vertical photons $separately$.
That is to say, we shall first write the state of the input light in the form
of $f(a_{1H}^\dagger,a_{1V}^\dagger,a_{2H}^\dagger,a_{2V}^\dagger)|00\rangle$
and then use
\begin{eqnarray}\begin{array}{l}
U_B(a_{1H}^\dagger, a_{1H},a_{2H}^\dagger,a_{2H})^TU_B^{-1}
=M_B(a_{1H}^\dagger, a_{1H},a_{2H}^\dagger,a_{2H})^T\\
U_B(a_{1V}^\dagger, a_{1V},a_{2V}^\dagger,a_{2V})^TU_B^{-1}=
M_B(a_{1V}^\dagger, a_{1V},a_{2V}^\dagger,a_{2V})^T\end{array}.
\end{eqnarray}
This is to say, we shall treat $H,V$ as different modes and the operators of
perpendicular polarizations $commute$. Here the creation operators of certain
polarization have the same property with those on a specific mode. For example
$$
a_{H}^\dagger |0\rangle  = |H\rangle =|1\rangle_H;\nonumber\\
{a_{H}^\dagger}^2 |0\rangle  = \sqrt 2|2H\rangle =\sqrt 2|2\rangle_H;
$$
and so on. Consider the case that each input beam contains one
photon, the polarization for beam 1 is horizontal and the
polarization for beam 2 is vertical. The input state is
$|\psi\rangle_{in}=|H\rangle_1\otimes |V\rangle_2 =a_{1H}^\dagger
a_{2V}^\dagger |00\rangle$. Given a 50:50 beam-splitter, the state
of the output beams is
\begin{equation}
|\psi\rangle_{out}=\frac{1}{2}(a_{1H}^\dagger +a_{2H}^\dagger)(a_{1V}^\dagger-a_{2V}^\dagger)|00\rangle.
\end{equation}
This is equivalent to the state
\begin{equation}
|\psi\rangle_{out}= |HV\rangle_1\otimes |0\rangle_2-|H\rangle_1\otimes |V\rangle_2 +|V\rangle_1\otimes |H\rangle_2 - |0\rangle_1\otimes |HV\rangle_2
\end{equation}
and the state $|HV\rangle_1=a_{1H}^\dagger a_{1V}^\dagger
|0\rangle$, which means mode 1 contains one horizontally polarized
photon and one vertically polarized photon. Given this formalism, we
can now show how a 50:50 beam-splitter may assist us to do the
(incomplete) Bell measurement with single-photon detectors. Suppose
each input mode  contains one and only one photon and we have placed
a single-photon detector at each side of the beam-splitter to detect
the output light beam of each mode. We want to see whether the input
beams are in the maximally entangled state of $|\psi^-\rangle
=\frac{1}{\sqrt 2}(|H\rangle_1\otimes |V\rangle_2-|V\rangle_1\otimes
|H\rangle_2)$ by watching the single-photon detectors. As we have
known already, there are 4 states in Bell basis. If we want to know
whether the input state is $|\psi^-\rangle$, we must be able to
exclude the other 3 states ($|\phi^\pm\rangle, |\psi^+\rangle$). The
detectors' status caused by the output beams of these 3 states must
be deterministically different from that of $|\psi^-\rangle$. For
such a purpose, we consider the consequence of the 4 Bell states one
by one. Given the input state $|\psi^+\rangle=\frac{1}{\sqrt
2}(|H\rangle_1\otimes |V\rangle_2+|V\rangle_1\otimes |H\rangle_2)$,
we can rewrite it into
\begin{equation}
|\psi^+\rangle=\frac{1}{\sqrt 2}(a_{1H}^\dagger a_{2V}^\dagger +
a_{1V}^\dagger a_{2H}^\dagger)|00\rangle.
\end{equation}
After passing through the 50:50 beam-splitter, the state is transformed into
\begin{equation}
|\psi\rangle_{out}=\frac{1}{2\sqrt 2}[(a_{1H}^\dagger +a_{2H}^\dagger)(a_{1V}^\dagger -a_{2V}^\dagger)
+(a_{1V}^\dagger +a_{2V}^\dagger)(a_{1H}^\dagger -a_{2H}^\dagger)]|00\rangle.
\end{equation}
According to Eq.(\ref{relation}) and Eq.(\ref{50:50}), this is
equivalent to
\begin{equation}
|\psi\rangle_{out}=\frac{1}{\sqrt 2}(a_{1H}^\dagger a_{1V}^\dagger-a_{2H}^\dagger a_{2V}^\dagger)|00\rangle
=\frac{1}{\sqrt 2}(|HV\rangle_1\otimes |0\rangle_2 - |0\rangle_1\otimes |HV\rangle_2).
\end{equation}
This shows,  if the input state is $|\psi^+\rangle$,  one mode of
the output must be vacuum. Thus one detector must be silent.
Similarly, given the input states of
$|\phi^{\pm}\rangle=\frac{1}{\sqrt 2}(|H\rangle_1\otimes
|H\rangle_2\pm |V\rangle_1\otimes |V\rangle_2)$, we find that the
output states are $\frac{1}{2\sqrt 2}(|2H\rangle_1 \otimes
|0\rangle_2+|0\rangle_1\otimes |2H\rangle_2)\pm |2V\rangle_1 \otimes
|0\rangle_2 \pm|0\rangle_1\otimes |2V\rangle_2)$. This also shows
that, given the input state $|\phi^+\rangle$ or $|\phi^-\rangle$,
one output beam must be vacuum thus one detector must be silent.

 But
what happens if the input state is $|\psi^-\rangle$ ? Given the
input of
\begin{equation}
|\psi\rangle_{in}=|\psi^-\rangle=\frac{1}{\sqrt 2}(a_{1H}^\dagger
a_{2V}^\dagger - a_{1V}^\dagger a_{2H}^\dagger)|00\rangle
\end{equation}
the output state must be
\begin{equation}
|\psi\rangle_{out}=\frac{1}{2\sqrt 2}[(a_{1H}^\dagger +a_{2H}^\dagger)(a_{1V}^\dagger -a_{2V}^\dagger)
-(a_{1V}^\dagger +a_{2V}^\dagger)(a_{1H}^\dagger -a_{2H}^\dagger)]|00\rangle
\end{equation}
which is equivalent to
\begin{equation}
|\psi\rangle_{out}=-\frac{1}{\sqrt 2}(a_{1H}^\dagger a_{2V}^\dagger-a_{1V}^\dagger a_{2H}^\dagger)|00\rangle
=-\frac{1}{\sqrt 2}(|H\rangle_1\otimes |V\rangle_2 -|V\rangle_1\otimes |H\rangle_2).
\end{equation}
This is to say, each output beam contains 1 photon.
 Therefore, {\em if both detectors click, the input beams must be
in $|\psi^-\rangle$}. One may go into \cite{bscm,inns} for the
historic development of collective measurement by a beam-splitter.

{\em Summary:} We can regard two perpendicular polarizations as two
modes and treat each polarization {\em separately}. A 50:50
beam-splitter can help us do an incomplete Bell measurement in
polarization space. Suppose
 each input beam contains one photon, if both detectors in the output space click, then we judge that
the input light must have been collapsed into $|\psi^-\rangle$.
\subsubsection{Bell measurement in quadrature space through homodyne detection.}
Now we consider the Bell measurement in a type of continuous basis,
say, in the coordinate basis of the eigenstates of position
difference $\hat x_1-\hat x_2$ and momentum sum $\hat p_1+\hat p_2$.
Since the operator $\hat x_1-\hat x_2$ and $\hat p_1+\hat p_2$
commute, the two measurements can be done simultaneously in
principle. To measure these two quantities, we only need a 50:50
beam-splitter as shown in figure \ref{bs0}. Given the 50:50
beam-splitter as defined by Eq.(\ref{50:50}), measuring the momentum
of the output mode 1'  is equivalent to a collective
 measurement of $\hat p_1+\hat p_2$ on the input beams and  measuring
the position of output mode 2' is equivalent to the collective
 measurement of $\hat x_1-\hat x_2$ on the input beams since the following
 transformation happens in the Heisenberg picture
 \begin{equation}
 \hat p_{1'} = U_B^\dagger \hat p_1 U_B = \frac{\hat p_1+\hat p_2}{\sqrt 2};~\hat x_{2'}= U_B^\dagger \hat x_2
 U_B=\frac{\hat x_1-\hat x_2}{\sqrt 2}.
 \end{equation}
Therefore, given a 50:50 beam-splitter, the Bell measurement to
two input beams is reduced to the local measurements of  position
and momentum on each output modes. The only technical problem
remaining now is how to measure the coordinate  of the
output beam 1' and and momentum of beam 2'. This can be done by the homodyne
detection\cite{homo1,homo101,homo102,homo2,homob}.

\begin{figure}
\epsffile{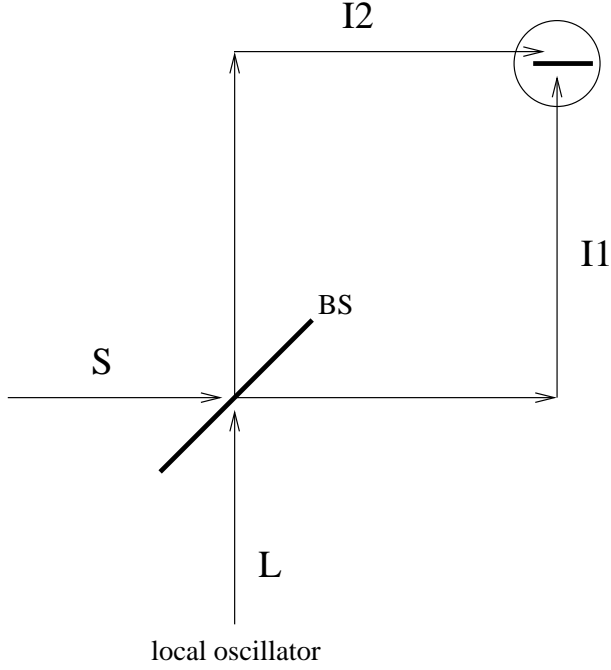} \caption{ Balanced homodyne detection with a
50:50 beam-splitter. By observing the current induced by the photon
difference of beam 1 and beam 2, one measures signal beam, beam S by
operator $\hat x(\theta)$ and $\theta$ is dependent on the state of local
oscillator, beam L.}\label{hdd0}
\end{figure}

As shown earlier, the states in Fock space can be represented in the
position-momentum basis, i.e., the eigen-states of $\hat
x(\theta)=\frac{1}{\sqrt 2}(a^\dagger e^{i\theta} +a e^{-i\theta})$,
$\{|x_\theta\rangle\langle x_\theta|,
x_\theta\in(-\infty,+\infty)\}$. However, the photon detector itself
only projects a state into the photon number states. To measure a
state in the basis of $\{|x_\theta\rangle\langle x_\theta|\}$ by
photon detectors, we need a bright classical light beam, which is
called as ``local oscillator''. In the measurement, the signal light
is ``phase locked" to the local oscillator. The local oscillator is
a coherent state $|\alpha\rangle_L$ and is denoted by ``L'' in
Fig.(\ref{hdd0}). We use the set-up to take quadrature measurement
on the signal beam, beam S. What is observed is actually the photon
number difference of two output modes, i.e.,
\begin{equation}
\langle\Delta N\rangle = \langle \psi_{out}|a_S^\dagger a_S
-a_L^\dagger a_L|\psi_{out}\rangle
=\langle \psi_{in}|U_B^{-1 }\left(a_S^\dagger a_S
-a_L^\dagger a_L\right)U_B|\psi_{in}\rangle.
\end{equation}
Based on the last equality of the equation above, the photon number difference
of the two output modes can be equivalently
regarded as the measurement outcome of
$U_B^{-1 }\left(a_S^\dagger a_S
-a_L^\dagger a_L\right)U_B=a_S^\dagger a_L+a_L^\dagger a_S$ on
{\em input state} $|\psi_{in}\rangle$, i.e., the state of light beams $S$ and $ L$. In this picture we have regarded the beamsplitter
itself as part of the measurement device.
Since the local oscillator is a very strong bright light in
coherent state ($e^{-|\alpha|^2/2+\alpha a_L^\dagger}|0\rangle$), operator
$a_L$ can be replaced by $\alpha$. Therefore the observed result of the photon
number difference of the output modes is actually the observation of
\begin{equation}
 |\alpha| (a^\dagger e^{i\theta}+a e^{-i\theta})=\sqrt 2 |\alpha|\hat x(\theta)
\end{equation}
to the input mode S, i.e., a measurement to input mode S in the
rotated position space.

{\em Remark:} Normally, $\theta$ value is unknown if beam L is
from a normal laser device. What we have known is that we have
observed $x(\theta)$ and $\theta$ is the phase parameter of beam
L. This is to say, we don't know the phase value but all pulses
have the same phase (phase locking).

As an application, we now demonstrate how to detect the position
fluctuation squeezing of a squeezed state. Given a single-mode
squeezed state $\hat S(\zeta) |0\rangle$ and $\hat S(\zeta)=\exp
(-\frac{\zeta}{2}{a^\dagger}^2 +\frac{\zeta^*}{2}{a}^2) $, $\zeta=
|\zeta| e^{i\phi}$, we find the quadrature fluctuation
\begin{equation}
\langle (\Delta x(\theta))^2\rangle
= \frac{1}{4}[\exp (-2|\zeta|)\cos^2(\theta-\phi/2)+\exp(2|\zeta|)\sin^2(\theta-\phi/2)]
\end{equation}
 We see that the fluctuation of $x$ will be less than 1/2 provided that
$|\theta-\phi/2|$ is not too large. In the experiment, the squeezed state is always phase-locked to the local
oscillator. Explicitly, one can produce many copies of input states and the local oscillator and
the value of $\phi$ and $\theta$ have a fixed relationship, e.g., $\phi=2\theta$.

\begin{figure}
\epsffile{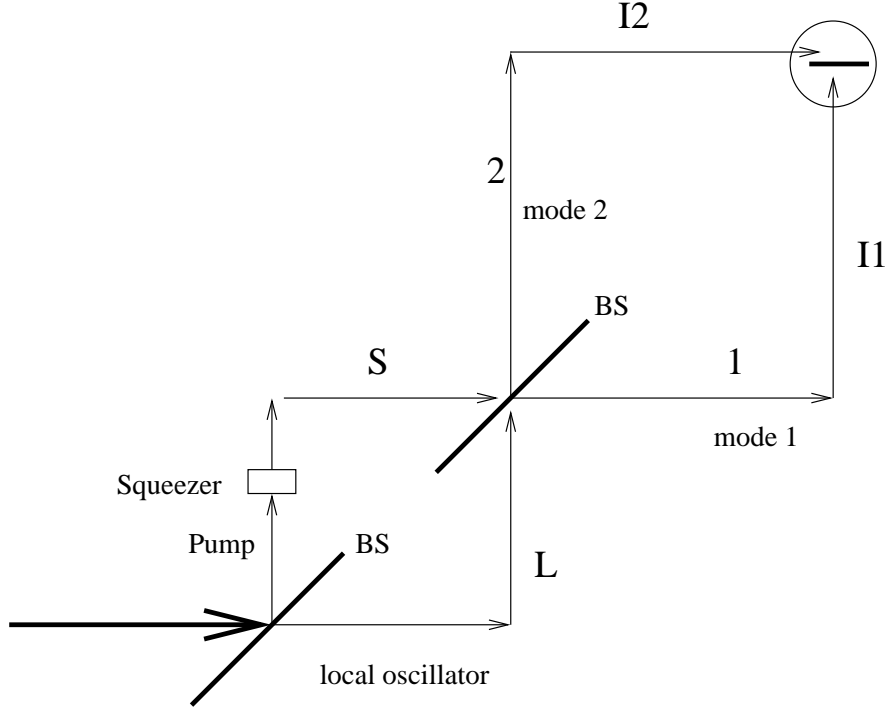} \caption{Verifying the position fluctuation
squeezing. The squeezed state is produced after the pump light
pass through the cavity (squeezer).}\label{msq0}
\end{figure}

The position squeezing is verified if one finds that the
fluctuation of $\hat x(\theta)$ is less than $1/4$, i.e.,
\begin{equation}\label{vsq}
\langle  x_{\theta i} ^2\rangle -\langle  x_{\theta i}\rangle^2<
1/2.
\end{equation}
here $x_{\theta i}$ is the observed value of the $i$th signal.
$\langle y_{ i} \rangle$ is the averaged value of all $y_i$.
 The set-up
is drawn in fig.(\ref{msq0}).
Since every squeezed state is phase-locked to the local oscillator,
we can set $\phi-\theta$ to be quite small in the experiment.  With
the observed position values of many copies of squeezed states, the
position fluctuation can be calculated by Eq.(\ref{vsq}).

Knowing how to measure the position, one can easily know how to do
the Bell measurement. We demonstrate it by $\hat x_1-\hat x_2$.
Consider figure \ref{bs0}. Measuring the quantity of $\hat x_1-\hat
x_2$ and $\hat p_1+\hat p_2$ of the light beam 1 and 2 is equivalent
to measure $\hat x$ and $\hat p$ respectively to beam $ 1'$ and $2
'$. The individual position measurement to beam 1' and momentum
measurement to beam 2' can be done by the homodyne detection as
shown already. Suppose the state of beam 1 and 2 is
$|\psi_{in}\rangle$ and the state of beam $1',2'$ is
$|\psi'\rangle$. Measurement outcome  of the position of beam $1'$
is
\begin{equation}
\langle \psi'|\hat x\otimes {\bf 1}|\psi'\rangle =\langle
\psi_{in}|U_B^{-1}\left(\hat x\otimes {\bf 1}\right) U_B
|\psi_{in}\rangle = \langle \psi_{in}|\left(\hat x_1-\hat x_2\right)
|\psi_{in}\rangle.
\end{equation}

 In short, using beam-splitters and strong reference light,
one can do the Bell measurement in position-momentum space, i.e.,
one can measure $\hat x_1-\hat x_2$ and $\hat p_1+\hat p_2$
simultaneously.

{\em Remark:} Different from the case of beam-splitter collective
measurement in the two-photon polarization space, here the Bell
measurement in position-momentum space is complete, deterministic
while the one in two-photon polarization space is incomplete (can
only judge whether the state is $|\psi^-\rangle$), non-deterministic
and normally with post selection only. This is an important
advantage of Gaussian states in QIP.
\subsection{Beam-splitter as an entangler}
A beam-splitter can work as a type of quantum entangler. For
example, given a 50:50 beam-splitter and the input beam of mode 1
containing one photon mode 2 containing nothing. The output state is
$\frac{1}{\sqrt 2}(|10\rangle +|01\rangle)$. Obviously, this is a
maximally entangled state in vacuum-and-one-photon space. However,
given another input state, the output state can be separable. For
example, if the input is vacuum or a coherent state, the output is
still vacuum or a (2-mode) coherent state. Actually, {\em the output
state is never entangled if the two-mode input state is classical}
(probabilistic mixture of coherent states). This can be regarded as
a theorem. The proof is very simple:  Any 2-mode coherent state
$|\alpha_1\rangle\otimes |\alpha_2\rangle$ will simply become
another 2-mode coherent state $|\alpha_1'\rangle\otimes
|\alpha_2'\rangle$  after the beam-splitter transformation.
Therefore, after passing the beam-splitter, the state will be
another probabilistic mixture of coherent states which is still a
classical state and must be separable (unentangled). Actually, no
matter how many beam-splitters are used, the final multi-mode state
must be separable given that the input light is in a classical
state. Details can be seen in Ref.\cite{btheorem}

Given the input of non-classical light, the output can be either entangled or
unentangled\cite{bwang}.
We now demonstrate the entanglement property when two one-mode squeezed states are used as the
input.
 Suppose the input states are the squeezed vacuum states in each mode, i.e.
\begin{equation}
\rho_{in}= S(\zeta_a) \otimes S(\zeta_b)|00\rangle\langle 00| \hat
S^\dagger(\zeta_a) \hat S^\dagger(\zeta_b),
\end{equation}
where $\hat S_a(\zeta_a),\hat S_b(\zeta_b)$ are squeezing operators
defined by $S(\zeta)$ in Eq.(\ref{jpsq348}). (Here $\zeta_a=r_a
e^{i\phi_a},~\zeta_b=r_be^{i\phi_b}$.)  The two-mode output state is
a pure state and can be entangled. The quantity of entanglement of
the output state is determined by the impurity of one mode. Suppose
the output state of mode $a$ is $\rho_{oa}$. The quantity of
entanglement of the output state is[1]
\begin{equation}
E(\rho_{oa})={\rm tr}(\rho_{oa}\ln\rho_{oa})\label{ed}.
\end{equation}
After passing the beam-splitter, the two-mode state is
\begin{equation}U_B S(\zeta_a)\otimes S(\zeta_b)|00\rangle={\mathcal U}|00\rangle.\end{equation}
 As stated earlier in theorem 3, the covariance matrix of the output state can be
derived based on the vacuum covariance matrix and the
transformation property of ${\mathcal U}$. Explicitly,
$$
{\mathcal U}^\dagger (\hat x_a,~\hat p_a,~\hat x_b,~\hat
p_b)^T{\mathcal U}=\left({\mathcal M}_{\zeta_a}\otimes {\mathcal
M}_{\zeta_b}\right)\cdot{\mathcal M}_B^{-1}(\hat x_a,~\hat
p_a,~\hat x_b,~\hat p_b)^T
$$
and ${\mathcal M}_{\zeta_a},~ {\mathcal M}_{\zeta_b}$ are defined
by Eq.(\ref{sqpmtran}), ${\mathcal M}_B$ is given in
Eq.(\ref{bbmpm}). Therefore the covariance matrix for the two-mode
state after passing through the beam-splitter is
\begin{equation}
\gamma'=\left[\left({\mathcal M}_{\zeta_a}\otimes {\mathcal
M}_{\zeta_b}\right)\cdot{\mathcal M}_B^{-1}\gamma_{in}({\mathcal
M}_B^{-1})^T\cdot \left({\mathcal M}_{\zeta_a}\otimes {\mathcal
M}_{\zeta_b}\right)^T\right]\label{bbet}
\end{equation}
and $\gamma_{in}$
is the characteristic matrix of the two-mode in-put state. This can be written in the $2\times 2 $ block form of $\left(\begin{array}{cc}\gamma_a'&\gamma_{ab}'\\
{\gamma_{ab}'}^T & \gamma_b' \end{array}\right)$ and
$\gamma_a'=\left(\begin{array}{cc}m_{11}& m_{12}\\
m_{21} & m_{22}\end{array}\right)$ is the covariance matrix of
mode $a$ of the output state. The matrix elements of $\gamma_a'$
can be explicitly calculated from Eq.(\ref{bbet}).
Since the entanglement quantity as defined by Eq.(\ref{ed}) is
unchanged by any local unitary operation, using an appropriate
symplectic transformation, we can first transform $\gamma_a'$ into
the diagonal form $\left(
\begin{array}{cc}\nu & 0\\0 & \nu\end{array}\right)$ and
\begin{equation}
\nu=\sqrt{m_{11}m_{22}-m_{12}^2}.\label{delta}
\end{equation}
This is the covariance matrix of  a thermal state.
Thus the quantity of entanglement for the output state is:
\begin{equation}
E(\rho_{oa})
=\ln\frac{2}{\nu+1}-
\frac{\nu-1}{2}\ln\frac{\nu-1}{\nu+1},
\end{equation}
 The above equation together with the previous equations for the
definition of $\nu$ formulates explicitly the entanglement quantity.
 The maximum entanglement is achieved through maximizing $\nu$.
This requests

\begin{equation}\label{cond}
2(\phi_1-\phi_0)-(\phi_b-\phi_a)=(2k+1)\pi,
\end{equation}
where $k$ is an arbitrary integer\cite{bwang}.

{\em Summary:} A beam-splitter can never produce entanglment with
classical light.  One can produce two-mode squeezed states given
one-mode squeezed states and a beam-splitter. The entanglement of
the output states can be optimized by adjusting the phase parameters
of the input squeezed states.

\section{Entanglement-based quantum tasks }\label{secebt}
Many important tasks in QIP are based on quantum entanglement. They
either directly need quantum entanglement resource, e.g., quantum
teleportation and dense coding, or the entanglement manipulation,
e.g., entanglement purification, or creating entanglement in
carrying out the tasks, e.g., cloning or quantum error correction,
and so on. We shall review most of these tasks with Gaussian states in
this section. But the issue of entanglement purification of strong
bipartite Gaussian states is not included in this section since  we
have placed it in another section which is focus on the mathematical
theory of quantum entanglement of Gaussian states.  We now start
with quantum teleportation.
\subsection{Teleportation with two-level states}
The idea of quantum teleportation was first proposed in
1993\cite{telepor1} with two-level systems. Suppose a photon with
Alice is in an unknown two-level quantum state in e.g., the
polarization space. Definitely, if Alice wants to send the unknown
quantum state to a remote party Bob, she can choose to simply send
the photon to Bob. However, with the help of quantum entanglement,
she has a better choice. Instead of sending the photon itself, she
can only move the unknown quantum state onto a photon at Bob's side
while not sending any physical particle to Bob, if she has
pre-shared an EPR pair with Bob. By this method, in order to send
any quantum information (unknown quantum state) to a remote party,
one does not have to send its physical carrier. This process is
called as {\em quantum teleportation}\cite{telepor1}. It is a bit
different from the classical teleportation: here once an unknown
state is teleported to a remote party, its original carrier has lost
all information of the state. Alice cannot choose to read the state
( measure the state) first and then announce the result, because the
original unknown state is changed once she reads it. Alice can
neither choose to copy the state first and then send one copy to
Bob, since the perfect cloning is forbidden by quantum non-cloning
theorem\cite{wooters}.
\subsubsection{Teleportation with two-level systems}
\begin{figure}
\begin{center}
\epsffile{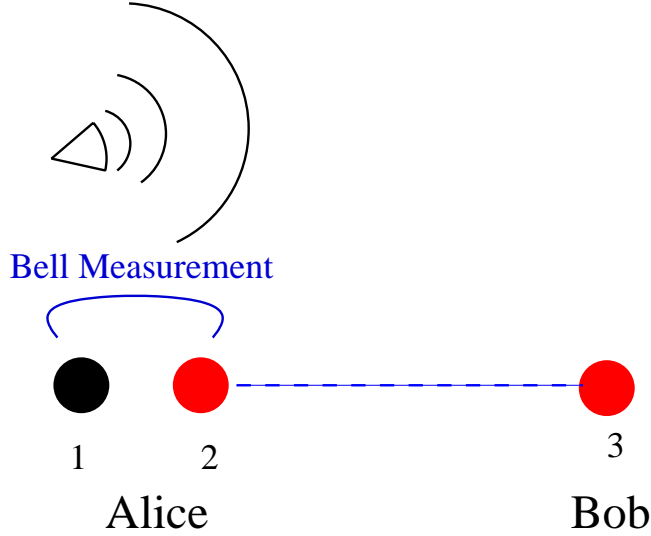}
\end{center}
\caption{ Quantum teleportation. Alice and Bob pre-share an EPR
pair ( photon 2 and 3). The unknown quantum state is initially
carried on photon 1. The Bell measurement outcome at Alice's side
is then broadcast. } \label{telefig2}
\end{figure}

The task of quantum teleportation\cite{telepor1} can be achieved if
they have pre-shared quantum entanglement. In figure \ref{telefig2},
Alice and Bob initially share a perfect EPR pair, particle 2 and 3.
Particle 2 is with Alice and particle 3 is with Bob. The two-level
state of particle 1 with Alice is unknown. Alice  measures particle
1 and 2 in the Bell basis first, then remotely instructs Bob taking
a local unitary transformation according to her measurement
outcome. After the local unitary transformation to particle 3 is
done, particle 3 is exactly in the initial unknown state of particle
1.
Now we demonstrate the idea by photon polarization.
Suppose the unknown polarization state of photon 1 is
\begin{equation}
\label{gen}
|\chi\rangle_{1}= \alpha | H \rangle_1 + \beta
| V \rangle_1
\end{equation}
and $\alpha, \beta$ are complex amplitudes satisfying
$|\alpha|^2+|\beta|^2=1$.
Photon 2 and 3 are in a maximally entangled state in polarization space, e.g.:
\begin{equation}
|\psi^{-}\rangle_{23}= \frac{1}{\sqrt{2}} \left( | H
\rangle_2 | V \rangle_3 - | V \rangle_2 | H
\rangle_3 \right) ,
\end{equation}
where $| H \rangle$ and  $| V\rangle$ represents the
horizontally- and vertically-polarized photon state, respectively.
The initial state of all three photons is
\begin{equation}
|\psi\rangle_{123}= |\chi\rangle_{1}|\psi^{-}\rangle_{23},
\end{equation}
which can be equivalently written in
\begin{eqnarray}
|\psi\rangle_{123}=
\frac{1}{2}\left[-|\psi^-\rangle_{12}|\chi\rangle_3
-|\psi^+\rangle_{12} (\alpha |H\rangle_3-\beta |V\rangle_3)\right.\nonumber \\
\left.+ |\phi^-\rangle_{12} (\alpha |V\rangle_3+\beta |H\rangle_3) +
|\phi^+\rangle_{12} (\alpha |V\rangle_3-\beta |H\rangle_3)\right].
\end{eqnarray}
One can easily see that after observing the outcome of Bell measurement
on photon 1 and 2, the corresponding local unitary transformation to
photon 3 will produce the initial state $|\chi\rangle$  on photon 3.

{\em Remark: }
Quantum teleportation does not offer any faster-than-light communication.
Bob can only produce state $|\chi\rangle$ on photon 3 {\em after} taking an appropriate local unitary transformation instructed
by Alice through classical communication which cannot be faster than light. Without Alice's message about the outcome
of her Bell measurement, the state of Bob's photon is not changed at all.

In fact, the initial unknown state is not necessarily a pure state.
It can be an arbitrary mixed state. Moreover, it can even be a part
of a composite state. This can be demonstrated in the concept of
entanglement swapping\cite{swapping}. For example, if initially
photon 1 is entangled with photon 0 and the state for the composite
system of photon 1 and 0 is an antisymmetric state
$|\psi^-\rangle_{01}$. The total state of photon 0,1,2, and 3 is
\begin{equation}
|\psi\rangle_{0123} =
\frac{1}{2}(|H\rangle_0|V\rangle_1-|V\rangle_0|H\rangle_1)
(|H\rangle_2|V\rangle_3-|V\rangle_2|H\rangle_3).
\end{equation}
After a Bell measurement on photon 1 and 2,
these two photons will be projected onto one of the 4 Bell states. To see what happens to photon 0 and photon 3,
we now recast the above state in
\begin{equation}
\frac{1}{2}(|\psi^+\rangle_{03}|\psi^+\rangle_{12}+
|\psi^-\rangle_{03}|\psi^-\rangle_{12}+
|\phi^+\rangle_{03}|\phi^+\rangle_{12}+
|\psi^-\rangle_{03}|\psi^-\rangle_{12}).
\end{equation}
This means, in all cases photon 0 and photon 3 becomes entangled despite the fact
that they have never been interacted. This is the so called entanglement swapping\cite{swapping}.

Quantum entanglement itself can also help to raise the channel
capacity of communication through quantum dense
coding\cite{bdensec}. If they pre-share an EPR pair, Alice at her
side can change the state of the whole pair into any one of the 4
Bell states. This helps her to transmit 2 bits information by
sending one qubit.

The task of quantum teleportation can also be done  in
position-momentum space with continuous variable states \cite{BL},
which is called as continuous variable quantum teleportation, CVQT.
One can use a two-mode Gaussian state as the entanglement resource
which is pre-shared between Alice and Bob. Alice can then teleport
an unknown state. If they have pre-shared perfect entanglement
resource, they can teleport an arbitrary continuous variable state.
Perfect entanglement means infinite squeezing which doesn't exist in
practice. Luckily, if the unknown state is limited to the family of
coherent states, one can teleport it with satisfactory fidelity
through the practically existing two-mode finitely squeezed states.
\subsection{CVQT with Gaussian states}
 Similar to that of the 2-level states, given the pre-shared two-mode
squeezed states, the CVQT can be performed to transfer the quantum
state of mode $1$ to the remote mode $3$. Suppose Alice and Bob have
three light beams, beam $1$ and beam $2$ are with Alice, beam $3$ is
with Bob. Beam $2$ and $3$ are in an entangled state, e.g., a two
mode squeezed state. Their goal is to transfer the state of beam $1$
into the remote mode $3$ through the entangled beams $2$ and $3$.
The task can be completed by the following 3 steps:
\begin{itemize} \item 1. Alice takes the Bell measurement on beam 1 and beam 2, i.e.,
measuring both $ \hat x_1- \hat x_2$ and $\hat p_1+\hat p_2$.
Suppose she has observed $\Delta$ and $\Sigma$ as the  outcome.
\item 2. Alice announces her measurement outcome $\Delta$ and
$\Sigma$. \item 3. Bob displaces beam 3 by $\Delta$ in position and
$\Sigma$ in momentum space.
\end{itemize} The state of beam 3 with Bob is now the outcome state of CVQT.
There are 3 main theoretical approaches to CVQT with two-mode
squeezed states: the Heisenberg picture \cite{cvqt1}, the
phase-space representation \cite{cvqt2} and the Schrodinger picture
\cite{hofmann,takeoka,vukics}.
\subsubsection{CVQT in the Heisenberg picture}
The first proposal of CVQT was presented in the Heisenberg picture
\cite{cvqt1}. For clarity, we start from the case with perfect
entanglement. Actually, the first proposal of CVQT assumed the
perfect entanglement\cite{cvqt1}. Say, initially, the light beams
of mode 1 and mode 2 are maximally entangled, i.e.,
\begin{equation}\label{teleh0}
\hat x_2-\hat x_3 = 0;~ \hat p_2+\hat p_3 =0.
\end{equation}
The light beam of mode 0 is the unknown field to be teleported.
Alice then performs Bell measurement i.e., simultaneous
measurement of $\hat x_1-\hat x_2$ and $\hat p_1+\hat p_2$ on mode
1 and mode 2. Suppose she has obtained
$$\hat x_1-\hat x_2=\Delta, ~ \hat p_1+\hat p_2=\Sigma$$ and $\Delta,~\Sigma$
are real numbers as Alice's measurement outcome. Given this outcome
and Eq.(\ref{teleh0}), the field of mode 3 can be expressed in
$$\hat x_3=\hat x_1-\Delta;~\hat p_3=\hat p_1-\Sigma.$$ After taking a displacement of $\Delta$ to
 $\hat x_3$ and $\Sigma$ to $\hat p_3$ Bob can obtain the following result for mode 3 as the outcome field
 of the quantum teleportation:
\begin{equation}
\hat x_3\longrightarrow \hat x_{tel}=\hat x_1;~ \hat
p_3\longrightarrow \hat p_{tel}=\hat p_1.
\end{equation}
The field of mode 3 after teleportation is exactly equal to the
initial field of mode 1.

In practice, the maximal entanglement resource is replaced by the
two-mode squeezed state. In such a case, according to
Eq.(\ref{eq80}), the initial field of the two-mode squeezed state
is:
\begin{equation}
\hat x_2-\hat x_3 = e^{-r}(\hat x_2^{(0)}-\hat x_3^{(0)});~ \hat
p_2+\hat p_3=e^{-r}(\hat p_2^{(0)}+\hat p_3^{(0)})
\end{equation}
and the superscript ``0" indicates the vacuum field. Given such an
imperfect entanglement, the outcome field of mode 2 after quantum
teleportation is
\begin{equation}
\hat x_3\longrightarrow \hat x_{tel}=\hat x_1+e^{-r}(\hat
x_2^{(0)}-\hat x_3^{(0)});~ \hat p_3\longrightarrow \hat
p_{tel}=\hat p_1+e^{-r}(\hat p_2^{(0)}+\hat p_3^{(0)}).
\end{equation}
The result is the same with the ideal one in the limit of
$r\longrightarrow +\infty$. However, if $r$ is finite, there will
be noise for the outcome field.

\subsubsection{CVQT in phase space representation} The
teleportation outcome can also be presented in the phase space
representation. In the original paper of phase space
representation CVQT, the Wigner function is used. Here we use the
the characteristic functions only.

There are three modes (light beams) in CVQT: mode 1 for the unknown
state which is to be teleported; mode { 2} and { 3} are for the
bipartite entangled state pre-shared by Alice and Bob. Also, there
are three steps in the CVQT: 1. Alice takes Bell measurement on beam
light 1 and 2; 2. Alice announces her measurement outcome of
$\Delta$ for $(\hat x_1-\hat x_2)$ and $\Sigma$ for $(\hat p_1+\hat
p_2)$. Bob displaces his light beam, beam 3 by $\Delta$ in
``position" space and $\Sigma$ in ``momentum" space.

The displacement operator for Bob in step 3 is $ D(\Delta, \Sigma)=
\exp(i \hat x_3 \Sigma-i\hat p_3 \Delta )$ as defined in
Eq.(\ref{pmdis}). It would make no difference to the final outcome
of the above stated CVQT if step 2 and step 3 above were replaced by
the following non-local operator in step 3
\begin{equation}\label{cvqtub}
G= \exp[-i \hat p_3 (\hat x_1-\hat x_2)]\exp [i\hat x_3 (\hat
p_1+\hat p_2) ].
\end{equation}
 This operator is non-local therefore cannot be implemented in
any real teleportation set-up. However, mathematically, the outcome
state of the standard CVQT procedure is identical to that from the
imagined one with such a non-local operator. We shall use this
virtual procedure  to deduce the general result of the outcome state
of CVQT {\em mathematically}. Here we consider the ensemble outcome
state of the CVQT with Bob, i.e., the state {\em averaged} over all
possible values of $\Delta$ and $\Sigma$.  If operator $G$ were
used, actually, Bob didn't need the classical information announced
by Alice in step 2, since we only care about the ensemble result of
CVQT outcome here.  Since the measurement operation in step 1 and
the displacement operator $G$ {\em commute}, if one exchanges the
order of  the Bell measurement and $G$ operation, the result should
be unchanged. Therefore the ensemble result of the transferred state
after the CVQT is
\begin{equation}\label{cvqtoce}
\rho_{tel}={\rm Tr_{1,2}} \left(G\rho_1\rho_{23}G^{-1}\right).
\end{equation}
Here $\rho_1$ and $\rho_{23}$ are the initial unknown state of mode
$1$ and the pre-shared entangled state of mode $2$ and $3$,
respectively. In this virtual protocol, there is no classical
communication between Alice and Bob. Therefore the delayed Bell
measurement at Alice's side can be actually omitted because the
outcome state at Bob's side after unitary operation $G$ will be
unchanged no matter whether Alice does the Bell measurement at her
side. Therefore to calculate the outcome state, we only need to
first take unitary operation $G$ and then take sub-trace of mode 1
and mode 2.

 Denote the initial characteristic function for the three-mode
 state to be
 \begin{equation}
 \chi_{o}(\xi)=\chi_1(\xi_1,\xi_2)\cdot\chi_{23}(\xi_3,\xi_4,\xi_5,\xi_6).
 \end{equation}
 $\chi_1(\xi_1,\xi_2)$ is the characteristic function of mode $1$,
 $\chi_{23}(\xi_3,\xi_4,\xi_5,\xi_6)$ is the characteristic function for the
 pre-shared entangled state (mode $2$ and $3$).
  Applying theorem 2 in section 1 to Eq. (\ref{cvqtoce}), we have
  the following
 characteristic function of the CVQT outcome state
 \begin{equation}
 \chi_{tel}(\xi_5,\xi_6) =\chi_o (M^T\xi)\mid_{\xi_1=\xi_2=\xi_3=\xi_4=0}
 \end{equation}
where $M$ is defined as
\begin{equation}
G^\dagger R G =MR
\end{equation}
and $R=(\hat x_1,\hat p_1,\hat x_2,\hat p_2,\hat x_3,\hat p_3)^T$.
Using Eq. (\ref{bch}), one can easily find:
\begin{eqnarray}
M =\left( \begin{array}{cccccc}
1&0&0&0&-1&0\\
0&0&0&1&0&-1\\
0&0&1&0&-1&0\\
0&1&0&0&0&1\\
1&0&-1&0&1&0\\
 0&1&0&1&0&1
\end{array}\right)
\end{eqnarray}
 Given this matrix, the characteristic function of the 3-mode state after unitary transformation $G$
 is
\begin{equation}
\chi_o'(\xi)=\chi_o(M^T\xi)=\chi_1(\xi_1+\xi_5,\xi_4+\xi_6)
\chi_{23}(\xi_3-\xi_5,\xi_2+\xi_6,-\xi_1-\xi_3+\xi_5,-\xi_2+\xi_4+\xi_6).
\end{equation}
According to theorem 2, the characteristic function for mode $3$
after teleportation is simply
\begin{equation}
\chi_{tel}=\chi_o(M^T\xi)|_{\xi_1=\xi_2=\xi_3=\xi_4=0}=\chi_1(\xi_5,\xi_6)\chi_{23}(-\xi_5,\xi_6,\xi_5,\xi_6)
\end{equation}
which is just
\begin{equation}\label{chtel}
\chi_{tel}(\xi_5,\xi_6)=\chi_1(\xi_5,\xi_6)\exp
\left[-\frac{1}{2}e^{-2r}(\xi_5^2+\xi_6^2)\right].
\end{equation}
Here we have used Eq. (\ref{chsq2}).
From this we can see that there will be excess noise in the
outcome state of teleportation, if $r$ is finite.

Similar to that of the two-level state, here one can also use the
two-mode squeezed state to teleport quantum
entanglement\cite{BL,cvswap1,cvswap2,cvswap3,vanpra}. Say,
initially, there are 2 two-mode squeezed states of beam 1,2 with
squeezing parameter $r_a$ and beam 3, 4 with squeezing parameter
$r_b$. After measuring beam 2,3 in the Bell basis, and an
appropriate displacement operation on beam 1 and beam 4, one can
obtained a two-mode squeezed state for beam 1 and beam 4. The
displacement can be carefully chosen so as to always obtain the same
two-mode squeezed state with squeezing parameter $r_{ab}$ in
 the form $e^{ \tanh r_{ab}}=e^{\tanh r_a\tanh r_b}$ no
matter what result of the Bell measurement on beam 2,3 is
obtained\cite{BL,vanpra}.

The CVQT result above is the ensemble outcome of the transferred state
averaged over all possible measurement outcome at Alice's side,
i.e., all possible values of $\Delta$ and $\Sigma$. One can obtain
the single-shot result through using the Shrodinger picture. This
can be done in  the position-momentum
representation\cite{takeoka}, the coherent-state
representation\cite{vukics} and so on. A very useful single-shot
result is given by Hoffman and coauthors in Schrodinger
picture\cite{hofmann}. There, it is found that if Alice obtains
her measurement outcome $\Delta,\,\Sigma$, after Bob takes the
displacement accordingly, the state transferred to Bob's side can
be described by the so-called transfer operator $T$:
\begin{equation}
|\psi\rangle_{tel} = \hat T (\Delta,\Sigma) |\psi_0\rangle
\end{equation}
where $|\psi\rangle_{tel}$ is the outcome state with Bob after the
CVQT, $|\psi_0\rangle$ is the initial unknown state at Alice's side
and
\begin{equation}
\hat T(\Delta,\Sigma)= \sqrt{\frac{1-\tanh r}{\pi}}\hat
D(\Delta,\,\Sigma) \exp (\frac{1}{2}\ln \tanh r a^\dagger a)\hat
D^{\dagger}(\Delta,\,\Sigma)
\end{equation}
and $ \hat D(\Delta,\Sigma)=\exp(-i\Delta \hat p+i \Sigma \hat x) $
as defined by Eq.(\ref{pmdis}).
Besides these, there are also results about the CVQT in phase-number
space\cite{numberphase,numberphase02,numberphase03}
\subsubsection{Teleportation fidelity.} As has been shown, if the
squeezing parameter $r$ of the pre-shared two-mode squeezed state is
finite, there is always noise in the outcome state. An interesting
question is whether we can obtain non-trivial outcome using finitely
squeezed states only. The answer is positive if the initial unknown
state is limited to an unknown {\em coherent state} only. If there
is no pre-shared entanglement resource, Alice and Bob may do the
task trivially: Alice uses an optimized measurement to detect the
state and then send her outcome to Bob through a classical channel.
In this way, the fidelity of the outcome state and the initial
unknown state cannot exceed $1/2$. However, as shown
below, in the CVQT with pre-shared two-mode squeezed states, the
fidelity of outcome state and the initial unknown state is always
larger than 1/2 provided that the squeezing factor is larger than 0.

If the initial unknown state is a coherent state of
$|\alpha=u+iv\rangle={\mathcal D}(\alpha)|0\rangle$, the
characteristic function given in Eq. (\ref{chtel}) is for a
displaced thermal state:
\begin{equation}
\rho_{tel}= {\mathcal D} (\alpha) \rho_{th}(\beta) {\mathcal
D}^\dagger (\alpha)
\end{equation}
where $\rho_{th}(\beta)=(1-e^{-\beta})e^{-\beta a^\dagger a}$ is a
thermal state and $e^{-\beta} = \frac{e^{-2r}}{1+e^{-2r}}$, which
is a direct consequence of Eq. (\ref{chthm}) and Eq.
(\ref{thtemp}). The teleportation fidelity is
\begin{equation}
F=\langle 0|{\mathcal D}^\dagger(\alpha) {\mathcal D} (\alpha)
\rho_{th}(\alpha) {\mathcal D}^\dagger (\alpha){\mathcal
D}(\alpha)|0\rangle = \langle 0|\rho_{th}(r)|0\rangle
\end{equation}
which is simply
\begin{equation}
F=1-e^{-\beta} = \frac{1}{1+ e^{-2r}}.
\end{equation}
This result is {\em independent} of the parameters $u,\, v$. This is
to say, all coherent states can be teleported with the same
fidelity. The fidelity is larger than $1/2$\cite{bfk} provided that $r>0$.
Since no classical scheme can  teleport an unknown coherent state
with the fidelity larger than $1/2$\cite{bfk}, the result shows that the CVQT
scheme can produce a non-classical result in teleporting an unknown
coherent state given any small squeezing for the pre-shared two-mode
squeezed state.

\subsection{Experiment}
\begin{figure}
\epsffile{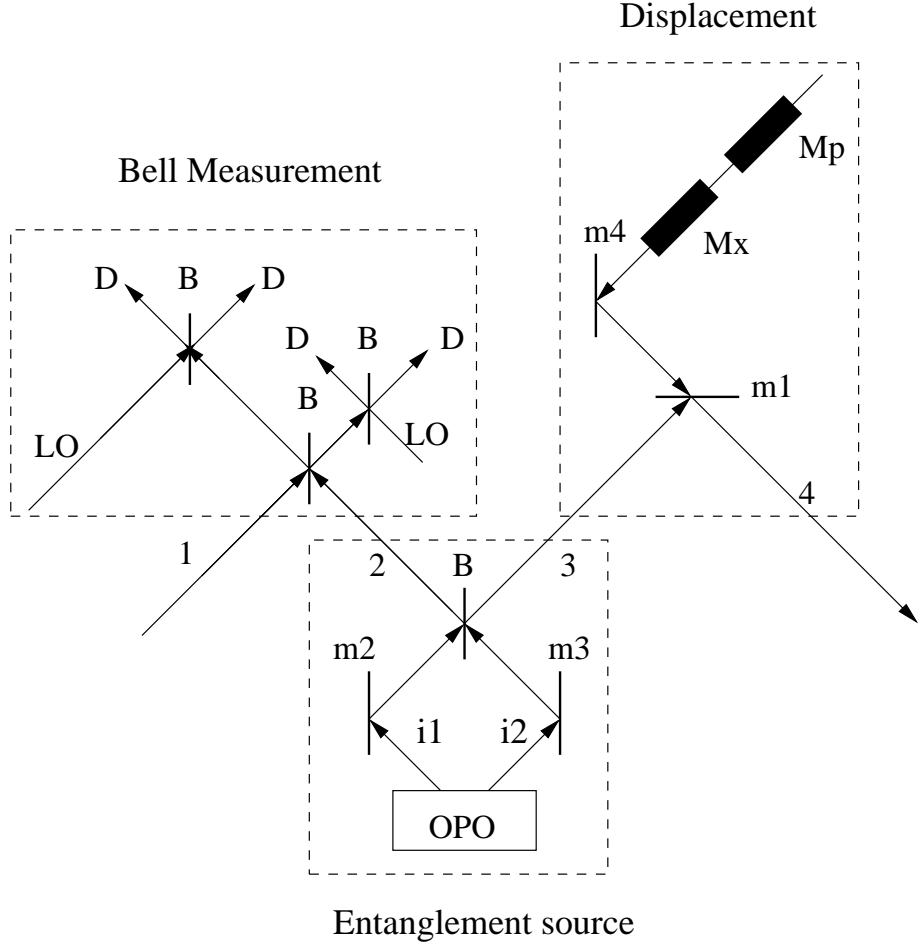}
\vskip 10pt
\caption{ Experimental set-up for the quantum teleportation with
Gaussian states. B: 50:50 beam splitter, m : mirror. Mx, Mp: displacement
operation. LO: local oscillator.
Beam 1 is the income unknown coherent light which is to
be teleported. Beams $i1,i2$ are two single-mode squeezed states.
Beam 2,3 are two-mode squeezed state which is
used as the entanglement source. Beam 4 is the outcome state which
is produced by taking appropriate displacements to beam 3.}\label{telee}
\end{figure}

There are 3 major steps in quantum teleportation: 1) Sharing
quantum entanglment.  The two-mode squeezed state
can be regarded as the pre-shared entanglement if the two modes
here are spartially separated. 2) The Bell measurement. This
requires to simultaneously measure the quantity of $\hat x_1-\hat
x_2$ $and$ $\hat p_1 + \hat p_2$. This can be done by the homodyne
detection. 3) The appropriate displacement operation dependent on
the Bell measurement outcome. One can see details in the figure
\ref{telee}.
The first teleportation experiment with two-mode squeezed states is
done in 1998\cite{telee0}. In the experiment\cite{telee0}, they have
teleported an unknown coherent state with intensity changing in a
large range. The averaged fidelity of the outcome state and the
input unknown state is around 0.58. On the other hand, no classical
method can reconstruct an unknown coherent state with a fidelity
larger than 1/2\cite{bfk}.

To keep the optical coherence, Alice and Bob must share the same
local oscillator or be phase locked to one strong reference
coherent light. The two mode squeezed states must also be  phase
locked to the local oscillator. Details of the coherence and the
validity have been discussed in\cite{barry,enk} . According to
Ref.\cite{enk}, the validity of existing experiment can be
explained in this way: The protocol requires Alice and Bob share
the same reference light. This can be regarded as the classical
communication for the calibration of measurement basis. Given the
shared reference light and the fact that all those two-mode
squeezed states are phase-locked to the reference light, the
quantum coherence has been kept very well, it is something like
multi-mode fields with same phase. Here, since they can only use
finitely squeezed states, the teleportation fidelity is limitted.
It is then interesting to ask the fidelity criterion for a good
quantum teleportation with 2-mode squeezed states. There are
several different viewpoints. A necessary condition for the a
non-classical teleportation of an unknown coherent state is
\begin{equation}
F_{av}\ge \frac{1}{2}
\end{equation}
and $F_{av}$ is the teleportation fidelity averaged over all coherent states.
This condition is required
since otherwise one can make it without entanglement resource\cite{bfk}.
Another viewpoint\cite{telecr2} sets the necessary condition to be
\begin{equation}
F_{av}\ge 2/3
\end{equation}
since one may clone an arbitrary coherent state with a fidelity
$2/3$\cite{clone3}. Besides these, there are also other
criteria\cite{bfk,telecr3}. Among all the existing experimental
results of teleporting unknown coherent
states\cite{telee0,telee2,telee3}, the highest average fidelity
achieved is around 0.64.

\subsection{Dense coding with two-mode squeezed states}
In the two-level-state case, one may implement quantum
dense-coding using the pre-shared EPR states.  In such a way, he
is able to realize two-bit classical communication through sending
one qubit only. The role of quantum entanglement and the advantage
to the simple qubit-sending is clear: if they don't pre-share an
EPR pair, one qubit can only carry one bit classical information.

Similar results also exist in CV state
case\cite{cvdense0,cvdense}. Here they need to pre-share a
two-mode squeezed state. Alice takes an arbitrary displacement,
$\hat D_1(\Delta,\Sigma)$ to her beam and then sends her beam to
Bob. By a Bell measurement $\hat x_1-\hat x_2$ and $\hat p_1+\hat
p_2$, Bob obtains information of $two$ parameters, $\Delta$ and
$\Sigma$. In the case of infinite squeezing of their pre-shared
state, Bob can obtain the two parameters exactly. However, we need
to know the result in the case they only pre-share a finitely
squeezed two-mode squeezed state. And we need to study the
advantage to the case of simply sending a light-beam with certain
common constraint, e.g., the same averaged photon number of the
states. After calculation\cite{cvdense}, it is found that the
channel capacity of a two-mode squeezed state can be larger than
that of a coherent state or a single-mode squeezed state. If they
use very bright light (i.e., the average photon-number of the
state is very large), the channel capacity of a two-mode squeezed
state is nearly 2 times of that of a coherent state or a single
mode squeezed state. Explicitly, the channel capacity of a
coherent state and that of a two-mode squeezed state can be found
in, e.g., \cite{cvdense,BL}. We shall present
 the details in section \ref{sec:Dense_Coding_with_Gaussian_Entanglement}.
Experimental works on dense coding with two-mode squeezed states
have also been done\cite{denseee}.

\subsection{Quantum error correction codes}\label{seqecc1}
The purpose of a quantum error correction code (QECC) is to
process quantum information robustly in the presence of noise.

Let's first consider a simple example in classical communication
through a noisy channel. Each time we want to transmit one bit
information which is either 0 or 1. Since the channel is noisy, it
flips the transmitted bit with probability $p$. This is to say,
sometimes 0 is changed into 1 and 1 is changed into 0 during the
transmission. The channel error rate here is $p$. However, we can
decrease the error rate by using error correction code of the
following:
\begin{equation}\label{qecc3}
0\rightarrow 000;~ 1\rightarrow 111.
\end{equation}
After the transmission, we recover the bit value of each code
by the majority rule. In such a way, the error rate is decreased to
$3p^2(1-p)+p^3$. If the channel noise ($p$ value) is small, the final
error rate drops drastically.

In the quantum information processing, there are something similar
to this, but there are also something different. We can still use
the same majority rule to correct bit-flip errors, but the initial
state is in general an unknown linear superposition of
$|0\rangle,|1\rangle$, the trivial repetition code doesn't work
and quantum entanglement has to be involved. Also, we need to add
something else to correct phase-flip errors.

Say, in general, we want to protect an unknown state of the form
\begin{equation}
|\psi_u\rangle =\alpha |0\rangle + \beta |1\rangle.
\end{equation}
A bit-flip error changes the state into
 \begin{equation}
\sigma_b |\psi_u\rangle =\alpha |1\rangle + \beta
|0\rangle.\label{unknown}
\end{equation}
A phase-flip error changes the state into
 \begin{equation}
\sigma_p |\psi_u\rangle =\alpha |0\rangle - \beta |1\rangle.
\end{equation}
A simple quantum code of
\begin{equation}
|0\rangle\otimes |00\rangle\rightarrow |000\rangle;~
|1\rangle\otimes |00\rangle\rightarrow |111\rangle \label{bfcode}
\end{equation}
can decrease the bit-flip errors. Using such a code, any unknown
state $|\psi_u\rangle=\alpha|0\rangle + \beta |1\rangle$ is encoded
by $\alpha |000\rangle + \beta |111\rangle$, which is normally an
entangled state. This code can not remove phase-flip errors. Say,
the channel noise can also change the state into $\alpha|0\rangle -
\beta |1\rangle$. To correct both types of errors, we can use a more
complete QECC\cite{qecc1,qecc2,qecc3,qecc4,qecc5,qecc6}. For
simplicity, we demonstrate the main idea here with Shor's 9-qubit
code[75].

Given any unknown two-level quantum state of equation (\ref{unknown}),
 we shall first encode it with the 3-qubit
phase-flip code, i.e., $|0\rangle\longrightarrow |\bar 0\bar 0\bar 0\rangle$ and
$|1\rangle\longrightarrow |\bar 1\bar 1\bar1\rangle$, and
\begin{equation}
|\bar 0\rangle = H |0\rangle =\frac{1}{\sqrt 2}(|0\rangle +
|1\rangle) ;~ |\bar 1\rangle = H |1\rangle =\frac{1}{\sqrt
2}(|0\rangle - |1\rangle) .
\end{equation}
Here $H$ is the Hadamard transform. We shall then encode each
qubit again by eq.(\ref{bfcode}) against the bit-flip errors. In
such a way we have a 9-bit code of the following:
\begin{eqnarray}
|0\rangle\rightarrow |0\rangle_{ec}=\frac{(|000\rangle +|111\rangle)(|000\rangle +|111\rangle)(|000\rangle +|111\rangle)}{2\sqrt 2}\nonumber\\
|1\rangle\rightarrow |1\rangle_{ec}=\frac{(|000\rangle -|111\rangle)(|000\rangle -|111\rangle)(|000\rangle -|111\rangle)}{2\sqrt 2}.
\end{eqnarray}
After sending out the quantum code $\alpha|0\rangle_{ec}+\beta
|1\rangle_{ec}$, if there is at most one bit-flip error and at most
one phase-flip error, we can always detect the positions of flip
precisely
 and recover the original code. Consider the 3 qubits in the first bracket. We first measure the parity of qubit 1 and 2
and obtain $z_{1\oplus 2}$, and then measure the parity of qubit 1
and 3 and obtain $z_{1\oplus 3}$. If both of them are 0 then there
is no bit-flip error. If both of them are 1, then qubit 1 must have
been flipped and we flip it back. Similarly, $z_{1\oplus
2}=0,z_{1\oplus 3}=1 $ means  qubit 3 is wrong and the opposite
result means qubit $2$ is wrong. Such type of parity measurement
does not destroy the code state itself. After we take such type of
parity check to all 3 blocks, we can locate and correct the only
bit-flip error, if there is one. We can also correct any phase-flip
error. Suppose there is only one phase flip. In each block, if any
qubit is phase flipped, the block state is changed from
$\frac{1}{\sqrt{2}}(|000\rangle\pm|111\rangle)$ into $\frac{1}{\sqrt
2}(|000\rangle\mp|111\rangle)$. We can take another type of syndrome
measurement by comparing the ``phase'' parity values of different
blocks. Say, $\frac{1}{\sqrt 2}(|000\rangle\pm|111\rangle)$
corresponds to the phase value $0,1$, respectively. We can first
measure the phase parity of block 1,2   and then measure that of
block 1,3 and we can determine the phase flipped block  if there is
at most one block that is phase flipped. Therefore, if there is not
more than one wrong qubit, we can always recover it to the perfect
code and the original quantum state is protected perfectly.

In making the 9-qubit code, we have used one important fact: to
correct the phase flip error, we need first do a basis
transformation and then construct the code in the conjugate basis,
i.e., from basis  $\{|0\rangle, |1\rangle\}$ to $\{|\bar
0\rangle,|\bar 1\rangle\}$. The similar idea can also be used to
construct the error-correction code for an unknown continuous
variable state\cite{qecc7}. Here we also have two conjugate basis,
$\{|x\rangle\}$ and $\{|\bar x\rangle \}$ which we shall call
position basis and momentum basis, respectively. To use the idea
of 9-bit code for the two level case, we need the relationship
between state $|x\rangle$ and state $|\bar x\rangle$.
 In the two level case, the two
conjugated bases are connected by Hadamard transform. In the CVQT
case, we have the fact that
\begin{equation}
|\bar x\rangle = H_c |x\rangle = \frac{1}{\sqrt \pi}\int e^{2i xy} |y\rangle {\rm d} y.
\end{equation}
where we have used the notation $|\bar x\rangle$ for a {\em
momentum} eigenstate with the value of the momentum being $x$,
i.e., $|\bar x\rangle = |p=x\rangle.$

Analogously to the 2-level QECC case, we can now  use the following code for any state in position space
\begin{equation}
|x_{encode}\rangle =\frac{1}{\pi^{3/2}}\int e^{2ix(y_1+y_2+y_3)}|y_1,y_1,y_1\rangle  |y_2,y_2,y_2\rangle |y_3,y_3,y_3\rangle
{\rm d} y_1{\rm d} y_2{\rm d} y_3.
\end{equation}
It can be easily examined that any displacement error or phase
error can be detected and corrected, if there is at most one
displacement or phase error in the 9 ``qubits''. It has been shown
that such a 9-qubit QECC can be implemented with linear optics and
squeezed lights\cite{qecc7}.

More generally, suppose ${\hat \epsilon_i}$ is any possible channel action. The sufficient and necessary condition
for quantum error correction is\cite{qecc4,qecc5}
\begin{equation}
\langle x'_{encode}|\hat \epsilon_i^\dagger \hat \epsilon _j | x_{encode}\rangle =\delta (x'-x)\lambda_{ij}.
\end{equation}
and $\lambda_{ij}$ is a complex number independent of encoded states.
It has been shown that similar to the case of 2-level states, an effective QECC code for the continuous variable
states can be constructed by fewer qubits, e.g., only 7 qubits of 5 qubits\cite{qecc8,qecc9}.

In the above, we have assumed the condition that there is at most
one displacement error or phase-shift error in the code. However, a
continuous variable state will always have some small errors. That
is to say, it is impractical to request that only one mode in the
quantum code has errors.  This will limit the application of the
above continuous variable QECC in practice. To avoid this
difficulty, one can also use Gaussian states to encode a discrete-level 
state for error correction\cite{qecc10}. We can construct an
encoded two-level quantum state based on the properties of the
following operators
\begin{eqnarray}
S_x=e^{2\sqrt{\pi}i\hat x};~ S_p=e^{-2\sqrt{\pi}i\hat p}
\end{eqnarray}
It can be easily examined that these two operators commute and there are two common eigenstates:
\begin{eqnarray}
|\tilde 0\rangle  \propto \sum_{-\infty}^{\infty} |x=2s\sqrt \pi\rangle
\propto \sum_{-\infty}^{\infty} |p=s\sqrt \pi\rangle;\nonumber\\
|\tilde 1\rangle  \propto \sum_{-\infty}^{\infty} |x=(2s+1)\sqrt \pi\rangle
\propto \sum_{-\infty}^{\infty} (-1)^s|p=s\sqrt \pi\rangle.
\end{eqnarray}
This code can protect an unknown state against errors that induce shifts in the value of $x$ and $p$.
Any unknown state of the type
\begin{equation}
|\tilde \psi \rangle = \alpha |\tilde 0\rangle +\beta |\tilde 1\rangle,
\end{equation}
is an eigenstate of operator $S_x$ and also an eigenstate of $S_p$. Therefore, it does not
destroy the state if we measure both values of $x$ modulo $\sqrt \pi$ and $p$ modulo $\sqrt \pi$.
After the measurements, we can correct small shift errors by a translation operation to displace $x$
and $p$ to their nearest integer multiples of $\sqrt \pi$. If the error
induced shifts of both
$x$ and $p$ are less than $\sqrt \pi/2$, the initial encoded state can be restored perfectly.

The encoded state here is a linear superposition of infinitely
squeezed states in $q$ and $p$. Now we consider the practical case
that we only have a finitely squeezed states. Suppose we encode
state $|0\rangle$ by
\begin{eqnarray}
|\tilde 0\rangle \approx \left(\frac{4}{\pi}\right)^{1/4}\int {\rm d}x |x\rangle  e^{-\frac{1}{2}(\Delta_p^2)x^2}
\times \sum_{-\infty}^{\infty}
e^{-\frac{1}{2}(x-2s\sqrt \pi)^2/\Delta_x^2}\nonumber\\
\approx \frac{1}{\pi^{1/4}}\int {\rm d}p |p\rangle
e^{-\frac{1}{2}(\Delta_x^2)p^2} \times \sum_{-\infty}^{\infty}
e^{-\frac{1}{2}(p-s\sqrt \pi)^2/\Delta_p^2}.
\end{eqnarray}
If $\Delta_x$ and $\Delta_p$ are small, then in principle these
shifts can be corrected with high probability. In the special case
that $\Delta_x=\Delta_p =\Delta$, it can be calculated that the
probability of uncorrected error caused by $\Delta$ is bounded
by\cite{qecc10}:
\begin{equation}
p_e \le \frac{2\Delta}{2}\exp (-\pi/4\Delta^2).
\end{equation}
Obviously, we can also concatenate a shift-resistant code with many Gaussian states.

\subsection{Gaussian cloning transformation}
It has been shown\cite{wooters} that the perfect quantum cloning
machine does not exist because it violates the linear superposition
principle in quantum mechanics. Say, given an arbitrary state
$|\psi\rangle$, map $C_0$ satisfying the following condition does
not exist:
\begin{equation}
C_0(|\psi\rangle) = |\psi\rangle|\psi\rangle.
\end{equation}
The idea for the proof is simple. Consider the 2-level system. If $C_0$ exists, it must satisfy
\begin{equation}
C_0(|0\rangle) = |00\rangle;~ C_0(|1\rangle) = |11\rangle.
\end{equation}
The linear superposition principle requires the same map to satisfy
\begin{equation}
C_0(|+\rangle) =\frac{1}{\sqrt 2} (|00\rangle +|11\rangle).
\end{equation}
However, if $C_0$ is a perfect cloning map, we request the conflicting result of $C_0(|+\rangle)=|++\rangle$.
This shows that $C_0$ cannot clone an arbitrary unknown 2-level state perfectly. If it clones
$|0\rangle$ and $|1\rangle$ perfectly, it cannot clone state $|+\rangle$ perfectly.

However, the quantum {\em approximate} cloning machines do exist\cite{clone2}.
We label the input unknown state by subscript
1 and the two ancilla by 2 and 3. The approximate cloning process can be mathematically described in:
\begin{equation}
C(|\psi\rangle_1 |0\rangle_2 |0\rangle_3) =|\psi'\rangle_{123}.
\end{equation}
We then discard qubit 3 and we want to have large values for the
following quantities (the cloning fidelity) :$ \langle \psi | \rho_1
|\psi\rangle $ and $\langle \psi | \rho_2 |\psi\rangle$ and
$\rho_1,\rho_2$ are states of qubit 1, 2 respectively, after tracing
out qubit 3 of the output state  $|\psi'\rangle_{123}$. This is the
so called $1\rightarrow 2$ cloning. The same concept can be extended
to $M\rightarrow N$ cloning. Various types of cloning machine have
been investigated for 2-level states\cite{clone2}, or even for
discrete $d-$level states\cite{clonenm}. The problem has then been
investigated with the continuous variable states \cite{clone3} by
Cerf, Ipe and Rottenberg (CIR). In particular, a type of Gaussian
cloning machine is constructed explicitly. CIR considered the
displacement invariant Gaussian cloning machine. Say, if two input
states are identical up to a displacement $ D (x',p') = e^{-ix'\hat
p}e^{ip'\hat x}$ as defined by Eq.(\ref{pmdis}), their respective
copies should be identical up to the same displacement.
Mathematically, the $1 \rightarrow 2$ cloning machine $C$ satisfies
\begin{equation}
C\left[ D(x',p')|\psi\rangle\langle\psi| D^\dagger (x',p')\right]
= D(x',p')^{\otimes 2}C\left[|\psi\rangle\langle\psi|\right]
D^\dagger (x',p')^{\otimes 2}.
\end{equation}
We now consider 3 modes. Mode 1 is for the unknown input state
denoted by $|\psi\rangle_1$. Mode 2 and 3 are ancilla. After the
cloning map, we discard mode 3 and mode 1,2 are the outcome of
$1\rightarrow 2$ cloning. Without loss of any generality, we set the
initial state of mode 2 and 3 to be
\begin{equation}
|\chi\rangle_{2,3}=\int\int_{-\infty}^{\infty}  f(x,p)|\Psi (x,-p)\rangle_{2,3}{\rm d}x{\rm d}p,
\end{equation}
where $f(x,p)$ is an arbitrary complex amplitude function and
\begin{equation}
|\Psi(x,p)\rangle =\frac{1}{\sqrt {2\pi}}\int_{-\infty}^{\infty} e^{ipy}|y\rangle|y+x\rangle {\rm d}y.
\end{equation}
The cloning transformation is
\begin{equation}
\hat U =e^{-i(\hat x_3-\hat x_2)\hat p_1} e^{-i\hat x_1(\hat p_2 +\hat p_3)}
\end{equation}
This is actually the analogy of qubit cloning
transformation\cite{clone2}. After applying $U$ to the state
$|\psi\rangle_1|\chi\rangle_{2,3}$, we obtain the following 3-mode
state:
\begin{equation}
\int\int_{-\infty}^{\infty}  f(x,p)\hat D(x,p) |\psi\rangle_1|\Psi (x,-p)\rangle_{2,3}{\rm d}x{\rm d}p,
\end{equation}
where mode 1 and mode 2 are taken as the two output states of the cloner. This class of cloning machines
are parameterized by
$f(x,p)$. If we choose $f(x,p)=e^{-(x^2+p^2)/2}/\sqrt \pi$, the cloner will provide two identical copies. For an arbitrary
input state $|\psi\rangle$, the noise of each output mode is $\sigma^2 =1/2$ and the state is
\begin{equation}
\rho_1=\rho_2 = \frac{1}{\pi}\int\int_{-\infty}^{\infty}e^{-x^2-p^2}\hat D(x,p)|\psi\rangle\langle \psi | \hat D^\dagger (x,p){\rm d}x{\rm d}p.
\end{equation}
In particular, if the input state is the coherent state $|\psi\rangle =|\alpha\rangle$,
the cloning fidelity is
\begin{equation}
\langle \alpha |\rho_1|\alpha\rangle=\langle \alpha |\rho_1|\alpha\rangle= \frac{1}{1+\sigma^2}= \frac{2}{3}.
\end{equation}
This cloning fidelity is independent of $\alpha$. It can be regarded
as the Gaussian-state analogy of the qubit cloner. Note that the
cloning noise here is larger than that of qubit cloner. The result
can also be extended to the $N\longrightarrow M$
cloning\cite{clone5}, in analogy with that of a
qubit-cloner\cite{clone6}. The cloning transformation here has been
proven to be the one that offers the largest fidelity in
principle\cite{clone3,clone6,clonegisin} because cloning
transformation can not help one to estimate an unknown state better
than the known result of optimized measurement. The result can be
extended to clone a squeezed state\cite{clone3}. Optical
implementations have been also proposed\cite{clone7}. One can also
do telecloning with quantum entanglement\cite{BL,paris}.
 An excellent review about quantum cloning has been presented very recently\cite{clonegisin}.

\subsection{Entanglement-based quantum tasks with weak Gaussian states: post-selection vs non-post-selection}
Most of the important results of entanglement-based quantum
information  for two-level systems have been experimentally
demonstrated, e.g.\cite{inns,ppuri}. The heart of the experimental
demonstration is the generation and manipulation of quantum
entanglement. Photon polarization is the most natural choice for the
entanglement space because it is easy to manipulate.  However, there
is no simple way to generate the perfect EPR state deterministically
in polarization space. In the existing experiments, probabilistic
EPR states are generated by spontaneous parametric down conversion
(SPDC). Actually, the entangled state generated by the SPDC process
is a type of mulit-mode weak Gaussian state in the form
\begin{equation}
|{\rm SPDC}\rangle = \exp [\beta(a^\dagger_Hb^\dagger_H
+a^\dagger_V b^\dagger_V)] |0\rangle,
\end{equation}
where the vacuum state $|0\rangle$ is the abbreviation of two-mode
vacuum, the subscripts of each creation operators indicate the
polarization and $\beta$ is a complex number with $|\beta|$ being
very small. Normally, we only need to consider the one-pair state
from the Taylor expansion given that $|\beta|$ is very small.
However, in the existing experiments, the photon detector normally
does not distinguish photon numbers exactly, one can only observe
whether it clicks. If the incident light beams contains one or more
than one photon, the detector may click. Given this fact, the higher
order terms in the weak Gaussian states take a non-trivial role and
in many experiments one may only choose the experimental results by
post selection. However, by carefully designing, in some cases one
can still obtain the non-post-selection results. In this subsection,
we shall first demonstrate how the higher order terms in the weak
Gaussian states from SPDC process can have non-trivial effects  and
what is the so called post-selection result by the example of
quantum teleportation. We then show how to convert the
post-selection result to non-post-selection result by two examples,
quantum teleportation and quantum entanglement concentration.
\subsubsection{Post-selection}
\begin{figure}
\epsffile{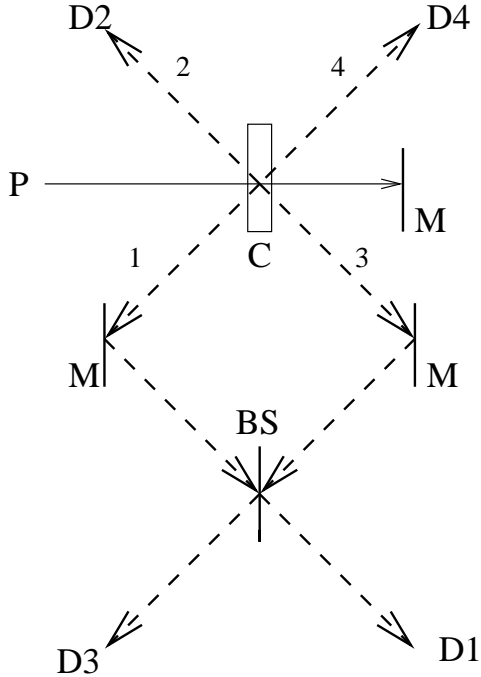}
\vskip 10pt
\caption{ The set-up of the first teleportation experiment in polarization space. C: nonlinear crystal, P: pump,
BS: 50:50 beamsplitter, M: mirror, Di: photon detector.
 }\label{inss0}
\end{figure}

Consider the set-up of the first teleportation quantum experiment in
figure \ref{inss0}. In the ideal case,  one  EPR pair is generated
at each side of the nonlinear crystal, i.e., beam 1, 2 is a pair and
beam 3, 4 is another pair. After the polarization of beam 2 is
measured, the polarization of beam 1 is known and therefore an
arbitrary polarization can be prepared on beam 1 by rotation. (The
rotation is not drawn in the figure.)
 Beam 1 and 3 are then sent to a beam-splitter for
an (incomplete) Bell measurement. As shown already, if there is
one and only one outcome photon on each side of the beam-splitter,
beam 1 and 3 have been collapsed to the anti-symmetric EPR state
and the original polarization state of beam 1 has been teleported
into the remote photon in beam 4, up to a fixed unitary rotation.
If detector D2 could distinguish the one-photon state and the
two-photon state, the set-up can be used as a teleporter without
post-selection.
  However,
 since the detectors there cannot distinguish
one photon and two photons, the result becomes more
complicated\cite{bkcomment}. What we can observe is whether the
detector clicks or not. It is also possible that initially beam 1, 2
contain two pairs while beam 3, 4 are vacuum. Suppose D2, D3 and D1
all click. There is considerable possibility that  beam 4 is
actually vacuum. That is to say, whenever we have observed the
three-fold coincident event of D1, D2 and D3, we are actually not
sure whether the
 teleportation has been done successfully, since D2 doesn't distinguish one photon or two photons. To ensure
a successful teleportation, we need to also observe beam 4. But once beam 4 is observed, the state is destroyed. We call
such type of results as post-selection-result because once we are sure of a successful result, the outcome state is destroyed already.
Obviously, if there is a sophisticated photon detector which distinguishes one-photon and two-photon
beams, the issue will be resolved\cite{kbcomment}. Or, we can use
strong entanglement in the position-momentum space as  shown earlier
in this section, because the strong two-mode squeezed state is
deterministically entangled in position-momentum space.  Actually,
this non-post-selection property is an important advantage of bright
Gaussian states for QIP tasks. Interestingly,  even with weak
Gaussian light, the entanglement-based experiments for two-level
systems can be done without post-selection in some cases.
\subsubsection{Non-post-selection teleportation}
 One can still use the same set-up, but  the pump light
after reflected back by the mirror in figure \ref{inss0} is
weakened\cite{kok2000,wgtele}. For example, we can weaken the
reflected pump light to 1/20 of its original intensity. In such a way,
once the event of 3-fold clicking of D1,D2, and D3 is observed, the
possibility of a successful teleportation is more than 20 times
larger than the posibility of nothing on beam 4\cite{wgtele}.
Mathematically, the initial Gaussian state is
\begin{equation}
|{\rm SPDC}\rangle = \exp \left[\frac{\beta}{20}(a_{1H}^\dagger
a_{2H}^\dagger+a_{1V}^\dagger a_{2V}^\dagger \right] \exp
\left[\beta (a_{3H}^\dagger a_{4H}^\dagger+a_{3V}^\dagger
a_{4V}^\dagger \right]|0\rangle.
\end{equation}
Doing Taylor expansion to the order of $\beta^2$, we have the
following approximate 4-photon state
\begin{eqnarray}
|{\rm SPDC}_4\rangle \approx \frac{\beta^2}{400} (a_{1H}^\dagger
a_{2H}^\dagger+a_{1V}^\dagger a_{2V}^\dagger )^2 |0\rangle
+\frac{\beta^2}{20}|\Phi^+\rangle_{12}\otimes |\Phi^+\rangle_{34}\nonumber\\
+\beta^2 (a_{3H}^\dagger a_{4H}^\dagger+a_{3V}^\dagger a_{4V}^\dagger )^2|0\rangle
\end{eqnarray}
and $|\phi^+\rangle_{ij}$ is a perfect EPR pair state for beam $i$
and $j$. Since we request detector D2 click, the state of the third
term is excluded. Once the event of 3-fold clicking of D1, D2 and D3
is observed, the initial state of $|SPDC\rangle$ could have been
collapsed to either the first term or the second term in the above
formula. But, the prior probability for the second term is 20 times
larger than that of the first term. Therefore, with very large
possibility the initial state contains one pair at each side of the
crystal. The experiment has been done sucessfully\cite{wgtele}. Here
the main technique used is to decrease the prior probability of
those unwanted terms. The similar technique can also be used for
non-post-selection entanglement swapping with SPDC state, in the
two-level space of vacuum and one-photon\cite{swap}. However,
weakening the probability of unwanted term is not the only way to
remove the post-selection condition. For different tasks we have
different ways.

\subsubsection{Non-post-selection entanglement concentration}
\begin{figure}
\begin{center}
\epsffile{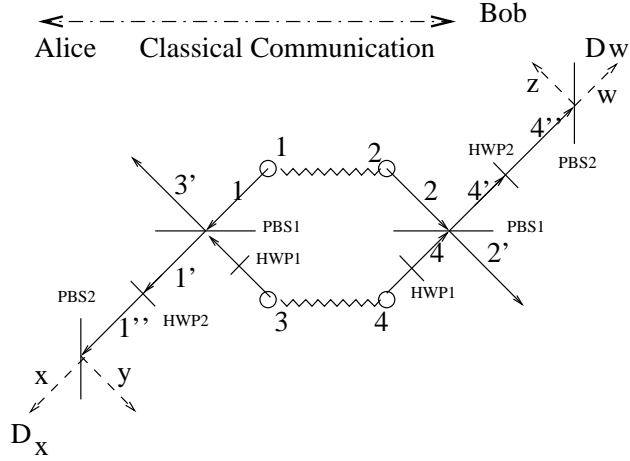}
\end{center}
\caption{Non-post-selection quantum entanglement concentration by
practically existing device of  linear optics. The two-fold
coincident event of both detector $D_x$ and detector $D_w$ being
clicked indicates that a maximally entangled state is produced on
beam 2' and 3'. PBS: polarizing beam-splitter.  Here HWP1 rotates
the polarization by $\pi/2$, HWP2 rotates the polarization by
$\pi/4$. }\label{distf1}
\end{figure}

The following raw state
\begin{equation}\label{key}
|r,\phi\rangle =\frac{1}{1+r^2}(|HH\rangle + r e^{i\phi}|VV\rangle)\otimes
(|HH\rangle + r e^{i\phi}|VV\rangle)
\end{equation}
can be probabilistically distilled into  the maximally entangled
state without post selection\cite{nonp}, even though the normal
yes-no single photon detectors are used. Note that in Eq.(\ref{key})
, both $r$ and $\phi$ are unknown parameters. The existing
experiments have demonstrated the post-selection
results\cite{yamamoto,ustc}. In particular, the special case $r=1$
is the one treated  in the recent experiment\cite{yamamoto}.
Consider the  schematic diagram, figure \ref{distf1}. (Earlier, a
similar scheme was given in\cite{panature,ins} for another
experiment.) The scheme requires the two fold coincidence event as
the indication that the maximally entangled state has been produced
on the outcome beams 2',3', i.e. whenever both detectors $D_x$ and
$D_w$ click, the two outcome beams, beam 2' and beam 3' must be in
the maximally entangled state:
\begin{equation}
|\Phi^+\rangle_{2',3'}=\frac{1}{\sqrt 2}(|H\rangle_{2'}|H\rangle_{3'}
|+V\rangle_{2'}|V\rangle_{3'}).
\end{equation}
Now we show the mathematical details for the claim above. The
polarizing beam-splitters transmit the horizontally polarized
photons
 and reflect
the vertically polarized photons. For clarity, we use the
Schr\"odinger picture. And we assume that the non-trivial time
evolutions to the light beams only take place while the light pass
through the optical device.

Consider Fig.\ref{distf1}.  Suppose initially two remote parties
Alice and Bob share two pairs of non-maximally entangled photons
as defined by Eq.(\ref{key}), denoted by photon pair 1,2 and
photon 3,4 respectively. The half wave plate HWP1 here is to
change the polarization between the horizontal and the vertical.
After photon 3 and 4 each pass through  HWP1, the state is evolved
to:
\begin{equation}
\frac{1}{1+r^2}(|HH\rangle_{12} + r e^{i\phi}|VV\rangle_{12})\otimes
(|VV\rangle_{34} + r e^{i\phi}|HH\rangle_{34}).
\end{equation}
Furthermore, after the  beams pass through the two horizontal polarizing
beamsplitters( denoted by PBS1), with perfect synchronization,
the state is
\begin{equation}
|\chi'\rangle = \frac{1}{1+r^2}(|H\rangle_{1'}|H\rangle_{2'}+ r e^{i\phi}
|V\rangle_{3'}|V\rangle_{4'})\otimes
(|V\rangle_1|V\rangle_2+ r e^{i\phi}
|H\rangle_3|H\rangle_4).
\end{equation}
This can be recast to the summation of three orthogonal terms:
\begin{equation}\label{chi}
|\chi'\rangle = \frac{1}{1+r^2}(|A\rangle + |B\rangle + |C\rangle),
\end{equation}
where
\begin{equation}
|A\rangle = r e^{i\phi}( |H\rangle_{1'}|H\rangle_{2'}|H\rangle_{3'}|H\rangle_{4'}
+|V\rangle_{1'}|V\rangle_{2'}|V\rangle_{3'}|V\rangle_{4'} );
\end{equation}
\begin{eqnarray}
|B\rangle = |H\rangle_{1'}|V\rangle_{1'}|H\rangle_{2'}|V\rangle_{2'}; \\
|C\rangle= r^2 e^{2i\phi} |H\rangle_{3'}|V\rangle_{3'}|H\rangle_{4'}|V\rangle_{4'}.
\end{eqnarray}
Obviously, term $|B\rangle$ means that there is no photon in beam
4' therefore this term never cause the requested two-fold
coincident event and hence it is ruled out. Similarly, term
$|C\rangle$ can also de disregarded  and only term $|A\rangle$
should be considered.

After the beams pass through HWP2,  state
$|A\rangle$ is evolved to
\begin{equation}
 |\Phi^+\rangle_{1''4''} |\Phi^+\rangle_{2'3'}+
 |\Psi^+\rangle_{1''4''} |\Phi^-\rangle_{2'3'}
\end{equation}
where $|\Phi^{\pm}\rangle_{ij}
=\frac{1}{\sqrt 2}( |H\rangle_i|H\rangle_j\pm |V\rangle_i|V\rangle_j$ and
 $|\Psi^{\pm}\rangle_{ij}
=\frac{1}{\sqrt 2}( |H\rangle_i|V\rangle_j\pm |V\rangle_i|H\rangle_j$.
After pass through the two PBS2 in the figure,  $|A\rangle$
is evolved to  state
\begin{equation}\label{xyzw}
|x\rangle|w\rangle |\Phi^+\rangle_{2'3'} + |y\rangle|z\rangle
|\Phi^+\rangle_{2'3'} +
(|x\rangle|z\rangle +|y\rangle|w\rangle)|\Phi^-\rangle_{2'3'}
\end{equation}
where $|s\rangle$ denote the state of one photon in beam $s$, $s$
can be $x,y,z$ or $w$.  From the above formula we can see that
only the first term will cause the requested two-fold coincident
event. This term indicates a maximally event-ready entangle state
between beam 2' and 3', i.e. $\frac{1}{\sqrt
2}(|H\rangle_{2'}|H\rangle_{3'}+ |V\rangle_{2'}|V\rangle_{3'})$.
Note that the result here is independent of the parameters
$r,\phi$.

To really produce the event-ready entanglement through the
concentration scheme here one need the deterministic supply of the
requested raw states. This  is rather challenging a task. However,
as it has been shown\cite{nonp}, even without such a deterministic
supply, one can still experimentally demonstrate that the scheme
$can$ produce the event-ready maximally entangled pair $if$ the
requested deterministic supply of raw state is offered.

{\em Remark:} The non-post-selection scheme for QIP is not limitted to teleportation and entanglement concentration,
e.g., there are non-post-selection schemes for the tsak of state Gaussification\cite{brownenp,eisertnp}.

\section{Quantum cryptography with weak coherent light}\label{sectomita}


\subsection{Introduction}
Security has been an important issue for the communication over the
networks these years. The internet traffic has been growing rapidly
to include important information, such as a huge amount of orders
for exchange markets or the decision of a government. Modern
economies and societies rely on the availability, confidentiality
and integrity of critical fiber optic network infrastructures to
function properly and efficiently. Recent deregulation improves the
accessibility to the networks and encourages new services. It also
results in a side effect that increases vulnerability against the
illegal tapping to steal confidential and commercially sensitive
data. Encryption, usually public-key and secret-key cryptosystems,
is one of the best means to keep such confidential from the
malicious eavesdroppers. However, those cryptosystem only provides
so called computational security, relies on the fact decryption
requires impractically long time for the current computers. The
encryption systems may be broken with the enormous calculation or
the sudden appearance of the polynomial algorithm, even before a
quantum computer runs Shor's algorithm to solves the RSA
cryptosystem. Also, some of the widely believed secure classical
cryptographic system have now proven to be indecure\cite{xywang}

It has been proven that the secret key cryptographic systems are
secure as long as the key is not used more than once (one-time-pad
or Vernum cipher.) This method needs the unrealistically vast pool
of the encryption key, thus in practice, one finite key is
repeating, causing the debilitation of the concealment. Quantum key
distribution (QKD) provides a protocol where  the sender (Alice) and
the legitimate receiver (Bob) can share the same bit sequence
successively without leaking any secret key information guaranteed
by the law of quantum mechanics. This unconditional security, which
cannot be achieved by conventional cryptography, clearly shows the
advantage of the quantum information technology. Unlike other
quantum information protocols proposed so far, QKD can be
implemented with current photonic technology. Therefore, number of
papers have reported the QKD transmission experiments.

In the following, we will consider a QKD protocol, so called BB84 (Bennett and Brassard 1984)\cite{bene} in particular. BB84 protocol has been proved to be unconditionally secure, if implemented with a perfect single photon source. It is, however, still difficult to provide a single photon source with a practically meaningful specification. Almost all the experiments towards practical implementation employ weak coherent light from semiconductor lasers instead. We will first briefly review the theoretical backgrounds of the unconditional security in BB84 protocol. Then we will discuss practical implementations of the protocol. Possible attacks to the QKD system using a weak coherent light will be examined. We will focus on the basic issues and the recent experimental progress. Readers who are interested in the detail of QKD should refer the review by Gisin et al. \cite{GRTZ02}, where they would find good introduction of the QKD protocol and state of art at 2001.

\subsection{QKD in practice}
\subsubsection{Requirement for single photon detectors}
In QKD systems with a single photon or a weak coherent light, we need to send and receive signals encoded in single photon states. One of the most important device in QKD transmission is single photon detector (SPD,) which limits both the transmission distance and the transmission rate. The SPDs should show high detection efficiency, low dark counts, and short response time. The ratio of the detection efficiency $\eta$ to the dark count probability $P_{d}$ determines the error rate $e_{B}$, as
\begin{eqnarray}
e_{B}  &  =\frac{1}{2}\frac{S\left(  1-v\right)  \eta+P_{d}}{P_{DET}%
}\nonumber\\
&  =\frac{1}{2}\frac{S\left(  1-v\right)  +P_{d}/\eta}{S\left(  1-P_{d}%
\right)  +P_{d}/\eta}, \label{QBER}%
\end{eqnarray}
where $v$ is the visibility of the interferometer, and $P_{DET}$ represents the detection probability per pulse. $P_{DET}$ is related to the probability $S$ that at least one photon arrives at the detector by
\begin{equation}
P_{DET}=S\eta+P_{d}-S\eta P_{d}. \label{PDET}%
\end{equation}
The probability $S$ is a function of the loss in the transmission line and the receiver. The photon loss in a $L$ km-long-fiber is given by $\alpha L$ [dB].When we assume the receiver loss is $\beta$ [dB], $S$ is given by
\begin{equation}
S=10^{-\left(  \alpha L+\beta\right)  /10} \label{Ss}%
\end{equation}
for a single photon source, and%
\begin{equation}
S=1-\exp\left[  -\mu10^{-\left(  \alpha L+\beta\right)  /10}\right]
\label{Sc}%
\end{equation}
for a coherent photon source with the average photon number of $\mu$. The error rate (\ref{QBER}) is given by the half of the inverse of signal-to-noise ratio $S/N$ (noise will give the error with the probability of 1/2.) Eq.(\ref{QBER}) shows that the ratio $P_{d}/\eta$ is a figure of merit of a SPD that determines the error probability, because $P_{d}$ and $1-v$ are small. Secure QKD can be achieved only when the error rate is kept lower than a threshold for security. The threshold varies according to the assumptions on the method of Eve's attack and error correction. Typical threshold value is around 11 $\%$\cite{Lutkenhaus00,Hamada03}. Then the ratio $P_{d}/\eta$ should be smaller than 10$^{-3}$ for 100 km fiber transmission in 1550 nm, where the commercial fibers take the lowest loss value (0.2 dB/km),  even with an ideal single photon source.

Response time of SPDs may limit the clock frequency of the system.  SPDs cannot detect photons during the recovery time (typically several hundreds nanoseconds) after one detection event. Another effect (sometimes more serious) is the afterpulse, false photon detections caused by residual electrons created by the previous detections. We can not send a photon pulse during the period of large afterpulse probability. The afterpulse effect remains typically 1 $\mu$s after the photon detection. This period may vary on the devices and the operating conditions. The afterpulse effect on error probability can be formulated as follows. We assume two detectors 1 and 2 to discriminate bit values 0 and 1, respectively. The probabilities $p_{1}(t_{n})$ that detector 1 fires and $p_{2}(t_{n})$ that detector 2 fires are given by the bit value $b(t_{n})=\{0,1\}$ at the $n$-th clock $t_{n}$ as
\begin{eqnarray}
p_{1}(t_{n}) &  =S\eta q(t_{n})+P_{d}+\sum_{i=-\infty}^{n-1}f(t_{n}%
-t_{i})p_{1}(t_{i})\\
p_{2}(t_{n}) &  =S\eta(1-q(t_{n}))+P_{d}+\sum_{i=-\infty}^{n-1}f(t_{n}%
-t_{i})p_{2}(t_{i})\label{aft1}%
\end{eqnarray}
where the function
\begin{equation}
q(t_{n})=v\left(  1-b(t_{n})\right)  +(1-v)b(t_{n})\label{aft2}%
\end{equation}
defines the fraction that a photon enters the detector 1, and the memory
function $f(t_{n}-t_{i})$ represents the afterpulse effect. A reasonable form
of $f$ might be $f(t)=A\exp[-\gamma t]$, but here we assume
\begin{equation}
f(t)=\left\{
\begin{array}
[c]{l}%
A\qquad(0\leq t\leq t_{M})\\
0\qquad(t<0,t>t_{M})
\end{array}
\right.  ,\label{memory}%
\end{equation}
for simplicity. The afterpulse probability $A$ remains constant during $M$
periods of the clock in this model. Then Eq. (\ref{aft1}) can be solved for
the asymptotic values. The error probability is given by%
\begin{equation}
\hat{e}_{B}=\frac{1-v+P_{d}/S\eta}{1+2P_{d}/S\eta}+\frac{v-1/2}{1+2P_{d}%
/S\eta}AM\approx e_{B}+\frac{1}{2}AM,\label{afterror}%
\end{equation}
if we neglect the events that both detectors fire simultaneously. Eq.(\ref{afterror}) shows that the afterpulse effect increase the error probability by $AM/2$. For example, typical values $A=10^{-3}$ and $M=100$ increase the error probability by 5 $\%$. The afterpulse effect can be reduced by neglecting the photon detection during the time $0\leq t\leq t_{M}$.

Devices for single photon detection must have a multiplication process inside the device: single photon detectors should yield many carriers from one photon, otherwise the amplifier noise will hide the signal. Avalanche photodiodes (APD) are widely used for the SPD, because of the high detection efficiency. A bias voltage higher than the break down is applied to the APDs to obtain large avalanche gain (order of $10^{6}$). The bias voltage should be decreased below the break down voltage after photon detection to quench the break down. This operation mode is called Geiger mode. Several methods are given to quench the APDs. The simplest one is called passive quench to put a high resistance in series to the APD. The break down current induce voltage drop in the resistance. More sophisticated method is active quench to use a circuit to detect the break down and decrease the bias. Si-APDs provides good performance for visible photon detection. Commercially available SPD modules detect single photons with the efficiency of $\eta \approx 50 \%$,
 while the dark count rate is kept at 100 counts/sec. However, the single photon detection in 1550 nm, which is suitable for fiber transmission, is still a big issue for the experiment.  The QKD experiments in 1550 nm have employed the SPDs using InGaAs/InP APDs as APDs with InGaAs absorption layer have sensitivity at this wavelength. However, the InGaAs/InP APDs in Geiger mode are suffered by large dark count probability and afterpulse, which cause errors in the qubit discrimination. The dark count probability and the afterpulse can be reduced by using gated-mode, where gate pulses combined with DC bias are applied to the APD. The reverse bias exceeds the break down voltage only in the short pulse duration.
Though the gated-mode works well, the short pulses produce strong spikes on the transient signals. High threshold in the discriminator is therefore necessary to avoid errors, at the cost of detection efficiency. High gate pulse voltage is also required to obtain large signal amplitude by increasing avalanche gain. Impedance matching helps to reduce the spikes to some extent\cite{YT01}. Bethune and Risk have introduced a coaxial cable reflection line to cancel the spikes\cite{BR00}. A unique observation has been made by Yoshizawa, et al. that photon detection changes the shape of the charge out spike. This change was shown to be useful for high speed photon detection.  Tomita and Nakamura\cite{TN02} reported a much simpler method: canceling the spikes by taking the balanced output of the two APDs required for the qubit discrimination. We describe the detail of the balanced APD photon detector in the following.

\begin{figure}[pts]
\begin{center}
\includegraphics[width=8cm]{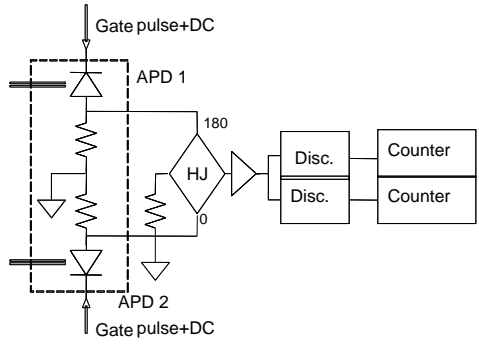}
\end{center}
\caption{Schematic of the photon detector. HJ and DISC stand for a hybrid junction,
 and discriminators.}
\label{SPDscheme}
\end{figure}

\begin{figure}[pts]
\begin{center}
\includegraphics[width=8cm]{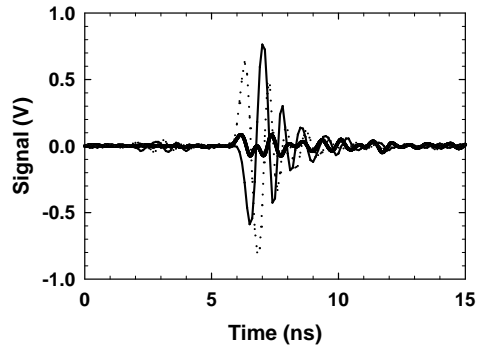}
\end{center}
\caption{Cancellation of the transient spike. Thin solid: APD 1, Dots:
APD 2, Thick solid: the differential output of the APD 1 and the APD 2.}
\label{spike}
\end{figure}

Figure \ref{SPDscheme}  depicts the schematic of the SPD. Two APDs (Epitaxx EPM239BA) and load resisters were cooled to between $-133^{\circ}$C and $-60^{\circ}$C by an electric refrigerator. Short gate pulses of 2.5 V p-p and 750 ps duration were applied to the APDs after being combined with DC bias by Bias-Tees. The output signals from the APDs were subtracted by a $180^{\circ}$ hybrid junction of 2-2000 MHz bandwidth. The differential signal was amplified and discriminated by two discriminators. Since the spikes were the common mode input for the $180^{\circ}$ hybrid junction, they would not appear at the output. The APD 1 provided negative signal pulses at the output, while the APD 2 provided positive pulses. One can determine which APD detects a photon from the sign of the output signals. Figure \ref{spike} shows the output signal of the amplifier without photon input. Almost identical I-V characteristics of the APDs enabled us to obtain a good suppression of the spikes. The lowest dark count probability of 7$\times10^{-7}$ per pulse with detection efficiency of 11 $\%$ at -96 $^{\circ}$C has been observed.
The ratio $P_{d}/\eta$ was as small as 6$\times10^{-6}$, which corresponds to 270 km QKD transmission with an ideal photon source. The detection efficiency and the dark count probability are increasing functions of the bias. The maximum
value of the detection efficiency is obtained when the DC bias is set to the
break down voltage. Larger values of the maximum detection
efficiency were observed at higher temperatures: the detection efficiency of 20 $\%$ at -60 $^{\circ}$C
with the dark count probability of 3$\times10^{-5}$ per pulse. Afterpulse probability was measured by applying two successive gate pulses to the APDs. Afterpulse is prominent at low temperatures. Afterpulse probability remained about 10$^{-4}$ for the 1 $\mu$s pulse interval at the temperatures higher than -96 $^{\circ}$C. This corresponds to 10$^{-5}$ error probability (per pulse) for 10 $\%$ detection efficiency.
Based on the observation on the dark count probability and the afterpulse probability, we conclude that the optimal operation temperature for the present APDs is around -96 $^{\circ}$C. The obtained afterpulse effect was shorter than the previous reports. This is probably due to the decrease of the gate pulse voltage. This is another advantage of the present SPDs. Recently, the dark count probability has been reduced to 2$\times $10$^{-7}$ per pulse at the detection efficiency of 10 $\%$\cite{KTNKN03}.
The $S/N$, or the ratio $P_{d}/\eta$ is improved about 50 times (17 dB) as much as the values previously reported.

\subsubsection{Implementation of BB84 protocol}
We first need to determine how to implement qubits to realize a protocol. Polarization is one of the straightforward way, because of the direct correspondence between the Bloch sphere for a spin-1/2 state and the Poincare sphere for the photon polarization state.  The four states used in BB84 protocol are given by
\begin{eqnarray}
 \left|0 \right\rangle &=& \left|H \right\rangle \nonumber \\
 \left|1 \right\rangle &=& \left|V \right\rangle \nonumber \\
 \left| \bar{0} \right\rangle &=& |F \rangle = \frac{1}{\sqrt{2}} \left( \left|H \right\rangle + \left|V \right\rangle \right) \nonumber \\
 \left| \bar{1} \right\rangle &=& |S \rangle = \frac{1}{\sqrt{2}} \left( \left|H \right\rangle - \left|V \right\rangle  \right),
\label{fourstates}
\end{eqnarray}
where $\left| H \right\rangle$,  $\left| V \right\rangle $, $\left|F \right\rangle $, and $\left| S \right\rangle $ denote the horizontal, vertical, $45^{\circ}$, and $135^{\circ}$ polarized states, respectively. We can choose the polarization states arbitrarily as long as $\left\langle 0 | 1 \right\rangle = \left\langle \bar{0} | \bar{1} \right\rangle =0 $ and $\left| \left\langle 0 | \bar{0} \right\rangle \right|^{2} =\left| \left\langle 0 | \bar{1} \right\rangle \right|^{2} = \left| \left\langle 1 | \bar{0} \right\rangle \right|^{2} = \left| \left\langle 1 | \bar{1} \right\rangle \right|^{2} = 1/2$. For example,  the circular polarization states $\left| L  \right\rangle$ and $\left| R \right\rangle $ will work; the BB84 protocol can be implemented with two of the three
 basis sets $\{ \left|H \right\rangle, \left| V \right\rangle \}$, $\{\left|F \right\rangle, \left|S \right\rangle\}$, and $\{ \left| L \right\rangle, \left| R \right\rangle \}$. As stated before, practical systems employ weak coherent light in place of a single photon. Polarization states of the coherent state light are given by
\begin{eqnarray}
\left|H \right\rangle _{coh} &=& \left| \alpha \right\rangle_{H} \nonumber \\
\left|V \right\rangle _{coh} &=& \left| \alpha \right\rangle_{V} \nonumber \\
\left|F \right\rangle _{coh} &=& \left| \alpha / \sqrt{2} \right\rangle_{H} \left| \alpha / \sqrt{2} \right\rangle_{V} \nonumber \\
\left|S \right\rangle _{coh} &=& \left| \alpha / \sqrt{2} \right\rangle_{H} \left| -\alpha / \sqrt{2} \right\rangle_{V},
\label{coherent polarization}
\end{eqnarray}
where the horizontally (vertically) polarized coherent state $\left| \alpha \right\rangle_{H(V)}$ is expressed by the number states as
\begin{equation}
\left|\alpha  \right\rangle _{H(V)} = e^{-\frac{\alpha^{2} }{2}} \sum_{n=0}^{\infty} \frac{\alpha^{n} }{n!} \left| n \right\rangle _{H(V)}
\end{equation}
The $45^{\circ}$ polarized coherent state is reduced to a superposition of single photon states in the weak light limit $(\alpha \ll 1)$:
\begin{equation}
\left|F \right\rangle _{coh} \approx \left|vac \right\rangle + \frac{\alpha }{\sqrt{2}} \left( \left|H \right\rangle + \left|V \right\rangle \right) + O(\alpha^{2})
\end{equation}
Similar expression holds for the $135^{\circ}$ polarized coherent state. Therefore, the weak coherent states (\ref{coherent polarization}) can approximate the four single photon states (\ref{fourstates}), when we neglect the vacuum states by postselection. This observation rationalizes the use of weak coherent states in the QKD systems.

The above polarization encoding is often employed in QKD systems for free-space transmission. However, the polarization encoding is not suitable for fiber transmission, because the polarization states fluctuate in fibers by the effects of environment:  birefringence induced by stress, and  polarization rotation by the geometric phase in twisted fibers. Nevertheless, it should be noted that the polarization coding may be still usable if we compensate the polarization changes after the transmission. This is due to the fact that the polarization fluctuation in the installed changes slow (in minutes or hours), and intermittent control with strong reference light will keep the polarization states.

\begin{figure}[pts]
\begin{center}
\includegraphics[width=8cm]{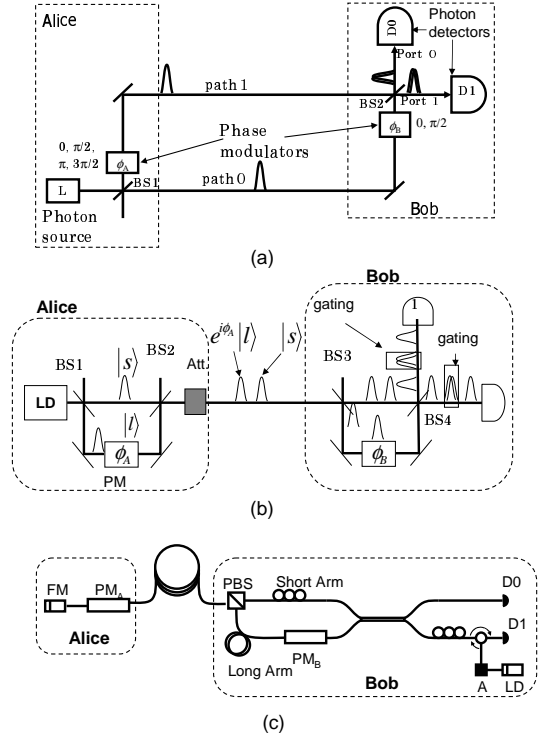}
\end{center}
\caption{A Mach-Zehnder interferometer set-up for implementing BB84 protocol with phase coding. (a): Basic scheme (b): Scheme consists of two asymmetric Mach-Zehnder interferometers and a single transmission channel (c) Plug and play system using a Faraday mirror for self-compensation}
\label{MZI}
\end{figure}

The alternate coding is phase coding. Suppose a Mach-Zehnder interferometer (MZI) as shown in Fig. \ref{MZI}(a). The amplitude of the photon is divided equally into the two arms of the interferometer (path P and path Q). Alice can prepare the four states for BB84 protocol by modulating the phase of the amplitude in one arm by one of the four values: $\phi_{A} = \left\{ 0^{\circ}, 180^{\circ} \right\}$, and $\left\{ 90^{\circ}, 270^{\circ} \right\}$. The modulation results in
 \begin{eqnarray}
 \left|0 \right\rangle &=& \frac{1}{\sqrt{2}} \left(\left|P \right\rangle + \left|Q \right\rangle \right) \nonumber \\
 \left|1 \right\rangle &=& \frac{1}{\sqrt{2}} \left(\left|P \right\rangle - \left|Q \right\rangle \right) \nonumber \\
 \left| \bar{0} \right\rangle &=& \frac{1}{\sqrt{2}} \left( \left|P \right\rangle + i \left|Q \right\rangle \right) \nonumber \\
 \left| \bar{1} \right\rangle &=&  \frac{1}{\sqrt{2}} \left( \left|P \right\rangle - i \left|Q \right\rangle  \right),
\label{fourphasestates}
\end{eqnarray}
where the wavefunction $\left|P \right\rangle$  ($\left|Q \right\rangle$) represents the state that a single photon exists in the path P (Q). The states given in Eq. (\ref{fourphasestates}) are equivalent to the states  $\{\left|F \right\rangle, \left|S \right\rangle, \left| L \right\rangle, \left| R \right\rangle \}$. Phase modulation in Bob's station refers to the basis selection. Bob obtains the correct bit values, when the difference between the phase modulations satisfies $\phi_{A}-\phi_{B} = 0^{\circ}$, or $180^{\circ}$.

MZI shown in Fig. \ref{MZI}(a) is however impractical, because it requires two fibers. Besides the economical reasons, the interferometer is sensitive to disturbance, because the photons in the different paths experience the independent fluctuations. A time division interferometer, which requires only one common fiber for transmission as shown in Fig.\ref{MZI}(b), will be more stable. Alice's asymmetric Mach-Zehnder interferometer (AMZ) divide the amplitude into double pulse, where one component passed through the shorter arm, and the other through the longer arm. Modulation of the relative phase between the pulse component results in the four states (\ref{fourphasestates}). After passing Bob's AMZ, the pulse contains three components. The first component originates from the amplitude that passed through the shorter arms in both interferometers. The middle component results from the two contributions: the shorter arm in Alice and the longer arm in Bob, and vice versa. The last component originates from the longer arms in both. Only the middle component shows interference, and contributes to the key transmission. Photon detection probabilities are 1/4, 1/2, and 1/4 for the first, middle, and last components. Therefore, the time division system loses a half of the transmitted photons. This time division phase coding was first implemented by Townsend et al \cite{TRT93} for QKD systems. They observed a clear interferometric fringe after the transmission in the fiber of 10 km-30 km long. The time division QKD system requires precise control of the path length. The two interferometers of Alice and Bob should be identical to obtain a high visibility. The path difference should be kept within 3 \%  of the wavelength to maintain the visibility higher than 0.98.

A simple method has been proposed to achieve stable interference without complicated active stabilization. This method, called 'Plug and Play' (P\&P)\cite{ZGGHMT97}, utilizes folded interferometer as shown in Fig. \ref{MZI}(c). Original proposal used three Faraday mirrors (FM)\cite{MHHTZG97}, but the simplified scheme works with one FM\cite{RG3Z98}. This scheme  combines the time division and polarization division interference. The output of the laser is divided into two arms by  beam splitter (BS). The one arm is set to be longer than the other. The outputs of the two arms are combined by polarization beam splitter (PBS) after the polarization rotation in  one arm by $90^{\circ}$. The transmitted light is reflected by the FM and phase modulation is applied to one of the time divided components. The light is sent back to Bob after attenuation to the single photon level. PBS divides the light into the two arms. Since the polarization is rotated $90^{\circ}$ by FM, the light component passed through the shorter arm in the outward transmission goes to the longer arm in the homeward transmission, and vice versa. The use of common AMZ guarantees the condition for the interference automatically. Moreover, the P\&P scheme is robust to the disturbance during the fiber transmission. To see this, we examine the role of the FM in P\&P systems. Since the reflected light propagates in the opposite direction, we need to be careful about the coordinate. Here we fix the direction of axes in the laboratory. The effect of FM in linear polarization basis reads $\sigma_{x}$ rotation. The effect of transmission line (fiber) on the polarization can be expressed by a unitary transform:
\begin{equation}
U=e^{i\alpha}R_{z}\left(  2\beta\right)  R_{y}(2\gamma)R_{z}\left(
2\delta\right)  \label{unitary}%
\end{equation}
where $R_{y}$ and $R_{z}$ stand for the rotation on $y$ axis%
\begin{equation}
R_{y}(2\gamma)=%
\left(\begin{array}{cc}
\cos\gamma & -\sin\gamma\\
\sin\gamma & \cos\gamma
\end{array}\right)
, \label{yrotation}%
\end{equation}
and the rotation on $z$ axis
\begin{equation}
R_{z}\left(  2\delta\right)  =%
\left(\begin{array}{cc}
e^{-i\delta} & 0\\
0 & e^{i\delta}%
\end{array}
\right)
, \label{zrotation}%
\end{equation}
respectively. The above unitary transform (\ref{unitary}) is general, as long as we can neglect depolarizing in the fiber. We can see that the total effect (not include global phase) of going-around the transmission line is just the transformation by the FM
\begin{equation}
R_{z}\left(  2\delta\right)  R_{y}(2\gamma)R_{z}\left(  2\beta\right)
\sigma_{x}R_{z}\left(  2\beta\right)  R_{y}(2\gamma)R_{z}\left(
2\delta\right)  = \sigma_{x}. \label{FM}%
\end{equation}
Eq. (\ref{FM}) implies that the outward and homeward polarizations are orthogonal, regardless of the disturbance at the transmission line and the initial polarization. This condition is essential for stable interference. Combination of an ordinary mirror and a quarter wave plate will rotate the polarization. However, this combination does not work for auto-compensation, because its effect is described by $R_{y} (\pi) \neq \sigma_{x}$. Even with FMs, the autocompensation becomes no longer perfect, if the transformation by the FM deviates from $\sigma_{x}$. Since the refractive index in materials depends on temperature, the rotation in FMs is no longer equal to $90^{\circ}$ as sifting temperature. It has been shown the combination of a phase modulation and a polarization rotation in a loop mirror yields $\sigma_{x}$ \cite{T5N04}. Temperature dependence can be easily compensated by changing apply voltage to the phase modulator, so that stable operation in a wide temperature range (from $-5^{\circ}$C to $75^{\circ}$C) has been obtained. It should be noted that we assume the disturbance in the homeward transmission is identical to the outward transmission.  We see this assumption reasonable, considering the fact that the round trip time is shorter (about 1 $\mu$s/km) than the time constant of the disturbance in the fiber (more than 1 s).

\subsubsection{Plug \& Play System}
Most of the successful QKD transmission experiments have been based on so called Plug-and-Play (P\&P) system, which contains autocompensation mechanism to achieve good interference performance with ease\cite{SGGRZ02}. In P\&P systems, the light source and the photon detectors are in the same side of the line; the pulses travel first from Bob to Alice, and then back to Bob. Back-scattered light will degrade the signal-to-noise ratio. The back-scattering is unavoidable due to non-uniformity in a fiber. The connections in practical fiber network will cause considerable reflection. Therefore, burst mode operation is necessary in practical P\&P systems, where a storage line is installed in Alice's station, and the length of the pulse train is set to be smaller than the storage line. The backward light will never cross the bright light in the transmission line.

Currently, two P\&P systems have appeared on market\cite{IDQ,MAGIQ}. QKD experiments on installed fibers have been also reported. One is the transmission over 67 km done by the group of Geneva University\cite{SGGRZ02}. They reported a successful key exchange at raw rate of 160 Hz with 5 \% QBER through a 67.1 km fiber from Geneva to Lausanne installed under the lake, where the measured fiber loss was 14.4 dB. Average photon density was set 0.2 photons/pulse in this experiment. Recently, Hasegawa et al\cite{HNIASM05}, have reported key exchange over 64 km and 96 km fiber transmission test-bed (JGNII). The fiber loss in 64 km single mode fiber was 13.29 dB. They could exchange key at 21.5 bps with 4.7 \% QBER for 64 km transmission and 8.2 bps with 9.9 \% QBER for 96 km transmission, respectively. The clock frequency of the system was 1 MHz, and average photon number was 0.1 photons/pulse. The fiber cable in the test bed contained ten cores, so that they used one core for quantum communication and another core for clock distribution.

So far, the transmission experiments were done in fairly stable environments, even in the installed fibers, the fibers were buried under the lake or well-maintained as a test-bed. In commercial fiber networks, which contain many connections and reflecting points, the loss and the back-scattering differ fiber to fiber. Access links for end-users sometimes use the fibers installed in the open-air. The fibers may experience mechanical vibration and temperature fluctuation. The system for quantum communications should be designed to be stable against the fluctuation of the environment. The QKD systems should have a clock synchronization system, which can trace the shift of fiber length due to thermal expansion and keep the optimum timing. A watch-dog system is also necessary to monitor key generation rate and error rate, and the transmission system should automatically re-set and calibrate itself on system errors. A QKD system fully equipped with the above functions has been developed. A 'hands-free' transmission was achieved through a 16 km commercial fiber link installed on electric poles with the system for fourteen days with the final key rate of 13 kbps\cite{NEC}.

\subsubsection{one-way transmission}
Although the  P\&P system works well for QKD systems using a weak pulse up to 100 km\cite{KTNKN03}, extending the transmission distance will be difficult even if a lower noise SPD is developed. This is because backscattering noise in the fiber dominates the detector noise, which is intrinsic to the bidirectional autocompensating system. Although the use of storage line and burst photon trains would reduce the backscattering, this would also reduce the effective transmission rate by one-third.

Unidirectional systems are free from the above problems. Such systems have also an advantage that they are compatible with photon sources of true single photons or quantum correlated photon pairs, which are believed to provide higher key rate after a long distance transmission. The difficulty in the unidirectional system has been the stabilization of two remote interferometers to achieve high visibility. The system should solve the conflict between stability and transmission distance. Recently, three systems have been reported that overcome such difficulty. One system employs an active control using fiber stretcher. The system performed a 122 km fiber transmission with QBER of 8.9 \%\cite{GYS04} and a continuous quantum key distribution session of 19 hours over a 20.3km installed telecom fiber\cite{YS05}. In the latter experiment strong reference pulses were sent between the quantum signals for the active control.
Another system installs an automatic compensation on polarization fluctuation by Faraday mirrors in the Michelson interferometers at the both sites. A 125 km QKD transmission from Beijin to Tianjin has been reported\cite{MZHGG04}. The third solution  is to use an integrated-optic interferometers based on planar lightwave circuit (PLC) technology\cite{NHN04}. The PLC technology provides a stable interference with a simple setting. The longest distance transmission over a 150 km fiber has been achieved\cite{KNHTKN04}. The detail of the PLC system will be described in the following.

The interferometers, asymmetric Mach-Zehnder Interferometers (AMZs)
 with a 5-ns delay in one of the arms were fabricated on a
silica-based PLC platform. Since the AMZs were fabricated using the same mask,
they had the same path length difference between the two arms.
The optical loss was 2 dB (excluding the 3-dB intrinsic loss at the coupler).
Polarization-dependent loss was negligible (0.32 dB).
One of the couplers was made asymmetric to
compensate for the difference in the optical loss between the two arms, so the
device was effectively symmetric. A Peltier cooler attached to the back of the
substrate enabled control of the device temperature with up to 0.01 $^{\circ
}$C precision. Polarization-maintaining fiber (PMF) pigtails aligned to the
waveguide optic-axis were connected to the input and output of the AMZ.

\begin{figure}[pts]
\begin{center}
\includegraphics[width=8cm]{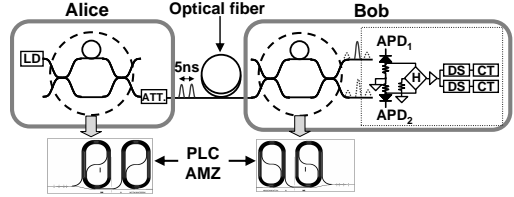}
\end{center}
\caption{Schematics of the integrated-optic interferometer system for quantum key distribution.
LD: laser diode, ATT: attenuator, APD: avalanche photodiode,
 DS: discriminator, CT: counter, H: $180^{\circ}$ hybrid junction.}
\label{PLC}
\end{figure}

Two AMZs were connected in series by optical fiber to produce a QKD
interferometer system (Fig. \ref{PLC}). Optical pulses that were 200 ps long and
linearly polarized along one of the two optic-axes were introduced into the
PMF pigtail of Alice's AMZ from a DFB laser at1550 nm. The input pulse was
divided into two coherent output pulses polarized along the optic-axis of the
output PMF, one passing through the short arm and the other through the long
arm. The two optical pulses were attenuated to the average photon number of
0.2. The two weak pulses propagated along the optical fiber and experience the
same polarization transformation. This is because the polarization in fibers
fluctuates much slower than the temporal separation between the two pulses.
After traveling through Bob's AMZ, these pulses created three pulses in each
of the two output ports. Among these three pulses, the middle presents the
relative phase between the two pulses. The interfering signal at the middle
pulses was discriminated by adjusting the applied gate pulse timing. The
system repetition rate was 1 MHz to avoid APD after-pulsing.

Precise control of the relative phase setting between the two AMZs and the
birefringence in the two arms of Bob's AMZ are necessary to obtain high
visibility. Both can be done by controlling the device temperatures. To set
the phase, it is sufficient to control the path length difference within
$\Delta L=\lambda/n$, where $n\sim1.5$ is the refractive index of silica. The
path length difference depends linearly on the device temperature with 5 $\mu
$m/K, due to the thermal expansion of the Si substrate. The birefringence in
the two arms can be balanced by controlling the relative phase shift between
two polarization modes, because the two arms have the same well-defined
optic-axes on the substrate. If the path length difference is a multiple of
the beat length $\Delta L_{B}=\lambda/\Delta n$, where $\Delta n$ is the modal
birefringence, the birefringence in the two arms is balanced and two pulses
interfere at the output coupler of Bob's AMZ no matter what the input pulse
polarization is. Since $\Delta n/n$ was the order of 0.01 for our device, the
birefringence was much less sensitive to the device temperature than the
relative phase. Therefore, we could easily manage both the phase setting and
the birefringence balancing simultaneously.

\begin{figure}[pts]
\begin{center}
\includegraphics[width=8cm]{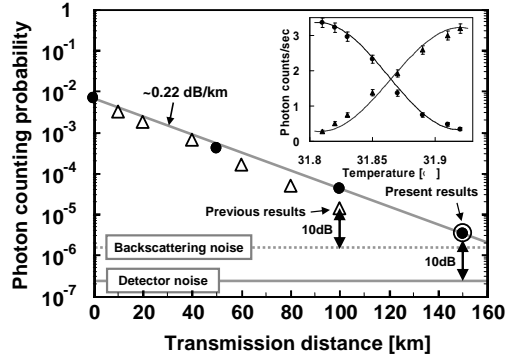}
\end{center}
\caption{Photon counting probability against transmission distance.
 Open triangles indicate the results in the P\&P system.
Inset: Fringe observed in photon count rate, obtained by changing the device temperature at 150 km.}
\label{PLCresult}
\end{figure}

The photon counting probability given by the key generation rate
divided by the system repetition rate is plotted  as a function of
transmission distance (Fig. \ref{PLCresult}). The measured data fit well with the upper
limit determined by the loss of the fiber used ( 0.22 dB/km). In Fig. \ref{PLCresult}, the base lines present the dark count probabilities. The interference fringe is shown in the inset. The visibility at 150 km was 82 $\%$ and 84 $\%$ for the two APDs\cite{KNHTKN04}, which corresponds to a quantum bit error rate (QBER) of 9 $\%$ and 8 $\%$, respectively. These satisfy the rule-of-thumb for secure QKD. The interference was stable for over an hour, which is good enough for a QKD system. The present system could achieve a much longer transmission distance than was attained in a previous experiment using the autocompensating system. The PLC system was shown to be stable for more
than three hours with only the temperature control at each local site.

A similar QKD implemmentation has been proposed to keep security over 146 km fiber transmission
by using strong reference light\cite{guang}.
\subsection{Eavesdropping and security}

\subsubsection{Attacks to weak coherent light BB84}
As we have observed, weak coherent states will work as a good approximation of single photon states. Nevertheless, coherent state pulses may contain multi-photon with small but non-negligible probability. The multi-photon states would give Eve an opportunity to obtain more information on the transmitted photon states than the single photon states. Two approaches have analyzed the security of the weak coherent light BB84 protocol. One assumes that any multi-photon state will provide the full information on the photon states, and tries to limit the probability of multi-photon states. This approach will yield the upper bound of the Eve's information and the conation of the unconditional security against all the possible attacks allowed by the laws of quantum mechanics. The other approach limits Eve's attack (often to the individual attacks) and estimates Eve's information for each attack\cite{Lutkenhaus00}. This approach will provide the security conditions useful to design practical QKD systems. We here concentrate ourselves to the latter approach. The former approach will be discussed in the following chapter.

Our assumptions in the present analysis are the following:
\begin{itemize}
\item Weak Coherent Light: QKD system uses coherent light whose average photon number per pulse is set less than one. The photon state may not be coherent state, as long as the photon number distribution obeys Poisson distribution. It will make the analysis easier to assume no phase correlation between the pulses. This assumption also prevents Eve to perform coherent attacks.

\item Individual Attack: Eve measures one qubit at a time. Eve can store a qubit in  a quantum memory till useful information is disclosed. Eve can also store the measurement results in conventional memories to analyze the results statistically. The assumption of individual attack implies that the information obtained by the measurement on $n$ qubits equals to the $n$ times the information obtained by the measurement on one qubit.

\item No PNS:  As we shall show it later, decoy states can detect a eavesdropping that alters the photon number distribution. Therefore, we don't have to worry about photon number splitting (PNS) attack\cite{cvqkd2,bra}, though it is a very strong eavesdropping method.

\item No control on Bob's equipment: Eve cannot control the performance of Bob's equipment. For example, Eve cannot change the dark count probability in Bob's photon detector. Eavesdropping affects the error rate originated in the transmission line. As a result, Eve's information on the key depends only on the error rate due to the imperfection of the transmission. Errors from the dark count only reduce the mutual information between Alice and Bob.
\end{itemize}
Eve tries to get the information by the optimal measurement under the above conditions. We will not assume no further restriction on Eve's ability. Some of Eve's ability may seem too good in the real situation. For example, Eve can replace the lossy fibers by lossless ones. Eve can measure the photon number contained in a pulse, and determine the optimal measurement. If we would stick with realistic eavesdropping under current technology, margin of the security would be much increased. However, this too limited approach may lose the advantage of QKD that guarantees the security whatever the technology may be developed.

Eve can improve her measurement with the classical information exchanged between Alice and Bob. Some information on the key bit are also announced. Security analysis should consider the information leakage. The classical information leakage in each step in the protocol is following:
\begin{itemize}
\item Bob show the position of the bits that he detected. Eve knows the position of the working bit, but obtains no information on the states.

\item All the basis are disclosed in sifting. Eve can measure the stored qubits with the optimal setting with the basis information.

\item In error rate estimation, some bit values are open. It makes no harm, however, because they are never used for cipher.

\item Partial information on bit values, parity of several bits, for example, are exchanged to detect and correct errors.
\end{itemize}

The simplest eavesdropping method is so called absorption-resend attack, where Eve measures the photon state and prepares the photon state to send Bob according to her measurement. Curty and  L\"{u}tkenhaus\cite{CL04,CL00} examined the attack for coherent light. As discussed for single photon, however, the absorption-resend attack is not optimal. The optimal individual attack for single photon state is given by cloning\cite{FGGNP97}. Eve sets her cloning machine to interact her qubit (ancilla) with the qubit sent by Alice and measure the ancilla to obtain the bit value. Eve may store the ancilla till the basis are open in the sifting. Eve can measure photon number in a pulse to prepare her cloning machine to be optimal. The photon number is in a different degree of freedom from that represents the qubits, so that photon number measurement doesnot affect the qubit.

In BB84, the four qubit states stay in a circle of the Bloch sphere. The optimal asymmetric cloning machine for these equatorial qubits is given for a single photon state, but is not known for $n$ photon state. We will consider a universal cloning machine here, which is suboptimal but the gap is expected to be small\cite{AGS04,CL04}. Note that the cloning machine should be universal, because Eve doesnot know the basis when she applies the machine. A $n \to n+1$ Asymmetric universal cloning machine is given by a unitary transform $U$ as follows:
\begin{eqnarray}
U\left| \phi\right\rangle ^{\otimes n}\left| 00\right\rangle
=\alpha\left| \phi\right\rangle ^{\otimes n}\left| \Phi^{\left(+\right)  }\right\rangle\nonumber\\ +
\beta\left[  \tilde{\sigma}_{z}\left|\phi\right\rangle ^{\otimes n}\left| \Phi^{\left(  -\right)}\right\rangle +
\tilde{\sigma}_{x}\left| \phi\right\rangle ^{\otimes n}\left| \Psi^{\left(  +\right)  }\right\rangle +
i\tilde{\sigma}_{y}\left| \phi\right\rangle ^{\otimes n}\left| \Psi^{\left(-\right)  }\right\rangle \right]
\label{ACM}
\end{eqnarray}
Eve forwards the first $n$ particles and keeps two particles.
Disturbance $D$ to the qubits sent to Bob is given by $D=2 \beta^{2}$. The coefficients $\alpha$ and $\beta$ satisfy
\begin{equation}
\alpha^{2} +n\left(  n+2\right)  \beta^{2} =1.
\end{equation}
In Eq. (\ref{ACM}), $\tilde{\sigma}_{k}$ is a superposition of the operations that applies a Pauli operator $\sigma_{k}$(i$k=x,y,z$)  to one of the $n$ particles.;
\begin{equation}
\tilde{\sigma}_{k}=\sigma_{k}^{\left(  1\right)  } \otimes 1^{\left(  2\right)}\otimes \cdots \otimes 1^{\left(  n\right)  }
+\cdots+1^{\left(  1\right)} \otimes \cdots \otimes 1^{\left(  n-1\right)  } \otimes \sigma_{k}^{\left(n\right)  }.
\end{equation}
This cloning machine is universal, because the result (\ref{ACM}) is independent of the input particle state $\left| \phi\right \rangle$. It is also optimal, in the sense that it yield the fidelity $F=\left(n^{2}+3n+1 \right)/\left(n^{2}+3n+2 \right)$ for the optimal universals cloning machine given by Gisin and Massar\cite{GM97}, when we assume identical fidelity for the first $n+1$ particles. However, it is not optimal as an $n \to n+2$ cloning machine, though it uses two ancilla, because the  fidelity of the last qubit is less than the optimal value.

Eve's error rate on the key is estimated from the distance between Eve's reduced density matrices as
\begin{equation}
E^{(n)}=\frac{1-\Delta^{(n)} }{2}
\end{equation}
where the distance can be measured by the fidelity:
\begin{equation}
\Delta^{(n)} = \sqrt{1-F\left( \rho_{|\phi\rangle}^{(n)},\rho_{|\phi' \rangle}^{(n)} \right)^{2}}. \label{Db}\\
\end{equation}
The reduced density matrix $\rho_{|\phi\rangle}^{(n)}$ is obtained by tracing out Bob's states from $U\left| \phi\right\rangle ^{\otimes n}\left| 00\right\rangle \left\langle 00 \right| \left\langle \phi \right|^{\otimes n}U^{\dagger}$ for each Alice's state $\left| \phi \right\rangle = |0 \rangle, |1\rangle, | \bar{0} \rangle, |\bar{1} \rangle$. Since Eve knows the basis when she measures the qubits, she needs to distinguish $\rho_{|0\rangle}^{(n)}$ from $\rho_{|1 \rangle}^{(n)}$ or $\rho_{|\bar{0}\rangle}^{(n)}$ from $\rho_{|\bar{1} \rangle}^{(n)}$.

\begin{figure}[pts]
\begin{center}
\includegraphics[width=8cm]{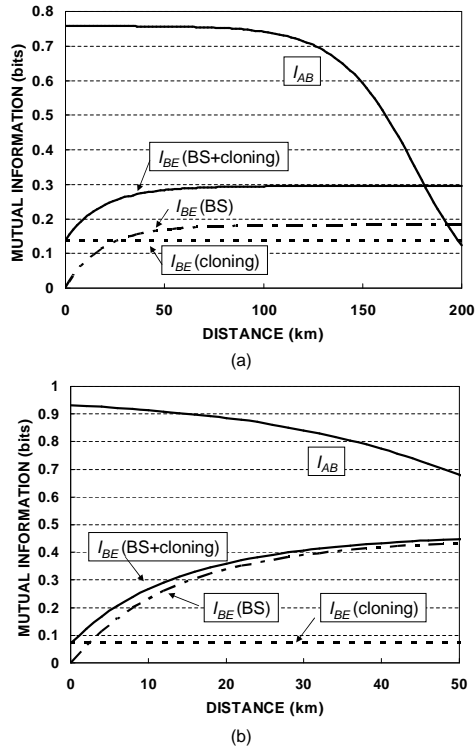}
\end{center}
\caption{Mutual information $I_{AB}$ and $I_{BE}$ as a function of transmission distance. (a)the parameters are taken from the long distance transmission experiment \cite{KNHTKN04}: $\alpha=0.22$dB/km, $\beta=5$dB, $P_{d}=2\times 10^{-7}$, $\mu=0.2$, $v=0.96$. (b) the parameters are taken from the short distance but high speed transmission experiment \cite{T5N04}: $\alpha=0.25$ dB/km, $\beta=1.5$ dB, $P_{d}=1 \times 10^{-4}$, $\mu=0.6$, $v=0.995$P}
\label{mutualinf}
\end{figure}

So far, we assume that Eve has an excellent technology to make an optimal cloning machine.Now Suppose Eve has a magic fiber that can transmit a photon without loss\cite{BBBSS92}. She can replace the fiber between Alice and Bob by her magic fiber and put a beam splitter before the fiber without any change in Bob's detection rate, if  she set the transmittance $T$ of the beam splitter equal to the inverse of the fiber loss. A photon will be detected by either Eve or Bob. Since Eve's detection does not correlated with Bob's, Eve will not gain any information, if a single photon was sent. However, coherent state light provide a finite probability that Bob and Eve detect a photon at the same time. If Eve keep the photon until the basis is open, she will obtain full information on the photon by measuring it according to the basis. The conditional probability of Eve's detection on the photons that Bob detected is given by $P_{BS}=(1-\exp[-R\mu])$, where $R=1-T$. This expression is the same as Eve's detection probability itself, because the Eve's detection process is independent of Bob's. The mutual information between Bob and Eve on the key bit therefore reads
\begin{equation}
I_{BE}^{BS}=P_{BS}. \label{BS}
\end{equation}
It is possible to apply the optimal cloning attack on the output light of the beam splitter. This combined attack may be the strongest among the individual attacks. Fig. \ref{mutualinf} show the mutual information $I_{AB}$ and $I_{BE}$ as a function of transmission distance, where the parameters are taken from the recent experiments for (a) long distance \cite{KNHTKN04} (b) short distance but high speed transmission \cite{T5N04}, as follows: the fiber loss $\alpha  = 0.22$ dB/km, the receiver loss $\beta = 5$ dB, dark count probability $P_{d}=2 \times 10^{-7} $/pulse, average photon number $\mu=0.2$, visibility of the interference $v=0.96$ in 150 km transmission, and $\alpha=0.25$ dB/km, $\beta=1.5$ dB, $P_{d}=1 \times 10^{-4}$, $\mu=0.6$, $v=0.995$ in 40 km transmission at raw key rate of 100 kbps, respectively. Visibility is usually a measure of the quality of the interferometer, but it is also a measure of the disturbance by eavesdropping in QKD experiments, where the disturbance $D$ is given by $D=1-v$. Impact of the parameters $P_{d}$, $D$, and $\mu$ is following. Dark counts, which result in Bob's error, reduce the mutual information $I_{AB}$. Disturbance increases Eve's information obtained by the cloning attack. Increasing average photon number raise the probability that the pulse contains more than two photons. Eve can obtain more information by the beam splitter attack. As shown in Fig. , the mutual information between Alice and Bob is larger than that between Alice and Eve with the present parameters. The results implies that the we can extract secure final keys from the transmitted raw keys in the experiments \cite{T5N04,KNHTKN04}.

\subsection{Ideas of unconditional security proof}
We have studied some typical types of Eavesdropping. However, the study of some types Eavesdropping itself does not make an unconditional
 security proof, because Eve may have many other choices. A security proof should not be dependent on any assumption that restrict
Eve's actions which do not violate laws of nature. Here we show some ideas of the security proof. Detais are presented in the next
section.

Since Bennett and Brassard proposed the first quantum key distribution protocol, the proof of the security has been one of the main issues.
Several authors have presented the proof of the unconditional security assuming a single photon source. It has been clarified that the role
of the quantum mechanics is to set a bound to Eve's knowledge on key.

Suppose Alice and Bob share a random variable $W$ ($n$ bits). Eve
tries to know the random bits without being detected. If Eve's
knowledge $V$ about $W$ is at most $t<n$: Renyi entropy is bounded
by, $R(W|V=v)\ge n-t$,  Alice and Bob can distill $r = n -t -s$ bits
of secure key $K$, Eve's mutual information on which satisfies
\cite{mayer1,BBCM95,LC99}
\begin{equation}
I(K;GV)=\frac{2^{-s}}{\ln2},
\end{equation}
with a random choice of universal hash function $G$ as $K=G(W)$. However, it is unknown so far whether the conclusion here for
classical communication also allpies for quantum case. Since in QKD, Eve may choose to store her ancillas first and then directly
attack the final key after the privacy aplification stage taken by Alice and Bob.

The first unconditional security proof is completed by Mayers which is based on the uncertainty principle and classical
error correction code (CSS code). As a result of Mayers' proof, Eve's information about the final key shared by Alice and
Bob must be exponentially close to 0, given whatever type of attacking schemes she may take. Mayers proof is rather complex.

Shor and Preskill\cite{shor2} simplified Mayers' proof from the
viewpoint of (virtual) entanglement distillation\cite{LC99}. Since
entanglement purification is equivalent to quantum error
correction(QEC), Alice and Bob can share nearly maximally entangled
pure states from degraded entangled photon pairs by performing QEC
on their sites. Shor and Preskill have shown that the entanglement
purification protocol can be reduced to BB84 protocol by the
following steps. The entangled photon source can be in Alice's site.
Suppose Alice measures one of the photon pair before Bob. The photon
state sent to Bob is same as a single photon state modulated
according to a random number, which corresponds to Alice's
measurement result. In this protocol, Alice sends random qubits
encoded in a random quantum code, and Bob obtains bit values after
performing QEC. The quantum code can be CSS code, so that this
protocol is called CSS code protocol. Bob measures qubits on the
computational basis $\{X,Z\}$, and phase does not affect the
measurement results. Therefore, QEC in the protocol needs not to
correct phase errors. Since measurement and QEC can commute,
\textit{i.e.}, Bob can measure the qubits first and  then correct
errors by a classical code corresponding to the CSS code. This
protocol is nothing but BB84. Performance of QEC would determine the
limit of error rate to guarantee the security. The limit, however,
will never exceed 25 \%, where the entanglement between Alice and
Bob can no longer be recovered. The security of QKD with an
imperfect source is then also shown.

\section{Security proofs and protocols of QKD with weak and strong Gaussian states}\label{secqkd}
This section contains the security proofs and protocols of QKD
with real setups.
 We shall first review the entanglement purification protocol given
by Bennett, DiVincenzo, Smolin and Wooters\cite{BDSW} (the BDSW
protocol) and then reduce it to the security proof of BB84 QKD
protocol with perfect single-photon source which is a difficult
technique. There are a number of useful methods for secure QKD
directly using laser light,  e.g.,  the decoy-state
protocol\cite{hwang,wang0d,lolo}, the SARG04 protocol\cite{scar},
the protocol with strong reference light\cite{srl} and the protocol
with bright squeezed light\cite{cvgp}. In this section, we shall go
into some of them after the security proof of standard BB84 protocol
with ideal source.
\subsection{Entanglement purification and security proof of QKD}\label{skey}
The first security proof of BB84 protocol is given by
Mayers\cite{mayer1}. The security proof for QKD can be simplified if
Alice and Bob have large quantum computers\cite{LC99}.  Later,
Mayers' security proof of BB84 protocol was simplified by Shor and
Preskill \cite{shor2} based on the simple idea of first doing
entanglement purification with a CSS code\cite{shor2,qecc1} and then
reducing it to a prepare-and-measure protocol (BB84) and  distilling
the data of measurement outcome by a {\em classical} CSS code. An
excellent tutorial of Shor and Preskill's proof has been presented
by Gottesman and Preskill\cite{cvgp}. Here we modify Shor and
Preskill's proof and we present an alternative approach which is
based on the idea of the (modified) BDSW entanglement purification
protocol\cite{BDSW}. By using our modified proof, the problems such
as how to construct a CSS code or the existence of good CSS codes is
circuvemented\cite{decouple}. (Mayers' proof does not have these
problems but it is rather complex.)

The first work to relate the security of QKD with quantum
entanglement was published in 1991\cite{Eke91}. It was then found
that actually one can use the idea of quantum entanglement as a {\em
mathematical} tool to prove the unconditional security. If Alice and
Bob share a number of perfect EPR pairs, they can measure them at
each side in $Z$ basis $(\{|0\rangle, |1\rangle\}) $ and use the
measurement outcome as the shared secure key. Say, outcome
$|0\rangle$ for bit value 0 and $|1\rangle$ for bit value 1. This
key will be perfect and no third party can have any information
about it. If the pre-shared EPR pairs are almost perfect
(exponentially close to the perfect ones), they can also use the
measurement outcome as the secure key since any third party's
information about the  key is exponentially close to 0\cite{LC99}.
If the pre-shared pairs are noisy but not too noisy, they can first
purify them into a smaller number of almost perfect pairs which are
exponentially close to perfect pairs and then obtain the secure
final key. This is the entanglement-purification-based protocol. We
shall use the modified BDSW protocol\cite{BDSW} for the entanglement
purification and then
 show how to reduce it to a prepare-and-measure one\cite{shor2,cvgp}.
\subsubsection{Main idea of entanglement purification}
Suppose initially Alice and Bob share $N+m$ raw pairs and each pair
is in one of the 4 Bell states. The goal of entanglement
purification is to distill out $K\le N$ pairs which are all in the Bell
state of $|\phi^+\rangle$. We can reach such a goal if most of the
initially shared raw pairs are in $|\phi^+\rangle$\cite{BDSW}. To
reach the goal, we need first to know the error rate of the raw
pairs and then we can locate the wrong pairs by hashing. By local
bit-flip or phase-flip operations, all the outcome pairs will be in
$|\phi^+\rangle$ state with a probability exponentially close to 1.
The original BDSW protocol\cite{BDSW} requests that each of the raw
pairs must be in one of the 4 Bell states, i.e., there is no
entanglement among different pairs. It is then shown by Lo and
Chau\cite{LC99} that actually the result of BDSW is unconditionally
correct given what ever initial state of the raw pairs. This makes
the base of the unconditional security of QKD. There are 2 main
steps: error test and hashing.
\subsubsection{Error test}
The initial $N+m$ shared raw pairs can be in any state, e.g., there
could be complex entanglement among different pairs. But here our
task is not to know the state of these pairs. We imagine that each
shared raw pair is measured in the Bell basis in the beginning but
no one reads the measurement outcome\cite{cvgp}. (As shown later,
the fidelity result of purification keeps unchanged if this virtual
Bell measurement in the beginning is omitted.) Therefore each pair
is now in one of the 4 Bell states and our task is to locate wrong
pairs, those pairs which are not in state $|\phi^+\rangle$. The goal
of error test is to see how many pairs are bit-flipped (i.e., in
state $|\psi^+\rangle$ or state $|\psi^-\rangle$) and how many of
them are phase-flipped (i.e., in the state $|\phi^+\rangle$ or state
$|\psi^-\rangle$).

 We use
notations $|0\rangle,|1\rangle$ for the eigenstates of operator $Z$
with eigenvalues $0,1$, $|\bar 0\rangle=\frac{1}{\sqrt 2}(|0\rangle+|1\rangle),
|\bar 1\rangle=\frac{1}{\sqrt 2}(|0\rangle-|1\rangle)$ for the
eigenstates of operator $X$ with eigenvalues $0,1$, and $Z_{A\oplus
B}$ for the basis of collective measurement of parity value in $Z$
basis for a pair. Given any state $\alpha
|a\rangle_A|b\rangle_B+\beta |a\oplus 1\rangle_A|b\oplus
1\rangle_B$, the measurement outcome in $Z_{A\oplus B}$ basis will
be $z_{A\oplus B}=a\oplus b$.  If Alice and Bob do individual
measurement at each side in $Z$ basis, there would be two outcome,
$z_A$ and $z_B$ at each side and we have
\begin{equation}
z_A\oplus z_B= z_{A\oplus B}.\label{eqind}
\end{equation}
\\{\em Remark:} Throughout this section, all summation symbols are for the bit wise summation, i.e.,
$a\oplus b =0$ if $a=b$ and  $a\oplus b=1$ if $a\not= b$.
\\ Similarly, we can also denote
the parity measurement in $X$ basis by $X_{A\oplus B}$ and its
measurement outcome is \begin{equation}x_{A\oplus B}=x_A\oplus
x_B.\label{eqindx}\end{equation}

The first question here is how the principles of classical
statistics works for a quantum system as stated in Ref.\cite{LC99}.
Since $Z_{A\oplus B}$ and $X_{A\oplus B}$ commute, one can in
principle measure any pair in both $Z_{A\oplus B}$ and $X_{A\oplus
B}$. A pair carries a bit-flip error if $z_{A\oplus B}=1$, a
phase-flip error if $x_{A\oplus B}=1$. Given $N+m$ raw pairs, we can
randomly choose $m$ pairs as {\em test pairs}. We want to know the
error rates (including bit-flip rate and phase-flip rate) of the
remaining $N$ pairs by examining the $m$ test pairs. This is called
the {\em error test}.

Since $X_{A\oplus B}$ and $Z_{A\oplus B}$ commute, we can imagine
that we had first measured each of the $m$ test pairs in both bases
and then estimate the error rates of the remaining $N$ pairs by
examining the measurement outcome of $m$ test pairs.
Further, we can divide the $m$ test pairs into two subsets and each
subset contains $m/2$ pairs. We can measure each pair in the first
subset in basis $Z_{A\oplus B} $ to see the bit-flip rate  and
measure each pair in the second subset in  $X_{A\oplus B}$ basis to
see the phase-flip rate. If $m$ is very large, the error rates
obtained in this way must be the same with the error rates obtained
by measuring each test pair in both  $Z_{A\oplus B} $ basis and
$X_{A\oplus B}$ basis. Therefore, after measuring one subset of the
test pairs in $Z_{A\oplus B} $ basis only and the other subset in
$X_{A\oplus B}$ basis only,  one can estimate the (upper bound)
values of both bit-flip rate and phase-flip rate of the remaining
$N$ pairs if these pairs were measured. Since the Bell measurement
and collective measurement $Z_{A\oplus B}$ or $X_{A\oplus B}$
commute, the operation of {`` Bell measurement plus $Z_{A\oplus
B}~(X_{A\oplus B})$" will give the same outcome of bit-flip rate
(phase-flip rate) with that of the measurement of $Z_{A\oplus
B}~(X_{A\oplus B})$, thus the virtual initial Bell measurement to
those $m$ test pairs can be omitted. Also, since all test pairs are
discarded, we can actually use {\em whatever} measurement to those
test pairs provided that the outcome tells us the value of
$z_{A\oplus B}$ of one subset and the value of $x_{A\oplus B}$ for
the other subset of the test pairs. Therefore we can replace
$Z_{A\oplus B}$ for the first subset by individual measurements in
$Z$ basis at each side, and $X_{A\oplus B}$ for the second subset by
individual measurements in $X$ basis at each side, because the
outcome of individual measurements determines the outcome of
collective measurements by Eq.(\ref{eqindx}). Further, it makes no
difference if Alice measures her halves of those test pairs before
entanglement distribution. This means, instead of sharing $m$ pairs
with Bob, she can simply send Bob $m$ single qubits with each state
being randomly chosen from $|0\rangle,|1\rangle$ and $|\bar
0\rangle,|\bar 1\rangle$ (i.e., BB84 states) for the future error
test.
\subsubsection{Hashing}
After the error test, Alice and Bob now share $N$ raw pairs. For
simplicity, we shall assume the non-local measurements $Z_{A\oplus
B},~X_{A\oplus B}$ and controlled-NOT gates for Alice and Bob at
this moment. However, these are only mathematical techniques for the
security proof. As shown later, all these assumed (virtual)
operations are not really necessary and the protocol is reduced to
the BB84 protocol. As assumed earlier in the `` error test" part,
Alice and Bob had measured each of the shared $N$ pairs in the Bell
basis but they don't look at the measurement outcome\cite{cvgp}.
Each pair must now be in one of the 4 Bell states. The task is to
find out the positions of those pairs which are not in
$|\phi^+\rangle$ and then correct the errors. Given
$N$ raw pairs,
 we can use two $N-$bit classical binary strings, the {\em bit-flip string}
$s_{b}$ and the {\em phase-flip string} $s_{p}$ to represent the
positions of bit-flips and phase-flips. For any $i$th pair, if it
bears a bit-flip error, the $i$th  element in string $s_{b}$ is 1,
otherwise it is 0; if it bears
 a phase-flip, the $i$th element  in string $s_{p}$ is 1, otherwise
it is 0.
For example, given 5 pairs
$|\phi^+\rangle|\phi^+\rangle|\psi^+\rangle|\phi^-\rangle
|\psi^-\rangle$, the two classical strings are
\begin{equation}
s_{b}= 00101;s_p=00011.
\end{equation}
The goal is to detect the positions of those pairs bearing a
bit-flip error or a phase-flip or both.  This goal can be achieved
by hashing, which consumes some of raw pairs and the shortened
bit-flip string and phase-flip string for those remaining pairs can
be determined. If the bit-flip rate and phase-flip rate are bounded by 
$t_b$, $t_{ph}$ respectively, the number of likely bit-flip string and
phase-string for $N$ raw pairs are bounded by
\begin{equation}
\omega_{b}=2^{N\cdot H(t_b)}  ;
\omega_{p}=2^{N\cdot H(t_{ph})}
\end{equation}
respectively and $H(x)=-x\log_2 x-(1-x)\log_2(1-x)$. (After the
error test, the values of $t_b,t_{ph}$  for the remaining $N$ pairs
are known already.)

To have a clear picture, we consider the result of classical hashing
first. Suppose we know that there are $\omega = 2^n$ candidates for
an unknown $N-$bit classical string. We can determine a shorter
sub-string  by revealing the parity values of $n+\delta N$ random
subsets of elements from the string. Say, each time we reveal the
parity value of a random subset and then discard one bit in the
subset. After we have revealed $n+\delta N$ parity values, we
examine each likely candidates for the remaining string whose length
is $N-n-\delta N$. There must be one candidate string $s_f$
satisfying all those revealed parity values. And the probability
that there is another (different) candidate string that also
satisfies those parity values is only $2^{-\delta N}$. Therefore,
the string for the remaining bits are determined. A detailed proof
of this is given in \cite{BDSW}.

(1) Bit-flip error correction. To use the classical result of
hashing here, we only need to know how to detect the parity value of
any subset of the bit-flip string $s_b$, i.e., the parity of a
shorter bit-flip string for any subset of those $N$ raw pairs.  This
can be done by the so-called biCNOT operations and  measurements in
$Z_{A \oplus B}$ basis.

 Suppose there are $u$ pairs in a subset $E$,
$E=\{e_1,e_2,\cdots,e_u\}$ is a subset that contains pair
$e_1,e_2,\cdots,e_u$. The bit-flip string for this subset is
$s_b(E)=c_{e_1}c_{e_2}\cdots c_{e_u}$ and the bit-flip string of $N$
pairs are $s_b=c_1c_2\cdots c_N$ and $N\ge u$, of course. We want to
know the parity value of string $s_b(E)$, i.e.,
\begin{equation}
\sum_{l\in E}c_l=c_{e_1}\oplus c_{e_2}\oplus\cdots\oplus c_{e_u}.
\end{equation}
We have the following fact
\begin{equation}
\sum_{l\in E} c_{l} =\sum_{l\in E}z_{Al\oplus Bl} \label{fact75}
\end{equation}
Here $z_{Al\oplus Bl}$ is the outcome data for pair $l$ if it is
measured in $Z_{A\oplus B}$ basis. However, as shown below in
obtaining $\sum_{l\in E}c_l$, we do NOT have to know each individual
value of $c_l$.

A CNOT gate in $Z$ basis is a gate that takes the following
transformation:
\begin{equation}
|z_1\rangle|z_2\rangle\longrightarrow |z_1\rangle|z_1\oplus
z_2\rangle.
\end{equation}
Here the first state is the control state, the second state is the
target state, $z_1,z_2$ can be any value from $\{0,1\}$. A biCNOT
gate contains two CNOT gate, one in Alice's side the other in Bob's
side. Explicitly, we first denote $|\phi^{+}\rangle,
|\phi^{-}\rangle, |\psi^{+}\rangle, |\psi^{-}\rangle$ by
$|\chi_{00}\rangle, |\chi_{01}\rangle,|\chi_{10}\rangle,
|\chi_{11}\rangle$, respectively. That is to say, for any pair state
$|\chi_{b,ph}\rangle$, the parity value $z_{A\oplus B}=b,~x_{A\oplus
B}=ph$. Given two pair state
$|\chi_{b_1,ph_1}\rangle|\chi_{b_2,ph_2}\rangle$, if we do a biCNOT in $Z$
basis on these two pairs with the second pair being the target, we
have
\begin{equation}
{\rm biCNOT_Z}(|\chi_{b_1,ph_1}\rangle|\chi_{b_2,ph_2}\rangle) =
|\chi_{b_1,ph_1\oplus ph_2}\rangle|\chi_{b_1\oplus b_2,ph_2}\rangle.
\label{backaction}\end{equation} This means, a biCNOT operation will
collect the parity value $z_{A1\oplus B1}\oplus z_{A2\oplus
B2}=b_1\oplus b_2$ into the target pair. Given
Eq.(\ref{backaction}), we can know the parity of string $s_b(E)$ by
the following way:  1. Gather $\sum_{l\in E}z_{Al+Bl}$ into pair
$d\in E$ by performing biCNOT gates between pair $d$ and other pairs
in set $E$, with pair $d$ being the target pair.  2. Measure pair
$d$ in basis $Z_{A\oplus B}$. 3. Discard pair $d$. For example, to
know the parity of bit-flip substring $s_b(E)=c_1c_4c_5$, they can
first do biCNOT operations in $Z$ basis as shown in Fig.
\ref{hashing}, then measure pair 5 in basis $Z_{A\oplus B}$ and then
discard pair 5.
\begin{figure}
\epsffile{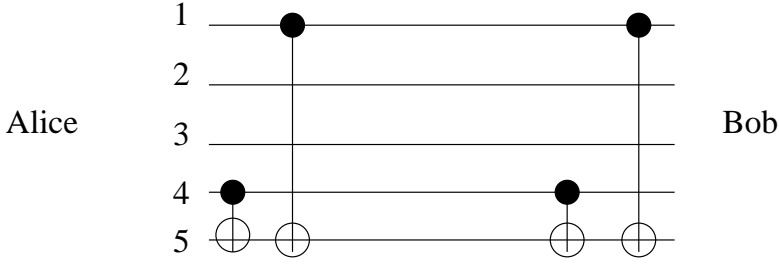}
\vskip 10pt
\caption{ They first do biCNOT operations in $Z$ basis and then measure the target pair (pair 5) in
 order to obtain  the parity value for the bit-flip string of a subset $E$ which contains
pair \{1,4,5\}.
 }\label{hashing}
\end{figure}

They repeat this for $q=NH(t_b) $ times and Bob can compute the
bit-flip string for the remaining $N-q$ pairs, i.e., he knows all
positions where the pair bears a bit-flip error. Bob flips those of
his qubits from bit-flipped pairs in $Z$ basis. The bit-flip error
correction is completed here.

However, the phase-flip string for the remaining $N-q$ pairs is not determined yet.
In doing the hashing for bit-flip string, the phase-flip string for the
remaining pairs changes.
However, the number of likely phase-flip string for the remaining pairs
cannot be larger than $\omega_p$, the number of likely phase-flip string
for the initial $N$ raw pairs. Because the initial phase-flip string of $N$ pairs
determines the phase-flip string of the remaining pairs after
bit-flip hashing.
Eq.(\ref{backaction}) shows that given the initial phase-flip string, the later phase-flip string for the remaining pairs is determined exactly.
Keeping this fact in the mind, they can then correct phase-flip errors:
\\(2) Phase-flip error correction.
They can also do hashing in $X$ basis. Eq.(\ref{backaction}) shows
that the biCNOT operation can collect the parity value $x_{A1\oplus
B1}\oplus x_{A2\oplus B2}=ph_1\oplus ph_2$ into pair 1, i.e., the
control pair. Therefore, given any set $E$, they can randomly choose
pair $d$ as the control pair and all the other pairs in set $E$ as
the target pair. They perform biCNOT gates between pair $d$ the
other pairs in set $E$. The parity value of $\sum_{l\in
E}x_{Al\oplus Bl}$ before biCNOT operation can be obtained by
measuring pair $d$ in $X_{A\oplus B}$ basis after the biCNOT
operations.  With this they can do hashing to correct phase-flip
errors.  Explicitly, at any step $j>q$, Alice announces $Ej$ which
is a random subset of all remaining pairs and pair $d_j\in Ej$. They
do biCNOT operations in $Z$ basis to collect the parity value
$\sum_{l\in Ej} \left(x_{Al}\oplus x_{Bl}\right)$ at pair $d_j$.
Pair $d_j$, the only control pair is discarded after it is measured
in $X_{A\oplus B}$ basis.  They repeat this for $p=NH(t_{ph})$ times
and Bob can compute the phase-flip string for the remaining $N-p-q$
pairs, i.e., he knows all positions where the pair bears a
phase-flip error. Bob correct those phase-flipped pairs by a taking
a phase shift operation to his own qubits. The purification is ended
here.
\\
{\em Remark:} Similar to the phase-flip hashing also needs the
biCNOT gates in $Z$ basis. But the target pair and control pair are
reversed.

\subsubsection{Reduction}
The above virtual entanglement purification protocol contains the
following elementary operations:
\begin{itemize}
\item{Bell measurement to each raw pair in the beginning}
\item{biCNOT gates}
\item{measurement in $Z_{A\oplus B}$ } basis and measurement in $X_{A\oplus B}$
basis.
\end{itemize}
The nonlocal measurements here are impossible tasks for Alice and
Bob who are spatially separated. The CNOT operations are difficult
tasks. We can get rid of all these in the real protocol by
reduction.

Since the Bell measurement commutes with biCNOT gates and any
measurements in $Z_{A\oplus B}$  basis and measurement in
$X_{A\oplus B}$ basis, we can delay the initial Bell measurement
until the end of the protocol and we shall obtain the same state
$\rho$ for the outcome $K$ pairs. We denote this protocol as
protocol V1. Suppose the fidelity $F=\langle
{\phi^+}^K|\rho|{\phi^+}^K\rangle=1-\epsilon$ for the outcome pairs
and $\epsilon$ is exponentially close to 0. Consider another
protocol V2 which simply skips the Bell measurement and performs
everything else the same as protocol V1. The outcome $K-$pair state
in protocol V2 is $\rho'$. obviously, if the Bell measurement were
performed to each outcome pair of protocol V2, the state of the
outcome pair of protocol V2 must be the same with that of protocol
V1. Therefore,
\begin{equation}
\rho=\sum_{\{b_y,ph_y\}}
\langle\Pi_{\{b_y,ph_y\}}|\rho'|\Pi_{\{b_y,ph_y\}}\rangle|\Pi_{\{b_y,ph_y\}}\rangle\langle
\Pi_{\{b_y,ph_y\}}|
\end{equation}
and
$|\Pi_{\{b_y,ph_y\}}\rangle=|\chi_{b_1,ph_1}\rangle\cdots|\chi_{b_K,ph_K}\rangle$,
and any $b_y$ or $ph_y$ can be 0 or  1. Also, as defined earlier, any
$|\chi_{a_yb_y}\rangle$ is one of the 4 Bell states  depending on
its subscript value. This leads to
\begin{equation}
\langle {\phi^+}^K|\rho'|{\phi^+}^K\rangle=\langle
{\phi^+}^K|\rho|{\phi^+}^K\rangle=1-\epsilon
\end{equation}
which means, although the outcome state of protocol V1 and V2 can be
different, the fidelity results of the two protocols are the same.
Since Eve's information upper bound is dependent on $F$ only\cite{LC99},
protocol V1 and V2 are the same secure. We shall only consider
protocol V2.

There are still some other non-local measurements of $Z_{A\oplus B}$
and $X_{A\oplus B}$ in the protocol. However, since all these are
performed on the pairs which are discarded in the protocol, it makes
no difference if they take any further operations to the discarded
pairs. In the step of bit-flip correction, they can measure the
discarded pairs in $Z$ basis at each side and obtain
 $\sum_{l\in E}z_{Al}$ and $\sum_{l\in E} z_{Bl}$ separately and then calculate
 the parity value by formula
 \begin{equation}\sum_{l\in E}c_l=\sum_{l\in E}z_{Al}+\sum_{l\in E} z_{Bl}
 .\end{equation}
The individual measurement in $Z$ basis at each side commutes with
the nonlocal measurement $Z_{A\oplus B}$ therefore they can exchange
the order of the two measurements. Furthermore, since the individual
measurements in $Z$ basis determines the outcome of the collective
measurement $Z_{A\oplus B}$, the collective measurement is
unnecessary. This is to say, in the protocol V2, all those nonlocal
measurements $Z_{Z\oplus B}$ can be simply
 replaced by the individual
measurement in $Z$ basis at each side.

The bit-flip error correction is now reduced to the following: At
any step $i$, Alice announces $Ei$ which is a random subset of all
remaining pairs and pair $d_i\in Ei$. They do biCNOT operations in
$Z$ basis to collect the parity value $\sum_{l\in Ei}
\left(z_{Al}\oplus z_{Bl}\right)$ at pair $d_j$. They measure the
target pair in $Z$ basis at each side. Alice announces her
measurement outcome, i.e.,    $\sum_{l\in Ei}z_{Al}$. Bob calculates
$\left(\sum_{l\in Ei} z_{Al}\right)\oplus \left(\sum_{l\in
Ei}z_{Bl}\right)$ and this is just the parity value of string
$s_b(Ei)$, the bit-flip string of pairs in subset $Ei$. They discard
pair $d_i$.

  The phase-flip correction can also be simplified. Due to the same reason as stated above,
  after the biCNOT gates in a subset $E_j$ is done, the non-local measurement $X_{A\oplus B}$ on the control pair
can be replaced by  measuring the control pair in $X$ basis in each
side because that pair will be discarded.
  Since their only purpose is to obtain a
secure final key, they need not really correct the phase-flip
errors. After Bob has known the positions of all phase-flips for the
remaining pairs, he does not need to really correct the phase errors
because this does not change the bit values of the final key and
actually no one knows whether he has performed the correction.
Therefore he even does not need to compute the positions of the
phase errors provided that he {\em can} do so. Consequently,
 the measurement in $X_{A\oplus B}$ basis on the control pair is unnecessary and they can directly discard the pair without any measurement.
This is to say, at each step of hashing for phase-flip string, they
{\em only} need to do some biCNOT operations in $Z$ basis
 and directly discard pair $d_j$.
Thus,
 all operations needed in both bit-flip
correction and phase-flip correction can now be done in $Z$ basis.
Therefore, Alice can choose to measure all her halves of $N$ pairs
in $Z$ basis before sending anything to Bob, i.e., she can directly
send Bob $N$ random qubits in $Z$ basis and Bob can measure each of
them before key distillation. The final key distillation becomes the
distillation of the data from the measurement outcome in $Z$ basis
at Bob's side.

We shall simply use {\em error correction} for the term {\em bit-flip error correction} because
the phase-flip error correction
 is now reduced to the {\bf Privacy amplification}:
At any step $j>q$, Alice's bits and Bob's bits are identical. Alice
announces a random subset $Ej$ and bit $d_j\in Ej$. For any $l\in
Ej$, they replace ${z_l}$ by $z_l\oplus z_{d_j}$ and they discard
bit $d_j$. Here $z_l$ is the bit value of the $l$th bit in $E_j$.
They need to repeat so by $p$ steps and obtain the final key.

If they don't use a quantum memory, Bob must measure each qubit once
he receives it. In such a case Bob has no way to know the right
basis of each individual qubit. Bob can randomly choose basis $Z$ or
$X$. They must discard those outcome from a wrong basis (basis
mismatch). We have the following prepare-and-measure protocol:

(0) Alice and Bob have agreed that states $|0\rangle,|\bar 0\rangle$ represent for bit value 0 and
states $|1\rangle,|\bar 1\rangle$   represent for bit value 1.
(1) Alice prepares $2(N+m)$ qubits. The preparation basis of each qubit is random,
with a prior probability of $P_z=\frac{2N+m}{2(N+m)}$ for $Z$ basis, and $P_x=\frac{m}{2(N+m)}$ for
$X$ basis.  The bit value of each individual qubit is randomly chose from 0 and 1.
(2) Alice sends these qubits to Bob and Bob measures each of them in a basis randomly chosen from
\{$X$, $Z$\}. Bob announces his measurement basis for each qubits and they
discard those outcome from a measurement basis that is different from
Alice's  preparation basis. Approximately, there should be $N+m$ classical bits
remaining
among which about $m/2$  are $X$ bits (outcome of measurement in $X$ basis)
and $N+m/2$ are $Z$ bits  (outcome of measurement in $Z$ basis).
(3) Bob announces the bit values of all $X$ bits and the same
number of $Z$ bits for error test. They discard all the announced bits and
 there are about $N$ bits remaining. After this error test, they know that
the bit-flip rate and phase-flip rate for the remaining $N$ bits are bounded by $t_b,t_{ph}$ respectively.
(4) They do bit-flip error correction and privacy amplification to the remaining $N$ classical data
as we have stated previously and obtain the final key.
The final key rate is
\begin{equation}
f=1-H(t_b)-H(t_{ph}).
\end{equation}
 In the
case $t_b=t_{ph}=t$, the noise threshold for BB84 protocol is $11\%$
where its key rate hits 0. However, if we use the encoded BB84
states and/or if the key distillation is done with two-way classical
communication, the noise threshold can be raised significantly, see
e.g., Ref.\cite{twogl,twochau,wangenc1,wangenc2}. Moreover, if the
channel noise is asymmetric, one can also raise the
efficiency\cite{qkdasym}.

{\em Remark}: Besides the approach of entanglement purification,
Koashi has presented another simple and clear picture based on the
uncertain principle\cite{koashin}.

\subsection{Secure  key distillation with a known fraction of tagged bits}
Although many QKD protocols such as the BB84\cite{bene} have been
proven to be unconditionally secure\cite{mayer1,LC99,shor2}, this
does not guarantee the security of QKD in practice, due to various
types of imperfections in real-life set-ups.
 In practical QKD, the source is often imperfect. Say, it may produce multi-photon
pulses with a small probability. Normally, weak coherent states are
used in practice. The probability of multi-photon pulses is around
$10\%$ among all non-vacuum pulses. Here we shall show how to
generate the secure final key even though the source is imperfect,
i.e., with a small probability of sending multi-photon pulses.
\subsubsection{Final key distillation with a fraction of tagged bits.}
For simplicity, let us imagine the following virtual case: Alice now
sends Bob the BB84 states (with most of them being prepared in $Z$
basis) from a perfect single-photon source. But, before sending out
any states, she randomly chooses a small fraction of the qubits and
tells Eve the right bases (and bit values) of them. This small
fraction of qubits are called {\em tagged qubits}. It was shown by
ILM-GLLP\cite{gllp} that we can also distill the secure final key by
a CSS code even with an imperfect source, if we know the bit-flip
rate, phase-flip rate and upper bound value of the fraction of
tagged bits, $\Delta$. Here we give a simple proof. 
Suppose after Bob has measured each qubits he
received and they discarded those outcome with basis mismatch, there
are $N+m$ classical bits remaining. About $m$ of them will be used
for error test, among which a half are $Z$ bits and a half are $X$
bits. The remaining $N$ bits are all $Z$ bits and they will be used
for the final key distillation. We shall call them as {\em
untested bits}. At this stage, any bit that is caused by a tagged
qubit is  a {\em tagged bit}. We assume Alice and Bob carry out the
protocol as if they didn't know which ones are tagged qubits but
they
 know the fraction of tagged bits,  $\Delta$.
They now distill the final key with those $N$ untested bits.
They need to know the number of likely bit-flip string and the number of likely phase-flip string for
those $N$ untested bits.
The error correction part is of no difference from that of the ideal protocol where there is no
tagged bits, i.e., after they
do the error test, they know that the upper bound
 of the bit-flip rate of those $N$ untested bits is $t_b$.
We shall use notations $t_{b,tag}$,
$t_{b,untag}$ for the upper bounds of   bit-flip rates
of the tagged bits and untagged bits from those
$N$ untested bits, respectively.
We have
\begin{equation}
\Delta t_{b,tag} +(1-\Delta)t_{b,untag}=t_b
\end{equation}
Here $t_b$ is the averaged bit-flip rate of all those untested bits.
The value of $t_b$ satisfies
\begin{equation}
t_b\le t_z +\delta_1
\end{equation}
and $t_z$ is the observed bit-flip rate of those test bits in $Z$
basis, $\delta_1$ is a small-value parameter due to the statistical
fluctuation.
 The number of
likely bit-flip string is bounded by
\begin{equation}
\omega_b=2^{\Delta N H(t_{b,tag})}\cdot 2^{(1-\Delta) NH(t_{b,untag})}\le 2^{NH(t_b)} .
\end{equation}
Since they don't know the refined error rates for tagged bits and untagged bits separately, they can
only use the value  $2^{NH(t_b)}$ as the number of likely bit-flip string for error correction.
What is a bit tricky is the
number of the likely phase-flip string. Also, there are two groups of
bits of the $N$ untested bits, tagged bits and untagged bits. If the number of phase-flip
string for the tagged bits is bounded by $\omega_{ph,tag}$ and the number of phase-flip string for
those untagged bits is bounded $\omega_{ph, untag}$ then the number of phase-flip string for all $N$ bits is
bounded by
\begin{equation}
\omega_{ph} =\omega_{ph, untag}\cdot \omega_{ph,tag}.
\end{equation}
Give any binary string of length $y$, in whatever case the number of
likely string is bounded by $2^{y}$. Therefore we have
\begin{equation}
\omega_{ph,tag}\le 2^{\Delta N}.
\end{equation}
To bound the value of $\omega_{ph, untag}$, we must bound the
phase-flip rate of those {\em untagged} untested bits.
Asymptotically, this value is given by the error rate of those {\em
untagged} test bits in $X$ basis. The observed value $t_x$ is the
{\em averaged} error rate of all those $X$ bits (all the $X$ bits
are used for test) among which only a fraction $(1-\Delta)$ are
untagged bits. Therefore, the separate error rate for the untagged
$X$ bits could actually
 be larger than the averaged one, $t_x$ but it must be bounded by
$t_x/(1-\Delta) $ which corresponds to the worst case that all errors in $X-$basis are carried by those untagged $X$ bits. Therefore
the phase-flip rate of those untagged untested bits are bounded by
\begin{equation}
t_{ph}\le \frac{t_x}{1-\Delta}+\delta_2.
\end{equation}
Here $\delta_2$ is a small-value parameter due to the statistical
fluctuation.
 Therefore the number of likely phase-flip string of all those untested bits is bounded by
\begin{equation}
\omega_{ph} \le 2^{\Delta N}\cdot 2^{(1-\Delta)N H(
\frac{t_x}{1-\Delta} +\delta_2)}.
\end{equation}
 If $N(1-\Delta)$ is very large, $\delta_1,~\delta_2$ are a very small.  Asymptotically, the final key rate
is
\begin{equation}
f=1- H(t_z)-\Delta-(1-\Delta) H( \frac{t_x}{1-\Delta} )\label{dllp}
\end{equation}
which confirms the result of ILM-GLLP\cite{gllp}.

{\em Remark:} The bit-flip and phase-flip for a tagged bit is rather
different. Consider the case with real entanglement. Once Alice
tells Eve in advance the bit value in $Z$ basis of certain pair,
Alice must have already measured it in $Z$ basis. The phase-flip is
immediately very large even Eve does not touch it in the future.
However, the bit-flip can still be 0 given a noiseless channel.

 The above model applies to the important situation that Alice uses the source of weak coherent light, i.e.,
 the
 weak light directly from a laser device.
Say, the source may sometimes produce the multi-photon pulses.
Obviously, to Alice and Bob, the situation here cannot be worse than
the situation of {\em tagged-bit} model where Alice announces some
of the basis (and bit values) therefore it must be secure here if
they use the model of {\em tagged bits} to treat the imperfect
source, provided that they know the value $\Delta$ for the raw bits.
In the case of using the coherent light as the source, if a bit is
created at Bob's side  due to  a multi-photon pulse from Alice, that
bit is regarded as a tagged bit. If the channel is lossy, it is not
a trivial task to know the tight upper bound of the fraction of
tagged bits, $\Delta$. Since we need to assume the channel to be
Eve's channel for security, the channel transmittance can be very
{\em different} for those single-photon pulses (untagged qubits) and
those multi-photon pulses (tagged qubits). Therefore the fraction of
multi-photon pulses of the source can be very different from the
fraction of tagged bits in Bob's raw key. In particular, if the
channel is quite lossy, since Alice doesn't know which pulses are
single-photon pulses, it is possible that the channel transmittance
for single-photon pulses is 0 and {\em all} bits in Bob's raw key
are tagged bits. This means, without a good method to upper bound
$\Delta$ tightly, the maximum secure distance for QKD with weak
coherent light of intensity not less than 0.1 is about only 20
kilometers given the existing detection technology\cite{gllp}.
 \subsubsection{PNS attack}
In practice, the channel can be very lossy. For example, if we want
to do QKD over a distance longer than 100 km using an optical fiber
and the light pulses of wavelength 1.55 ${\rm \mu m}$ , the overall
transmittance can be in the magnitude order of $10^{-3}$ or even
$10^{-4}$. (Suppose the detection efficiency is around 10\%.) This
opens a door for the Eve by the so called photon-number-splitting
(PNS) attack\cite{cvqkd2,bra}
 as shown
in figure (\ref{pns0}). Suppose at a certain time Alice sends out a
multi-photon pulse. Every photon inside the same pulse is in the
same state in coding space, e.g., the polarization space. Eve can
keep one photon from the pulse and sends other photons of the pulse
to Bob through a transparent channel. This action will not cause any
noise in the coding space but Eve may have full information about
Bob's bit: After the measurement basis is announced by Alice or Bob,
Eve will be always able to measure the photon she has kept in the
correct basis. Therefore, here we regard all those bits caused by
multi-photon pulses from Alice as the {\em tagged bits}. According
to the model given by ILM-GLLP\cite{gllp} as we have studied, if the
fraction of tagged bits in Bob's raw key is not too large, we can
still obtain a secure final key by equation (\ref{dllp}). In using
this result, we must first know the value $\Delta$.
Alice does not know which pulse contains more than one photons, she
only knows a distribution over different photon numbers for all
pulses. As we have mentioned, the fraction of multi-photon pulses
from the source can be very  {\em different} from the fraction of
tagged bits in Bob's raw key, for, Eve's channel transmittance can
be dependent on the photon number of the pulse from Alice. Naively,
one can assume the worst case to estimate $\Delta$: the channel
transmittance for multi-photon pulses is 1 and we check how many raw
bits are generated. If the number is larger than the number of
multi-photon pulses, there must be some untagged bits. However, in
such a way, the light intensity must be rather weak in order to
have some untagged bits. Suppose the channel transmittance is $\eta$
and the light pulse intensity is $x$. The density operator of a
phased randomized coherent state of intensity $x$ is
\begin{equation}
\hat \rho_x =e^{-x}\sum \frac{x^n}{n!}|n\rangle\langle
n|.\label{source0}
\end{equation}
Consider the normal case that there is no Eve, the channel
transmittance is $\eta$ to every photon. If Alice sends $N_0$
pulses, Bob will find $N_0(1-e^{-\eta x})$ counts at his side.
However, for security, they have to assume this to be Eve's channel.
Using the naive worst-case estimation, we require
\begin{equation}
1-e^{-\eta x}> e^{-x} x^2/2,
\end{equation}
i.e.,
\begin{equation}
x e^{-x}< 2\eta.
\end{equation}
This shows, to guarantee that not all raw bits are tagged bits,
the efficiency must be bounded by
 $\eta^2$.  Given that $x\ge 0.1$, we request $\eta\ge 4.5\%$ to obtain non-zero secure key.
 For security, we have to assume that Eve can also control the instantaneous detection efficiency
 of Bob's detector. Thus, if Bob's detection efficiency is 10\% and the the light 
intensity loses a half
 over every 15 km, the cut off distance is not larger than 20 km.
 To
obtain the secure final key with a meaningful key rate, we must have
a better way to verify $\Delta$, upper bound of fraction of tagged
bits in raw key.

{\em Remark.} Bob cannot verify the tagged bits at his side by
measuring the photon number in each coming pulses. Suppose he finds
certain pulse contains only one photon. The bit caused by that pulse
could be still a tagged bit because the pulse could have contained
two photons when Alice sent it. {\em It is the photon number in the
pulse at Alice's side that only matters for the security here.}

\begin{figure}
\begin{center}
\epsffile{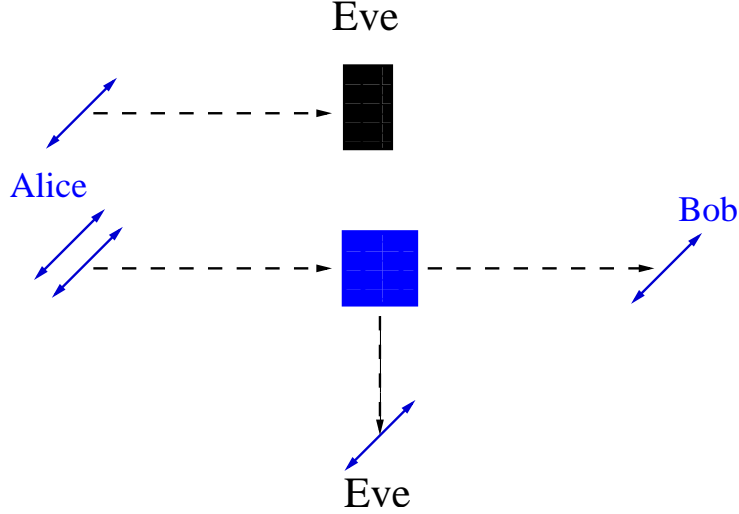}
\end{center}
\caption{ Schematic diagram for photon-number-splitting attack. Eve
is in the middle and controls the channel. If Alice sends a
single-photon pulse (untagged qubit), Eve may absorb it. If Alice
sends a multi-photon pulse, Eve may split it and keeps one photon.
Of courses, Eve has many other choices. } \label{pns0}
\end{figure}
Earlier, the PNS attack has been investigated where Alice and Bob
monitor only how many non-vacuum signals arise, and how many errors
happen. However, it was then shown\cite{kens1} that the
simple-minded method does not guarantee the final security. It is
shown\cite{kens1}  that in a typical parameter regime nothing
changes if one starts to monitor the photon number statistics as Eve
can adapt her strategy to reshape the photon number distribution
such that it becomes Poisonian again. Finding a faithful and tight
upper bound for $\Delta$ value is strongly non-trivial because Eve can produce
whatever type photon-number-dependent  lossy channel that 
does not violate laws of the nature.
Although  some types of {\em specific} PNS attack, e.g., the
beamsplitter attack could be detected by simple method such as
tomography at Bob's side, a method to manage {\em whatever} lossy channel
 is strongly non-trivial.

In short, given ILM-GLLP\cite{gllp} result, the remaining task is to verify the fraction of tagged bits in Bob's raw key faithfully.
A reliable tight verification is non-trivial.
 The central task for the decoy-state method is to
make a tight verification of $\Delta$, the upper bound of
 the fraction of tagged bits or equivalently,
$\Delta_1$, the lower bound of the fraction of untagged bits in Bob's raw key.

{\em Remark:} In the above, in showing that certain old protocol is
{\em insecure}, we have used some specific attacking schemes. Of
course Eve can have many choices in the attack, e.g. methods in
Ref\cite{kens1}. Definitely, in showing that certain protocol is
{\em secure}, one should not assume any specific attacking scheme.
We are now going to present the decoy-state method which does not
assume any specific attacking scheme. The security of this method is
only based on principles of quantum mechanics and classical
statistics therefore is {\em unconditional}.

\subsection{Decoy-state method}

The first idea of decoy-state method and the first protocol
   is given by Hwang\cite{hwang}. Hwang proposed to do the non-trivial verification by changing the intensity of pulses.
In Hwang's first protocol, two intensities are used. The intensity
of signal pulses is set to be around 0.3 and the intensity of
decoy-state pulses is set to be 1. By watching the counting rate of
decoy pulses, one can deduce the upper bound of the fraction of
tagged bits among all those bits caused by signal pulses. For the
conceptual clarity and the mathematical simplicity, here we give up
Hwang's original statement and derivation and we shall directly use
the technique of density matrix convex\cite{wang0d} where only a few
parameters are involved\cite{wang0d}.

\subsubsection{Basic idea and  protocol}
We start from the classical statistics.

{\em Proposition 1.} Given a large number of identical and
independent pulses,  the averaged value per pulse of any physical
quantity  for some randomly sampled pulses  must be (almost) equal
to that of the remaining pulses, if both the number of sampled
pulses and the number of the remaining pulses are sufficiently
large.

In a standard QKD protocol with a perfect single-photon source, this
proposition is used for the error test: They check the error rate of
a random subset, and use this as the error rate of the remaining
bits. Also, this proposition can be used for estimation of the
averaged value of any other physical quantities, such as the {\em
counting rate}.
 In the protocol, Alice sends pulses to Bob. Given a lossy channel,
after a pulse is sent out from Alice,
Bob's detector may click or not click during a certain time window.
If his detector clicks, a raw bit is generated. Counting rate is the ratio
of the number
of Bob's clicks and the number of pulses sent out from Alice.
More specifically, if source $x$ sends out $N_x$ pulses and
Bob's detector clicks $n_x$ times in the appropriate time windows,
the counting rate for pulses from  source $x$ is
\begin{equation}
S_x=\frac{n_x}{N_x}.
\end{equation}
Obviously, this quantity for any real source can be directly
observed in the protocol itself. In the QKD protocol, Alice controls
the source. We shall use the concept of a {\em mixed source}. Source
$X$ and source $Y$ together make a {\em mixed source} if the
following conditions are satisfied: (1)  Each individual pulse has a
probability $p_X$ to be produced from source $X$ and probability
$p_Y$ to be produced from $Y$. (And $p_X + p_Y=1$.) (2) The state of each
individual pulses are independent. (3) Except for the states in the
photon-number space, pulses from source $X$ and pulses from source
$Y$ are indistinguishable by any other physical quantities, e.g.,
the wavelength, the polarization, the transmission path and so on.
In particular, if $X$ and $Y$ make a {\em mixed source} and states
of pulses from each source are identical in photon-number space,
then the counting rate of source $X$ must be equal to that of source
$Y$, since  in such a cases all pulses are identical and pulses from
$X$ can be regarded as the sampled pulses and pulses from $Y$ can be
regarded as the remaining pulses in using {\em proposition 1}.

{\em Proposition 2.} Given that the light pulses from a  mixed
source that contains source $X$ and source $Y$, if $X$ and $Y$
produce the same states in photon-number space, the counting rate
for pulses from source $X$ must be equal to that of source $Y$,
provided that the number of pulses from each source is sufficiently
large.

One intensity we shall use in the protocol is $\mu$. Imagine that
all pulses of this intensity is produced by source $A_\mu$. The state
can be re-written in the following equivalent convex form:
\begin{equation}
\hat \rho_{\mu}= e^{-\mu}|0\rangle\langle0|+\mu
e^{-\mu}|1\rangle\langle 1| +c\rho_c\label{oo}
\end{equation}
and $c=1-e^{-\mu}-\mu e^{-\mu}>0$,
\begin{equation}
\rho_c=\frac{1}{c}\sum_{n=2}^\infty P_\mu(n)|n\rangle\langle
n|\label{coherent}
\end{equation}
and $P_\mu(n)=\frac{e^{-\mu} \mu^{-n}}{n!}$. This convex form shows
that the source sends out 3 types of pulses: sometimes sends out
vacuum, sometimes sends out $|1\rangle\langle 1|$, sometimes sends
out pulses of state $\rho_c$. Therefore, source $A_\mu$ can be
equivalently regarded as 3 sources, $A_{\mu 0}$ producing those
vacuum pulses, $A_{\mu 1}$ producing those single-photon pulses and
$A_{\mu c}$ producing those pulses in state $\rho_c$. Nobody outside
Alice's lab can tell whether Alice has actually used these 3 sources
or  $A_\mu$, the real source. Those of Bob's bits due to the counts
caused by the pulses in state $\rho_c$ from Alice are regarded as
$tagged$ bits. Since we know explicitly the probability of pulses
$\rho_c$ for the source, we shall know the fraction of tagged bits
if we know $s_c$, the {\em counting rate} of state $\rho_c$. The
{\em counting rate} of any state $\rho$ is the probability that
Bob's detector counts whenever Alice sends out a pulse in state
$\rho$. Suppose that Alice has another source, $A'$ which always
produces state $\rho_c$ and Alice sometimes uses $A_\mu$ sometimes
uses $A'$. This is to say, Alice uses a mixed source that contains
$A_\mu$ and $A'$. Since $A_\mu$ itself can be regarded as three
sources, we can also say that Alice has used source $A_{\mu 0}$,
$A_{\mu 1}$ and a mixed source that contains $A_{\mu c}$ and $A'$.
Since the state of pulses from $A_{\mu c}$ is identical to that of $A'$
in photon-number space, according to our {\em proposition 2,} pulses
from source $A_{\mu c}$ and pulses from source $A'$ must have the
same counting rate. This is to say, by watching the counting rate of
pulses from source $A'$, Alice can judge the counting rate of all
those multi-photon pulses from source $A_\mu$. We can have the
following artificial protocol: Alice sends many pulses to Bob with
two sources, $A_\mu,A'$. Among all these pulses, he knows which $N'$
pulses are produced by source $A'$. After Alice sends out all
pulses, Bob announces which time his detector has counted and which
time has not. Alice finds that among all those counts at Bob's side,
$n'$ of them are due to those pulses from source $A'$. Then Alice
knows the counting rate of source $A'$ is $n'/N'$. This is also the
counting rate of source $A_{\mu c}$, which is just the counting rate
of all those multi-photon pulses from source $A_\mu$. Note that Eve
cannot treat the pulses from source $A'$ and the pulses from source
$A_{\mu c}$  differently.

In the above artificial model, we have used source $A'$ 
that only produces state $\rho_c$. In practice we don't have such a source.
But we can have another coherent light source, $A_{\mu'}$ which produces a coherent 
state with averaged photon number
(intensity) $\mu'$ ($\mu'>\mu, \mu'e^{-\mu'}>\mu e^{-\mu}$).
We now consider the case that Alice uses a real mixed source that contains source $A_\mu$ 
which produces  the coherent
state with averaged photon number (intensity) $\mu$ and source $A_{\mu'}$.
 Alice can use only one laser
device to produce such a mixed source by randomly switching the intensity between $\mu,\mu'$.
Since $\mu'>\mu$ and $\mu' e^{-\mu'}>\mu e^{-\mu}$, the state for source
$A_{\mu'}$ can be written in the convex form:
\begin{equation}
\hat\rho_{\mu'}=e^{-\mu'}|0\rangle\langle0|+\mu'
e^{-\mu'}|1\rangle\langle 1| +c\frac{\mu'^2 e^{-\mu'}}{\mu^2
e^{-\mu}}\rho_c + d\rho_d\label{dd}.
\end{equation}
Here $d\ge 0$ and $\rho_d$ is a density operator. We don't need the
explicit formula for $\rho_d$, we shall only need the fact that
state $\rho_{\mu'}$ can be written in the above convex form. Source
$A_{\mu'}$ can be equivalently regarded as  4  virtual sources:
$A_{\mu'0}$ which contains all vacuum pulses from $A_{\mu'}$,
$A_{\mu'1}$ which contains all single-photon pulses from $A_{\mu'}$,
$A_{\mu' c}$ which contains all $\rho_c$ pulses of  source
$A_{\mu'}$ and $A_{\mu' d}$ that contains all $\rho_d$ pulses of
source $A_{\mu'}$. Also, source $A_\mu$ can be equivalently regarded
as 3 sources as we have mentioned before. Of course, $A_{\mu' c}$
and source $A_{\mu c}$ make a mixed source and they produce
identical states in photon number space. This means, by proposition 2, $s_c(\mu)$, the counting
rate of state $\rho_c$ from source $A_\mu$ is equal to $s_c(\mu')$, the counting
rate of state $\rho_c$ from source $A_{\mu'}$, i.e.,
\begin{equation}
s_c(\mu)= s_c(\mu')=s_c.
\end{equation}
 By the same reason we have a more general formula for counting rate
\begin{equation}\label{asyeq19}
s_\alpha (\mu) = s_\alpha (\mu') = s_\alpha.
\end{equation}
Here the subscript $\alpha$ represents for a state.  We shall use
$\alpha = 0,1,c$ for (the counting rates of) vacuum state,
single-photon state and state $\rho_c$ respectively.
\\
{\em Remark:}  Source $A_\mu$ and $A_{\mu'}$ are not identical and
Eve can treat pulses from these two sources differently. But source
$A_{\mu\alpha}$ and $A_{\mu'\alpha}$ are identical and Eve
cannot treat pulses from these two sources differently.

Here Alice has only used two real sources, $A_{\mu},A_{\mu'}$.
Counting rate of any real source can be observed directly in the
protocol. But counting rate of a virtual source e.g., $A_{\mu c}$
cannot be observed directly. We must deduce it mathematically based
on the {\em observed} results of the protocol.

Alice has no way to know which pulses are from source $A_{\mu' \alpha}$ or source $A_{\mu \alpha}$ . But she knows
which pulses are from source $A_{\mu'}$ and which ones from $A_\mu$. 
We shall show that she can know an
{\em upper bound} of the fraction of tagged bits from source $A_{\mu c}$
by watching the counting rate of source  $ A_{\mu'}$.

We denote the counting rates for source $A_{\mu'}$ and  source $A_\mu$ by  $S_{\mu'}, S_\mu$ respectively.
These values can be observed directly:
\begin{equation}
S_\mu =\frac{n_\mu}{N_\mu}  ,  ~S_{\mu'} =\frac{n_{\mu'}}{N_{\mu'}}.
\end{equation}
$N_\mu,N_{\mu'}$ are number of pulses sent out from source $A_\mu,A_{\mu'}$ respectively; $n_{\mu},n_{\mu'}$ are
number of clicks  of Bob's detector corresponding to pulses  from $A_\mu,A_{\mu'}$ respectively.
According to equation (\ref{dd}),
we have the following equation:
\begin{equation}
S_{\mu'}= e^{-\mu'}s_0+\mu' e^{-\mu'}s_1 + c\frac{\mu'^2 e^{-\mu'}}
{\mu^2 e^{-\mu}}s_c +ds_d.\label{origin}
  \end{equation}
Here we denote  $s_0,~s_1$ for the counting rates of vacuum pulses
and single-photon pulses from $A_{\mu'}$,  i.e. source $A_{\mu' 0}$
and $A_{\mu' 1}$ , respectively, $s_c$ for counting rate of $\rho_c$
pulses from $A_{\mu'}$, i.e., source $A_{\mu' c}$ and $s_d$ for
counting rate of $\rho_d$ pulses from $A_{\mu'}$, i.e., source
$A_{\mu' d}$. Given the fact $s_0\ge 0$, $s_1\ge 0$ and $s_d\ge 0$,
we transform eq(\ref{origin}) into an inequality for the upper bound
of $s_c$:
\begin{equation}
s_c \le
\frac{\mu^2e^{-\mu}}{c\mu'^2e^{-\mu'}}S_{\mu'}\label{crude}.
\end{equation}
 This is the bound value for counting rate of source $A_{\mu' c}$. This
is {\em also} the bound value for pulses from any source that
produces $\rho_c$ states only, including the source $A_{\mu c}$,
i.e., those $\rho_c$ pulses from source $A_\mu$.  Therefore we have
the following upper bound for the fraction of tagged bits for source
$A_{\mu}$:
\begin{equation}
\Delta \le \frac{\mu^2e^{-\mu}S_{\mu'}}{\mu'^2e^{-\mu'}S_{\mu}},
\end{equation}
and we have used
\begin{equation}
\Delta=c\frac{s_c}{S_{\mu}}\label{new}.
\end{equation}
In the normal case that there is no
Eve's attack, Alice and Bob will find
$
S_{\mu'}/S_{\mu}=\frac{1-e^{-\eta\mu' }}{1-e^{-\eta\mu}}= \mu'/\mu
$ in their protocol therefore they can verify
$
 \Delta \le \frac{\mu e^{-\mu}}{\mu' e^{-\mu'}}
,$
which is just eq.(13) of Hwang's work\cite{hwang}.

The above is the main result of Hwang's work. We have simplified the
original derivation given by Hwang\cite{hwang}. In short, Hwang's
protocol works in this way:  By watching the counting rate of decoy
state (intensity $\mu'$), we can obtain $\Delta$ value for the
signal state (intensity $\mu$).

{\em Remark}: Here a tricky point is that  Eve cannot treat the
pulses from $A_{\mu \alpha}$  and the pulses from $A_{\mu' \alpha}$
($\alpha = 0,1,c$) differently although she can treat the pulses
from $A_\mu$ and the pulses from $A_{\mu'}$ somehow differently.
Lets consider a similar classical story. There are professional
basketball players and football players. We mix them and let them
 pass through a gate controlled by Eve. Eve knows that each one
must be either a basketball player or a football player. Eve can
partially distinguish each one's profession by his height: the
taller ones are more likely to be basketball players. She can then
produce a different transmission rate for the two types of players.
For example, if she wants football players' transmission rate to be
higher, she only needs to block all those taller guys. Consider a
subgroup of football players $G_f$ and subgroup of basketball
players $G_b$. People in these subgroups are all in the same hight,
1.80 m. Eve cannot treat people in these two subgroups differently
according to his profession. Say, given whatever strategy Eve may
have used, if $x\%$ of guys in group $G_f$ are blocked, there must
be also $x\%$ guys in subgroup $G_b$ being blocked. Here, all
football players and all basketball players are in the same role of
pulses from source $A_\mu$ and pulses from source $A_{\mu'}$ in the
decoy-state protocol. Players in subgroup $G_b$ or subgroup $G_f$
only are in the same role of pulses from source $A_{\mu \alpha}$ or
source $A_{\mu' \alpha}$ in the decoy-state protocol.
\subsubsection{The issue of unconditional security}
The security of Hwang's method is a direct consequence of the
separate prior art result of ILM-GLLP\cite{gllp} therefore a
separate security proof is not necessary. ILM-GLLP\cite{gllp} have
offered methods to distill the unconditionally secure final key from
raw key if the upper bound of fraction of tagged bits is known,
given whatever imperfect source and channel. Decoy-state method
verifies such an upper bound for coherent-state source. We can
consider an analogy using the model of pure water distillation: Our
task is to distill pure water by heating from raw water that may
contain certain poison constitute. Suppose it is known that the
poison constitute will be evaporated by heating. We want to know
how long the heating is needed to obtain the pure water for certain.
If we blindly heat the raw water for too long, all raw water will be
evaporated and we obtain nothing. If we heat the raw water for too
short a period, the water could be still poisonous.
 ``ILM-GLLP'' finds an explicit formula for the heating time which is  dependent on
 the upper bound of the fraction
of poison constitute. They have proven that one can always obtain pure water if we use that 
formula for the heating time.
However, the formula itself does not tell us how to examine the fraction of poison constitute. 
``Decoy-state'' method is a method to verify a
tight
upper bound of the fraction of the poison constitute. It is guaranteed by the classical statistical principle that the
verified upper bound by ``decoy-state method'' is (always) larger than the true value. Using this analog, the next question is how to
obtain a tighter upper bound: if the verified value over estimates too much, it is secure but it is inefficient.

Hwang's result is a large step towards the efficient and secure QKD
with existing setup. However, the result can be further improved for efficient 
application in practice. The estimated $\Delta$ value here is still too large.
Given such a value, one cannot obtain a meaningful key rate in
practice for long distance QKD with existing setups. We want a
faithful and {\em tighter} estimation. We want a way to obtain a
value that is only a bit larger than the true value in the normal
case that there is no Eve (for efficiency), {\em and} it is  {\em
always}  larger than the true value in whatever case (for security).
The improved result of a tightened estimation of $\Delta$ including
both the analytical formula in the asymptotical case and the
numerical calculation in the non-asymptotic case are then given by
Wang\cite{wang0d} and later confirmed and further studied by others\cite{lolo1}.
\subsubsection{Improved decoy-state method}
The improvement is possible because Hwang's method has not
sufficiently used the information of counts from different
intensities. Actually, in doing the verification, Alice has only
used the counting rate of one intensity, the source $A_{\mu' }$. It
should be interesting to consider the case that Alice uses more
intensities and considers the observed results {\em jointly}.

The decoy-state method is not immediately useful in practice until
the more tightened estimation is given\cite{wang0d}.
There\cite{wang0d}, the counting rates of different intensities are
treated jointly with non-trivial inequalities through the density
operator convex technique. Quantitative results including both
explicit formulas and numerical results are also given
\cite{wang0d}. There\cite{wang0d}, for the first time an
explicit formula for the tight upper bound of $\Delta$ is given and
the statistical fluctuation is studied with detailed numerical
results are presented. Earlier, a review on Hwang's result with some
rough ideas were presented\cite{lo4}. However, there is no
quantitative result though it proposed to use vacuum to test dark
count and very weak coherent state to test single-photon counting
rate.
Naturally, one can expect a higher key rate
if one uses more intensities. The key rate in the limit of using infinite
intensities is studied in \cite{lolo}.

Here we are most interested in a protocol that is {\em practically}
efficient. Obviously, there are several criteria for a practically
efficient protocol. 1. The protocol must be clearly stated. For
example, there should be {\em quantitative} description about the
intensities used and {\em quantitative} result about the
verification. Because we need the $explicit$ information of
intensities in the implementation and the $explicit$ value of
$\Delta$ for key distillation. 2. The result of verified value
$\Delta$ should be tight in the normal case when there is no Eve.
This criterion is to guarantee a good final key rate. 3. It should
only use a few different intensities. In practice, it is impossible
to switch the intensity among infinite number of different values.
4. It should be robust to possible statistical fluctuations. Note
that the  counting rates are very small parameters. The effects of
possible statistical fluctuations can be very important because the
repetition rate of any real system is limited hence we cannot assume too
large the number of pulses.

Concerning the above criteria, tightened estimation of $\Delta$
value is then obtained\cite{wang0d} through jointly using the
information of counting rates of 3 intensities, vacuum, $\mu$ and
$\mu'$. We call this protocol as 3-intensity protocol\cite{wang0d}.
For convenience, we shall always assume
\begin{equation}
\mu'>\mu; \mu' e^{-\mu'} > \mu e^{-\mu} \label{condition}
\end{equation} in this paper.
Since we randomly change the intensities among 3 values, we can
regard it as a mixture of 3 sources. Source $A_0$ that produces
vacuum pulses, source $A_\mu$ that produces coherent-state pulses of
intensity $\mu$ and source $A_{\mu'}$ that produces coherent-state
pulses of intensity $\mu'$. States from source $A_\mu$ and
$A_{\mu'}$ are given by Eq.(\ref{oo}) and Eq.(\ref{dd}),
respectively. In the protocol, they can directly watch the counting
rates of each source of $A_0,A,A_{\mu'}$ as we have stated before,
i.e., $S_\mu,S_{\mu'}$ are shall be regarded as {\em known}
parameters. Suppose they find $S_0,S_\mu,S_{\mu'}$ for each of them.
In the asymptotic case, we have the following equations:
\begin{equation}
S_{\mu}= e^{-\mu}S_0 + \mu e^{-\mu}s_1 + cs_c \label{originmu}
  \end{equation}
\begin{equation}
S_{\mu'}= e^{-\mu'}S_0 + \mu' e^{-\mu'}s_1 + c\frac{\mu'^2
e^{-\mu'}} {\mu^2 e^{-\mu}}s_c +ds_d \label{originp}
  \end{equation}
In the above we have used the same notations $S_0,s_1,s_c$ in both
equations. This is because we have assumed that the counting rates
of the same state from different sources are equal.
$S_0,S_\mu,S_{\mu'}$ are known, $s_1$ and $s_d$ are unknown, but
they are never less than 0. Therefore setting $s_d, s_1$ to be zero
we can obtain the following crude result by using Eq.(\ref{originp})
alone.
\begin{equation}
cs_c\le \frac{\mu^2e^{-\mu}}{\mu'^2e^{-\mu'}}\left(S_{\mu'}-
e^{-\mu'}S_0-\mu' e^{-\mu'}s_1\right). \label{origin8}
\end{equation}
Since $s_1\ge 0$, we obtain a crude result for the upper bound of $s_c$
\begin{equation}
cs_c\le \frac{\mu^2e^{-\mu}}{\mu'^2e^{-\mu'}}\left(S_{\mu'}-
e^{-\mu'}S_0\right). \label{origin9}
\end{equation}
However, we can further tighten the verification by using
Eq.(\ref{originmu}). Having obtained the crude result above,
 we now show that the verification can be done
more sophisticatedly and one can further tighten the bound significantly.
In obtaining inequality (\ref{origin9}), we have dropped terms $s_1$ and $s_d$, since
we only have trivial knowledge about $s_1$ and $s_d$ there,
i.e., $s_1\ge 0$ and $s_d\ge 0$ then. Therefore, inequality(\ref{origin9}) has no advantage to Hwang's result at that moment.
 However, after we have obtained
the crude upper bound of $s_c$, we can have a larger-than-0 lower
bound for $s_1$ by Eq.(\ref{originmu}), provided that our crude
upper bound for $s_c$ given by Eq.(\ref{origin9}) is not too large.
 Combining the crude upper bound for $s_c$ given
by Eq.(\ref{origin9}) and Eq.(\ref{originmu}), we have the
non-trivial lower bound for $s_1$ now:
\begin{equation}
s_1 \ge S_{\mu}-e^{-\mu}S_0 - c s_c
 > 0. \label{new1}
\end{equation}
With this new lower-bound of $s_1$, we can further tighten the
upper-bound of $s_c$ by Eq.(\ref{origin8}) and obtain a more
tightened $s_c$. With the new $s_c$, we can again raise the lower
bound of $s_1$ by Eq.(\ref{originmu}). The final bound values are
determined by infinite  iterations. Therefore tight values for $s_c$
and $s_1$ can be obtained by solving the simultaneous constraints of
equation (\ref{originmu}) and  inequality (\ref{origin8}). We obtain
\begin{equation}\label{link}
s_1\ge \frac{a_c'{\mathcal S}- a_c {\mathcal
S'}}{a_1a_c'-a_1'a_c},~s_c\le\frac{a_1'{\mathcal S}-a_1{\mathcal
S'}}{a_1'a_c-a_1a_c'}
\end{equation}
and ${\mathcal S}= S_\mu-e^{-\mu}S_0,~ {\mathcal S'
=S_{\mu'}-e^{-\mu'}S_0},~ a_1=\mu e^{-\mu}, ~a_1'=\mu'
e^{-\mu'},~a_c=c,~a_c'=\frac{c\mu^2e^{-\mu'}}{\mu'^2e^{-\mu}}.$
Given these, we can easily find the lower bounds and upper bounds
for the fraction of untagged bits and tagged bits for both intensity
$\mu$  and $\mu'$. In general, raw bits caused by both intensities
can be used for the final key distillation. For the intensity $\mu$,
we have
\begin{equation}
\Delta_1 = \frac{\mu e^{-\mu} s_1}{S_\mu}, ~ \Delta =
\frac{cs_c}{S_\mu}
\end{equation}
for the fraction of untagged bits and tagged bits, respectively; for
the intensity $\mu'$, we have
\begin{equation}\label{mains}
\Delta_1' = \frac{\mu' e^{-\mu'} s_1}{S_{\mu'}}, ~ \Delta' =
\frac{a_c's_c}{S_{\mu'}}
\end{equation}
for the fraction of untagged bits and tagged bits, respectively.
 For example,
\begin{equation}
\Delta \le
\frac{\mu}{\mu'-\mu}\left(\frac{\mu e^{-\mu} S_{\mu'}}{\mu' e^{-\mu'} S_{\mu}}-1\right)
+\frac{\mu e^{-\mu}s_0 }{ \mu' S_\mu }.\label{assym}
\end{equation}
Also, through the fact of
\begin{equation}
\Delta_0+\Delta_1+\Delta=1
\end{equation}
we have \begin{equation} \Delta_1=
1-\Delta-e^{-\mu}s_0/S_{\mu}\label{fractionsingle}
\end{equation}
and $\Delta_0$ is the fraction of vacuum counts.
 In the case of $s_0<<\eta$, if there is no Eve,
$S_\mu'/S_\mu=\mu'/\mu$. Alice and Bob must be able to verify
\begin{equation}
\Delta = \left. \frac{\mu \left(e^{\mu'-\mu}-1\right)}{\mu'-\mu}\right|_{\mu'-\mu\rightarrow 0}=\mu
\end{equation}
in the protocol.
Of course, other parameters such as $\Delta_1',\Delta'$ can also be
calculated explicitly by Eq.(\ref{link},\ref{mains}). Given these,
we can distill the final key from both intensities. The values of
$\mu,\mu'$ should be chosen in a reasonable range, e.g., from 0.2 to
0.6. Actually, we can distill more final bits from raw bits created
by intensity $\mu'$ than that of $\mu$, we shall call pulses of
intensity $\mu'$ as the signal pulses and pulses of intensity $\mu$
as decoy pulse for simplicity.

{\em Summary:} The 3-intensity decoy-state protocol is stated by the following:
Alice switches the intensity of each pulses randomly among 3 values, 0, $\mu\sim 0.2$ and $\mu'\sim 0.5$.
Suppose she sends out $N_0$ pulses of intensity 0, $N_\mu$ pulses of intensity $\mu$ and $N_{\mu'}$ pulses of
intensity $\mu'$. Bob then announces which pulses have caused a click. According to Bob's announcement, Alice knows
that the pulses of intensities 0, $\mu$ and $\mu'$ have caused $n_0,n_\mu,n_{\mu'}$ clicks at Bob's side.
Alice uses $S_0=\frac{n_0}{N_0}$, $S_\mu= \frac{n_\mu}{N_\mu}$ and $S_{\mu'}=\frac{n_{\mu'}}{N_{\mu'}}$ as the input
of eq.(\ref{assym}) and obtain the value $\Delta$. She can also obtain the value of other parameters such as 
 $\Delta'$, $\Delta_1$, and $\Delta_1'$ based on this. Raw bits caused by both intensities of $\mu$ and $\mu'$ can be used 
 for the final key distillation, if the key rate is larger than 0.
\subsubsection{Statistical fluctuations}
The results above are only for the asymptotic case. Before applying
the decoy-state method in practice, we must first resolve the very
important security problem related to the statistical fluctuation.
In practice, the number of pulses are finite thus there are
statistical fluctuations, i.e., Eve has non-negligibly small
probability to treat the pulses from different sources a little bit
differently, even though the pulses are in the same state. This
problem was first proposed and solved in \cite{wang0d} and then
further studied\cite{lolo1,wang2d}. Mathematically, this can be
stated by
\begin{equation}
s_{\alpha}(\mu)=(1+r_\alpha)s_\alpha(\mu')
\end{equation}
and the real number $r_\alpha$ is the relative statistical
fluctuation for $s_\alpha$, $\alpha=0,1,c$.  It is $insecure$ if we
simply use the asymptotic result in practice. Since the actual
values are actually different from what we have estimated from the
observed data. Our task remaining is to verify a tight upper bound
of $\Delta$ and the probability that the real value of $\Delta$
breaks the verified upper bound is exponentially close to 0.

The counting rate of any state $\rho$ from different sources
now can be slightly different with a non-negligible
probability. We shall use the primed
notation for the counting rate of any state from source $A_{\mu'}$ and the original notation
for the counting rate 0f any state from source $A$.
Explicitly, constraints of Eq(\ref{crude},\ref{new1})
are now converted to
\begin{equation}
  \left\{ \begin{array}{l}
e^{-\mu}s_0 + \mu e^{-\mu}s_1 + c s_c=S_{\mu},
 \\ cs'_{c}\le \frac{\mu^2e^{-\mu}}{\mu'^2e^{-\mu'}}\left(S_{\mu'}
- \mu' e^{-\mu'} s'_1
- e^{-\mu'}s'_0\right) .
  \end{array}
  \right) \label{couple}
 \end{equation}
with setting \begin{equation}s_\alpha = (1+r_\alpha)s_\alpha'\end{equation} for $\alpha=0,1,c$. For security, we need to seek
the worst-case solution of $s_1'$ or $s_c'$ of the above equations over all possible $r_\alpha$.
As shown in ref[161], the maximum values of $\{r_\alpha\}$ lead to the smallest solution of $s_1'$ or
largest solution of $s_c'$ in the equations above. What is the reasonable largest value of $r_\alpha$ ?
We can figure out the issue by classical statistics. We regard $r_{\alpha M}$ as the upper bound of $r_\alpha$
if the probability that $r_\alpha>r_{\alpha M}$ is exponentially close to 0.

Given $N_1+N_2$ copies of state $\rho$,  suppose the counting rate
for $N_1$ randomly chosen states is $s_{\rho}$ and the counting rate
for the remaining states  is $s'_{\rho}$, the probability that
$|s_\rho-s'_\rho|>\delta_\rho$ is less than
$\exp\left(-O{\delta_\rho}^2N_0/s_\rho\right)$ and $N_0 ={\rm
Min}(N_1,N_2)$. Now we consider the difference of counting rates for
the same state from different sores, $A$ and $A_{\mu'}$.
 To make a faithful estimation
with exponential certainty, we require
${\delta_\rho}^2N_0/s_\rho =100$. This causes a relative fluctuation
\begin{equation}
r_\rho=\frac{\delta_\rho}{s_{\rho}}\le 10\sqrt{\frac{1}{s_{\rho}N_0}}\label{statis}.
\end{equation}
The probability of violation is less than $e^{-O(100)}$.
 To formulate the relative fluctuation $r_1,r_c$
by $s_c$ and  $s_1$, we only need to check the number of pulses in
states $\rho_c$, $|1\rangle\langle 1|$  in each sources in the
protocol. That is, using eq.(\ref{statis}), we can replace $r_1,r_c$
in eq.(\ref{couple}) by $10e^{\mu/2}\sqrt{\frac{1}{\mu s_1N}}$,
$10\sqrt{\frac{1}{c s_cN}}$, respectively and $N$ is the number of
pulses in source $A$.
 With these inputs, eq.(\ref{couple}) can now be solved
numerically for the largest value of $s_c$ in the likely range of
statistical fluctuation, i.e., the fluctuation beyond the assumed
range is in the magnitude order of $e^{-O(100)}$. Good numerical
results for a tighten estimation of $\Delta$ value have
obtained\cite{wang0d} by many parameter settings based on existing
technology\cite{GYS04,YS05}. For example, given
$\mu=0.3,\mu'=0.43$,and $\eta=10^{-3}$, we obtain $\Delta=34.4\%$
which is greatly less than Hwang's asymptotic result $60.4\%$,
though it is still a bit larger than the true value, $25.9\%$.
Definitely, we can also replace $s_\alpha$ by
$s_\alpha'(1+r_\alpha)$ ($\alpha=0,1,c$) in Eq.(\ref{couple}) and
then find out the worst-case result for $s_1',~s_c'$ numerically.
The fraction of multi-photon counts for pulses of intensity $\mu'$
have also been tightly verified\cite{wang0d}.

\subsubsection{Robustness with respect to small errors}
We now study how robust the method is.
In the protocol, we use different intensities.
In practice, there are both statistical fluctuations and small operational errors in switching
the intensity.
We shall show that, by using the counting rates of 3 intensities, one can still verify
tight bounds even we take all theses errors and fluctuations into consideration.

There are small operational errors inevitably. Say, in setting the
intensity of any light pulse, the actual intensity can be slightly
different from the one we have assumed. There are also fluctuation
to the photon number distribution for each intensities\cite{harry}.

At any time Alice decides to set the intensity of the pulse to be
$\mu$ or $\mu'$, the actual intensity could be $\mu_i$, $\mu_i'$
which can be a bit different from $\mu$ or $\mu'$. The intensity
errors of different pulses can be {\em correlated}. Due to this
possible correlation, neither the decoy pulses nor the signal pulses
are {\em independent}, the state of decoy pulses or signal pulses
cannot be simply represented by a {\em single-pulse} density
operator as described by Eq.(\ref{source0}).
With the correlated intensity error, the pulses from class $A_{\mu
\alpha}$ and class $A_{\mu' \alpha}$ ($\alpha=0,1,c$) are actually
{\em not} randomly mixed in the protocol as shown below  therefore
the conditions for the propositions of decoy-state method are not
satisfied. For example, the intensity can be dependent on the
temperature. In a certain interval, all pulses can be brighter or
darker than the supposed value. Consider an extreme example. Suppose
the actual intensity of each pulse is $10\%$ larger than the
supposed one in the first half of quantum-state transmission, $10\%$
lower than the supposed one in the second half of the transmission.
 If Eve's channel transmittance is $4t$ during the first half of
pulse transmission and $t$ during the second half of pulse
transmission, $s_1$, the counting rates of pulses from source
$A_{\mu1}$ and $s_1'$, that of class $A_{\mu'1}$ can be calculated
by: $ \frac{4t\times 1.1 x e^{-1.1x}+ t\times 0.9x e^{-0.9x}}{ 1.1 x
e^{-1.1x}+ 0.9x e^{-0.9x}} $ with the setting of $x=\mu, \mu'$.  For
$\mu=0.2,~\mu'=0.6$, we find $s_1=1.023s_1'$ rather than $s_1=s_1'$.
   This has clearly shown that
 Proposition 2 cannot be blindly  used because here because $A_{\mu1}$ and $A_{\mu'1}$ are {\em not}
 randomly mixed, although they contain the same quantum states.
 The ``randomly mixing" condition for two sources $A_{\mu \alpha},~A_{\mu' \alpha}$
  requires the following condition:
  For any pulse sent out from Alice, if it belongs to $A_{\mu \alpha}\cup A_{\mu' \alpha}$, the
probability that the pulse belongs to $A_{\mu \alpha}$ and the
probability that the pulse belongs to $A_{\mu' \alpha}$ must be
constant throughout the quantum communication stage of QKD. But the
correlated intensity error can violate this requirement.

 The problems listed above can be overcome in various ways\cite{incontrol1,incontrol2,incontrol3}.
First, there is a theory\cite{incontrol1} for whatever error pattern
provided that the largest intensity error of a single pulse is not
too large. But in such a case, the final key rate drops drastically
with the intensity error. As is shown in Ref.\cite{incontrol2}, if
the intensity errors are random and independent, even though there are large errors,
such as 20\% fluctuation, the key rate is almost the same with the
ideal case where the intensity is controlled exactly. The efficiency
becomes quite good here because the linear terms of the fluctuation
disappear and only the quadratic terms of the fluctuation take
effect in the protocol. To make sure that the intensity error of
each pulse is random, we can use, for example, the feed forward
control demonstrated under another topic\cite{zeinature}. If we are
not sure of the error pattern, we can also use the method proposed
in Ref.\cite{incontrol3}, where we request that at each time, a
father pulse is produced and then attenuated by a two-value
attenuator to create a decoy pulse or a signal pulse. This method
works for whatever error pattern and also the key rate is quite
good. Say, if the intensity error is bounded by $5\%$, to reach the
same key rate of the ideal protocol, the distance given by the
method of Ref.\cite{incontrol3} is only 1 km shorter than that of
the ideal protocol where pulse intensities are controlled exactly.

Due to the small operational error, the intensity of light pulses in
source $A_0$ could be slightly larger than 0. This doesn't matter
because a little bit over estimation of the vacuum count will only
decrease the efficiency a little bit but not at all undermine the
security\cite{incontrol2}. (Eq.(\ref{assym}) shows that
overestimation of dark counts will lead to overestimation of
$\Delta$ value.)
 Therefore
we don't care about the operational error of this part. Say, given
$n_0$ counts for all the pulses from source $A_0$, asymptotically,
we can simply assume the tested vacuum counting rate to be
$s_0=n_0/N_0$, though we know that the actual value of vacuum
counting rate is less than this.
\subsubsection{Final key rate and further studies}
Given the methods to verify $\Delta$, the upper bound of fraction of
tagged bits or $\Delta_1$, the lower bound of fraction of untagged
bits, we can calculate the final key rate by Eq.(\ref{dllp}).
However, there are more efficient formulas. As it is pointed out in
Ref\cite{lolo}, actually, one need only correct the phase-flip
errors of single-photon counts and remove tagged bits for privacy
amplification. In particular, the following formula is
recommended\cite{lolo,lolo1} for key rate of any intensity $x$:
\begin{equation}
Q_1(x)=\Delta_1(x)+\Delta_0(x)-H(E(x))-\Delta_1(x)H(e_1).
\end{equation}
Here $\Delta_1(x)$ and $\Delta_0(x)$ are fractions of single-photon counts and vacuum counts
respectively for the
intensity $x$ which can be either $\mu$ or $\mu'$; $E(x)$ is the
observed error rate of bits caused by source $A_x$ and $e_1$ is the
error-rate of counts caused by single-photon pulses. This formula
gives a higher key rate than that of Eq.(\ref{dllp}). Given $\Delta$
value or $\Delta_1(x)$ value,  $e_1$ can be estimated
efficiently. We only need to derive the $e_1$ value. We
can do so by using the weaker source\cite{lolo1}, intensity $\mu$.
Obviously, if the observed total error rate is $E(\mu)$ for source
$\mu$, then $e_1$ can be calculated by
\begin{equation}
e_1= \frac{E(\mu)-\frac{e^{-\mu}S_0}{2S_\mu}}{\Delta_1(\mu)}.
\end{equation}

After the major works presented in\cite{hwang,wang0d,lolo}, the decoy-state method has been further studied[158-171,277].
For example, Harrington {\em et al.}\cite{harry} numerically studied the effect of fluctuation of the state itself. Ref.\cite{wang2d}
proposed a 4-state protocol: using 3 of them to make optimized
verification and using the other one $\mu_s$ as the main signal pulses. This is because, if we want to optimize the verification
of $\Delta$ value, $\mu,\mu'$ cannot be chosen freely. Therefore we use another intensity $\mu_s$ to optimize the final key rate.
It is shown numerically on how to choose the intensity for the main signal pulses ($\mu_s$) and good key rates are obtained in a number
of specific conditions.
The results of final key rates\cite{wang0d,wang2d} show that good
key rate can be obtained even the channel transmittance is around
$10^{-4}$. There are even improved formulas for a higher key
rate\cite{lonothing,koashid}. This corresponds to a distance of
120-150 kilometers for practical QKD with coherent states. Using
two-way classical communication\cite{twogl} in the final key
distillation can further increase the QKD distance\cite{ma2w}. The
theory of decoy-state has now been extensively demonstrated by a
number of experiments\cite{Lo06,peng,ron,yuane}. The decoy-state
method also applies to other types of source, e.g., the
parametric-down-conversion source\cite{decoypdc}.
\subsubsection{Summary}
 Given the result of ILM-GLLP\cite{gllp}, one knows how to distill the
secure final key if he knows the fraction of tagged bits. The
purpose of decoy state method is to do a tight verification of the
the fraction of tagged bits. The main idea of decoy-state method is
to use different intensities of source light and one can verify the
fraction of tagged bits of certain intensity by watching the the
counting rates of pulses of different intensities\cite{hwang}. With
the mathematical technique of density operator convex and jointly
treating the counting rates of different intensities with
non-trivial inequalities\cite{wang0d}, the upper bound of fraction
of tagged bits or the lower bound of the fraction of untagged bits
can be verified so tightly\cite{wang0d} that the decoy-state method
is immediately implementable with the existing matured technology.
Since the counting rates are small quantities, and in any real
setup, the number of different intensities and pulses are limited,
the effect of statistical fluctuation is very
important\cite{wang0d}. It has been shown that the decoy-state
method  can work in practice even with the fluctuations and other
errors\cite{wang0d}. If one uses infinite number of different
intensities and each intensity consists of infinite pulses, one can
actually verify the $\Delta$ value perfectly\cite{lolo}. The
decoy-state method has promised a distance of 120-150 kilometers for
practical QKD\cite{wang0d,wang2d}. To further raise the distance, we
need to improve the existing technologies, this includes decreasing
the dark counts, raising the detection efficiency and the system
repetition rate.

{\em Remark:} Although decoy-state method is promising for the
practical quantum key distribution, it is not the only choice. Other
promising methods for practical QKD include the method using strong
reference light\cite{srl}, the mixed B92 protocol\cite{scar} (i.e.,
SARG04 protocol ),  and so on. For those earlier protocols and
implementations, one may refer to the excellent review on QKD
presented by Gisin, Ribordy, Tittle and Zbinden\cite{GRTZ02}. For a
review of most recent developments of QKD theory, one may refer to
Ref\cite{norbertrc}.
\subsection{SARG04 protocol}
This protocol wa proposed by Sacrani, Aci, Rigbord and Gisin in
2004\cite{scar}, and is called SARG04 protocol. The protocol uses
BB84 states but the bit value of each qubit is represented by its
preparation basis, say, $Z$ basis for 0 and $X$ basis for one.
Consider the B92\cite{srl} protocol first. Alice may send Bob either
state $|0\rangle$ in $Z$ basis or state $|+\rangle$ in $X$ basis.
Bob will measure  each qubit in either $X$ basis or $Z$ basis and he
has a probability 1/4 to obtain a conclusive result. For example,
suppose Alice has sent him a state $|0\rangle$. If Bob measures it
in $Z$ basis, he will be surely obtain $|0\rangle$ and they will
discard this data, because both $|0\rangle$ and $|+\rangle$ can lead
to this result and Bob does not know whether the bit value is 0 or
1. However, if he measures it in $X$ basis, he could obtain a
conclusive result. If he obtain $|+\rangle$, they will discard the
data because Bob does not know the bit value. But if he obtain
$|-\rangle$ which is orthogonal to $|+\rangle$, Bob concludes that
the qubit must have been prepared in state $|0\rangle$ by Alice. In
summary, Bob changes his measurement bases between $Z$ and $X$
randomly and those
 measurement results of $|1\rangle$ and $|-\rangle$ indicate
conclusive bit values. They will only use those conclusive bit values.

There is a security drawback of B92 protocol given a lossy channel, though it is unconditionally secure if the lossy rate is lower than
certain threshold\cite{tkn92}. Obviously, if the channel transmittance is less than 1/4, it cannot be secure because Eve may
measure each qubit in the middle in either $Z$ or $X$, with 1/4 probability, she obtains a conclusive result and she re-produces
Alice's state and sends it to Bob through a transparent channel; with 3/4 probability the result is inconclusive, she blocks it and
pretends her action to be channel loss.

SARG04 has developed the B92 protocol and avoided the above
drawback. SARG04 uses 4 states which are just the BB84 states. Bob
will just measure each qubits in either $Z$ basis or $X$ basis.
Then, Alice announces a set of two states for each qubits. There are
4 possible sets, $S_{0+}=\{|0\rangle,|+\rangle\}$,
$S_{0-}=\{|0\rangle,|-\rangle\}$, $S_{1+}=\{|1\rangle,|+\rangle\}$,
$S_{1-}=\{|1\rangle,|-\rangle\}$. Each BB84 qubit must belong to at
least one set in the above. Alice just randomly chooses a set that
contains the state of that qubit. After Alice's announcement of a
specific set for each qubit, Bob knows which of his data corresponds
to the conclusive results. Explicitly, whenever Bob obtains an
outcome state that is orthogonal to one of the two states in Alice's
announced set, he has obtained a conclusive bit. And they discard
the other data.

One obvious advantage of the SARG04 protocol is that it does not
change the physical set-up of BB84 at all therefore can be
immediately used in practice. And, not only has it overcome the
drawback of B92 protocol itself, but also it is secure under the PNS
attack even we use weak coherent light with a very lossy channel.
For example, we consider the case that the pulse contains 2 photons.
In this case Eve can split the pulse and keep one photon. Suppose
the actual state is $|0\rangle$ and later Alice announces a set
$S_{0+}$. Suppose Bob has happened to obtain a conclusive result. In
such a case, if Eve also measures it in a basis randomly chosen from
$X,Z$, Eve only has a possibility of $1/4$ to obtain a conclusive
result. In  the normal realization of BB84 with one intensity of
coherent light, Eve can obtain all information of Bob's raw key by
the PNS attack. But the same attacking scheme may only allow Eve to
be sure of  $1/4$ of Bob's raw key in the SARG04 protocol. Of
course, Eve may have many other choices\cite{sarg04}. However, as it
has been shown by Koashi\cite{sargproof}, given whatever Eve may
take, the SARG04 protocol with weak coherent light is always secure
with a net final key rate at least in the magnitude order of
$O(\eta^{3/2})$. Moreover, if we increase the number of possible
bases, the key rate can be further raised\cite{sargproof}.

\subsection{QKD in position-momentum space}
In principle, BB84 protocol of quantum key distribution is secure because it uses 2 bases and measurements in these two bases do
not commute. By the uncertainty principle, if Eve looks at the transmitted qubits in whatever basis, she must cause noise to
 thoses qubits prepared in another basis. The key point here for security is the noncommunity of two bases. Therefore, secure
QKD should not be limited to quantum states in two-level space such
as polarization and phase coding.  It should also allow protocols
with quadrature measurements of continuous variables, for example,
the position and momentum. For simplicity, we shall call any
quadrature observable as position $\hat x$ and momentum $\hat p$
provided that they satisfy the same commutation relation.

Basically, Alice can encode the bit values in either position or momentum. If Eve looks at the bit values in position basis,
she must cause noise to the information in momentum basis, and vice versa. There have been a number of protocols for QKD in
 position-momentum space\cite{cvqkd1,cvqkd2,cvqkd3,cvqkd4,cvqkd5,cvqkd6,cvqkd7,cvqkd8,cvqkd9,cvqkd10,cvqkd11,cvqkd12}.
All these protocols encode the bit values by Gaussian states, more
specifically, coherent states or squeezed states.  Most of the
studies about the security are limited to the  individual attack
while the effects of collective attack is rarely
discussed\cite{mheid}. The only unconditional security proof uses
one-mode or two-mode squeezed states based on the idea of purifying
the encoded two-level entangled pairs\cite{cvgp}. Here we shall
first review the protocols with coherent states and squeezed states,
and then go into the unconditional security proof.
\subsubsection{Protocols of QKD with Gaussian states}
A. Protocols with coherent states.
\\
 We can take two types of
displacement on a coherent state, say, position displacement $e^{-i
x \hat p}$ and momentum displacement $e^{ip \hat x}$. A possible
protocol works as follows\cite{cvqkd3}. Alice generates two random
numbers and modulates these numbers on a coherent state by taking
displacement operations in position and momentum respectively. Bob
detects either position or momentum of each light pulse. By a public
channel, they compare their results of those times Bob has measured
the momentum. If Bob's results agree with Alice's in an acceptable
rate, they would use the position value as the raw key, i.e., the
results of the times Bob has measured in position basis. Otherwise,
they discard the protocol.

Consider the most intuitive Eve's attacking. To each pulse, Eve could guess a basis that would be used by Bob and she measures in this
basis herself and then reproduce a coherent state with the same displacement in that basis and sends it to Bob. However, in average,
there is 50\% probability Eve makes wrong guess and Bob will find large noise at those times Eve has used a wrong basis therefore
Eve's presence will be detected according to the protocol itself.
This is similar to the case of original BB84 protocol.

However, Eve may do it more sophisticatedly given a lossy channel.
Suppose the channel transmittance is only 50\%. But Eve may have a
channel with perfect transmittance. Eve may then split each pulse by
a 50:50 beam-splitter. She keeps one daughter pulse and sends the
other one to Bob through a perfect channel. Bob can not detect Eve's
presence  because Eve has just pretended her action to be the
channel loss. After Bob announces the measurement basis for each
pulse, Eve measure her pulses accordingly. In such a way, Eve's
amount of information about Alice's key is the same with that of
Bob. Therefore, a necessary condition for such a QKD protocol with
coherent state is that the channel loss should be less than a
half\cite{cvqkd7}. However, such a necessary condition is only for
the case of key distillation with one-way classical communication.
In this case, Bob accepts all remaining bits if the protocol passes
the error test. However, if they use two-way classical
communication, they may discard those bits with larger error rate
and only keep those bits with smaller error rate\cite{cvqkd8}. This
is a kind of advantageous distillation by post-selection. If they
use this post-selection, the  channel loss does not have to be less
than a half for a secure QKD. With the method of post-selection,
Alice and Bob can do QKD even more efficiently\cite{cvqkd11}. Bob
actually doesn't have to  switch his measurement bases. Alice draws
two random numbers $x_A,p_A$ and sends Bob a coherent state of
$|\alpha\rangle =|x_A+ip_A\rangle$. Bob splits it by a 50:50
beamsplitter and measures one daughter pulse in position basis and
the other in momentum basis. They use post-selection\cite{cvqkd8} to
reverse any initial ``information advantage'' a potential Eve may
have obtained and distill a final key.  This type of QKD without
bases switching has been demonstrated experimentally with a channel
loss up to 90\%\cite{cvqkd11}.
 Actually, if Eve only uses individual attack by a beamsplitter, the secure distance for QKD with coherent states is un-limitted, if a
sufficiently large post-selection threshold is set.
Some experiments of QKD with position-momentum measurement have been done by using the homodyne detection\cite{cvqkd6,cvqkd11,cvqkd12}.

Beyond the framework of BB84 and its variants, Yuen's group proposed a key expansion scheme\cite{y00} with mesoscopic coherent
states by making use of the noise of the states (Y-00 protocol). The security issue of Y-00 has been actively
discussed\cite{y001,y002}.

B. Protocols with squeezed states. \\ There are also proposals for
QKD with displaced squeezed states of one-mode or two-mode, e.g.,
\cite{cvqkd4}. If we use squeezed states, the source noise is
compressed. A coherent state has the same  statistical
fluctuations in the two bases. But a squeezed state can have a
smaller noise in one basis and largere noise in another basis. For
example\cite{cvqkd4}, Alice can send displaced squeezed vacuum
state to Bob, which is squeezed in either position basis or
momentum basis. Bob measures each pulse in a basis randomly chosen
from position or momentum.  Again, the security here is a result
of uncertainty relationship.

Consider the case of infinite squeezing. Alice may either produce a
random position eigenstate or a random momentum eigenstate. In this
case, if Bob measures a state in the right basis, he obtains a value
that is identical to Alice's value. Also, a position eigenstate
distributes uniformly in momentum space, and vice versa. This means,
if the state is has been measured in a wrong basis, it would very
likely produce a wrong result if it is measured latter by Bob in the
right basis. Alice and Bob will only use those data obtained from
the right bases. If Eve has intercepted the state and measured it in
a basis as she guessed and then reproduced a state according to her
measurement result and sent it to Bob, there is 50\% probability
that she has caused large noise.

 In practice, infinite squeezing is
not available. However, finitely squeezed states also have the
similar property. If it is squeezed in position basis and its
averaged position (momentum) is $x_A$, then with high probability
the measurement result falls inside a small interval centered by
$x_A$, if one measures it in position basis. But, if anybody
measures it in momentum basis, the result would be in a large range.
This means that Eve will cause large noise to the state squeezed in
position basis if she measures it in momentum (position) basis. In
the QKD protocol with finite squeezing, Alice will only use a set of
discrete values as the averaged position and momentum for squeezed
states in each basis. The states in the same basis with different
discrete values are {\em almost} orthogonal, but the overlap of
states between different bases are considerable. Explicitly,
consider the squeezed vacuum state $S(r)|0\rangle$ and operator
$S(r)=\exp[r{a^\dagger}^2-ra^2]$. If we take a displacement
operation $D(x_A,p_A)$, this state will be centered at $x_A,p_A$ in
position space and  momentum space respectively. If we measure the
position, the probability that the result falls inside the interval
$(x_A-\delta/2,x_A+\delta/2)$ is
\begin{equation}
p_{\delta} = \int_{x_A-\delta/2}^{x_A+\delta/2} \langle x |\rho (x_A,p_A,r)| x\rangle {\rm d} x
\end{equation}
and $\rho (x_A,p_A,r)$ is the density operator of displacement
squeezed vacuum state, which is squeezed in position basis  and
centered at $x_A$ and $p_A$ respectively in position space and
momentum space, respectively. After calculation one
finds\cite{cvqkd4}
\begin{equation}
p_{\delta} ={\rm erf}\left(\frac{\delta}{2\sqrt \nu}\right)
\end{equation}
and $\nu = 1/2 e^{-2r}$,
\begin{equation}
{\rm erf}(x) = \frac{2}{\sqrt \pi}\int_0^x e^{-t^2}{\rm d} t.
\end{equation}
If we choose $\delta$ to be $1/8$ and $r=3.3$,  the probability of
obtaining a wrong value (a value that is outside the assumed
interval of $\delta$ centered at $x_A$) is less than $10^{-3}$.
However, if the measurement is done in momentum basis, the likely
interval will be $e^r/2$ which is more than 100 times larger than
$\delta$. This is to say, whenever Eve measures a pulse in a wrong
basis, she must cause large noise to the state therefore her
presence will be detected. If they use the coherent states, the
interval $\delta$ for the right basis will become larger and the
interval in wrong basis will be smaller.

However, in practice, the channel can be very lossy. As it has been
calculated\cite{cvqkd3}, the effects of channel loss is rather
severe. Even Alice starts from the infinitely squeezed state, if we
want the same error probability for the same interval as shown
above, the transmission distance in the normal optical fiber can be
only 1 km, with a loss rate of $1.2\times 10^{-6}/m $. Eve may
attack the protocol more effectively by pretending her action to be
the channel loss. Again, here she can use beam-splitter attack. As
it has been shown\cite{cvqkd4}, the negative effect of loss can be
reduced by amplifiers and the method is secure against the
individual beam-splitting attack.

There are also proposals of QKD with 2-mode squeezed
light\cite{cvqkd3,cvqkd5}. Given 2-mode squeezed states, it is
possible to make a pre-determined key\cite{cvqkd5}. The protocol
inspects eavesdropper by checking quantum correlation of two beams,
one with Alice and the other sent with Bob. This is the analog of
Ekert 91 protocol\cite{Eke91} with Gaussian entanglement.
\subsection{Security proof of QKD with squeezed states}
So far we have not considered the issue of {\em unconditional} security. It has been shown by Gottesman and Preskill\cite{cvgp} that one can do
unconditionally secure  quantum key distribution with one-mode (or 2-mode) squeezed states. To make the proof, they first construct
the encoded $2\times 2$ EPR pair state\cite{qecc10}  and then use the similar reduction technique
as used in the
case of qubit-QKD by Shor and Preskill\cite{shor2}. (Definitely, one can also use the method in section \ref{skey} for final key distillation.)
As a result, QKD with one-mode squeezed state is unconditionally secure under whatever type
of eavesdropping.

\subsubsection{Security proof of QKD in position-momentum space.}
As we have mentioned in section \ref{seqecc1}, we can use encoded $2\times 2$ EPR state in position-momentum space
\begin{equation}
|\tilde \phi^+\rangle =\frac{1}{\sqrt 2}(|\tilde 0\tilde 0\rangle + |\tilde 1\tilde 1\rangle)
\end{equation}
and
\begin{eqnarray}
|\tilde 0\rangle  \propto \sum_{-\infty}^{\infty} |x=2s\sqrt \pi\rangle
\propto \sum_{-\infty}^{\infty} |p=s\sqrt \pi\rangle;\nonumber\\
|\tilde 1\rangle  \propto \sum_{-\infty}^{\infty} |x=(2s+1)\sqrt \pi\rangle
\propto \sum_{-\infty}^{\infty} (-1)^s|p=s\sqrt \pi\rangle.
\end{eqnarray}
Here we have bit-flip operator
\begin{equation}
\tilde \sigma_x =e^{-i\sqrt \pi \hat p}
\end{equation}
which causes a bit-flip of $|\tilde 0\rangle\rightarrow |\tilde
1\rangle  ;  |\tilde 1\rangle\rightarrow |\tilde 0\rangle$;
phase-flip operator
\begin{equation}
\tilde \sigma_z =e^{i\sqrt \pi \hat x}
\end{equation}
which causes a phase-flip of $ |\tilde 0\rangle\rightarrow |\tilde
0\rangle  ; |\tilde 1\rangle \rightarrow -|\tilde 1\rangle$. If
Alice starts with such states and then do entanglement purification
as if they were two-level states, they can obtain the secure final
key. More explicitly, Alice sends half of each pair to Bob. They
will measure $e^{i2\sqrt \pi\hat x}$ and $e^{-i2\sqrt \pi \hat p}$
to determine the values of $x$ or $p$ modulo $\sqrt \pi$ and then
add a displacement to the state to adjust it to the nearest integer
multiples of $\sqrt \pi$. The noisy channel could have shifted a few
states to an extent that it has a wrong nearest integer multiples of
$\sqrt \pi$. Such errors can be removed by entanglement purification
in the encoded space space, if the error rate is not too high. In
particular, they can measure $e^{i\sqrt \pi \hat x}$ to see whether
the position of the state is an even (bit value 0) or odd (bit value
1) multiples of $\sqrt\pi$,  or measure $e^{-i\sqrt \pi \hat p}$ to
see whether the momentum of the state is an even (bit value 0) or
odd (bit value 1) multiples of $\sqrt\pi$. Furthermore, they can
directly measure $\hat x$ or $\hat p$ to conclude the bit value
equivalently. In short, to do the secure QKD in position-momentum
space with
 the encoded EPR states, Bob shifts his state according to the value announced by Alice and then
 together with Alice purifies the entanglement
in the decoded space.\\
 Similarly, Alice may produce any displaced EPR state in the form
\begin{equation}
|\tilde \phi^+(d_x,d_p)\rangle =\left[I \otimes D(\sqrt \pi
d_x,\sqrt \pi d_p)\right]\frac{1}{\sqrt 2}(|\tilde 0\tilde 0\rangle
+ |\tilde 1\tilde 1\rangle).
\end{equation}
Here the displacement operator $D$ is defined in Eq.(\ref{pmdis}).
After she sends the displaced parts to Bob, Bob measures $\hat x$ or
$\hat p$ modulo $\sqrt \pi$. Alice then announces the displacement
$d_x$ or $d_p$ and Bob takes the reverse displacement to his states.
On the other hand, it makes no difference if Alice measures her part
in the beginning, before she sends anything to Bob. Thus, Alice need
only prepare displaced encoded BB84 states in $ D(\sqrt \pi
d_x,\sqrt \pi d_p)(|\tilde 0\rangle, \tilde 1\rangle)\}$ space and
sends them to Bob. Suppose she prepares $N+m$ states among which $N$
are prepared in $\{|\tilde 0\rangle,|\tilde 1\rangle\}$ basis and
$m$ are prepared in $D(\sqrt \pi d_x,\sqrt \pi d_p)\{\frac{1}{\sqrt
2}(|\tilde 0\rangle\pm |\tilde 1\rangle)\}$ basis.
 Bob measures each of them either in position or momentum
space and they will discard the data with basis mismatch. Also, Bob
will revise his data by classical information of $d_p$ or $d_x$ from
Alice. In particular, if Bob measures $\hat x$, he does not need
phase information, i.e., $d_p$, therefore Alice does not have to
tell him this. This means Alice can actually choose a random value
of $0\le d_p\le 2\sqrt \pi$. Thus, if Alice decides to send Bob a
displaced $|\tilde 0 \rangle$, she can actually sends him
\begin{equation}
 \int_0^{2\sqrt \pi} e^{i\sqrt \pi  y \hat x}e^{-i\sqrt \pi  d_x \hat p}|\tilde 0\rangle\langle\tilde 0|e^{i\sqrt \pi  d_x \hat p}
e^{-i\sqrt \pi  y \hat x}{\rm d}y.
\end{equation}
If we write it in position space, the factors for off diagonal terms are all 0 after integration and the state is actually
\begin{equation}
\rho(d_x, z=0)=\sum_s|x=(2s+d_x)\sqrt \pi\rangle\langle x=(2s+d_x)\sqrt\pi |.
\end{equation}
Similarly,  if Alice wants send Bob a state in position basis with bit value 1 and displacement $d_x$, she only needs to directly prepare the following state
for Bob\begin{equation}
\rho(d_x, z=1)=\sum_s|x=(2s+1+d_x)\sqrt \pi\rangle\langle x=(2s+1+d_x)\sqrt\pi |.
\end{equation}
Averaged over all possible displacement $0\le d_x\le 2\sqrt \pi$, Alice only needs to send random position states !
Due to the same reason, for those states in basis of states displaced from $\frac{1}{\sqrt 2}(|\tilde 0\rangle\pm|\tilde 1\rangle)$, she can actually
uses arbitrary momentum states.
Therefore, the protocol actually only requires
Alice to prepare random position states and random momentum states.
A state with explicit position  or momentum  is not available in practice because
it requires infinite energy. But one can prepare (finitely) squeezed states in either position space or momentum space. Such an imperfect source means excessive
channel noise even they have a noiseless channel. However, as we shall show it later\cite{cvgp}, if the initial state is sufficiently squeezed, the
 bit-flip error rate can be still less than $11\%$ and one can distill some final key.

{\em Remark}: In the security proof above, the EPR state in the
encoded subspace is used as a mathematical tool. However, such
type of encoded EPR state is unnormalizable. It should be
interesting to seek alternative proofs which circumvents such
unnormalizable states.
\subsubsection{Realization with squeezed states}
  As we have pointed
it out already, the infinitely squeezed states (position or momentum
eigenstates) are unavailable since these states are un-normalizable.
In practice, highly squeezed states would be quite technically
demanding. An interesting question  is the minimum squeezing
demanded for the protocol with squeezed states. Say, if the source
is not perfectly squeezed, there are errors due to the source
itself. Also, the position $x$ and momentum $p$ should be always
finite. A related question is how to choose the probability
distribution of position and momentum. The perfect encoded BB84
states cannot be used here because this means that with large
probability that $p,x$ should be very large. We want to reduce the
protocol to the realistic case that the probability of using large
position or momentum state  will be negligibly small for large. This
is indeed the case by a further examination of the security with a
noisy source.

In the entanglement distillation, it is only the error rate of the
{\em shared} raw pairs that matters. The final key rate does not
depend on how the errors of the shared raw pairs had been generated
e.g., source error or transmission error, it {\em only} depends on
the value of the error rate of those shared  raw pairs. Alice and
Bob need test the error rate of the shared pairs and then carry out
the distillation protocol. Moreover, for the purpose of QKD, it
doesn't matter if Alice measures her qubits in the very begining,
i.e., the purification protocol is equivalent to a
prepare-and-measure protocol with (noisy) bipartite states.

As shown in Ref\cite{cvgp}, we can make a secure protocol with
two-mode squeezed states or directly produce single-mode Gaussian
state which can also be produced from a bipartite state {\em in
principle}.

We consider the case that Alice initially creates a two-mode Gaussian state in the form
\begin{eqnarray}
|\psi(\Delta)\rangle_{AB}=\frac{1}{\sqrt \pi }\int{\rm d}x_A{\rm d}x_B\exp\left[-\frac{1}{2}\Delta^2\left(\frac{x_A+x_B}{2}\right)^2\right]\nonumber\\
\times \exp\left[-\frac{1}{2}\left(\frac{x_A-x_B}{2}\right)^2/\Delta^2\right]|x_A,x_B\rangle\nonumber\\
=\frac{1}{\sqrt \pi }\int{\rm d}p_A{\rm d}p_B\exp\left[-\frac{1}{2}\Delta^2\left(\frac{p_A-p_B}{2}\right)^2\right]\nonumber\\
\times \exp\left[-\frac{1}{2}\left(\frac{p_A+p_B}{2}\right)^2/\Delta^2\right]|p_A,p_B\rangle
\end{eqnarray}
where $\Delta$ is real and positive. In the case Alice measures
position and obtain $x_A$, she has prepared the following one-mode
squeezed state for Bob:
\begin{equation}
|\psi(x_A)\rangle_B =\frac{1}{(\pi\tilde\Delta^2)^{1/4}}\int {\rm d} x_B
 \exp\left[-\frac{1}{2}(x_B-x_{B0})^2/\tilde\Delta^2\right]|x_B\rangle
\end{equation}
and \begin{equation}
x_{B0}= \left({{1 - {1\over 4} \Delta^4}\over{ 1+ {1\over 4}\Delta^4}}\right) ~x_A
=\left( 1 - \tilde \Delta^4\right)^{1/2}x_A~,
\end{equation}
and
\begin{equation}
\label{tilde_delta}
\tilde \Delta^2= {\Delta^2\over 1+ {1\over 4}\Delta^4}~.
\end{equation}
The probability distribution for the outcome of Alice's measurement can be expressed as
\begin{equation}
\label{pqa}
P(x_A)={\tilde \Delta\over \sqrt{\pi}}~ \exp\left(-\tilde \Delta^2 x_A^2\right) ~,
\end{equation}
and we can easily see that if Alice and Bob both measure $x$, then the difference of their outcomes is governed by the probability distribution
\begin{equation}
{\rm Prob}(x_A-x_B)= {1\over \sqrt {\pi \Delta^2}}\exp[ - (x_A-x_B)^2/\Delta^2]~.
\end{equation}
Also, if Alice measures her pulse in momentum space, she will have a
probability distribution of
\begin{equation}
\label{pqb} P(p_A)={\tilde \Delta_p\over \sqrt{\pi}}~
\exp\left(-\tilde \Delta_p^{2} p_A^2\right)
\end{equation}
where $\tilde\Delta_p^2 = \frac{4\Delta^2}{1+4\Delta^4}$.
 If she obtains $p_A$, she has prepared the state
\begin{equation}
|\psi(p_A)\rangle_B =\frac{1}{(\pi\tilde\Delta^{-2})^{1/4}}\int {\rm
d} x_B
 \exp\left[-\frac{1}{2}(p_B-p_{B0})^2\tilde\Delta^2\right]|p_B\rangle
\end{equation}
and \begin{equation} p_{B0}= \left( 1 - \tilde
\Delta_p^4\right)^{1/2}p_A.
\end{equation}
 We now ignore the noise due to channel
and take a look at the the error rate due to the finite squeezing of
the source. If $|x_A-x_B|< \sqrt \pi/2$, Alice and Bob still share
the same bit-value since the bit-value is determined by the nearest
integer of multiples of $\sqrt \pi$. However, if $|x_A-x_B|> \sqrt
\pi/2$, there will be a bit-flip error. Therefore the bit-flip error
rate caused by the imperfect source is
\begin{equation}
P(|x_A-x_B|>\sqrt \pi/2) \le
\frac{2\Delta}{\pi}\exp(-\pi/4\Delta^2).
\end{equation}
If $\Delta < 0.486$, the error rate from the source is less than
$1\%$ which is significantly below the threshold value $11\%$. We
also have the similar result for bit-flip error rate in momentum
space. The effect of channel loss and other imperfections can also
be studied in a similar way\cite{cvgp}.

Also, it makes no difference if Alice directly prepares the
single-pulse states for transmission rather than first preparing
bipartite states and then measuring her pulse in BB84 basis, if the
directly created single-pulse state is identical to that of half of
bipartite state. This is to say, Alice can obtain secure final key
even she uses a noisy single-pulse source which produces
$\rho_{0,1,+,-}$ as the noisy BB84 states, provided that {\em in
principle } there {\em exists} a bipartite $\rho_{AB}$ which has the
following property: If we measure photon $A$ of $\rho_{AB}$ in $Z$
basis and obtain $|0\rangle$ or $ |1\rangle$, state for photon $B$
will become $\rho_{0}$ or $\rho_1$ accordingly; if we measure photon
$A$ in $X$ basis and obtain $|+\rangle$ or $|-\rangle$, state for
photon $B$ will become $\rho_{+}$, $\rho_-$ accordingly. Explicitly,
we have the following final protocol for
 QKD with squeezed states\cite{cvgp}:\\

\begin{itemize}
\setlength{\itemsep}{-\parskip}

\item [\bf 1:]
Alice sends a number of displaced squeezed states to Bob. Most of
her states are squeezed in position. For those position squeezed
state, Alice uses a probability distribution $P(x_0)=\frac{\tilde
\Delta}{\sqrt\pi}\exp\left(-\frac{\tilde\Delta^2}{(1-\tilde\Delta^4)^{1/2}}x_0^2\right)$
to produce a wavepacket of
$\psi(x)=\frac{1}{(\pi\tilde\Delta^2})^{1/4}\int {\rm d}x
\exp\left[-\frac{1}{2}(x-x_0)^2/\tilde\Delta^2\right]$. For those
momentum squeezed state, Alice uses a probability distribution
$P(p_0)=\frac{\tilde
\Delta_p}{\sqrt\pi}\exp\left(-\frac{\tilde\Delta_p^2}{(1-\tilde\Delta_p^4)^{1/2}}p_0^2\right)$
to produce a wavepacket of
$\psi(p)=\frac{1}{(\pi\tilde\Delta_p^2})^{1/4}\int {\rm d}x
\exp\left[-\frac{1}{2}(p-p_0)^2/\tilde\Delta_p^2\right]$. She then
sends  Bob these states.
\item [\bf 2:] After receives the states, Bob measures each one in either position basis or momentum basis.
\item [\bf 3:] Bob announces his measurement basis for each state. Alice asks him to discard those data from a wrong basis.
\item [\bf 4:] Alice announces the value $x$, modulo $\sqrt{\pi}$. Bob subtracts Alice's value from his own measured value,
 and corrects to the nearest integer multiple of $\sqrt{\pi}$. Bob and Alice extract their shared bit according to whether the integer
is an even number  or an odd number.
\item [\bf 5:] They then do the error test in two level space.
If the error rates are not too large they go on to do the final key
distillation, otherwise, they abort the protocol. Bob announces bit
values of all those bits obtained from measurement in momentum basis
and the same number of bit values from position basis. The remaining
bits are all from position basis. They distill the final key by the
method in section \ref{skey}, including error correction and privacy
amplification.
\end{itemize}

\section{Mathematical theory of quantum entanglement with Gaussian states}\label{secmet}

In this section, we review the entanglement properties of Gaussian
states of radiation fields. Firstly, we introduce a notion of
quantum entanglement with its basic properties - separability and
distillability in
Section~\ref{sec:General_Properties_of_Quantum_Entanglement},
where we note that the partial transposition plays a key role in
the entanglement theory. Secondly,  in
Section~\ref{sec:Entanglement_Properties_of_Gaussian_States}, we
review the entanglement properties of Gaussian states focusing on
the qualitative aspects. These properties include the description
of partial transposition of Gaussian states, necessary and
sufficient conditions for the separability and distillability of
bipartite Gaussian states, and classification of tripartite
Gaussian states. We also point out that Gaussian states cannot be
distilled by Gaussian operations with classical communication.
This section is ended with a concluding remark.

\subsection{General Properties of Quantum Entanglement} \label{sec:General_Properties_of_Quantum_Entanglement}

First of all, we introduce the notion of quantum entanglement \cite{TWD03} and briefly review its general properties.
The notion of quantum entanglement comes from the nonlocal properties of a wave function in the Hilbert space composed of local Hilbert spaces,
$\mathcal{H}=\mathcal{H}_{A}\otimes \mathcal{H}_{B}\otimes \cdots $.
A notable example is the wave function
$\left| \psi _{-}\right\rangle =\left( \left|
0\right\rangle _{A}\left| 1\right\rangle _{B}-\left| 1\right\rangle
_{A}\left| 0\right\rangle _{B}\right) / \sqrt{2} $
in the Hilbert space $\mathcal{H}=\mathcal{H}_{A}\otimes \mathcal{H}_{B}$ with
$\dim \mathcal{H}_{A}= \dim \mathcal{H}_{B}=2$.
Here,
$\{\left| 0\right\rangle _{A(B)},\left| 1\right\rangle _{A(B)}\}$ is the orthonormal basis of $\mathcal{H}_{A(B)} $.
This wave function cannot be written as a separable form
$\left| \chi \right\rangle _{A}\otimes \left| \chi \right\rangle _{B}$
for any local wave functions $\left| \chi \right\rangle _{A} $ and
$\left| \chi \right\rangle _{B} $.
A state described by such a wave function is called an entangled (pure) state.
A statistical mixture of pure states is called a mixed state.
Whether such a state is entangled or separable is defined as follows \cite{Wer89}.

\smallskip

\noindent{\em Definition 1.--}
A state described by a density operator $\rho $ (or simply a state $\rho $) is called separable
if it can be written as the convex sum of the tensor products of local density operators,
\begin{equation} \label{eq:separable}
\rho =\sum_{i=1}^{k}\lambda _{i}\rho _{A}^{i}\otimes \rho _{B}^{i} ,
\end{equation}
where $0\leq \lambda _{i}\leq 1$ and $\sum_{i=1}^{k}\lambda _{i}=1$.
Otherwise it is called an entangled state.

\smallskip

To make the following arguments as plain as possible,
we confine ourselves to the bipartite case here;
we consider only the states on the Hilbert space composed of two local Hilbert spaces.

To distinguish whether a given state is separable or entangled is one of the most difficult problems in the entanglement theory and much work has been devoted on this subject \cite{LBC00,Ter02,Bru02,BCH02}.
The following theorem is fundamental on the separability problem \cite{Per96}.

\smallskip

\noindent{\em Theorem 4.--} Let $\rho $ be a density operator on
the composite Hilbert space $\mathcal{H}=\mathcal{H}_{A}\otimes
\mathcal{H}_{B}$. If $\rho $ is separable, then $\rho ^{T_{A}}\geq
0$ and $\rho ^{T_{B}}\geq 0$.

\smallskip

Here, $\rho ^{T_{A}}$ is defined as $\rho
_{ij,kl}^{T_{A}}=\left\langle v_{j}v_{k}\right| \rho \left|
v_{i}v_{l}\right\rangle =\rho _{jk,il} $, where $\left|
v_{i}v_{j}\right\rangle = \left|v_{i}\right\rangle _{A}\otimes
\left| v_{j}\right\rangle _{B} $ denotes an orthonormal basis
vector in $\mathcal{H}$ with $\left\{ \left| v_{i}\right\rangle
_{A(B)}\right\} $ being the orthonormal basis of
$\mathcal{H}_{A(B)}$. That is, $\rho ^{T_{A}}$ is a partial
transposition of $\rho $ on the subsystem $A$. $\rho ^{T_{B}}$ is
also defined similarly. We say that a state $\rho $ is PPT
(Positive Partial Transpose) if $\rho ^{T_{A}}\geq 0$ and $\rho
^{T_{B}}\geq 0$. Otherwise it is called NPPT (Non-Positive Partial
Transpose). Theorem 4 assets that separable states are always PPT.
It is proven that a PPT state in
$\mathcal{H}=\mathcal{H}_{A}\otimes \mathcal{H}_{B}$ with $\dim
\mathcal{H}_{A}=2$ and $\dim \mathcal{H}_{B}=2, 3$ is separable
\cite{HHH96}. However, there exists a PPT but entangled state in
higher dimensions \cite{Hor97}.

Distillation of quantum entanglement is an operation that extracts a pure maximally entangled state from a several copies of a given state by local quantum operations supplemented by classical communication (LOCC).
If there exist an LOCC operation $\Lambda _{LOCC} $ and $n,m\in \mathbb{N}$
such that
\[
\mathrm{Tr}[(\left| \psi _{-}\right\rangle \left\langle \psi _{-}\right| )^{\otimes
m}\Lambda _{LOCC}(\rho ^{\otimes n})] < 1-\varepsilon
\]
for arbitrary $\varepsilon $, then the state $\rho $ is called distillable \cite{Hor01b}.
Here, $\rho ^{\otimes n}$ is the abbreviation of
\[
\underbrace{\rho \otimes \cdots \otimes \rho}_{\mbox{$n$}}.
\]
Because only a maximally entangled pure state provides a reliable task in quantum information processing such as quantum teleportation \cite{telepor1} and quantum cryptography \cite{Eke91}, the distillation is of practical significance and the distillability is an important characterization of an entangled state.

All physically admissible operations are allowed for local quantum operations; they include local measurements/transformations of the state, adding/removing an auxiliary local system (ancilla), etc.
Only classical information on the local operations can be exchanged through classical channels (Fig.\ref{fig:1}).
The direct transfer of the local states is strictly forbidden.
Evidently, entanglement cannot be created by LOCC; it is inevitably degraded during the LOCC processes.
A distillation procedure maximizes the amount of entanglement of a part of given states while destroying the rest of them.

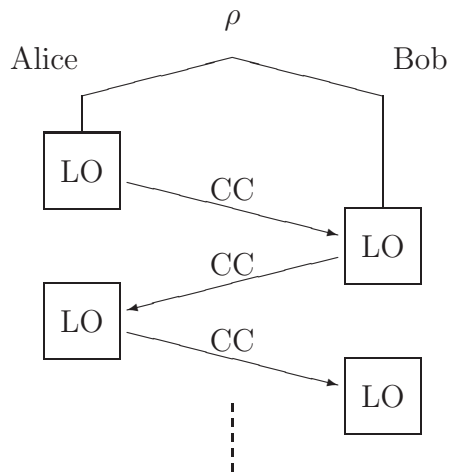
\begin{figure}[h]
\setlength{\unitlength}{1mm}
\begin{center}
\begin{picture}(70,65)(0,0)

\multiput(35,0)(0,2){5}{\line(0,1){1}}
\put(50, 5){\framebox(10,10){LO}}
\put(10,15){\framebox(10,10){LO}}
\put(50,25){\framebox(10,10){LO}}
\put(10,35){\framebox(10,10){LO}}
\put(30,15){\makebox(10,5){CC}}
\put(30,25){\makebox(10,5){CC}}
\put(30,35){\makebox(10,5){CC}}
\put(21,18.5){\vector( 4,-1){28}}
\put(49,28.5){\vector(-4,-1){28}}
\put(21,38.5){\vector( 4,-1){28}}
\put(25,55){\makebox(20,10){$\rho $}}
\put(15,45){\line(0,1){ 5}}
\put(55,35){\line(0,1){15}}
\put(15,50){\line( 4,1){20}}
\put(55,50){\line(-4,1){20}}
\put( 5,50){\makebox(10,10){Alice}}
\put(55,50){\makebox(10,10){Bob}}
\end{picture}
\end{center}

\caption{Procedure of LOCC.}
\label{fig:1}
\end{figure}

If $\rho$ is separable, evidently $\rho$ is not distillable.
However, an entangled state $\rho$ is not always distillable.
One of the most useful criteria for distillability is the following.

\smallskip

\noindent{\em Theorem 5.--} A PPT state is not distillable.

\smallskip

Although it is conjectured that the converse of this theorem is
not the case in general, the following partial result is known
\cite{DCL00,HHH97}. Let $\rho $ be a density operator on the
composite Hilbert space$\mathcal{H}_{A}\otimes \mathcal{H}_{B}$
with $\dim \mathcal{H}_{A}=2$ and $\dim \mathcal{H}_{B}\geq 2$.
Then $\rho $ is distillable if it is NPPT. As mentioned
beforeCPPT but entangled states exists, but Theorem 5 asserts
that such entangled states are still undistillable. Such states
that are entangled but undistillable is called a bound entangled
state \cite{Hor97,HHH98} whereas distillable states are called
free entangled states. It is an open question whether or not NNPT
states are always distillable (Fig.~\ref{fig:2}).

\begin{figure}[h]
\begin{center}
\includegraphics[scale=.5]{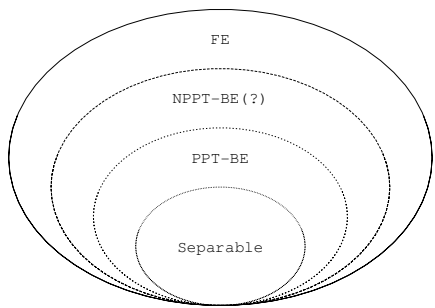}
\end{center}
\caption{States on a bipartite Hilbert space.
Separable, PPT-BECNPPT-BECand FE stand for separable states, PPT bound entangled states,
NPPT bound entangled states, and free entangled states, respectively.
}
\label{fig:2}
\end{figure}

There are several distillability criteria other than the PPT criterion \cite{NK01,VW02a,Hir03,HH99,CAG99}.
Among them, the most frequently used is the following reduction criterion \cite{HH99,CAG99}.

\smallskip

\noindent{\em Theorem 6.--} If
\begin{equation} \label{eq:RC0}
\mathrm{Tr}_{B}\rho \otimes \mathbf{I}_{B}-\rho \ngeq 0 ,
\end{equation}
then the state $\rho $ is distillable.

\smallskip

For example, it can be shown that a phase-damped two-mode squeezed
state \cite{Hir001} is always distillable by Theorem 6
\cite{WMT01}. We use this reduction criterion in the proof of the
distillability of NPPT bipartite Gaussian states in
Section~\ref{subsec:Distillability}.
\subsection{Entanglement Properties of Gaussian States} \label{sec:Entanglement_Properties_of_Gaussian_States}

Quantum entanglement of the radiation field is a quantum mechanical correlation between modes of the field.
In the following we focus on the fundamental properties of entanglement of Gaussian states, i.e., the separability and distillability \cite{GKD01,KGL03} and review the results obtained so far in depth.
Although we deal with bipartite Gaussian states in most part of the following arguments, we also give a brief overview on the separability of mutipartite Gaussian states in Section~\ref{subsec:Separability}.

Let $\rho $ be a bipartite Gaussian state shard by Alice and Bob and $n_{A} (n_{B})$ modes be in the possession of Alice (Bob).
We call the state $\rho $ the state of $n_{A} \times n_{B}$ modes (or $n_{A} \times n_{B}$ state) by convention (The total number of modes is $n=n_{A}+n_{B}$).

In Section~\ref{sec:General_Properties_of_Quantum_Entanglement}, we observed that the partial transposition is an important tool to characterize the entanglement properties of a given state.
So the question is how to describe the partial transposition for radiation field systems.
In order to answer this, we note the fact that the transposition of the density operator $\rho $ is given by $\rho ^{T}=\rho ^{*}$.
Hence, the canonical variable $P=-i\partial /\partial Q$ changes its sign while the canonical variable $Q$ remains unchanged under the transposition.
As for the covariance, the partial transposition causes the replacement
$\gamma _{jk} \rightarrow -\gamma _{jk} $ if one of the two indices $j$ and $k$ refers to the modes undergoing the transposition.
For a $n_{A}\times n_{B}$ Gaussian state $\rho $,
the covariance matrix of $\rho ^{T_{A}}$ is given by $\widetilde{\gamma }=F\gamma F$,
where $F=F_{A}\oplus F_{B}$ with $F_{A}=\oplus _{j=1}^{n_{A}}\mathrm{diag}(1,-1)$.
Hence, the necessary and suffcient condition for $\rho ^{T_{A}}$ describing a physical state, or equivalently, $\rho $ being PPT is $\widetilde{\gamma }+iJ\geq 0$ by Theorem 4.
Here, $J_{A} \oplus J_{B}$ with $J_{A(B)}=\oplus _{j=1}^{n_{A(B)}}J_{1}$.
This condition is also rewritten as $\gamma +i\widetilde{J}\geq 0$,
where $\widetilde{J}=FJF=(-J_{A})\oplus J_{B}$.
Therefore, we have the following.

\smallskip

\noindent{\em Theorem 7.--}
A Gaussian state is PPT if and only if its covariance matrix $\gamma $ satisfies $\gamma +i\widetilde{J}\geq 0$
with
$\widetilde{J}=(-J_{A})\oplus J_{B}$.

\smallskip

Note that the displacement in the characteristic function or Wigner function can be set to zero by local canonical transformations.
This means that the displacement is irrelevant to the entanglement properties of the states, so we can assume that it is zero without loss of generality.

First of all, we discuss the separability of a $1\times 1$ mode Gaussian state,
which is a starting point in the following discussion.
For a $1\times 1$ Gaussian state,
it is possible to make its covariance matrix to take a simple form by local canonical transformations on local canonical variables.
Note that the entanglement properties are completely preserved under such local transformations.
Furthermore, a physical state is always physical under these transformations.

The covariance matrix of a Gaussian state of $1\times 1$ mode is written as
\[
\gamma =\left(
\begin{array}{cc}
A & C \\
C^{T} & B
\end{array}
\right) .
\]
Here, $A$, $B$, and $C$ are $2 \times 2$ real matrices, and $A=A^{T}>0$, and $B=B^{T}>0$.
From the Williamson's theorem (Theorem 6), there exist symplectic matrices $S_{A}$ and $S_{B}$ such that
$S_{A}AS_{A}^{T}=\mathrm{diag}(n_{a},n_{a})$ and
$S_{B}BS_{B}^{T}=\mathrm{diag}(n_{b},n_{b})$, where $n_{a},n_{b}>0$.
By the symplectic matrix $S_{A} \otimes S_{B}$, we have
\[
(S_{A}\oplus S_{B})\gamma (S_{A}\oplus S_{B})^{T}=\mathrm{diag}%
(n_{a},n_{a},n_{b},n_{b})+\left(
\begin{array}{cc}
0 & \widetilde{C} \\
\widetilde{C}^{T} & 0
\end{array}
\right)
\]
with $\widetilde{C}=S_{A}AS_{B}^{T}$.
Next, we choose orthogonal matrices $O_{A},O_{B}\in O(2)$ such that $K=O_{A}\widetilde{C}O_{B}^{T}$ is the singular value decomposition of $\widetilde{C}$:
$K=\mathrm{diag}(k_{x},k_{p})$
with $(k_{x} \geq k_{p} \geq 0)$.
If $O_{A(B)}$ is not symplectic, $\widetilde{O}_{A(B)}=O_{A(B)}\mathrm{diag}(1,-1)$ turns out to be symplectic.
According to the replacement $O_{A}(O_{B})\rightarrow \widetilde{O}_{A}(\widetilde{O}_{B})$,
we have $K\rightarrow \mathrm{diag}(k_{x},\pm k_{p})$, which is still diagonal.
Namely, by choosing $O_{A},O_{B} \in Sp(2,\mathbb{R})\cap O(2)$,
we have $O_{A}\widetilde{C}O_{B}^{T}=\mathrm{diag}(k_{x},k_{p})$ with
$(k_{x}\geq \left| k_{p}\right| )$.
Note that $O_{A(B)}\mathrm{diag}(n_{a(b)},n_{a(b)})O_{A(B)}^{T}=\mathrm{diag}%
(n_{a(b)},n_{a(b)})$.
Putting all things together, the covariance matrix of the original state is transformed as
\begin{equation} \label{eq:standard_form_of_cm}
\gamma \rightarrow (S_{A}^{\prime }\oplus S_{B}^{\prime })\gamma
(S_{A}^{\prime }\oplus S_{B}^{\prime })^{T}=\left(
\begin{array}{cccc}
n_{a} & 0 & k_{x} & 0 \\
0 & n_{a} & 0 & k_{p} \\
k_{x} & 0 & n_{b} & 0 \\
0 & k_{p} & 0 & n_{b}
\end{array}
\right)
\end{equation}
by the local symplectic transformation $S_{A}^{\prime }\oplus S_{B}^{\prime }$
with $S_{A(B)}^{\prime }=O_{A(B)}S_{A(B)}$.
The matrix in the right-hand side of Eq.~(\ref{eq:standard_form_of_cm}) is called the standard form of the ($1 \times 1$ mode) covariance matrix.
From Theorem 1, we have the following.
A $1\times 1$ Gaussian state is a physical state if and only if the four parameters in its covariance matrix in the standard form satisfies
\begin{equation} \label{eq:physical_condition_g1}
d_{x}\geq 1
\end{equation}
and
\begin{equation} \label{eq:physical_condition_g2}
d_{x} d_{p} +1 \geq n_{a}^{2}+n_{b}^{2}+2k_{x}k_{p} .
\end{equation}
Here, $d_{x(p)} = n_{a} n_{b} - k_{x(p)}^2$.
Furthermore, from Theorem 7 we observe that the stats is PPT if and only if
\begin{equation} \label{eq:PPT_condition_g}
d_{x} d_{p} +1 \geq n_{a}^{2}+n_{b}^{2}-2k_{x}k_{p}
\end{equation}
in addition to Eqs.~(\ref{eq:physical_condition_g1}) and (\ref{eq:physical_condition_g2}).
Equivalently, the state is NPPT if and only if
\begin{equation} \label{eq:NPPT_condition_g}
d_{x} d_{p} +1 < n_{a}^{2}+n_{b}^{2}-2k_{x}k_{p}
\end{equation}
in addition to Eqs.~(\ref{eq:physical_condition_g1}) and (\ref{eq:physical_condition_g2}).

Since $\det S_{A}=\det S_{B}=1$, $\det A$, $\det B$, $\det C$, and $\det \gamma $ are invariant under the transformation [Eq.~(\ref{eq:standard_form_of_cm})], and these are given by
$\det A=n_{a}^{2}$, $\det B=n_{b}^{2}$, $\det C=k_{x}k_{p}$, and
$\det \gamma =(n_{a}n_{b}-k_{x}^{2})(n_{a}n_{b}-k_{p}^{2})$.
If $\det A=\det B$, we call the Gaussian state symmetric.
When the state is asymmetric, the fluctuations of canonical variables are not balanced between mode A and B as seen from the definition of the covariance [Eq.~(\ref{eq:covariance})].
For example, if $n_{a}>n_{b}$,
$\left\langle Q_{A}^{2}\right\rangle =\left\langle P_{A}^{2}\right\rangle > \left\langle Q_{B}^{2}\right\rangle =\left\langle P_{B}^{2}\right\rangle $,
or equivalently, the photon number of mode $A$ is larger than that of mode $B$;
$\left\langle a_{A}^{\dagger }a_{A}\right\rangle > \left\langle a_{B}^{\dagger }a_{B}\right\rangle $.

The Wigner correlation matrix also acquires the standard form by local canonical transformations;
\begin{equation} \label{eq:standard_form_of_Wigner_cm}
\gamma ^{-1}=\left(
\begin{array}{cccc}
N_{a} & 0 & K_{x} & 0 \\
0 & N_{a} & 0 & K_{p} \\
K_{x} & 0 & N_{b} & 0 \\
0 & K_{p} & 0 & N_{b}
\end{array}
\right) ,
\end{equation}
where $K_{p}\geq |K_{x}|\geq 0$ .
A $1\times 1$ Gaussian state is a physical state if and only if the four parameters in its Wigner correlation matrix in the standard form satisfies
\begin{equation} \label{eq:physical_condition_W1}
D_{x} \leq 1
\end{equation}
and
\begin{equation} \label{eq:physical_condition_W2}
D_{x} D_{p} +1 \geq N_{a}^{2}+N_{b}^{2}+2K_{x}K_{p} .
\end{equation}
Here, $D_{x(p)} = N_{a} N_{b} - K_{x(p)}^2$.
Furthermore, the stats is NPPT if and only if
\[
D_{x} D_{p} +1 < N_{a}^{2}+N_{b}^{2}-2K_{x}K_{p}
\]
in addition to Eqs.~(\ref{eq:physical_condition_W1}) and (\ref{eq:physical_condition_W2}).

Contrary to the covariance matrix, if $N_{a}>N_{b}$,
the photon number of mode $A$ is smaller than that of mode $B$;
$\left\langle a_{A}^{\dagger }a_{A}\right\rangle < \left\langle a_{B}^{\dagger }a_{B}\right\rangle $.

\subsubsection{Separability} \label{subsec:Separability}

Firstly, we discuss the separability of Gaussian state of $1\times
1$ mode. To begin with, we note the following fact. As discussed in
Section 1.2, a bipartite Gaussian state admits the following $P$
representation:
\[
\rho =\int d^{2}\alpha d^{2}\beta P(\alpha ,\beta )\left| \alpha ,\beta \right\rangle \left\langle
\alpha ,\beta \right| .
\]
If the $P$ functions $P(\alpha ,\beta )$ are positive,
the state is a statistical mixture of separable states
$\left| \alpha ,\beta \right\rangle \left\langle \alpha ,\beta \right|
=\left| \alpha \right\rangle \left\langle \alpha \right| \otimes \left|
\beta \right\rangle \left\langle \beta \right| $
so that the state is separable by Definition 1.
For a bipartite Gaussian state, the $P$ functions are calculated as
\begin{eqnarray*}
P(\alpha ,\beta ) &=&\frac{1}{4\pi ^{4}}\int d\xi _{1}\cdots d\xi _{4} \\
&&\times \exp (i\sqrt{2} \mathrm{Im} \alpha \xi _{1}+i\sqrt{2} \mathrm{Re} \alpha \xi _{2}+i\sqrt{2} \mathrm{Im} \beta \xi _{3}+i\sqrt{2} \mathrm{Re} \beta \xi _{4})
\\
&&\times \exp \left[ -\frac{1}{4}\xi ^{T}(\Gamma -\mathbf{I}_{4})\xi \right] .
\end{eqnarray*}
The the $P$ functions $P(\alpha ,\beta )$ are positive if $\Gamma \geq \mathbf{I}_{4}$ or $\gamma \geq \mathbf{I}_{4}$.
This observation leads to the following lemma \cite{Sim00}.

\smallskip

\noindent{\em Lemma 1.--}
Let $\gamma $ be a covariance matrix in the standard form [Eq.~(\ref{eq:standard_form_of_cm})] for a $1 \times 1$ Gaussian state.
If $k_{x} k_{p} \geq 0$, then the state is separable.

\smallskip

\noindent{\em Proof.}
If $k_{x} k_{p} \geq 0$, $k_{x}\geq k_{p}\geq 0$.
Firstly, let us consider the case $k_{x}>0$.
We perform two successive local canonical transformations
$S_{1}=\mathrm{diag}(x,x^{-1},x^{-1},x)$ with $x>0$ and
$S_{2}=\mathrm{diag}(y,y^{-1},y,y^{-1})$ with $y>0$ to the canonical variables.
These symplectic transformations correspond to local squeezing operations [see Eq.~(\ref{eq:squeezer})].
The covariance matrix $\gamma $ is changed to
\[
\gamma ^{\prime }=S_{2}S_{1}\gamma S_{1}^{T}S_{2}^{T}=\left(
\begin{array}{cccc}
y^{2}x^{2}n_{a} & 0 & y^{2}k_{x} & 0 \\
0 & y^{-2}x^{-2}n_{a} & 0 & y^{-2}k_{p} \\
y^{2}k_{x} & 0 & y^{2}x^{-2}n_{b} & 0 \\
0 & y^{-2}k_{p} & 0 & y^{-2}x^{2}n_{b}
\end{array}
\right) .
\]
This matrix is decomposed into the matrix on the $Q_{1},Q_{2}$ plane and that on the $P_{1},P_{2}$ plane.
The matrix on the $Q_{1},Q_{2}$ plane is
\[
\gamma _{Q}=y^{2}\left(
\begin{array}{cc}
x^{2}n_{a} & k_{x} \\
k_{x} & x^{-2}n_{b}
\end{array}%
\right) ,
\]
while the matrix on the $P_{1},P_{2}$ plane is
\[
\gamma _{P}=y^{-2}\left(
\begin{array}{cc}
x^{-2}n_{a} & k_{p} \\
k_{p} & x^{2}n_{b}
\end{array}
\right) .
\]
Because $n_{a},n_{b}>0$, $k_{x}>0$, and $k_{p}\geq 0$, we can choose $x$ such that $\gamma _{Q} \gamma _{P} = \gamma _{P} \gamma _{Q}$:
$x=(k_{x}n_{a}+k_{p}n_{b})^{1/4}/(k_{p}n_{a}+k_{x}n_{b})^{1/4}$.
Since the two commuting matrices $\gamma _{Q}$ and $\gamma _{P}$ can be diagonalized simultaneously by a common orthogonal matrix, $\gamma ^{\prime }$ can be diagonalized as
$\gamma ^{\prime \prime }=S\gamma ^{\prime }S^{T}=\mathrm{diag}(\kappa
_{+},\kappa _{+}^{\prime },\kappa _{-},\kappa _{-}^{\prime }) $
by the orthogonal matrix of the form
\[
S=\left(
\begin{array}{cccc}
\cos \theta  & 0 & -\sin \theta  & 0 \\
0 & \cos \theta  & 0 & -\sin \theta  \\
\sin \theta  & 0 & \cos \theta  & 0 \\
0 & \sin \theta  & 0 & \sin \theta
\end{array}%
\right) .
\]
Here,
\[
\kappa _{\pm }=\frac{1}{2}y^{2}\left\{ x^{2}n_{a}+x^{-2}n_{b}\pm \sqrt{%
(x^{2}n_{a}-x^{-2}n_{b})^{2}+4k_{x}^{2}}\right\}
\]
and
\[
\kappa _{\pm }^{\prime }=\frac{1}{2}y^{-2}\left\{ x^{-2}n_{a}+x^{2}n_{b}\pm
\sqrt{(x^{-2}n_{a}-x^{2}n_{b})^{2}+4k_{p}^{2}}\right\} .
\]
It is easy to see that $SJS^{T}=J$, i.e., the orthogonal matrix $S$ is symplectic.
Since $SS_{2}S_{1}$ is symplectic, the transformed state is still a physical state
so that the covariance matrix $\gamma ^{\prime \prime }$ satisfies
$\gamma ^{\prime \prime }+iJ\geq 0$ by Theorem 4.
This means $\kappa _{-}\kappa _{-}^{\prime }\geq 1$.
If we choose $y$ such that $\kappa _{-}=\kappa _{-}^{\prime }$,
$\kappa _{+},\kappa _{+}^{\prime }\geq \kappa _{-}=\kappa _{-}^{\prime }=1$, i.e.,
$\gamma ^{\prime \prime }=S\gamma ^{\prime }S^{T}\geq \mathbf{I}_{4}$.
From this we have $\gamma ^{\prime }\geq \mathbf{I} _{4}$ by noting that $S$ is orthogonal,
and therefore $\gamma ^{\prime }$ is a covariance matrix for a separable state.
Since the separable state with the covariance matrix $\gamma ^{\prime }$ is obtained by local canonical transformations from the original state with the covariance matrix $\gamma $, the original state is also separable.
Secondly, we consider the case $k_{x}=0$.
In this case, $k_{x}=k_{p}=0$ and $\gamma =\mathrm{diag}(n_{a},n_{a},n_{b},n_{b})$.
By Theorem 4, we have $n_{a},n_{b}\geq 1$.
Hence, $\gamma \geq \mathbf{I} _{4}$ and $\gamma $ is a covariance matrix for a separable state.\hfill $\Box$

Now let us prove the following Simon' theorem \cite{Sim00}.
This result has been also obtained by Duan {\it et al.} independently \cite{DGC00a}.

\smallskip

\noindent{\em Theorem 8.--}
A Gaussian state of $1\times 1$ mode is separable if and only if it is PPT.

\smallskip

\noindent{\em Proof.}
Since the separable states are always PPT by Theorem 1, it suffices to prove that the PPT condition implies the separability.
Firstly, let us assume that $k_{x}k_{p}<0$ in the covariance matrix of $1\times 1$ Gaussian state $\rho $ written in the standard form [Eq.~(\ref{eq:standard_form_of_cm})].
If $\rho $ is PPT, the partially transposed state $\rho ^{T_{A}}$ is a physical state; $\rho ^{T_{A}} \geq 0$.
The covariance matrix of $\rho ^{T_{A}}$ is given by Eq.~(\ref{eq:standard_form_of_cm}) with $k_{p}\rightarrow -k_{p}$, so that $\rho ^{T_{A}}$ is separable by Lemma 1.
Therefore, the original state $\rho $ is also separable.
Secondly, we consider the case $k_{x}k_{p}\geq 0$.
The state is separable by Lemma 1 and Eq.~(\ref{eq:PPT_condition_g}) is always satisfied due to Eq.~(\ref{eq:physical_condition_g2}).
Therefore, we conclude that the PPT condition implies the separability.
\hfill $\Box$

For general bipartite Gaussian states, we have the following necessary and sufficient condition for separability \cite{WW01,GKL01a}.

\smallskip

\noindent{\em Theorem 9.--}
A Gaussian state with covariance matrix $\gamma $ is separable if and only if there exist covariance matrices $\gamma _{A}$ and $\gamma _{B}$ such that
\begin{equation} \label{WW_separability_criterion}
\gamma \geq \gamma _{A}\oplus \gamma _{B}.
\end{equation}

\smallskip

\noindent{\em Proof.}
Suppose a bipartite Gaussian state $\rho $ is separable:
$\rho =\sum_{k}\lambda _{k}\rho _{A}^{k}\otimes \rho _{B}^{k}$.
The covariance of $\rho $ is calculated as
\begin{equation} \label{eq:cm_in_WW_th_1}
\gamma _{\alpha \beta }=\sum_{k}\lambda _{k}\left( \gamma _{A}^{k}\oplus
\gamma _{B}^{k}\right) _{\alpha \beta }+2\sum_{k}\lambda _{k}m_{\alpha
}^{k}m_{\beta }^{k}-2m_{\alpha }m_{\beta } ,
\end{equation}
where $\gamma _{A(B)}^{k}$ is the covariance matrix of $\rho _{A(B)}^{k}$ and
$m_{\alpha }=\mathrm{Tr}\left( \rho R_{\alpha }\right) =\sum_{k}\lambda _{k}\mathrm{Tr}\left(
\rho _{A}^{k}\otimes \rho _{B}^{k}R_{\alpha }\right) \equiv \sum_{k}\lambda
_{k}m_{\alpha }^{k} $ .
Here we define a matrix
$\Delta \equiv \gamma -\sum_{k}\lambda _{k}\left( \gamma _{A}^{k}\oplus \gamma _{B}^{k}\right)$.
By noting $\sum_{k}\lambda _{k}=1$ and using Eq.~(\ref{eq:cm_in_WW_th_1}),
it is easy to see that
$\xi ^{T}\Delta \xi =\sum_{k,l}\lambda _{k}\lambda _{l}(s_{k}-s_{l})^{2}\geq 0$
for every vector $\xi \in \mathbb{R}_{2f}$, where
$s_{k}=\sum_{\alpha =1}^{2f}\xi _{\alpha }m_{\alpha }^{k}$.
That is, the matrix $\Delta $ is positive (semi-)definite.
Hence, we have $\gamma \geq \gamma _{A}\oplus \gamma _{B}$
by choosing $\gamma _{A(B)}=\sum_{k}\lambda _{k}\gamma _{A(B)}^{k}$.
Conversely, let us suppose that $\gamma \geq \gamma _{A}\oplus \gamma _{B}$.
That is, $\gamma $ is written as $\gamma =\gamma _{A}\oplus \gamma _{B}+P$ with $P\geq 0$.
Let $\sigma (d) $ be a density operator with the covariance matrix $\gamma _{A}\oplus \gamma _{B}$ and the displacement $d$:
\[
\sigma (d)\propto \int d^{2n}\xi
\exp \left[ -\frac{1}{4}\xi ^{T}J^{T}(\gamma _{A}\oplus
\gamma _{B})J\xi +i(Jd)^{T}\xi \right] \mathcal{W}(-\xi ).
\]
Note that $\sigma (d) $ is separable.
It is easy to observe
\[
\int d^{2n} d \sigma (d)\exp \left[ -d^{T}P^{-1}d\right]
\propto \int d^{2n} \xi \exp \left[ -\frac{1}{4}\xi ^{T}J^{T}\gamma J\xi
\right] \mathcal{W}(-\xi ) \propto \rho .
\]
That is, $\rho $ is a statistical mixture of separable density operators $\sigma (d)$ with positive weight $\exp \left[ -d^{T}P^{-1}d\right] $.
Namely, $\rho $ is separable by Definition 1.
\hfill $\Box$

Now let us define a minimal PPT state as follows \cite{WW01}.
If $\gamma +iJ\geq 0$ and $\gamma +i\widetilde{J}\geq 0$,
$\gamma $ is called a PPT covariance matrix.
Furthermore, the PPT covariance matrix $\gamma $ is called minimal if
$\gamma=\gamma ^{\prime } $ for any PPT covariance matrix $\gamma ^{\prime }$ such that $\gamma \geq \gamma ^{\prime }$.
A Gaussian state with minimal PPT covariance matrix is called a minimal PPT state.
Under this definition, we have the following \cite{WW01}.

\smallskip

\noindent{\em Lemma 2.--}
A PPT covariance matrix $\gamma $ is minimal if and only if
\begin{equation} \label{eq:minimal_PPT_condition}
\mathrm{supp}(\gamma +iJ)\cap \mathrm{supp}(\gamma +i\widetilde{J}%
)=\varnothing .
\end{equation}

\smallskip

Here, $\mathrm{supp}M=\{\Phi |M\Phi \neq 0\}$ denotes the support of $M$.

\smallskip

\noindent{\em Proof of Lemma 2.}
Let us suppose that the PPT covariance $\gamma $ is minimal.
If Eq.~(\ref{eq:minimal_PPT_condition}) is violated,
then there exists a vector $\xi \in \mathbb{C}^{2f}$ such that
$\xi ^{\dagger } (\gamma +iJ) \xi >0$, i.e.,
$\gamma +iJ\geq \varepsilon \xi \xi ^{\dagger }\equiv \varepsilon \Delta $
for some $\varepsilon >0$.
We also have $\gamma +i\widetilde{J}\geq \varepsilon \Delta $.
Now let $\gamma ^{\prime }=\gamma -\varepsilon \Delta $.
Since $\gamma ^{\prime }+iJ\geq 0$ and $\gamma ^{\prime }+i\widetilde{J}\geq 0$,
$\gamma ^{\prime }$ is a PPT covariance matrix.
However, it is not minimal because $\gamma \geq \gamma ^{\prime }$ and $\gamma \neq \gamma ^{\prime }$.
This contradicts our assumption that the PPT covariance matrix $\gamma $ is minimal.
Therefore, Eq.~(\ref{eq:minimal_PPT_condition}) must be satisfied.
Conversely, let us suppose that Eq.~(\ref{eq:minimal_PPT_condition}) holds.
If the PPT covariance matrix $\gamma $ is not minimal, there exists a PPT covariance matrix $\gamma ^{\prime }$ such that $\gamma \geq \gamma ^{\prime }$ and $\gamma \neq \gamma ^{\prime }$.
It is always possible that
$\Delta =\gamma -\gamma ^{\prime }$
be a matrix of rank one:
$\Delta =\xi \xi ^{\dagger }$.
Since $\gamma ^{\prime }+iJ\geq 0$, we have
$\xi ^{\dagger }(\gamma +iJ)\xi \geq \xi ^{\dagger }\Delta \xi >0$.
We also have
$\xi ^{\dagger }(\gamma +i\widetilde{J})\xi >0$.
This means that
$\xi \in \mathrm{supp}(\gamma +iJ)\cap \mathrm{supp}(\gamma +i%
\widetilde{J})$,
which contradicts Eq.~(\ref{eq:minimal_PPT_condition}).
Therefore, $\gamma $ must be a minimal PPT covariance matrix.
\hfill $\Box$

Note that Eq.~(\ref{eq:minimal_PPT_condition}) is equivalent to
$\mathrm{Re} \mathcal{N} \cup \mathrm{Re}\widetilde{\mathcal{N}} = X$.
Here $\mathcal{N} = \mathrm{Ker}(\gamma +iJ)$ and
$\widetilde{\mathcal{N}} = \mathrm{Ker}(\gamma +i\widetilde{J})$ and
$\mathrm{Re} V$ denotes the real restriction of the complex vector space $V$; it consists of all real parts of vectors in $V$.
This is a subspace of the phase space $X$.
We have $\dim \mathrm{Re} \mathcal{N}\leq 2f$ since $\dim \mathrm{Re} V=2\dim V$ by definition of the real restriction.

Now, let us prove the following theorem \cite{WW01} that is the extended version of Theorem 8.

\smallskip

\noindent{\em Theorem 10.--}
A Gaussian state of $1\times n _{B}$ modes is separable if and only if it is PPT.

\smallskip

\noindent{\em Proof.}
Since the separable states are always PPT by Theorem 1, it suffices to prove that the PPT condition implies the separability.
In the following, we confine ourselves to the separability of minimal PPT states.
If every minimal PPT state is shown to be separable, every PPT state is also separable by Theorem 9.
Firstly, let us suppose that $\mathcal{N}\cap \widetilde{\mathcal{N}} \neq \{0\}$.
Then, there exists a vector $\Phi $ such that $\Phi \in \mathcal{N}\cap \widetilde{\mathcal{N}}(\Phi \neq 0)$.
Since $\Phi $ satisfies $(\gamma +iJ)\Phi =(\gamma +i\widetilde{J})\Phi =0$,
we have $(J-\widetilde{J})\Phi =0$.
If we write the vector $\Phi $ as
$\Phi =(x_{1},x_{2},\cdots ,x_{2f})^{T}$,
$(J-\widetilde{J})\Phi =(x_{2},-x_{1},\cdots )^{T}=0$.
This means that the first two component (Alice's part) of $\Phi $ are zero;
$\Phi =(0,0)^{T}\oplus \Phi _{B}$ $(\Phi _{B} \neq 0)$.
Here let $X_{B}$ denote Bob's phase space ($\dim X_{B} =2f_{B}$) and $X_{C}$ be a real restriction of $\mathcal{N}_{B}=\mathrm{Ker}(\gamma _{B}+iJ_{B})$:
$X_{C}=\mathrm{Re} \mathcal{N}_{B}$.
This is a subspace of the phase space $X_{B}$ and is defined as the phase space of the subsystem $C$ ($\dim X_{C}=2f_{C}$).
For a vector $\Phi _{B}\in \mathcal{N}_{B}$,
we write its restriction on the subsystem $C$ as $\Phi _{C}$.
Then, $\Phi _{C}$ satisfies $(\gamma _{C}+iJ_{C})\Phi _{C}=0$.
Hence, $\dim \mathcal{N}_{B}\leq \dim \mathcal{N}_{C}$.
Here, $\mathcal{N}_{C}=\mathrm{Ker}(\gamma _{C}+iJ_{C})$.
Therefore,
$\dim X_{C}=\dim \mathrm{Re} \mathcal{N}_{B}
=2\dim \mathcal{N}_{B}\leq 2\dim \mathcal{N}_{C}\leq \dim X_{C}$, i.e.,
$\dim X_{C}=2\dim \mathcal{N}_{C}=\dim \mathrm{Re} \mathcal{N}_{C}$
so that the state on the subsystem $C$ is pure (see the comment below Theorem 5).
This means that the state on the total system is written as
$\rho _{A,B\backslash C,C}=\rho _{A,B\backslash C}\otimes \rho _{C}$.
So we can focus on the separability problem of the states $\rho _{A,B\backslash C}$ of the smaller system and assume $\mathcal{N}\cap \widetilde{\mathcal{N}}=\{0\}$ in the following.
Let us suppose that $\Phi (\neq 0)\in \mathcal{N}$ and
$\widetilde{\Phi }(\neq 0)\in \widetilde{\mathcal{N}}$.
Note that $\Phi \notin \widetilde{\mathcal{N}}$.
Since $\gamma $ is Hermitian,
$\left\langle \widetilde{\Phi },\gamma \Phi \right\rangle =\left\langle
\gamma \widetilde{\Phi },\Phi \right\rangle $.
By noting $\gamma \Phi =-iJ\Phi $, $\gamma \widetilde{\Phi }=-i\widetilde{J}\widetilde{\Phi }$,
and $(i\widetilde{J})^{\dagger }=i\widetilde{J}$,
we have
$\left\langle \widetilde{\Phi },-iJ\Phi \right\rangle =
\left\langle -i\widetilde{J}\widetilde{\Phi },\Phi \right\rangle =
\left\langle \widetilde{\Phi },-i\widetilde{J}\Phi \right\rangle $.
Namely, $\widetilde{\Phi }$ is orthogonal to $(J-\widetilde{J})\Phi$:
$\left\langle \widetilde{\Phi },(J-\widetilde{J})\Phi \right\rangle =0$.
Here, the vector $(J-\widetilde{J})\Phi $ must be non-zero.
Otherwise we would have $(\gamma +iJ)\Phi =(\gamma +i\widetilde{J})\Phi =0$,
contradicting our assumption $\Phi \notin \widetilde{\mathcal{N}}$.
Since $J-\widetilde{J}=(2J_{A})\oplus 0$, $(J-\widetilde{J})\Phi \neq 0$ means that Alice's part of $\Phi $ is non-zero.
Alice's part of $\widetilde{\Phi }$ is also shown to be non-zero.
If $\dim \mathcal{N}\geq 2$,
$\dim \mathcal{N}_{A}\geq \dim \mathcal{N}\geq 2$
so that there exist two mutually orthogonal vectors
$\Phi _{1}=\Phi _{A}^{(1)}\oplus (0,\cdots ,0) \in \mathcal{N}$ and
$\Phi _{2}=\Phi _{A}^{(2)}\oplus (0,\cdots ,0) \in \mathcal{N}$
with $\Phi _{A}^{(1)}\neq \Phi _{A}^{(2)}$.
Now, for a vector $\widetilde{\Phi }=\widetilde{\Phi }_{A}\oplus (*,\cdots ,*)$,
both $(J-\widetilde{J})\Phi _{1}^{\prime }=\Phi _{A}^{(1)}$ and
$(J-\widetilde{J})\Phi _{2}^{\prime }=\Phi _{A}^{(2)}$
must be orthogonal to $\widetilde{\Phi } $.
This means that three two-dimensional vectors,
$\Phi _{A}^{(1)}$, $\Phi _{A}^{(2)}$, and $\widetilde{\Phi }_{A}$ must be orthogonal to each other.
This is impossible and we must conclude $\dim \mathcal{N}=1$.
It is also shown that $\dim \widetilde{\mathcal{N}}=1$ in a similar manner.
So we have $\dim \mathrm{Re} \mathcal{N}=\dim \mathrm{Re} \widetilde{\mathcal{N}}=2$.
Since $\gamma $ is a minimal PPT covariance matrix,
$\mathrm{Re} \mathcal{N}\cup \mathrm{Re} \widetilde{\mathcal{N}}$
is identical to the phase space $X$ (see the comment below Proof of Lemma 2).
Hence, $\dim X\leq \dim \mathrm{Re} \mathcal{N}+\dim \mathrm{Re} \widetilde{\mathcal{N}}=4$.
This means that $\dim X=4$ because $\dim X\geq 4$.
That is, $X$ is a phase space of the $1\times 1$ mode system.
Therefore, the state is separable by Theorem 8 because it is PPT.
\hfill $\Box$

Can we remove the restriction on the mode in the above theorem?
Unfortunately and expectedly, we cannot do so.
There exists an entangled PPT Gaussian states, i.e., a bound entangled bipartite Gaussian state.
Such examples ($2 \times 2$) are shown explicitly by Werner and Wolf \cite{WW01}.

Theorem 9 gives a necessary and sufficient condition for the separability of all bipartite Gaussian states.
However, it is a difficult task to check whether Eq.~(\ref{WW_separability_criterion}) holds or not for a covariance matrix of a given state.
The following operational criterion devised by Giedke {\it et al.} make it possible to determine whether a given state is separable or not by a series of simple computations.

We write a covariance matrix of a bipartite Gaussian state as
\[
\gamma _{0}=\left(
\begin{array}{cc}
A_{0} & C_{0} \\
C_{0}^{T} & B_{0}
\end{array}
\right) .
\]
Starting from this, we construct a sequence of matrices $\{\gamma _{N}\}_{N=0}^{\infty }$
according to the following nonlinear map:
If $\gamma _{N}-iJ\geq 0$,
$A_{N+1}=B_{N+1}=A_{N}-\mathrm{Re}(X_{N}) $ and $C_{N+1}=-\mathrm{Im}(X_{N}) $.
Otherwise, $\gamma _{N+1}=0 $.
Here, $X_{N}=C_{N}(B_{N}-iJ)^{-1}C_{N}^{-1}$ and
\[
\gamma _{N}=\left(
\begin{array}{cc}
A_{N} & C_{N} \\
C_{N}^{T} & B_{N}
\end{array}
\right) .
\]
Now, we have the following theorem.

\smallskip

\noindent{\em Theorem 11.--}
\begin{enumerate}
\item
If for some $N\in \mathbb{N}$, we have $A_{N}-iJ_{A}\ngeq 0$, then $\rho $ is not separable.
\item
If for some $N\in \mathbb{N}$, we have $A_{N}-||C_{N}||_{op}\mathbf{I}-iJ_{A}\geq 0$, then the state $\rho $ is separable.
Here, $||\cdot ||_{op}$ denotes the operator norm.
\end{enumerate}

\smallskip

In order to check the separability of a Gaussian state, we just apply the nonlinear map $\gamma _{N}\rightarrow \gamma _{N+1}$ starting from $\gamma _{0} $ until (1) or (2) in Theorem 11 is met.

The separability problem of bipartite Gaussian states has been thus completely solved.

For mutipartite Gaussian states \cite{LB00,LB01,GKL01b}, the problem is not simple; there are many different ways in which each subsystem is entangled/disentangled to other subsystems.
At present the classification problem of multipartite Gaussian states is solved only for the simplest system, a $1\times 1\times 1$ tripartite Gaussian state \cite{GKL01b}.

Let us consider a Gaussian state composed of three subsystems A, B, and C, each of which holds only one mode.
This tripartite state is also considered as a bipartite state if two subsystems A and B are grouped together for instance.
We write such a bipartite state AB-C and say that the total system is divided into two subsystems AB and C by a bipartite cut between AB and C.
Since a bipartite state AB-C is a $2 \times 1$ mode state, Theorem 10 can be applied.
According to the number of bipartite states with different bipartite cuts exhibiting NPPT/PPT, we have the four different classes.

\begin{description}
\item[Class 1:]
Fully inseparable states that is NPPT under all three bipartite cuts.
\item[Class 2:]
Partially inseparable states that is NPPT under two bipartite cuts but is PPT under the remaining one.
\item[Class 3:]
Partially inseparable states that is NPPT under one bipartite cut but is PPT under the remaining two.
\item[Class 4:]
States that is PPT under all three bipartite cuts.
\end{description}
Continuous-variable analogues to Greenberger-Horne-Zeilinger states considered in \cite{LB01} belong to Class 1.
Note that PPT under all bipartite cuts does not imply the separability.
The last Class 4 is further divided into two subclasses.
\begin{description}
\item[Class 4a:]
Entangled states $\rho _{ABC}$ that cannot be written as the statistical mixture of product states as
\begin{equation} \label{eq:triseparable}
\rho _{ABC}=\sum_{i}\lambda _{i}\rho _{A}^{i}\otimes \rho _{B}^{i}\otimes \rho _{C}^{i} .
\end{equation}
\item[Class 4b:]
Fully separable states that can be written as Eq.~(\ref{eq:triseparable}).
\end{description}


\begin{figure}[t]
\setlength{\unitlength}{1mm}
\begin{center}
\begin{picture}(100,82.5)(0,0)

\put(20,35.0){\circle*{2}}
\put(30,20.0){\circle*{2}}
\put(40,35.0){\circle*{2}}
\put(20,35.0){\thicklines\line( 2,-3){10}}
\put(40,35.0){\thicklines\line(-2,-3){10}}
\put(20,35.0){\thicklines\line( 1, 0){20}}
\put(16,27.5){\line( 1,0){28}}
\put(20,20.0){\line( 2,3){15}}
\put(40,20.0){\line(-2,3){15}}
\put( 0,22.5){\makebox(16,10)[c]{PPT}}
\put(12,10.0){\makebox(16,10)[c]{PPT}}
\put(32,10.0){\makebox(16,10)[c]{NPPT}}
\put(20, 0.0){\makebox(20,10)[ct]{(c) Class 3}}

\put(60,35.0){\circle*{2}}
\put(70,20.0){\circle*{2}}
\put(80,35.0){\circle*{2}}
\put(60,35.0){\thicklines\line( 2,-3){10}}
\put(80,35.0){\thicklines\line(-2,-3){10}}
\put(60,35.0){\thicklines\line( 1, 0){20}}
\put(56,27.5){\line( 1,0){28}}
\put(60,20.0){\line( 2,3){15}}
\put(80,20.0){\line(-2,3){15}}
\put(84,22.5){\makebox(16,10)[c]{PPT}}
\put(52,10.0){\makebox(16,10)[c]{PPT}}
\put(72,10.0){\makebox(16,10)[c]{PPT}}
\put(50, 0.0){\makebox(40,10)[ct]{(d) Class 4a and 4b}}

\put(20,75.0){\circle*{2}}
\put(30,60.0){\circle*{2}}
\put(40,75.0){\circle*{2}}
\put(20,75.0){\thicklines\line( 2,-3){10}}
\put(40,75.0){\thicklines\line(-2,-3){10}}
\put(20,75.0){\thicklines\line( 1, 0){20}}
\put(16,67.5){\line( 1,0){28}}
\put(20,60.0){\line( 2,3){15}}
\put(40,60.0){\line(-2,3){15}}
\put( 0,62.5){\makebox(16,10)[c]{NPPT}}
\put(12,50.0){\makebox(16,10)[c]{NPPT}}
\put(32,50.0){\makebox(16,10)[c]{NPPT}}
\put(20,40.0){\makebox(20,10)[ct]{(a) Class 1}}

\put(60,75.0){\circle*{2}}
\put(70,60.0){\circle*{2}}
\put(80,75.0){\circle*{2}}
\put(60,75.0){\thicklines\line( 2,-3){10}}
\put(80,75.0){\thicklines\line(-2,-3){10}}
\put(60,75.0){\thicklines\line( 1, 0){20}}
\put(56,67.5){\line( 1,0){28}}
\put(60,60.0){\line( 2,3){15}}
\put(80,60.0){\line(-2,3){15}}
\put(84,62.5){\makebox(16,10)[c]{PPT}}
\put(52,50.0){\makebox(16,10)[c]{NPPT}}
\put(72,50.0){\makebox(16,10)[c]{NPPT}}
\put(60,40.0){\makebox(20,10)[ct]{(b) Class 2}}

\end{picture}
\end{center}
\caption{Classification of $1 \times 1 \times 1$ Gaussian states.}
\label{fig:3}
\end{figure}
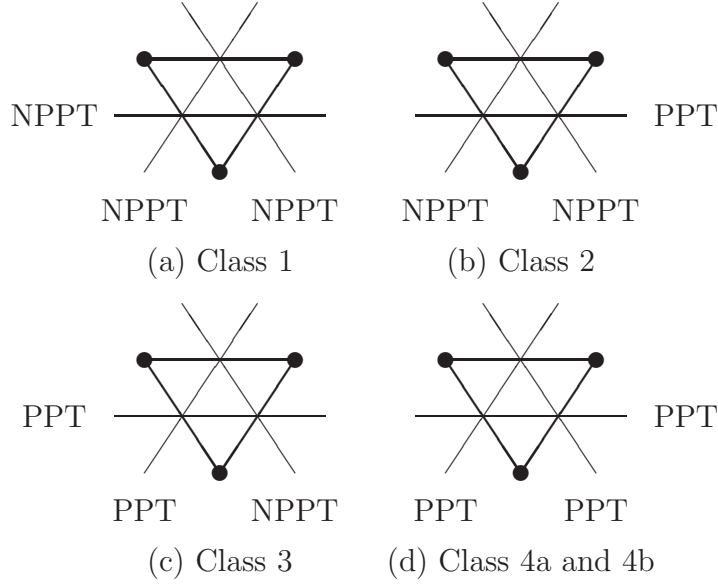

Namely, we have five different entanglement classes for tripartite $1 \times 1 \times 1$ mode Gaussian states in total (Fig.~\ref{fig:3}).
States of Class 4a are tripartite bound entangled states.
The complete criterion to distinguish between Class 4a and 4b has been also obtained by Giedke {\it et al.} \cite{GKL01b}.

\subsubsection{Distillability} \label{subsec:Distillability}.

The classification of bipartite Gaussian states by distillability is completely solved
and it is established that the NPPT condition is the necessary and sufficient condition for the distillability by Giedke {\em et al.} \cite{GDC01}

\smallskip

\noindent{\em Theorem 12.--}
A bipartite Gaussian state is distillable if and only if it is NPPT.

\smallskip

Although, we cannot rule out the possibility of existence of undistillable (bipartite) NPPT states in the general setting, the family of Gaussian states does not harbor such peculiar states.

To prove this, we need three lemmas.

\smallskip

\noindent{\em Lemma 3.--}
An NPPT $1\times 1$ symmetric Gaussian state is distillable.

\smallskip

\noindent{\em Proof.}
According to the reduction criterion (Theorem 3),
if there exists a state vector $\left| \psi \right\rangle $ such that
\begin{equation} \label{eq:RC1}
\left\langle \psi \right| \mathrm{Tr}_{B}\rho \otimes \mathbf{I}_{B}-\rho \left| \psi
\right\rangle <0 ,
\end{equation}
then the state $\rho $ is distillable.
The covariance matrix of $\rho $ is written in the standard form (\ref{eq:standard_form_of_cm}) without loss of generality:
\[
\gamma =\left(
\begin{array}{cc}
\gamma _{A} & C \\
C^{T} & \gamma _{B}
\end{array}
\right) .
\]
Here, $\gamma _{A}=\gamma _{B}=\mathrm{diag}(n,n)$ and $C=\mathrm{diag}(k_{x},k_{p})$.
By noting $\mathrm{Tr}_{B}\mathcal{W}(-\xi _{B})=2\pi \delta ^{(2)}(\xi _{B})$,
we can take the partial trace of $\rho $;
\begin{eqnarray} \label{eq:rho_A}
\mathrm{Tr}_{B}\rho  &=&\frac{1}{(2\pi )^{2}}\int d^{4}\xi \exp \left( -%
\frac{1}{4}\xi ^{T}J^{T}\gamma J\xi \right) \mathrm{Tr}_{B}\mathcal{W}(-\xi )
\nonumber \\
&=&\frac{1}{2\pi }\int d^{2}\xi _{A}\exp \left( -\frac{1}{4}\xi
_{A}^{T}J_{A}^{T}\gamma _{A}J_{A}\xi _{A}\right) \mathcal{W}(-\xi _{A}).
\end{eqnarray}

Now, let us take a two-mode squeezed state as the state vector $\left| \psi \right\rangle $;
$\left| \psi \right\rangle =(\cosh r)^{-1}\sum_{n=0}^{\infty }\tanh
^{n}r\left| n\right\rangle _{A}\otimes \left| n\right\rangle _{B}$.
The covariance matrix of the pure state $\left| \psi \right\rangle \left\langle \psi \right| $
takes the form,
\[
\delta =\left(
\begin{array}{cc}
\delta _{A} & D \\
D^{T} & \delta _{B}
\end{array}
\right) .
\]
Here, $\delta _{A}=\delta _{B}=\mathrm{diag}(\cosh 2r,\cosh 2r)$ and
$D=\mathrm{diag}(\sinh 2r,-\sinh 2r)$.

$\mathrm{Tr}_{B}\left| \psi \right\rangle \left\langle \psi \right| $ takes the form of the right-hand side of Eq.~(\ref{eq:rho_A}) with the replacement $\gamma _{A}\rightarrow \delta _{A}$.
Now, we have
\begin{eqnarray*}
&&\left\langle \psi \right| \mathrm{Tr}_{B}\rho \otimes \mathbf{I}_{B}\left| \psi
\right\rangle  =\mathrm{Tr}_{A}(\mathrm{Tr}_{B}\rho \mathrm{Tr}_{B}\left|
\psi \right\rangle \left\langle \psi \right| ) \\
&=&\frac{1}{(2\pi )^{2}}\int d\xi _{A}\int d\xi _{A}^{\prime }\exp \left( -%
\frac{1}{4}\xi _{A}^{T}J_{A}^{T}\gamma _{A}J_{A}\xi _{A}-\frac{1}{4}\xi
_{A}^{\prime T}J_{A}^{T}\delta _{A}J_{A}\xi _{A}^{\prime }\right)  \\
&&\times \mathrm{Tr}_{A}[\mathcal{W}(-\xi _{A})\mathcal{W}(-\xi _{A}^{\prime
})].
\end{eqnarray*}
After integration, the result is $\left\langle \psi \right|
\mathrm{Tr}_{B}\rho \otimes \mathbf{I}_{B}\left| \psi \right\rangle
=2[\det (\gamma _{A}+\delta _{A})]^{-1/2}$. Similar calculations
yield $\left\langle \psi \right| \rho \left| \psi \right\rangle
=\mathrm{Tr}(\rho \left| \psi \right\rangle \left\langle \psi
\right| )=4[\det (\gamma +\delta )]^{-1/2}$. Hence,
Eq.~(\ref{eq:RC1}) is written as $[\det (\gamma _{A}+\delta
_{A})]^{-1/2} < 2[\det (\gamma+\delta )]^{-1/2} $, which turns out
to be $(n-k_{x})(n+k_{p})<1$ for $r\rightarrow \infty $. Here, we
observe that the NPPT condition [Eq.~(\ref{eq:NPPT_condition_g})]
yields $(n-k_{x})(n+k_{p})<1$ by noting $k_{x}k_{p}<0$ for
inseparable states (Lemma 1). Therefore, Eq.~(\ref{eq:RC1}) holds
for the two-mode (infinitely) squeezed state $\left| \psi
\right\rangle $ so that $\rho $ is distillable. \hfill $\Box$

\smallskip

\noindent{\em Lemma 4.--}
An NPPT $1 \times 1$ Gaussian state can be transformed by LOCC into a symmetric NPPT $1\times 1$ state.

\smallskip

\noindent{\em Proof.}
If the state is not symmetric, the photon number of one mode is larger than that of the other mode.
So it is expected that the symmetrization can be achieved by equalizing the photon number by mixing the mode of larger photon numbers with a vacuum mode of physically minimal photon numbers.
In the following, we show that this local operation can be done successfully under the NPPT condition.
We write the Wigner correlation matrix $\gamma _{AB}^{-1} $ of the state in the standard form [Eq.~(\ref{eq:standard_form_of_Wigner_cm})] with $N_{a} > N_{b} $.
The Wigner correlation matrix of the vacuum mode used as an ancilla is given by
$\gamma _{anc}^{-1}=\mathrm{diag}(1,1)$.
The displacement of the vacuum mode is always zero.
Now, let us make a local compound system B+ancilla to pass through a beam splitter.
In the language of phase space, we apply a symplectic transformation $S=\mathbf{I}_{A}\oplus S_{bs}$ on
the Wigner correlation matrix of the total system,
$\gamma _{AB+anc}^{-1}=\gamma _{AB}^{-1}\oplus \gamma _{anc}^{-1}$.
Here, the symplectic matrix $S_{bs}$ is given by Eq.~(\ref{eq:beam_splitter}).
The Wigner correlation matrix $\gamma _{AB+anc}^{-1}$ is changed to
\begin{equation} \label{eq:Wigner_cm_1_in_distillation_lemma_2}
\gamma _{AB+anc}^{\prime -1}=S^{T}\gamma _{AB+anc}^{-1}S=\left(
\begin{array}{cc}
M_{AB} & C \\
C^{T} & M_{anc}
\end{array}
\right) ,
\end{equation}
where,
\begin{equation} \label{eq:matrix_1_in_distillation_lemma_2}
M_{AB}=\left(
\begin{array}{cccc}
N_{a} & 0 & cK_{x} & 0 \\
0 & N_{a} & 0 & cK_{p} \\
cK_{x} & 0 & c^{2}N_{b}+s^{2} & 0 \\
0 & cK_{p} & 0 & c^{2}N_{b}+s^{2}
\end{array}
\right) ,
\end{equation}
\begin{equation} \label{eq:matrix_2_in_distillation_lemma_2}
C=\left(
\begin{array}{cc}
sK_{x} & 0 \\
0 & sK_{p} \\
sc(N_{b}-1) & 0 \\
0 & sc(N_{b}-1)
\end{array}
\right) ,
\end{equation}
and
\begin{equation} \label{eq:matrix_3_in_distillation_lemma_2}
M_{anc}=\mathrm{diag}(c^{2}+s^{2}N_{b},c^{2}+s^{2}N_{b})
\end{equation}
with $c=\cos \theta $ and $s=\sin \theta $.
The Wigner function for the state of the total system is given by
\begin{eqnarray} \label{eq:Wigner_function_in_distillation_lemma_2}
&&W_{AB+anc}(x_{A},p_{A},x_{B},p_{B},x_{anc},p_{anc}) \nonumber \\
&=&\frac{1}{\pi ^{3}\sqrt{\det \gamma _{AB+anc}^{\prime }}}\exp \left[
-(x_{A},\cdots ,p_{anc})\gamma _{AB+anc}^{\prime -1}(x_{A},\cdots
,p_{anc})^{T}\right] .
\end{eqnarray}
If we measured the $P$-quadrature of the ancilla with the measured result $p_{0}$, the $Q$-quadrature of the ancilla would be completely uncertain and the Wigner function would take the form
\[
W_{AB}(x_{A},p_{A},x_{B},p_{B})=\int dx_{anc}W_{AB+anc}(x_{A},\cdots
,x_{anc},p_{anc}=p_{0}) .
\]
Substituting Eq.~(\ref{eq:Wigner_function_in_distillation_lemma_2}) with
Eqs.~(\ref{eq:Wigner_cm_1_in_distillation_lemma_2}),
(\ref{eq:matrix_1_in_distillation_lemma_2}),
(\ref{eq:matrix_2_in_distillation_lemma_2}),
(\ref{eq:matrix_3_in_distillation_lemma_2}) into the above equation
and performing Gaussian integration, we obtain
\[
W_{AB}(x_{A},\cdots ,p_{B})\propto \exp \left[ -(x_{A},\cdots ,p_{B})%
\widetilde{M}_{AB}(x_{A},\cdots ,p_{B})^{T}\right] ,
\]
where,
\begin{equation} \label{eq:Wigner_cm_2_in_distillation_lemma_2}
\widetilde{M}_{AB}=\left(
\begin{array}{cc}
\widetilde{M}_{A} & \widetilde{C} \\
\widetilde{C}^{T} & \widetilde{M}_{B}
\end{array}
\right)
\end{equation}
with
\[
\widetilde{M}_{A}=\mathrm{diag}\left( N_{a}-\frac{s^{2}K_{x}}{s ^{2} N_{b}+c ^{2}}
,N_{a}\right)
\]
and
\[
\widetilde{M}_{B}=\mathrm{diag}\left( c^{2}N_{b}+s^{2}-\frac{%
s^{2}c^{2}(N_{b}-1)^{2}}{s ^{2} N_{b}+c ^{2}},c^{2}N_{b}+s^{2}\right) .
\]
In order to put the state with the Wigner correlation matrix Eq.~(\ref{eq:Wigner_cm_2_in_distillation_lemma_2}) symmetric,
$\det \widetilde{M}_{A}=\det \widetilde{M}_{B}$ must hold.
This yields
\begin{equation} \label{eq:symmetrization_condition}
\tan ^{2}\theta =\frac{N_{a}^{2}-N_{b}^{2}}{N_{b}-D_{x}N_{a}} .
\end{equation}
This equation has a real solution for $\theta $ only when $N_{b}-D_{x}N_{a}\geq 0$
since $N_{a}>N_{b}$.
Here, we note that $N_{b}-D_{x}N_{a}\geq 0$ is equivalent to $(N_{b}-D_{x}N_{a})(N_{a}-D_{p}N_{b})\geq 0$, which turns out to be
$(N_{b}-D_{x}N_{a})(N_{a}-D_{p}N_{b})=(N_{a}K_{x}+N_{b}K_{p})^{2}\geq 0$
by Eq.~(\ref{eq:physical_condition_W2}).
Therefore, there always exists a $\theta $ satisfying Eq.~(\ref{eq:symmetrization_condition}).
The proof is thus completed.
\hfill $\Box$

\smallskip

\noindent{\em Lemma 5.--}
An NPPT $n_{A}\times n_{B}$ Gaussian state can be transformed by LOCC into a $1\times 1$ NPPT state.

\smallskip

\noindent{\em Proof.}
Let $\gamma $ denote the covariance matrix for an NPPT Gaussian state of $n_{A}\times n_{B}$ modes.
Since the state is NPPT, there exists a vector $\xi \in \mathbb{C}^{2(n_{A}+n_{B})}$ such that
\begin{equation} \label{eq:NPPT_1_in_distillation_lemma_3}
\xi ^{\dagger } (\gamma +i\widetilde{J}) \xi \leq -\varepsilon <0
\end{equation}
for some $\varepsilon >0 $ by Theorem 7.
$\widetilde{J} $ is again given by $\widetilde{J}=(-J_{A})\oplus J_{B}$.
Here, we write $\xi = \xi ^{(A)} \oplus \xi ^{(B)} $ and
$\xi ^{(A,B)} =\xi _{r}^{(A,B)} + i \xi _{i}^{(A,B)} $
($\xi _{r,i}^{(A)}\in \mathbb{R}^{2n_{A}}$ and
$\xi _{r,i}^{(B)}\in \mathbb{R}^{2n_{B}}$).
It can be assumed that two vectors $\xi _{r}^{(A,B)} $ and $\xi _{i}^{(A,B)} $
satisfy $\xi _{r}^{(A,B)T }J\xi _{i}^{(A,B)}\neq 0$.
If $\xi _{r}^{(A,B)T}J\xi _{i}^{(A,B)}=0$,
we can replace $\xi _{i}^{(A,B)}$ by
$\xi _{i}^{(A,B)\prime }=\xi _{i}^{(A,B)}+\delta ^{(A,B)} J\xi _{r}^{(A,B)}$ with
$\delta ^{(A,B)} \neq 0$, so that $\xi _{r}^{(A,B)T}J\xi _{i}^{(A,B)\prime }\neq 0$ and
$\xi ^{\dagger }(\gamma +iJ)\xi \rightarrow \xi ^{\dagger }(\gamma +iJ)\xi
+O(\delta ^{(A,B)})$.
Therefore, Eq.~(\ref{eq:NPPT_1_in_distillation_lemma_3}) still holds if $|\delta ^{(A,B)}|$ is sufficiently small.
Now we construct local canonical transformations $S_{A}$ as follows.
We choose $f_{j}^{(A)} $ $(j=1,2,\cdots ,2n_{A})$ such that
$f_{2j}^{(A)T}J_{A}f_{2k-1}^{(A)}=\delta _{jk}$,
$f_{2j-1}^{(A)T}J_{A}f_{2k-1}^{(A)}=f_{2j}^{(A)T}J_{A}f_{2k}^{(A)}=0$
$(j,k=1,2,\cdots ,n_{A})$
with
$f_{1}^{(A)} = \xi _{r}^{(A)} $ and
$f_{2}^{(A)}=-(\xi _{r}^{(A)T }J\xi _{i}^{(A)})^{-1}\xi _{i}^{(A)}$.
This is always possible and $\{f_{j}^{(A)}\}_{j=1}^{2n_{A}}$
constitutes a basis of the local phase space $X_{A}$ \cite{Arn89}.
Here, we define $S_{A}^{T} e_{j}^{(A)} = f_{j}^{(A)} $ $(j=1,2,\cdots ,2n_{A})$ for
$e_{1}^{(A)} =(1,0,0,\cdots ,0)^{T},
e_{2}^{(A)} =(0,1,0,\cdots ,0)^{T},\cdots \in \mathbb{R}^{2n_{A}}$.
The matrix $S_{A}$ (and $S_{A}^{T}$) thus defined is shown to be symplectic.
The matrix $S_{B}$ is also defined in a similar manner.
A vector defined as
$\eta _{A}=(S_{A}^{T})^{-1}\xi ^{(A)}$ takes the form
$a_{A}e_{1}^{(A)}+b_{A}e_{2}^{(A)}$ by definition of $S_{A}$, where
$a_{A},b_{A}\in \mathbb{C}$.
In a similar manner, a vector $\eta _{B}=(S_{B}^{T})^{-1}\xi ^{(B)}$ is written in the form
$a_{B}e_{1}^{(B)}+b_{B}e_{2}^{(B)}$, where $a_{B},b_{B}\in \mathbb{C}$.
Now, we apply the local symplectic transformation $S_{A}\oplus S_{B} $ on $\gamma $;
\begin{equation} \label{eq:canonical_in_distillability_lemma_3}
\gamma ^{\prime }=(S_{A}\oplus S_{B})\gamma (S_{A}\oplus S_{B})^{T} .
\end{equation}
By noting $(S_{A}\oplus S_{B})\widetilde{J}(S_{A}\oplus S_{B})^{T}=\widetilde{J}$,
Eq.~(\ref{eq:NPPT_1_in_distillation_lemma_3}) is rewritten as
\begin{equation} \label{eq:NPPT_2_in_distillation_lemma_3}
\eta ^{\dagger }(\gamma ^{\prime}+i\widetilde{J})\eta <0 ,
\end{equation}
where $\eta = \eta _{A} \oplus \eta _{B} $.
Since all components of vectors $\eta _{A} $ and $\eta _{B} $ other than their first two are zero,
Eq.~(\ref{eq:NPPT_2_in_distillation_lemma_3}) is further rewritten as
\begin{equation} \label{eq:NPPT_3_in_distillation_lemma_3}
\eta ^{\prime \dagger }(\gamma _{red}+i\widetilde{J}_{red})\eta ^{\prime }<0 ,
\end{equation}
where $\eta ^{\prime } =(a_{A},b_{A})^{T}\oplus (a_{B},b_{B})^{T}$,
\[
\gamma _{red}=\left(
\begin{array}{cccc}
\gamma _{1,1}^{\prime } & \gamma _{1,2}^{\prime } & \gamma
_{1,n_{A}+1}^{\prime } & \gamma _{1,n_{A}+2}^{\prime } \\
\gamma _{2,1}^{\prime } & \gamma _{2,2}^{\prime } & \gamma
_{2,n_{A}+1}^{\prime } & \gamma _{2,n_{A}+2}^{\prime } \\
\gamma _{n_{A}+1,1}^{\prime } & \gamma _{n_{A}+1,2}^{\prime } & \gamma
_{n_{A}+1,n_{A}+1}^{\prime } & \gamma _{n_{A}+1,n_{A}+2}^{\prime } \\
\gamma _{n_{A}+2,1}^{\prime } & \gamma _{n_{A}+2,2}^{\prime } & \gamma
_{n_{A}+2,n_{A}+1}^{\prime } & \gamma _{n_{A}+2,n_{A}+2}^{\prime }
\end{array}
\right) ,
\]
and $\widetilde{J}_{red}=(-J_{1}) \oplus J_{1} $.
$\gamma _{red} $ is the covariance matrix of a $1\times 1$ state that is obtained by local canonical transformations [Eq.~(\ref{eq:canonical_in_distillability_lemma_3})] followed by tracing out the local modes other than the first one at each local side.
Since the sequence of these operations is LOCC, it partially destroys the entanglement of the state.
However, the resulting state is still NPPT owing to Eq.~(\ref{eq:NPPT_3_in_distillation_lemma_3}).
\hfill $\Box$

\noindent{\em Proof of Theorem 12.}
Since distillable states are always NPPT by Theorem 2,
it suffices to show that the NPPT condition implies the distillability.
If a $n_{A} \times n_{B}$ Gaussian state is NPPT, it can be transformed into a NPPT $1 \times 1$ Gaussian state by LOCC (Lemma 5).
This state can be further transformed into a symmetric $1 \times 1$ Gaussian state by LOCC (Lemma 4).
The resulting state is distillable by Lemma 3.
\hfill $\Box$

\smallskip

Several distillation protocols for Gaussian states have been proposed so far \cite{eisertnp,DGC00b,DGC00c,FMF03}.
For example, the protocol of Duan {\it et al.} utilizes the local quantum nondemolition measurement and distills Gaussian states to finite-dimensional maximally entangled pure states \cite{DGC00b,DGC00c}.
This scheme is physically feasible but the experimental realization may be a long way off.
So the natural question is:
Can we make a distillation protocol with local quantum operations that are much more tractable experimentally?
One such a candidate for local quantum operations is a Gaussian operation or Gaussian channel that maps every Gaussian state into a Gaussian state.
The Gaussian operation is shown to be implementable by linear optics with homodyne detection \cite{GC02,ESP02,Fiu02} that are experimentally accessible with current technology.
The recent research work, however, have dashed this hope.
It has revealed the following discouraging fact: Gaussian states cannot be distilled by local Gaussian operations and classical communication \cite{GC02,ESP02,Fiu02}.
This means that a distillation protocol for Gaussian states must involve non-Gaussian operations at some stage.
In the protocol proposed by Eisert {\it et al.} \cite{eisertnp}, a non-Gaussian operation is required only in the initial step and a series of Gaussian operations of \cite{BES03} is applied in the subsequent steps.
The non-Gaussian operation does not require too much; it consists of a single measurement that distinguishes between vacuum states and non-vacuum ones.

\subsection{Conclusions}

In this section, we have reviewed the entanglement properties of Gaussian states starting from the fundamental level.
We have focused on the qualitative characterization of Gaussian state entanglement, i.e., the separability and distillability and have given the complete proofs on several results.
The important point is that the complete characterization by separability and distillability has been established for bipartite Gaussian states.

For the quantitative description of entanglement, some entanglement measures \cite{Hor01a} are required.
Among them, the logarithmic negativity $E_{N}$ (or negativity) is computable and is widely used in the theory of Gaussian state entanglement, although it is not a true entanglement measure in a rigorous sense \cite{VW02b}.
The explicit formula of $E_{N}$ for a $1 \times 1$ Gaussian state can be written in terms of parameters of the covariance matrix in the standard form.
General formula of other entanglement measures such as entanglement of formation ($E_{F}$), relative entropy of entanglement, and the distillable entanglement for general Gaussian states are not known yet.
However, Giedke {\it et al.} give a closed formula for $E_{F}$ of a $1 \times 1$ symmetric Gaussian state \cite{GWK03}.
One of the biggest problem related to the entanglement measures is the additivity question of $E_{F}$, i.e., whether or not $E_{F}(\rho _{1}\otimes \rho _{2})=E_{F}(\rho _{1})+E_{F}(\rho _{2})$.
Is this additivity holds for Gaussian states?
The answer is not known yet.

We do not intend that this section is a comprehensive guide to literatures.
So, many important literatures have been omitted in the references.
Some of them are, however, found in the excellent review articles \cite{EP03,BL}.

\section{Classical capacities of Gaussian channels}\label{seccc}
In this section, we review the classical capacities of Gaussian channels and their properties.
Firstly, we introduce basic ideas of classical information transmission via a quantum channel in Section~\ref{sec:Classical_Capacity_of_Quantum_Channels},
which includes the concept of mutual information, Shannon capacity, Holevo-Schumacher-Westmoreland capacity, and Holevo capacity.
The dense coding scheme is also introduced.
Secondly, we define the Gaussian channels and show the additivity property of a special case of Gaussian channel -- a lossy or attenuation channel in Section~\ref{sec:Gaussian_Channels}.
The explicit formula of the von Neumann entropy for Gaussian states is given and its properties are discussed.
Thirdly, we discuss the properties of Gaussian channels with Gaussian inputs in Section~\ref{sec:Gaussian_Channels_with_Gaussian_Inputs},
where we note that the Gaussian Holveo capacity is greatly simplified due to the special properties of von Neumann entropy of Gaussian states discussed in Section~~\ref{sec:Gaussian_Channels}.
We mention the additivity properties of some Gaussian channels with Gaussian inputs.
Finally, in Section~\ref{sec:Dense_Coding_with_Gaussian_Entanglement}, we review the dense coding with Gaussian state entanglement.

\subsection{Classical Capacities of Quantum Channels} \label{sec:Classical_Capacity_of_Quantum_Channels}

In this section, we introduce basic ideas of classical information transmission via a quantum channel \cite{nielsen,Hay06}.
Suppose that the sender (Alice) has some messages $X=\{x_{1},x_{2},\cdots ,x_{n}\}$ that she wants to send to the receiver (Bob) by using a noisy quantum channel $\Phi $.
To this end, Alice encodes her message using a quantum state $\rho _{x}$ ($x\in X$) with a priori probability $p_{x}$.
The input state $\rho _{x}$ is changed to $\Phi (\rho _{x}) $ at the output of the channel,
where Bob decode the message by measuring the received state $\Phi (\rho _{x}) $ by a suitable positive operator-valued measures (POVM) $\{E_{y}\}=\{E_{1},E_{2},\cdots ,E_{m}\}$ with the probability $\mathrm{Tr}(E_{y}\Phi (\rho _{x}))$.
The channel is completely characterized by the conditional probability $p_{y|x}$ for the probability of obtaining $y\in Y$ given that Alice sent a message $x\in X$ if the channel is memoryless and the signal states $\rho _{x}$ and the POVM $\{E_{y}\}$ are fixed.
In the memoryless channel, the signal states $\rho _{x}$ are independent of earlier or later usage of the channel.
The probability of obtaining $y\in Y$ is given by $p_{y}=\sum_{x}p_{y|x}p_{x}$ and the joint probability of $x\in X$ and $y\in Y$ is computed as $p_{x,y}=p_{y|x}p_{x}$.
The amount of information gained by Bob is given by the following mutual information,
\begin{equation} \label{eq:mutual_information_1}
H(X:Y)=H(X)+H(Y)-H(X,Y)=H(Y)-H(Y|X),
\end{equation}
where $H(X)=-\sum_{x}p_{x}\ln p_{x}$
and
$H(X,Y)=-\sum_{x}p_{x,y}\ln p_{x,y}$
are Shannon entropies and
\[
H(Y|X)=\sum_{x}p_{x}H(Y|X=x)
\]
denotes the conditional entropy with $H(Y|X=x)=-\sum_{y}p_{y|x}\ln p_{y|x}$.
Note that the mutual information [Eq.~(\ref{eq:mutual_information_1})] is written as
\begin{equation} \label{eq:mutual_information_2}
H(X:Y)=\sum_{x,y}p_{x,y}=p_{y|x}p_{x}\ln \frac{p_{y|x}}{p_{y}}.
\end{equation}

One of the most important bounds on the mutual information is the following Holevo bound.
\[
H(X:Y)\leq S(\rho )-\sum_{x}p_{x}S(\rho _{x}),
\]
where $\rho =\sum_{x}p_{x}\rho _{x}$.

When the set of signal states $\{\rho _{x}\}$ and the POVM $\{E_{y}\}$ are given, the achievable classical capacity of a quantum channel is calculated as the supremum of the mutual information over all probability distributions of input signal states;
\begin{equation} \label{eq:Shannon_capacity}
C_{S}(\Phi )=\sup_{p_{x}}H(X:Y).
\end{equation}
This is the quantum version of Shannon's noisy channel coding theorem and we call the capacity given by Eq.~(\ref{eq:Shannon_capacity}) the Shannon capacity.

Yet, a further optimization over the quantum states to be sent and the POVM yields the ultimate information transmission rate.
This defines the usual classical capacity of a quantum channel.
By noting entangled signal states are allowed for $n$ invocations of the channel, the classical capacity of the quantum channel $\Phi$ is calculated as
\begin{equation} \label{eq:HSW_capacity}
C_{HSW}(\Phi )=\lim_{n\rightarrow \infty }\frac{1}{n}\chi (\Phi ^{\otimes n})
\end{equation}
with
\begin{equation} \label{eq:Holevo_capacity}
\chi (\Phi )=\sup_{\{p_{x},\rho _{x}\}}\left[ S\left( \Phi (\rho )\right)
-\sum_{x}p_{x}S\left( \Phi (\rho _{x})\right) \right],
\end{equation}
where $\rho =\sum_{x}p_{x}\rho _{x}$ and
the supremum is taken over all probability distributions and sets of states.
This result is the content of the cerebrated Holveo-Schmacher-Westmoreland (HSW) theorem \cite{Hol98a,SW97}.
Hereafter we call the capacity given by Eq.~(\ref{eq:HSW_capacity}) the HSW capacity.
The quantity defined by Eq.~(\ref{eq:Holevo_capacity}) is called the Holevo capacity.

If the input $\rho _{x}^{(n)}$ is a product states
$\rho _{x}^{(n)}=\rho _{x,1}\otimes \rho _{x,2}\otimes \cdots \otimes \rho_{x,n}$ for
$n$ invocations of channel $\Phi$, then
$\chi (\Phi ^{\otimes n})=n\chi (\Phi )$ so that $C_{HSW}(\Phi )=\chi (\Phi )$.
That is, the Holevo capacity $\chi (\Phi )$ gives the single-shot HSW capacity
$C_{HSW}^{(1)}(\Phi )$.
By definition, $C_{HSW}^{(1)}(\Phi )\leq C_{HSW}(\Phi )$.
However, it is conjectured that
\begin{equation} \label{eq:additivity_of_Holevo_capacity}
\chi \left( \bigotimes_{j=1}^{n}\Phi _{j}\right) =\sum_{j=1}^{m}\chi \left(
\Phi _{j}\right)
\end{equation}
holds.
This is called the additivity conjecture of the Holevo capacity \cite{AHW00}.
If $\Phi _{j} $ in Eq.~(\ref{eq:additivity_of_Holevo_capacity}) are identical for all $i$,
we have $C_{HSW}^{(1)}(\Phi )=C_{HSW}(\Phi )$.
Even this weak form of additivity question is still open for general quantum channels.

The classical capacity can be increased if there is an additional resource in the form of an entangled state shared between Alice and Bob.
Such an augmented capacity is called the entanglement-assisted classical capacity \cite{BSS02}.
The dense coding \cite{bdensec} is the fundamental scheme of entanglement-assisted quantum communication.
The transmission of classical information via a quantum channel in the dense coding scheme goes as follows.
Initially, Alice and Bob share an entangled state $\rho _{AB} $.
Alice performs a unitary transformation $U_{m} $ [or generally a completely positive trace preserving (CPTP) map] on her state to encode the message and send the transformed state to Bob via the quantum channel.
Upon receiving this transformed state,
Bob combines it with his initially shared state and retrieve the message by POVM on the state
$(U_{m}\otimes \mathbf{I}_{B})\rho _{AB}(U_{m}^{\dagger }\otimes \mathbf{I}_{B})$.
If the quantum channel is noiseless and the initially shared state $\rho _{AB} $ is on the Hilbert space $\mathcal{H}_{A}\otimes \mathcal{H}_{B}$ with $\dim \mathcal{H}_{A}=\dim \mathcal{H}_{B}=d$,
the optimal capacity of the dense coding is given by \cite{Hir01}
\begin{equation} \label{eq:dense_coding_capacity}
C_{d}=\ln d+S(\rho _{B})-S(\rho _{AB}),
\end{equation}
where $\rho _{B}=\mathrm{Tr}_{A}\rho _{AB}$ and we have assumed the unitary coding.
If $\rho _{AB}$ is the maximally entangled state,
$\rho _{AB}=\left| \psi _{+}\right\rangle \left\langle \psi _{+}\right| $
with
$\left| \psi _{+}\right\rangle =d^{-1/2}\sum_{i=1}^{d}\left| i\right\rangle
_{A}\otimes \left| i\right\rangle _{B}$,
then $S(\rho _{B})-S(\rho _{AB})=\ln d$ so that $C_{d}=2\ln d$.
For $d=2$, 2 bits of classical information can be transmitted by sending a single qubit.
The capacity of Eq.~(\ref{eq:dense_coding_capacity}) is based on the HSW theorem.
If the signal states and the decoding scheme are given, the dense coding capacity is computed via the Shannon capacity [Eq.~(\ref{eq:Shannon_capacity})].

\subsection{Gaussian Channels} \label{sec:Gaussian_Channels}

A Gaussian channel $\Phi $ is a CPTP map that transforms the Weyl
operator $\mathcal{W}(\xi )=\exp (i\xi ^{T}JR)$' as follows.
$\mathcal{W}(\xi )=\exp (i\xi ^{T}JR)$ as follows.
\[
\mathcal{W}(\xi )\mapsto \mathcal{W}(X\xi )\exp \left( -\frac{1}{2}\xi
^{T}Y\xi \right),
\]
where $X$ and $Y$ are $2n\times 2n$ real matrices $Y$ is positive and symmetric ($Y = Y^{T} \geq 0$).
Here, we have assumed that both the input and output Gaussian are $n$ mode state.
The complete positivity of the channel is expressed in terms of these matrices as \cite{Lin00}
\[
Y+iJ_{n}-iX^{T}J_{n}X\geq 0.
\]

Among quantum channels for continuous variable systems, a Gaussian channel has its own significance.
Namely, it corresponds to the so-called Gaussian operations that can be implemented by current experimental techniques such as beam splitters, phase shifters, squeezers, and homodyne measurements.
Optical fibers in optical systems are well modeled by Gaussian channels.
The covariance matrix is transformed according to
\begin{equation} \label{eq:covariance_matrix_transformation_of_Gaussian_channel}
\gamma \mapsto \phi (\gamma )=X^{T}\gamma X+Y,
\end{equation}
Hereafter, we write a Gaussian channel by a capital Greek letter and the corresponding transformation on the covariance matrix by the corresponding lower case Greek letter.

The von Neumann entropy plays a fundamental role in the analysis of various capacities of quantum channels.
The von Neumann entropy of a Gaussian state is given in terms of the symplectic eigenvalues of the state.
Here, we derive the formula for the von Neumann entropy of a Gaussian state following \cite{HSH00,SEW05}.
We note that the von Neumann entropy can be computed as
\begin{equation} \label{eq:entropy_by_p-norm}
S(\rho )=-\lim_{p\rightarrow 1+}\frac{d}{dp}\left\| \rho \right\| _{p},
\end{equation}
where
$\left\| \rho \right\| _{p}=\left( \mathrm{Tr}\left| \rho \right| ^{p}\right) ^{1/p}$
is the Schatten $p$-norm ($p\geq 1$) and $\left| A\right| =\sqrt{A^{\dagger }A}$.
Without loss of generality, we can assume that the displacement of the Gaussian state is zero because of the unitary invariance of the von Neumann entropy, by noting that the the displacement of the Gaussian state can be always set to be zero by a unitary transformation.
By virtue of Williamson theorem, the covariance matrix is written as
\begin{equation} \label{eq:Williamson_standard_form_of_covariance}
\gamma \mapsto S\gamma S^{T}=\bigoplus_{j=1}^{n}\nu _{j}\mathbf{I}_{2}
\end{equation}
with $\nu _{j}\in [1,\infty )$ and $S\in Sp(2n,\mathbb{R})$.
This symplectic transformation corresponds to a unitary transformation in the underlying Hilbert space.
Therefore, a Gaussian state is unitarily equivalent to the state with covariance matrix given by the right-hand side of Eq.~(\ref{eq:Williamson_standard_form_of_covariance}) and vanishing displacement, which is written as
\begin{equation} \label{eq:thermal_state_in_Fock_space}
\rho =\bigoplus_{j=1}^{n}\frac{2}{\nu _{j}+1}\sum_{k=0}^{\infty }\left(
\frac{\nu _{j}-1}{\nu _{j}+1}\right) ^{k}\left| k\right\rangle
_{jj}\left\langle k\right|
\end{equation}
in the Fock space representation, where
\[
\left| k\right\rangle _{j}=\frac{1}{\sqrt{k!}}(a_{j}^{\dagger })^{k}\left|
0\right\rangle, \qquad k = 0,1,\cdots
\]
denotes the number state of the $j$th mode.
Equation.~(\ref{eq:thermal_state_in_Fock_space}) describes the state in a thermal equilibrium (the thermal state) with the average photon number of the $j$th mode being
$\left\langle n\right\rangle _{j}=(\nu _{j}-1)/2$.
From Eq.~(\ref{eq:thermal_state_in_Fock_space}), we have
\[
\mathrm{Tr}\rho ^{p}=\prod_{j=1}^{n}\frac{2^{p}}{f_{p}(\nu _{j})}
\]
with $f_{p}(x)=(x+1)^{p}-(x-1)^{p} $.
Therefore, form Eq.~(\ref{eq:entropy_by_p-norm})
\begin{equation} \label{eq:entropy_of_Gaussian_state}
S(\rho )=\sum_{j=1}^{n}g\left( \frac{\nu _{j}-1}{2}\right)
\end{equation}
with
\begin{equation} \label{eq:g-function_in_entropy_of_Gaussian_state}
g(x)=(x+1)\ln (x+1)-x\ln x.
\end{equation}
Equation~(\ref{eq:entropy_of_Gaussian_state}) with
Eq.~(\ref{eq:g-function_in_entropy_of_Gaussian_state}) gives
the von Neumann entropy of a Gaussian state $\rho $.

Since
\[
\left\langle k\right| \left( \frac{\nu _{j}-1}{\nu _{j}+1}\right)
^{a_{j}^{\dagger }a_{j}}\left| k\right\rangle =\left( \frac{\nu _{j}-1}{\nu
_{j}+1}\right) ^{k},
\]
the state [Eq.~(\ref{eq:thermal_state_in_Fock_space})] is also written as
\[
\rho =\bigotimes_{j=1}^{n}\frac{2}{\nu _{j}+1}\left( \frac{\nu _{j}-1}{\nu
_{j}+1}\right) ^{a_{j}^{\dagger }a_{j}}.
\]
It is evident from this equation that $\ln \rho $ depends linearly on $n_{j}=a_{j}^{\dagger }a_{j}$.
This observation leads to the following important theorem \cite{HSH99}.

\smallskip

\noindent{\em Theorem 1. --}
Among all states with fixed covariance matrix, the Gaussian state is one which maximizes the von Neumann entropy.

\noindent{\em Proof.}
Let $\widetilde{\rho }$ be a Gaussian state with fixed covariance matrix.
There exists a state $\rho $ with the same covariance matrix.
Without loss of generality, we assume that the displacements of $\widetilde{\rho }$ and $\rho $ are zero.
We thus have
\[
S(\widetilde{\rho })-S(\rho )=\mathrm{Tr}\rho (\ln \rho -\ln \widetilde{\rho }%
)+\mathrm{Tr}(\rho -\widetilde{\rho })\ln \widetilde{\rho }.
\]
The first term on the right-hand side is the nonnegative relative entropy and the second term vanishes since
$\mathrm{Tr}\left( \rho a_{j}^{\dagger }a_{j}\right) =\mathrm{Tr}\left( \widetilde{\rho }%
a_{j}^{\dagger }a_{j}\right) $
and $\ln \widetilde{\rho }$ is a linear function of $n_{j}=a_{j}^{\dagger }a_{j}$.
Therefore, we have $S(\widetilde{\rho })\geq S(\rho )$.
This completes the proof.
\hfill $\Box$

\smallskip

Important examples of Gaussian channels are classical channels and thermal noise channels including lossy or attenuation channels.

In the classical noise channel,
a classical Gaussian noise is added to the input states.
Since
$\mathcal{W}(\xi )R_{j}\mathcal{W}^{\dagger }(\xi )=R_{j}+\xi _{j}$,
the classical noise channel is described by
\begin{equation} \label{eq:classical_noise_channel}
\Phi ^{cl}(\rho _{(\gamma ,m)}) =\frac{1}{\pi ^{n}\sqrt{\det Y}}
\int d^{2n}\xi \exp (-\xi ^{T}Y^{-1}\xi )\mathcal{W}(\xi )\rho _{(\gamma ,m)}
\mathcal{W}^{\dagger }(\xi )
=\rho _{(\gamma +Y,m)}
\end{equation}
with $Y\geq 0$.
In Eq.~(\ref{eq:classical_noise_channel}),
$\rho _{(\gamma ,m)}$ stands for a Gaussian state
with covariance matrix $\gamma $ and displacement $m$.
Namely, the transformations of covariance matrix is given by
\[
\phi ^{cl}(\gamma )=\gamma +Y.
\]

In the thermal noise channel, the signal Gaussian states interact
with an environment that is in thermal equilibrium. This channel is
modeled by beam splitters that couple the input Gaussian state and
the thermal reservoir. Let $a_{j}$ and $b_{j}$ be annihilation
operators of the $j$th mode of the singnal state $\rho $ and the
thermal state $\rho _{th}$ that acts as a thermal reservoir. The
action of the beam splitter is described by the transformations,
$a_{j}\mapsto \cos \theta _{j} a_{j}+\sin \theta _{j} b_{j} $ and
$b_{j}\mapsto -\sin \theta _{J} a_{j}+\cos \theta _{j} b_{j} $.
Accordingly, the corresponding symplectic transformation $S_j$ takes
the form of Eq.(\ref{eq:beam_splitter}) with $\theta=\theta_j$,
Therefore, the output Gaussian state has the covariance matrix,
\begin{equation} \label{eq:thermal_noise_channel_covariance_1}
\phi ^{th}(\gamma )=\mathrm{Tr}_{th}[S^{-1}(\gamma \oplus \gamma _{th})(S^{-1})^{T}],
\end{equation}
where $S=\bigoplus_{j=1}^{n}S_{j}$, $\gamma
_{th}=\bigoplus_{j=1}^{n}\left( 2\left\langle n\right\rangle
_{j}+1\right) \mathbf{I}_{2}$ denotes the covariance matrix of the
thermal state $\rho _{th}$ with $\left\langle n\right\rangle _{j}$
being the average photon number of the $j$th mode, and
$\mathrm{Tr}_{th}$ describes the trace over the thermal state.  The
right-hand side of Eq.~(\ref{eq:thermal_noise_channel_covariance_1})
is calculated as
\begin{equation} \label{eq:thermal_noise_channel_covariance_2}
\phi ^{th}(\gamma )=X^{T}\gamma X+Y,
\end{equation}
where
\begin{equation} \label{eq:thermal_noise_channel_covariance_X}
X=\bigoplus_{j=1}^{n}\sqrt{\eta _{j}}\mathbf{I}_{2}
\end{equation}
and
\begin{equation} \label{eq:thermal_noise_channel_covariance_Y}
Y=\bigoplus_{j=1}^{n}\left( 2\left\langle n\right\rangle _{j}+1\right) (1-\eta _{j})\mathbf{I}_{2}
\end{equation}
with $\eta _{j}=\cos ^{2}\theta _{j}$ being the transmittivity of the beam splitter.
At zero temperature ($\left\langle n\right\rangle _{j}=0$),
the thermal noise channel is reduced to the lossy or attenuation channel \cite{HW02,EW05}.

At present, the additivity of Holevo capacity is proven and the HSW
capacity is computed exactly only for the lossy channel among
Gaussian channels\cite{GGL04b}, which is shown in the following. In
the following discussion, we need the following theorem
\cite{Bek81}.

\smallskip

\noindent{\em Theorem 2. --}
Among all single mode states (that are not necessarily Gaussian)
with fixed average photon number,
the thermal state is the one that maximizes the von Neumann entropy.

\noindent{\em Proof.}
The average photon number of the mode of frequency $\omega $ and therefore
the energy $E=\mathrm{Tr}(\rho H)$ of the mode is fixed with $H=\omega (a^{\dagger }a+1/2)$.
The problem is to find the state maximizing the von Neumann entropy
$S=-\mathrm{Tr}(\rho \ln \rho )$ under the normalization condition $\mathrm{Tr}\rho =1 $.
This is given by the solution of the variational problem
\[
\delta \left[ -\mathrm{Tr}(\rho \ln \rho )-\mu \mathrm{Tr}(\rho H)-\nu \mathrm{Tr}\rho \right] =0,
\]
where $\mu $ and $\nu $ are Lagrange multipliers.
Direct computations yield
\[
\rho =\frac{1}{Z(\beta )}\exp (-\beta H)
\]
with $Z(\beta )=\mathrm{Tr}\exp (-\beta H)$ and $\beta =(S/E)_{\max }$.
That is, the state $\rho $ is a thermal state that is in thermal equilibrium
with temperature $\beta ^{-1}$.
\hfill $\Box$

\smallskip

Let $\Phi$ be a lossy channel that maps an $m$ mode Gaussian state to an $m$ mode Gaussian state.
By definition,
\begin{equation} \label{eq:lossy_channel_capacity_1}
C_{HSW}(\Phi ) \geq \chi (\Phi )
=\sup_{\{p_{x},\rho _{x}\}}\left[ S\left( \Phi (\overline{\rho })\right)
-\sum_{x}p_{x}S\left( \Phi (\rho _{x})\right) \right],
\end{equation}
where $\overline{\rho }=\sum_{x}p_{x}\rho _{x}$.
Since the states $\rho _{x}$ are infinite-dimensional states,
the right-hand side of Eq.~(\ref{eq:lossy_channel_capacity_1}) becomes arbitrarily large
if we do not impose some constraint on the signal states.
Here, we take the energy constraint
\begin{equation} \label{eq:energy_constraint_1}
\sum_{k=1}^{m}\omega _{k}N_{k}=\mathcal{E}
\end{equation}
and write $C_{HSW}(\Phi ,\mathcal{E})$ instead of $C_{HSW}(\Phi )$ as the HSW capacity for the channel $\Phi $.
In Eq.~(\ref{eq:energy_constraint_1}),
$\omega _{k}$ denotes the frequency of $k$th mode
and $N_{k}$ stands for the average photon number of the $k$th mode of the average input state;
Eq.~(\ref{eq:energy_constraint_1}) describes the constraint
that the energy (excluding the zero point oscillation energy) of the average input state $\overline{\rho }$
should be $\mathcal{E}$.

Now we take the tensor product of coherent states
\begin{eqnarray} \label{eq:coherent_state_input_for_lossy_channel}
\rho (m_{1},m_{2},\cdots ,m_{m}) &=&\rho _{(\gamma ,m_{1})}\otimes \rho
_{(\gamma ,m_{2})}\otimes \cdots \otimes \rho _{(\gamma ,m_{m})} \nonumber \\
&=&\left| \alpha _{1}\right\rangle \left\langle \alpha _{1}\right| \otimes
\left| \alpha _{2}\right\rangle \left\langle \alpha _{2}\right| \otimes
\cdots \otimes \left| \alpha _{m}\right\rangle \left\langle \alpha
_{m}\right|
\end{eqnarray}
for the input signal state and assume that the probability distribution of the displacement $m_{k}$ of the $k$th mode takes the form,
\begin{eqnarray} \label{eq:probability_distribution_for_lossy_channel}
P_{k}(m_{k})=\frac{1}{2\pi N_{k}}\exp \left[ -\frac{m_{k}^{2}}{2N_{k}}%
\right]
\end{eqnarray}
instead of optimizing the signal ensemble in the right-hand side of
Eq.~(\ref{eq:lossy_channel_capacity_1}).
In Eq.~(\ref{eq:coherent_state_input_for_lossy_channel}), $\gamma =\mathrm{diag}(1,1,\cdots ,1,1)$ and $m_{k}=\sqrt{2}\alpha _{k}$ ($k=1,2,\cdots ,m$).

The output state $\Phi (\rho _{x})$ is still a coherent state \cite{Hel81}.
This is easily seen by Eqs.~(\ref{eq:thermal_noise_channel_covariance_2}), (\ref{eq:thermal_noise_channel_covariance_X}), and (\ref{eq:thermal_noise_channel_covariance_Y}).
The average output state is calculated as
\[
\Phi (\overline{\rho }) =\int \prod_{k=1}^{m}dm_{k}P_{k}(m_{k})\rho (m_{1},m_{2},\cdots ,m_{k})
=\rho _{(\gamma _{1}0)}\otimes \rho _{(\gamma _{2},0)}\otimes \cdots
\otimes \rho _{(\gamma _{m},0)},
\]
where $\gamma _{k}=\mathrm{diag}(1+2\eta _{k}N_{k},1+2\eta _{k}N_{k})$.
That is, the state $\Phi (\overline{\rho })$ is a tensor product of thermal states with average photon numbers $\eta _{k}N_{k}$ ($k=1,2,\cdots,m$).
Since the state $\Phi (\rho _{x})$ is pure, $S\left( \Phi (\rho _{x})\right) =0$ so that
\[
S\left( \Phi (\overline{\rho })\right) -\sum_{x}p_{x}S\left( \Phi (\rho
_{x})\right)
=S\left( \Phi (\overline{\rho })\right) =\sum_{k=1}^{m}g(\eta _{k}N_{k}).
\]
Therefore, we have
\begin{equation} \label{eq:HSW_capacity_of_lossy_channel_lower_bound}
C_{HSW}(\Phi ,\mathcal{E})\geq \max_{N_{k}}\sum_{k=1}^{m}g(\eta _{k}N_{k}),
\end{equation}
where maximum in the right-hand side of Eq.~(\ref{eq:HSW_capacity_of_lossy_channel_lower_bound}) is taken for all $N_{k}$ satisfying the energy constraint (\ref{eq:energy_constraint_1}).

Next, let $\{p_{x}^{*},\rho _{x}^{*}\}$ be the optimal signal ensemble, which gives the capacity $C_{HSW}(\Phi)$.
Since $S\left( \Phi ^{\otimes n}(\rho _{x}^{*})\right) \geq 0$, we have
\[
C_{HSW}(\Phi ,\mathcal{E})\leq \lim_{n\rightarrow \infty }\frac{1}{n}S\left( \Phi ^{\otimes n}(%
\overline{\rho }^{*})\right)
\]
with $\overline{\rho }^{*}=\sum_{x}p_{x}^{*}\rho _{x}^{*}$.
By the subadditivity of the von Neumann entropy \cite{OP93}, we have
\[
S\left( \Phi ^{\otimes n}(\overline{\rho }^{*})\right) \leq
\sum_{l=1}^{n}\sum_{k=1}^{m}S\left( \Phi (\rho _{k}^{(l)})\right),
\]
where $\Phi (\rho _{k}^{(l)})$ is the reduced density operator of the $k$th mode in the $l$th realization of the channel, which is obtained from $\Phi ^{\otimes n}(\overline{\rho }^{*})$
by tracing over all the other modes and over the other $n-1$ channel realizations.
Now, let $N_{k}^{(l)}$ be the average photon number of the state $\rho _{k}^{(l)}$ so that
the average photon number of the output state $\Phi (\rho _{k}^{(l)})$ is $\eta _{k}N_{k}^{(l)}$.
Since among all states with this average photon number, the thermal state has the maximal von Neumann entropy (Theorem 2), we have
\[
S\left( \Phi (\rho _{k}^{(l)})\right) \leq g(\eta _{k}N_{k}^{(l)}).
\]
Therefore
\[
C_{HSW}(\Phi ,\mathcal{E})\leq \lim_{n\rightarrow \infty }\frac{1}{n}\sum_{l=1}^{n}%
\sum_{k=1}^{m}g(\eta _{k}N_{k}^{(l)}),
\]
where the summation in the right-hand side is taken under the constraint
\[
\sum_{k=1}^{m}\omega _{k}\sum_{l=1}^{n}\frac{N_{k}^{(l)}}{n}=\mathcal{E}.
\]
Note that $n^{-1}\sum_{l=1}^{n}N_{k}^{(l)}=N_{k} $ is the average photon number of the $k$th mode of the average input state.
Since $d^{2}g/dx^{2}=-1/[x(x+1)]<0$, $g$ is concave so that
\[
\frac{1}{n}\sum_{l=1}^{n}\sum_{k=1}^{m}g(\eta _{k}N_{k}^{(l)}) \leq
\sum_{k=1}^{m}g\left( \sum_{l=1}^{n}\eta _{k}\frac{N_{k}^{(l)}}{n}\right)
\leq \max_{N_{k}}\sum_{k=1}^{m}g(\eta _{k}N_{k}).
\]
Therefore,
\begin{equation} \label{eq:HSW_capacity_of_lossy_channel_upper_bound}
C_{HSW}(\Phi ,\mathcal{E})\leq \max_{N_{k}}\sum_{k=1}^{m}g(\eta _{k}N_{k}),
\end{equation}
where maximum in the right-hand side is taken for all $N_{k}$ satisfying the energy constraint (\ref{eq:energy_constraint_1}) again.
From Eqs.~(\ref{eq:HSW_capacity_of_lossy_channel_lower_bound}) and (\ref{eq:HSW_capacity_of_lossy_channel_upper_bound}), we have
\[
C_{HSW}(\Phi ,\mathcal{E}) = \max_{N_{k}}\sum_{k=1}^{m}g(\eta _{k}N_{k}).
\]
The right-hand side is the energy-constrained classical capacity of the lossy channel.
Note that the energy-constrained Holevo capacity is weakly additive;
$\chi \left( \Phi ^{\otimes n}\right) =n\chi \left( \Phi \right) $
and the optimal signal ensemble is given by Eqs.~(\ref{eq:coherent_state_input_for_lossy_channel}) and (\ref{eq:probability_distribution_for_lossy_channel}).

\subsection{Gaussian Channels with Gaussian Inputs} \label{sec:Gaussian_Channels_with_Gaussian_Inputs}

It is conjectured that the optimal signal states for Gaussian channels are Gaussian states \cite{HS04}.
However, it is not proven yet so that we have
\begin{equation} \label{eq:Gaussian_HSW_capacity_inequality}
C_{HSW}(\Phi ,\mathcal{E})\geq C_{G}(\Phi ,\mathcal{E})
\end{equation}
in general.
$C_{G}(\Phi ,\mathcal{E})$ denotes the HSW capacity of the Gaussian channel $\Phi $ for Gaussian state inputs under the energy constraint specified by $\mathcal{E} $, which is given in terms of the energy-constrained Holevo capacity for Gaussian state inputs or the Gaussian Holevo capacity \cite{Hol98b},
\begin{equation} \label{eq:Gaussian_Holevo_capacity}
\chi_{G}(\Phi ,\mathcal{E})=\sup_{\mu ,\rho _{(\gamma ,m)}}\left[ S(\Phi (\overline{\rho }%
))-\int \mu (d\gamma ,dm)S(\Phi (\rho _{(\gamma ,m)}))\right],
\end{equation}
where
\[
\overline{\rho }=\int \mu (d\gamma ,dm)\rho _{(\gamma ,m)}
\]
is the average signal state.
In Eq.~(\ref{eq:Gaussian_Holevo_capacity}),
the supremum is taken over all possible probability measures $\mu $
and signal states $\rho _{(\gamma ,m)}$ constituting the signal ensemble under the energy constraint
\begin{equation} \label{eq:energy_constraint_2}
\mathrm{Tr}\overline{\rho }H=\mathcal{E}
\end{equation}
with
\[
H =\sum_{k=1}^{n}\omega _{k}\left( a_{k}^{\dagger }a_{k}+\frac{1}{2}\right)
=\frac{1}{2}\sum_{k=1}^{n}\omega _{k}\left( R_{2k-1}^{2}+R_{2k}^{2}\right).
\]
The right-hand side of Eq.~(\ref{eq:Gaussian_HSW_capacity_inequality}) is given by
\[
C_{G}(\Phi ,\mathcal{E})=\lim_{n\rightarrow \infty }\frac{1}{n}\chi
_{G}\left( \Phi ^{\otimes n},\mathcal{E}^{\otimes n}\right).
\]

Some properties of the capacity of Gaussian channels are greatly simplified if we restrict the signal states to Gaussian ones.
In the rest part of this section, we discuss the additivity properties of Gaussian Holevo capacity.
Let us recall that the von Neumann entropy of a Gaussian state depends only on the covariance matrix, and that channel $\Phi $ affects only the covariance matrix.
Therefore, if we find a single state $\rho _{(\gamma ^{*},m)}$ that minimizes $S(\Phi (\rho ))$,
all possible Gaussian states $\rho $ with the covariance matrix $\gamma ^{*}$ also minimizes
$S(\Phi (\rho ))$.
This observation indicates that the optimal signal ensemble
that attains the Gaussian Holevo capacity consists of Gaussian states
with the common covariance matrix $\gamma ^{*}$ and a certain probability distribution
of the displacement $m$.
If we restrict the signal ensemble to that described above,
it suffices to take a Gaussian probability distribution for the probability measure
$\mu (d\gamma ,dm)=\mu (dm)$.
This is shown as follows \cite{HSH99}.
If $\mu (dm)$ is a Gaussian distribution;
\[
\mu (dm)=\frac{1}{\pi ^{n}\sqrt{\det Y_{\mu }}}\exp (-m^{T}Y_{\mu }^{-1}m)dm
\]
with $Y_{\mu } > 0$,
the average input signal state is calculated as
\begin{equation} \label{eq:average_input_signal_state}
\overline{\rho }=\int \mu (dm)\rho _{(\gamma ,m)}=
\rho _{(\gamma +Y_{\mu},0)}.
\end{equation}
That is, $\overline{\rho }$ is also a Gaussian state with the covariance matrix
$\overline{\gamma }=\gamma +Y_{\mu }$ and has the vanishing displacement.
Equation~(\ref{eq:average_input_signal_state}) even holds for $Y_{\mu } \geq 0$.
Since the displacement of $\overline{\rho }$ is zero,
\[
\mathrm{Tr}\overline{\rho }H =\frac{1}{2}\sum_{k=1}^{n}\omega _{k}\left(
\left\langle R_{2k-1}^{2}\right\rangle +\left\langle R_{2k}^{2}\right\rangle
\right)
=\frac{1}{4}\sum_{k=1}^{n}\omega _{k}\mathrm{Tr}\overline{\gamma }_{[k]},
\]
where
\[
\overline{\gamma }_{[k]}=\left(
\begin{array}{cc}
\overline{\gamma }_{2k-1,2k-1} & \overline{\gamma }_{2k-1,2k} \\
\overline{\gamma }_{2k,2k-1} & \overline{\gamma }_{2k,2k}
\end{array}
\right),
\]
so that the energy constraint [(\ref{eq:energy_constraint_2})] is written as
\begin{equation} \label{eq:energy_constraint_3}
\sum_{k=1}^{n}\omega _{k}\mathrm{Tr}(\gamma +Y_{\mu })_{[k]}=4\mathcal{E}.
\end{equation}
Since such a signal ensemble described above is not always optimal, we have
\begin{equation} \label{eq:Gaussian_Holevo_capacity_lower_bound_1}
\chi_{G}(\Phi ,\mathcal{E})\geq \sup_{\gamma, Y_{\mu }(\geq 0)}
\left[ S(\phi (\gamma +Y_{\mu
}))-S(\phi (\gamma ))\right],
\end{equation}
where the supremum is taken under the constraint (\ref{eq:energy_constraint_3}).
In Eq.~(\ref{eq:Gaussian_Holevo_capacity_lower_bound_1}), $S(\phi (\gamma ))$ stands for the von Neumann entropy of $\rho _{\phi (\gamma )}$, the Gaussian state with covariance matrix $\phi (\gamma )$.
Hereafter, we have occasions to write the von Neumann entropy as $S(\gamma )$ instead of $S(\rho _{\gamma })$ when we are dealing with Gaussian states.
Note that the right-hand side of Eq.~(\ref{eq:Gaussian_Holevo_capacity_lower_bound_1}) is written as\begin{equation} \label{eq:Gaussian_Holevo_capacity_lower_bound_2}
\sup_{\gamma }S(\phi (\gamma ))-\inf_{\gamma }S(\phi (\gamma ))
=\sup_{\gamma }S(\phi (\gamma ))-S_{\min }(\Phi ),
\end{equation}
where the supremum is taken under the constraint
\begin{equation} \label{eq:energy_constraint_4}
\sum_{k=1}^{n}\mathrm{Tr}\omega _{k}\gamma _{[k]}=4\mathcal{E}.
\end{equation}
In Eq.~(\ref{eq:Gaussian_Holevo_capacity_lower_bound_2}),
\[
S_{\min }(\Phi )=\inf_{\rho \in \mathcal{G}}S(\Phi (\rho ))
\]
defines the Gaussian mimimum output entropy and $\mathcal{G}$ denotes the set of all Gaussian states.
By Theorem 1, we can see that for Gaussian signal state $\rho _{(\gamma ,m)}$ and
the probability measure $\mu (d\gamma ,dm)$,
the quantity within the brackets of the right-hand side of Eq.~(\ref{eq:Gaussian_Holevo_capacity})
cannot exceed the value of the right-hand side of (\ref{eq:Gaussian_Holevo_capacity_lower_bound_1}).
Therefore, the equality holds in the inequality (\ref{eq:Gaussian_Holevo_capacity_lower_bound_1});
\[
\chi_{G}(\Phi ,\mathcal{E})=\sup_{\gamma}
S(\phi (\gamma))-S_{\min }(\Phi ).
\]
Again, the supremum is taken under the constraint [Eq.~(\ref{eq:energy_constraint_4})].


Let $\mathcal{E}$
be the value for the energy constraint [Eq.~(\ref{eq:energy_constraint_1})] for
the Gaussian channel $\Phi _{j}$ and $\mathcal{E}=\sum_{j=1}^{m}\mathcal{E}_{j}$.
From the definition, the Gaussian Holevo capacity of the tensor product channel
is greater than or equal to the supremum of the sum of the Gaussian Holevo capacity
of individual channels;
\begin{equation} \label{eq:Gaussian_Holevo_capacity_lower_bound_3}
\chi_{G}(\Phi ,\mathcal{E})\geq
\sup_{\{\mathcal{E}_{j}\},\sum_{j=1}^{m}\mathcal{E}_{j}=\mathcal{E}}
\sum_{j=1}^{m}\chi_{G}(\Phi _{j},\mathcal{E}_{j}).
\end{equation}
Here, the supremum is taken over all possible combinations of $\mathcal{E}_{j}$
under the constraint $\sum_{j=1}^{m}\mathcal{E}_{j}=\mathcal{E}$.
If the equality holds in the inequality (\ref{eq:Gaussian_Holevo_capacity_lower_bound_3}),
we say that the energy-constrained Gaussian Holevo capacity is additive
for Gaussian channels $\Phi _{j}$.
Now let $\rho $ be a Gaussian state on the composite Hilbert space
$\mathcal{H}_{1}\otimes \cdots \otimes \mathcal{H}_{m}$
and define
$\rho _{j}=\mathrm{Tr}_{\mathcal{H}_{1}\otimes \cdots
\otimes \mathcal{H}_{j-1}\otimes \mathcal{H}_{j+1}\otimes \cdots
\otimes \mathcal{H}_{m}}\rho $.
By noting the subadditivity of von Neumann entropy \cite{OP93},
$S(\rho )\leq \sum_{j=1}^{m}S(\rho _{j})$,
we have
\[
S(\phi (\gamma ))\leq \sum_{j=1}^{m}S(\phi _{j}(\gamma _{j})),
\]
where $\gamma _{j}$ denotes the covariance matrix of the Gaussian state $\rho _{j}$.
Therefore, if the minimal output entropy $S_{\min }(\Phi )$ is additive
for the channels $\Phi _{j}$;
\[
S_{\min }\left( \bigotimes_{j=1}^{m}\Phi _{j}\right) =\sum_{j=1}^{m}S_{\min
}\left( \Phi _{j}\right),
\]
then
\[
\chi_{G}(\Phi ,\mathcal{E})\leq
\sup_{\{\mathcal{E}_{j}\},\sum_{j=1}^{m}\mathcal{E}_{j}=\mathcal{E}}
\sum_{j=1}^{m}\chi_{G}(\Phi _{j}, \mathcal{E}_{j}).
\]
This implies the additivity of the energy-constrained Gaussian Holevo capacity;
\[
\chi_{G}(\Phi ,\mathcal{E})=\sup_{\{\mathcal{E}_{j}\},\sum_{j=1}^{m}\mathcal{E}_{j}=\mathcal{E}}
\sum_{j=1}^{m}\chi_{G}(\Phi _{j},\mathcal{E}_{j}).
\]

Due to Eq.~(\ref{eq:entropy_by_p-norm}) the additivity of the Gaussian mimimum output entropy is implied by the multiplicativity of the Gaussian maximal output $p$-norm for $p\rightarrow 1+ $.
\[
\lim_{p\rightarrow 1+}\xi _{p}\left( \bigotimes_{j=1}^{m}\Phi _{j}\right)
=\lim_{p\rightarrow 1+}\prod_{j=1}^{m}\xi _{p}\left( \Phi _{j}\right),
\]
where
\[
\xi _{p}(\Phi )=\sup_{\rho \in \mathcal{G}}\left\| \Phi (\rho )\right\| _{p}
\]
defines the Gaussian maximal output $p$-norm.

Serafini {\it et al.} \cite{SEW05} proved that
the Gaussian maximal output $p$-norm of a tensor product of
identical single mode Gaussian channels and that of single mode channels described by
$X_{i}$ and $Y_{i}$ [(\ref{eq:covariance_matrix_transformation_of_Gaussian_channel})]
such that $\det X_{i}$ are identical and $Y_{i}>0$ for all $i$,
are multiplicative for Gaussian state inputs for $p > 1$.
Consequently, the Gaussian minimal output entropy and energy-constrained Gaussian Holevo capacity
are additive for such tensor product channels.
Hiroshima \cite{Hir05} also proved the multiplicativity of the Gaussian maximal output $p$-norm of a tensor product of classical noise channels of arbitrary modes and that of thermal noise channels of arbitrary modes.
Therefore, the Gaussian minimal output entropy and energy-constrained Gaussian Holevo capacity
are additive for such tensor product channels.

The additivity properties of Gaussian channels under Gaussian state
inputs are shown to be equivalent to the additivity of Gaussian
entanglement of formation $E_{G}$ defined by Wolf {\it et al.}
\cite{WGKWC04}, which is an entanglement monotone under Gaussian
local operations and classical communication (GLOCC) [216]. By
definition, $E_{G}$ is an upper bound for the true entanglement of
formation $E_{F}$ for Gaussian states. However, for symmetric
two-mode Gaussian states, $E_{G}$ coincides with $E_{F}$
\cite{WGKWC04}. Furthermore, $E_{G}$ is additive for symmetric
states \cite{WGKWC04}.

As mentioned earlier, only lossy channels within the class of
Gaussian channels are proven to exhibit additivity properties --
additvity of energy-constrained HSW capacity and the minimum output
entropy. For general quantum channels, the additivity question is
still an open problem. Despite many efforts devoted to the
additivity problems of quantum channels, the additivity properties
have been proven for a few examples, such as entanglement breaking
channels \cite{Shor02}, unital qubit channels \cite{King02},
depolarizing channels \cite{King03}, and contravariant channels
\cite{MY04}.
\subsection{Dense Coding with Gaussian Entanglement} \label{sec:Dense_Coding_with_Gaussian_Entanglement}

The dense coding scheme in continuous variable systems was discussed by Ban \cite{cvdense0} and Braunstein-Kimble \cite{cvdense}.
In this section, we introduce the Braunstein-Kimble's scheme of dense coding.

Suppose that Alice and Bob share initially the two-mode squeezed state
$\rho _{12}=U(r)\left| 0\right\rangle \left\langle 0\right| U^{\dagger }(r)$,
where $U(r)=\exp [-r(a_{1}^{\dagger }a_{1}-a_{2}^{\dagger }a_{2})]$
is the two-mode squeezing operator with $r$ being the squeezing parameter ($r > 0$).
The mode 1 (2) is in the possession of Alice (Bob).
The characteristic function of the shared state $\rho _{12}$ is given by
\[
\chi (\eta _{1},\eta _{2})=\mathrm{Tr}[\rho _{12}\exp (\eta _{1}a_{1}^{\dagger }-\eta
_{1}^{*}a_{1})\exp (\eta _{2}a_{2}^{\dagger }-\eta _{2}^{*}a_{2})]
\]
so that the Wigner function takes the form,
\begin{eqnarray*}
W(\alpha _{1},\alpha _{1}) &=&\frac{1}{\pi ^{4}}\int d^{2}\eta _{1}\int
d^{2}\eta _{2}
\exp (\eta _{1}^{*}\alpha _{1}-\eta _{1}\alpha _{1}^{*})\exp (\eta
_{2}^{*}\alpha _{2}-\eta _{2}\alpha _{2}^{*})\chi (\eta _{1},\eta _{2}) \\
&=&\frac{4}{\pi ^{4}}\exp \left[ -e^{-2r}(\alpha _{1}-\alpha
_{2})_{R}^{2}-e^{2r}(\alpha _{1}-\alpha _{2})_{I}^{2}\right.  \\
&&\left. -e^{2r}(\alpha _{1}+\alpha _{2})_{R}^{2}-e^{-2r}(\alpha _{1}+\alpha
_{2})_{I}^{2}\right],
\end{eqnarray*}
where the subscript $R(I)$ refers to the real (imaginary) part of $\alpha _{1}\pm \alpha _{2}$.
Now, Alice performs a unitary transformation,
\[
D(\alpha _{in})=\exp (\alpha _{in}a_{1}^{\dagger }-\alpha _{in}^{*}a_{1})
\]
on the mode 1 to encode her message characterized by the displacement $\alpha _{in}$.
The state $\rho _{12} $ is transformed as
\begin{equation} \label{eq:signal_state_in_dense_coding}
\rho _{12}\rightarrow D(\alpha _{in})\rho _{12}D^{\dagger }(\alpha
_{in})=\rho _{12}^{\prime }.
\end{equation}
The characteristic fucntion of $\rho _{12}^{\prime }$ is computed as
\[
\chi ^{\prime }(\eta _{1},\eta _{2})=\exp (-\eta _{1}^{*}\alpha _{in}+\eta
_{1}\alpha _{in}^{*})
\]
so that the Wigner function of $\rho _{12}^{\prime }$ takes the form,
\[
W^{\prime }(\alpha _{1},\alpha _{2}) = W(\alpha _{1}-\alpha _{in},\alpha _{2}).
\]
After Alice send her state (mode 1) to Bob via a noiseless channel, Bob receive the state and he has the state of Eq.~(\ref{eq:signal_state_in_dense_coding}) in his hand.
Bob's task is to decode Alice's original signal $\alpha _{in}$.
To this end, Bob performs a unitary transformation on $\rho _{12}^{\prime }$ represented by a 50-50 beam splitter;
$\alpha _{1}\rightarrow (\alpha _{1}+\alpha _{2})/\sqrt{2}$ and
$\alpha _{2}\rightarrow (\alpha _{1}-\alpha _{2})/\sqrt{2}$.
If he measured $\mathrm{Re}\alpha _{1}$ and $\mathrm{Im}\alpha _{2}$ by an ideal balanced homodyne detection with the measured results $\alpha _{1R}$ and $\alpha _{2I}$ respectively,
$\mathrm{Im}\alpha _{1}$ and $\mathrm{Re}\alpha _{2}$ would be completely uncertain and the Wigner function of the state would be reduced to
\[
\int d\mathrm{Im}\alpha _{1}\int d\mathrm{Re}\alpha _{2}W^{\prime }\left(
\frac{\alpha _{1}+\alpha _{2}}{\sqrt{2}},\frac{\alpha _{1}-\alpha _{2}}{%
\sqrt{2}}\right) =\frac{2e^{r}}{\pi }\exp \left( -2e^{r}\left| \alpha -\frac{%
\alpha _{in}}{\sqrt{2}}\right| ^{2}\right) =W^{\prime \prime }(\alpha ),
\]
where $\alpha =\alpha _{1R}+i\alpha _{2I}$.
This represents a highly peaked distribution about the displacement $\alpha _{in}/\sqrt{2}$ and gives the conditional probability $P(\alpha |\alpha _{in})$ of obtaining $\alpha $ given an original message $\alpha _{in}$.
Let $\alpha _{in}$ be distributed as
\[
P(\alpha _{in})=\frac{1}{\pi \sigma ^{2}}\exp \left( -\frac{\left| \alpha
_{in}\right| ^{2}}{\sigma ^{2}}\right).
\]
Thus, the probability of obtaining $\alpha $ is computed as
\[
P(\alpha ) =\int d^{2}\alpha _{in}P(\alpha |\alpha _{in})P(\alpha _{in})
=\frac{2}{\pi (\sigma ^{2}+e^{-2r})}\exp \left[ -\frac{2\left| \alpha
\right| ^{2}}{\sigma ^{2}+e^{-2r}}\right]
\]
and the mutual information that is accessible to Bob is calculated by Eq.~(\ref{eq:mutual_information_2}) as
\begin{equation} \label{eq:mutual_information_in_dense_coding}
H^{d}(A :B)=\int d^{2}\alpha _{in}d^{2}\alpha P(\alpha |\alpha
_{in})P(\alpha _{in})\ln \left( \frac{P(\alpha |\alpha _{in})}{P(\alpha )}%
\right)
=\ln (1+\sigma ^{2}e^{2r}).
\end{equation}
Since the average photon number of mode 1 is given by
\[
\overline{n}(\alpha _{in})=\mathrm{Tr}\left( \rho _{12}^{\prime }a_{1}^{\dagger
}a_{1}\right) =\left| \alpha _{in}\right| ^{2}+\sinh ^{2}r,
\]
the average photon number of the average input signal state is computed as
\[
\overline{n}=\int d^{2}\alpha _{in}P(\alpha _{in})\overline{n}(\alpha
_{in})=\sigma ^{2}+\sinh ^{2}r.
\]

For fixed $\overline{n}$, the mutual information [Eq.~(\ref{eq:mutual_information_in_dense_coding})] is maximized when $\sigma =\sqrt{\sinh r\cosh r}$ ($\overline{n}=e^{r}\sinh r$),
yielding the dense coding capacity of
\[
C_{d}=\ln (1+\overline{n}+\overline{n}^{2})\sim 4r, \qquad r\rightarrow \infty .
\]

Now, we recall that the thermal state maximizes the von Neumann entropy if the average photon number $\overline{n}$ of the mode is fixed (Theorem 2).
Thus, the signal ensemble with pure signal states and the probability distribution such that the average input signal state is a thermal state achieves the HSW capacity of noiseless quantum channel under the fixed photon number;
\begin{equation} \label{eq:HSW_capacity_of_noiseless_channel_for_fixed_photon_number}
C_{HSW}=g(\overline{n}).
\end{equation}
Substituting $\overline{n}=e^{r}\sinh r$ into Eq.~(\ref{eq:HSW_capacity_of_noiseless_channel_for_fixed_photon_number}),
we find $C_{HSW}\sim 2r$ for $r\rightarrow \infty$.
That is, the dense coding sheme presented here allows twice as much as information to be encoded;
$C_{d}\sim 2C_{HSW}$ for $r\rightarrow \infty $.

\subsection{Entanglement measure}
\section{Estimation theory for Gaussian states}\label{sechayashi}
In this section, we treat state estimation theory in the three methods, the
Bayesian method, the group covariant method, and the unbiased method.
For the preparation of this topic, we first discuss information quantities
for one-mode bosonic quantum system in Subsection \ref{12-s-1}.
Next, we treat the measurement theory for this system in Subsection
\ref{12-s-2}.
In particular, the heterodyne measurement is explained in this section.
In Subsection \ref{12-s-3}, we give the mathematical formulation of the
estimation theory. In Subsection \ref{12-s-4}, we consider the case many
copies of the unknown state.
Then, we show that the estimation problem with $n$ copies can be resulted
in the estimation with that with the single copy when the unknown parameter
is the expectation of the filed operator of the unknown Gaussian state.
We treat the estimation theory by the Bayesian method, the group covariant
method, and the unbiased method in Subsections \ref{12-s-5}, \ref{12-s-6},
and \ref{12-s-7}, respectively.
In these subsections, we show the optimality of the heterodyne measurement.
Finally, we briefly treat the simple hypothesis testing in the Gaussian
case.
For the estimation theory for other state families
and a more deep analysis of the simple hypothesis testing,
see Hayashi \cite{Hay06}.
Since our analysis treats the estimation of displacement parameter,
it can be extended to the estimation of displacement parameter
for squeezed Gaussian states family whose elements are given by
the Gaussian mixture of the squeezed states.
For this extension, it is sufficient to replace the
annihilation operator $a$ by
the operator $\cosh r a - e^{-i\phi} \sinh r a^{\dagger}$.

\subsection{Information quantities}\label{12-s-1}
The essence of information theory is describing the operational bound based
on the information quantities. Hence, as a preliminary, we need to give
concrete expressions of these quantities in the case of Gaussian states
$\rho_{\zeta,\overline{N}}:=
\frac{1}{\pi \overline{N}}
\int_{\complex}|\alpha \rangle \langle \alpha|
e^{-\frac{|\alpha-\zeta|^2}{\overline{N}}}
d \alpha$.
In order to express the difference between two states, we often treat the
relative entropy as
\begin{align*}
D(\rho\|\sigma):=
\Tr \rho (\log \rho -\sigma).
\end{align*}
Since the equation $D(\rho\|\sigma)=D(\sigma\|\rho)$ does not necessarily
hold, the relative entropy does not satisfies the axiom of the distance.
However, this value express how large the difference between two state.
Hence, it can be regarded as a kind of distance between two states.
In the Gaussian case, it can be calculated
\begin{align}
D(\rho_{\zeta,\overline{N}}\|\rho_{\zeta',\overline{N}})=
D(\rho_{0,\overline{N}}\|\rho_{\zeta-\zeta',\overline{N}})=
|\zeta-\zeta'|^2\log \frac{\overline{N}+1}{\overline{N}}.\label{12-29-1}
\end{align}
As another quantity,
Bures' fidelity
\begin{align*}
F(\rho,\sigma):=
\Tr \sqrt{\sqrt{\rho}\sigma\sqrt{\rho}}=
\Tr \sqrt{\sqrt{\sigma}\rho\sqrt{\sigma}}
\end{align*}
is known\cite{twamley,scutaru,wang2,multi}. When the state $\rho$ is
a pure state $|\phi\rangle \langle \phi|$, the its square equals the
overlap probability between two states, {\it i.e.,}
\begin{align*}
F^2(|\phi\rangle \langle \phi|,\sigma)=
\langle \phi|\sigma |\phi\rangle
\end{align*}
Using this value,
Bures' distance $d_b(\rho,\sigma)$ is defined as
\begin{align*}
d_b(\rho,\sigma):= \sqrt{1- F^2(\rho,\sigma)}.
\end{align*}
It satisfies the axiom of distance.
In the Gaussian case,
this value is calculated as
\begin{align*}
F(\rho_{\zeta,\overline{N}},\rho_{\zeta',\overline{N}})=
\exp \left(-
\frac{|\zeta-\zeta'|^2}
{8(2\overline{N}+1)}
\right).
\end{align*}
In order to treat the geometric aspect of the set of parametric set of
states
$\{\rho_\theta| \theta \in \Theta\subset \real^d \}$,
we often focus on the metric of the family.
This is because the metric expresses the infinitesimal distance between two
states.
In the case of the set of probability distributions
$\{p_\theta(\omega)|\theta \in \Theta\subset \real\}$,
Fisher information
\begin{align*}
J_\theta:=
\int
\left(\frac{d \log p_\theta(\omega)}{d\theta}\right)^2
p_\theta(\omega)d \omega
\end{align*}
is used for this purpose in the one-parametric case.
In the multi-parametric case, the metric defined from the Fisher
information matrix:
\begin{align*}
J_{\theta;k,l}:=
\int
\frac{\partial \log p_\theta(\omega)}{\partial\theta^k}
\frac{\partial \log p_\theta(\omega)}{\partial\theta^l}
p_\theta(\omega)d \omega
\end{align*}
is used. This is because it is always equal to the constant times of Fisher
information if a metric satisfies monotonicity for quantum operation.

Hence, it is thought that Fisher information presents the information
quantities.
Moreover, using this property, we can characterize the bound of the
accuracy of estimation by Fisher information.

However, in the quantum state case, there are several information
quantities satisfying this invariance.
Hence, there is arbitrariness of the quantum version of Fisher information.
This arbitrariness is caused by the arbitrariness of the quantum version of
the logarithmic derivative. In the one-parametric case
$\{\rho_\theta| \theta \in \Theta \subset \real\}$,
three quantum versions are known.
When $\rho_\theta$ is non-degenerate,
one is the symmetric logarithmic derivative (SLD)
$L_\theta$\cite{Helstrom:1967}:
$\frac{1}{2}\left(
L_\theta \rho_\theta
+ \rho_\theta L_\theta
\right)=
\frac{d \rho_\theta}{d \theta}$.
Other are the right logarithmic derivative (RLD)
$\hat{L}_\theta$\cite{HolP}:
$\rho \hat{L}_\theta= \frac{d \rho_\theta}{d \theta}$,
and
the Kubo-Mori-Bogoljubov (KMB) logarithmic derivative
$\tilde{L}_\theta$\cite{Na,Petz3}:
$\tilde{L}_\theta= \frac{d \log \rho_\theta}{d \theta}$.
In the above definition,
the SLD
$L_\theta$ and the KMB logarithmic derivative $\tilde{L}_\theta$
are Hermitian while
the RLD $\hat{L}_\theta$ is not necessarily
Hermitian.
Based on these logarithmic derivatives, we can define three quantum
analogues of Fisher informations, SLD Fisher information
$J_\theta$\cite{Helstrom:1967},
RLD Fisher information $\hat{J}_\theta$\cite{HolP}, and KMB Fisher
information $\tilde{J}_\theta$\cite{Na,Petz3}:
\begin{align}
J_\theta:= \Tr \rho_\theta L_\theta ^2,\quad
\hat{J}_\theta:= \Tr \rho_\theta \hat{L}_\theta \hat{L}_\theta ^\dagger ,
\quad
\tilde{J}_\theta:= \Tr \rho_\theta \tilde{L}_\theta^2.\label{12-28-8}
\end{align}
These information quantities are closely related to differences between two
states.
The SLD Fisher information $J_\theta$ is characterized as the limit of
Bures' distance\cite{Ma-mas,Uhlmann:1993}:
\begin{align*}
J_\theta=
\lim_{\epsilon \to 0}
\frac{4}{\epsilon^2}d_b^2(\rho_\theta,\rho_{\theta+\epsilon}),
\end{align*}
and the KMB Fisher information $\tilde{J}_\theta$ is by relative
entropy\cite{Na}:
\begin{align*}
\tilde{J}_\theta=
\lim_{\epsilon \to 0}
\frac{2}{\epsilon^2}D(\rho_\theta\|\rho_{\theta+\epsilon})
=
\lim_{\epsilon \to 0}
\frac{1}{\epsilon^2}D(\rho_{\theta+\epsilon}\|\rho_\theta).
\end{align*}
Moreover, these Fisher informations satisfy the relation\cite{Petz3}:
\begin{align*}
J_\theta\le \tilde{J}_\theta \le \hat{J}_\theta.
\end{align*}
In the multi-parametric case, based on the the partial derivative for the
$k$-th parameter $\frac{\partial \rho_\theta}{\partial \theta^k}$, we can
similarly define the logarithmic derivatives for the $k$-th parameter
$L_{\theta;k}, \hat{L}_{\theta;k}$, and $\tilde{L}_{\theta;k}$.
Hence, the SLD Fisher information matrix $J_{\theta;k,l}$\cite{Hel}, the
RLD Fisher information matrix $\hat{J}_{\theta;k,l}$\cite{HolP}, and the
KMB Fisher information matrix $\tilde{J}_{\theta;k,l}$\cite{Na} are defined
as
\begin{align*}
J_{\theta;l,k}:= \Tr \rho_\theta L_{\theta;k}\circ L_{\theta;l},\quad
\hat{J}_{\theta;l,k}
:= \Tr \rho_\theta \hat{L}_{\theta;k} \hat{L}_{\theta;l} ^\dagger
,\quad
\tilde{J}_{\theta;l,k}
:= \Tr \rho_\theta \tilde{L}_{\theta;k}\tilde{L}_{\theta;l}.
\end{align*}
In the Gaussian case, {\it i.e.},
$\{\rho_{(\theta_1+i \theta_2)/\sqrt{2},\overline{N}}|
\theta \in \real^2\}$,
these matrices
${\bf J}_{\theta}=[J_{\theta;k,l}]$\cite{Hel},
${\bf \hat{J}}_{\theta}=[\hat{J}_{\theta;k,l}]$\cite{HolP}, and
${\bf \tilde{J}}_{\theta}=[\tilde{J}_{\theta;k,l}]$
are calculated as
\begin{align}
{\bf J}_{\theta}&=
\frac{1}{\overline{N}+1/2}
\left(
\begin{array}{cc}
1& 0 \\
0 & 1
\end{array}
\right),\quad
{\bf \tilde{J}}_{\theta}=
\log \frac{1+\overline{N}}{\overline{N}}
\left(
\begin{array}{cc}
1 & 0 \\
0 & 1
\end{array}
\right), \nonumber\\
{\bf \hat{J}}_{\theta}&=
\frac{1}{\overline{N}(\overline{N}+1)}
\left(
\begin{array}{cc}
\overline{N}+1/2 & i/2 \\
-i/2 & \overline{N}+1/2
\end{array}
\right),\quad
{\bf \hat{J}}_{\theta}^{-1}=
\left(
\begin{array}{cc}
\overline{N}+1/2 & i/2 \\
-i/2 & \overline{N}+1/2
\end{array}
\right).\label{12-28-22}
\end{align}

Further, if the state family is parameterized by a complex number as
$\rho_{z},z \in \complex$, we often use the following Fisher
information\cite{YL}.
\begin{align*}
\overline{J}_{z} := \Tr \rho_z \overline{L}_z
\overline{L}_z^\dagger,\quad
\frac{\partial \rho_z}{\partial z}=
\rho_z \overline{L}_z.
\end{align*}
In the Gaussian case\cite{YL},
we have
\begin{align}
\frac{\partial \rho_{\theta,\overline{N}}}{\partial \theta}
=
\frac{1}{\overline{N}}
\rho_{\theta,\overline{N}}
(a^{\dagger}-\overline{\theta})
=
\frac{1}{\overline{N}+1}
(a^{\dagger}-\overline{\theta})
\rho_{\theta,\overline{N}}, \quad
\overline{J}_{\theta} &=\frac{1}{\overline{N}+1}.\label{12-28-10}
\end{align}
Here, we use the formula
\begin{align}
\rho_{\theta,\overline{N}}
(a^{\dagger}-\overline{\theta})
=
\frac{\overline{N}}{\overline{N}+1}
(a^{\dagger}-\overline{\theta})
\rho_{\theta,\overline{N}}.
\label{12-28-3}
\end{align}

\subsection{Measurement theory}\label{12-s-2}
In the quantum system, any measurement can be described by positive
operator valued measure (POVM) $M$.
If the set of the measuring data $\Omega$ is discrete-valued, the POVM $M$
can be given by the set of positive semi-definite operators
$\{M_{\omega}\}_{\omega \in \Omega}$ satisfying
\begin{align*}
\sum_{\omega \in \Omega}
M_{\omega}= I
\left(
\hbox{ or }
\int_\Omega M_{\omega}
d \omega
= I
\right),
\end{align*}
where the identity operator is described by $I$.
When the state of the system is given as the density $\rho$, the
measurement data obeys the distribution
\begin{align*}
P^M_\rho (\omega)= \Tr M_\omega \rho.
\end{align*}
If any element $M_\omega$ is a projection, it is called a projection valued
measure (PVM).

In the single-mode bosonic system $L^2(\real)$, the number detection is
described by the POVM $\{|n\rangle \langle n|\}_{n=0}^\infty$.
When the state is given as the coherent state $|\alpha\rangle$, the data
obeys the Poisson distribution
$|\langle n | \alpha \rangle|^2= e^{-|\alpha|^2}\frac{|\alpha|^2n}{n!}$.

When the set $\Omega$ of measuring data $\omega$ has continuous values, the
POVM has the integral form $M(d \omega)$.
That is, for any integrable subset $B \subset \Omega$, the operator $\int_B
M(d \omega)$ is positive semi-definite, and the total operator $\int_\Omega
M(d \omega)$ is equal to the identity operator $I$.
In particular, if any partial integral $\int_B M(d \omega)$ is a
projection,
it is called a projetion valued measure (PVM).

In the single mode bosonic system $L^2(\real)$, the measurement of the
position operator $Q$ is described by its spectral measure $E_Q(d q)$, and
that of the momentum operator $P$ is by its spectral measure of $E_P(d p)$.
The simultaneous spectral measure of $P$ and $Q$ is impossible.
However, when we focus on the two mode system $L^2(\real)^{\otimes 2}
= L^2(\real^2)$,
we can simultaneously measure the operators
$\frac{1}{\sqrt{2}}(P_1+P_2)$ and $\frac{1}{\sqrt{2}}(Q_1-Q_2)$
because they are commutative with each other.
Using this property, we can jointly measure the operators $P$ and $Q$ as
follows.
Assume that we prepare the main target bosonic system $L^2(\real)$ with
arbitrary state $|\phi\rangle \langle \phi|$, and the (additional) ancilla
bosonic system $L^2(\real)$ with the state
$|\phi_0\rangle \langle \phi_0|$.
When we perform the measurement corresponding to
the simultaneous spectral measure of
$\frac{1}{\sqrt{2}}(P_1+P_2)$ and
$\frac{1}{\sqrt{2}}(Q_1-Q_2)$,
the data $p$ and $q$ obey the distribution
\begin{align*}
&P(p,q)=
|\langle \psi\otimes \psi_0|
\frac{1}{\sqrt{2}}(P_1+P_2)= p,
\frac{1}{\sqrt{2}}(Q_1-Q_2)=q\rangle|^2\\
=&|
\frac{1}{\sqrt{2\pi}}\int_{-\infty}^\infty
e^{i pq'}
\psi (\frac{q'+q}{\sqrt{2}})
\psi_0(\frac{q'-q}{\sqrt{2}})
d q'|^2\\
=&|
\frac{1}{\sqrt{2\pi}}
\int_{-\infty}^\infty
e^{i \sqrt{2}p q''}
\psi (q'')
\psi_0(q''- \sqrt{2} q)
\sqrt{2} d q''|^2\\
=&
\frac{1}{\pi}
\left|
\langle\overline{\psi_0}|D(\sqrt{2}q,\sqrt{2}p)|\psi\rangle\right|^2.
\end{align*}
where $q''= \frac{q'+q}{\sqrt{2}}$.
Note that
$D(s,t)= e^{i(s Q- tP)}=D(\alpha)
=e^{i \alpha a + i {\alpha}^* a^\dagger}$,
where $\alpha= \frac{s +it}{\sqrt{2}}$.
Therefore,
if we regard this protocol as the measurement for the main target bosonic
system $L^2(\real^2)$, it is described by
the POVM
\begin{align*}
\frac{1}{\pi}
D(\alpha)|\overline{\psi_0}\rangle
\langle\overline{\psi_0}|D(\alpha)^\dagger
d q dp,
\end{align*}
where $\alpha= p+ i q$.
When $\langle \psi_0 |Q_2 | \psi_0 \rangle =
\langle \psi_0 |P_2 | \psi_0 \rangle =0$,
the average of $p,q,p^2,q^2$, and $pq$ are calculated as
\begin{align*}
\int_{\real^2}
p \Tr \rho
\frac{1}{\pi}
D(\alpha)|\overline{\psi_0}\rangle
\langle\overline{\psi_0}|D(\alpha)^\dagger
d p dq
&= \Tr \frac{1}{\sqrt{2}}(P_1+P_2)
(\rho\otimes | \psi_0 \rangle \langle \psi_0|)\\
&=\frac{1}{\sqrt{2}}\Tr P_1 \rho\\
\int_{\real^2}
q \Tr \rho
\frac{1}{\pi}
D(\alpha)|\overline{\psi_0}\rangle \langle\overline{\psi_0}|
D(\alpha)^\dagger
d p dq
&=
\frac{1}{\sqrt{2}}\Tr Q_1 \rho,
\end{align*}
and
\begin{align*}
\int_{\real^2}
p^2 \Tr \rho
\frac{1}{\pi}
D(\alpha)|\overline{\psi_0}\rangle \langle\overline{\psi_0}|
D(\alpha)^\dagger
d p dq
&= \Tr \left(\frac{1}{\sqrt{2}}(P_1+P_2) \right)^2
(\rho\otimes | \psi_0 \rangle \langle \psi_0|)\\
&=\frac{1}{2}((\Tr P_1^2 \rho)+
\langle \psi_0 |P_2^2 | \psi_0\rangle) \\
\int_{\real^2}
q^2 \Tr \rho
\frac{1}{\pi}
D(\alpha)|\overline{\psi_0}\rangle \langle\overline{\psi_0}
|D(\alpha)^\dagger
d p dq
&=
\frac{1}{2}((\Tr Q_1^2 \rho)+
\langle \psi_0 |Q_2^2 | \psi_0\rangle )\\
\int_{\real^2}
q p \Tr \rho
\frac{1}{\pi}
D(\alpha)|\overline{\psi_0}\rangle
\langle\overline{\psi_0}|D(\alpha)^\dagger
d p dq
&=
\frac{1}{2}((\Tr Q_1 \circ P_1 \rho)+
\langle \psi_0 |Q_2 \circ P_2 | \psi_0\rangle ),
\end{align*}
where $Q_1 \circ P_1:= \frac{Q_1 P_1 + P_1 Q_1}{2}$.
Hence, the covariance matrix is given by
\begin{align}
\frac{1}{2}
\left(
\begin{array}{cc}
\Tr P_1^2 \rho & \Tr Q_1 \circ P_1 \rho \\
\Tr Q_1 \circ P_1 \rho &\Tr Q_1^2 \rho
\end{array}
\right)
-
\frac{1}{2}\left(
\begin{array}{cc}
(\Tr P_1\rho)^2 & (\Tr P_1 \rho)(\Tr Q_1 \rho) \\
(\Tr P_1 \rho)(\Tr Q_1 \rho) &(\Tr Q_1 \rho)^2
\end{array}
\right)\nonumber\\
+
\frac{1}{2}\left(
\begin{array}{cc}
\langle \psi_0 |P_2^2 | \psi_0\rangle
&  \langle \psi_0|Q_2 \circ P_2 | \psi_0\rangle ) \\
\langle \psi_0|Q_2 \circ P_2 | \psi_0\rangle &
\langle \psi_0 |Q_2^2 | \psi_0\rangle
\end{array}
\right). \label{12-3}
\end{align}
That is, the accuracy of this measurement depends on
the choice of the ancilla state $\psi_0$.
For example, if we choose the vacuum state $|0\rangle$ as the ancilla state

$\psi_0$,
the last term equals
$\frac{1}{2} I_2$.
In this case, this measurement
has the form
$\frac{1}{\pi}|\alpha \rangle \langle \alpha |d \alpha$, and is called
the heterodyne measurement.
If we perform this measurement to
the system with the coherent state $|\alpha\rangle \langle \alpha|$,
the data obeys the
Gaussian distribution
$\frac{1}{\pi}e^{-|\alpha-\zeta|^2}$.
Hence, if the state of the system is
Gaussian state $\rho_{\zeta,\overline{N}}
= \int_\complex |\beta\rangle \langle \beta |
e^{-\frac{|\alpha-\zeta|^2}{N}}d \beta$,
the data obeys the
Gaussian distribution
$\frac{1}{\pi(1+\overline{N})}
e^{-\frac{|\alpha-\zeta|^2}{1+\overline{N}}}$.
If the squeezed state $|\zeta,0\rangle$
is choosed as the ancilla state $\psi_0$,
the last term of (\ref{12-3}) is equal to
a real symmetric matrix $S$ with determinant $\frac{1}{4}$.
In addition,
when $\rho$ is the Gaussian state $\rho_{\zeta,\overline{N}}$,
the above matrix is equal to
\begin{align}
(\overline{N}+\frac{1}{2}) I_2 + S \label{12-28-20}.
\end{align}

Indeed,
the heterodyne measurement
$\frac{1}{\pi}|\alpha \rangle \langle \alpha |d \alpha$
not only satisfies the
condition
\begin{align}
a = \int_\complex
\frac{\alpha}{\pi}|\alpha \rangle \langle \alpha |d \alpha,
\label{12-28-11}
\end{align}
but also
can be regarded
as the measurement of the annihilation operator $a$
because of the following reason.
Assume that a POVM $M(d \alpha)$  satisfies
the condition
\begin{align*}
a = \int_\complex
\alpha M(d \alpha).
\end{align*}
Then,
we obtain
\begin{align}
a a^\dagger
\ge \int_\complex
M(d \alpha) \label{12-1}
\end{align}
because
\begin{align*}
a a^\dagger
-
 \int_\complex
|\alpha|^2 M(d \alpha) =
\int_\complex
(a - \alpha)
M(d \alpha)
(a - \alpha)^\dagger
d \alpha\ge 0.
\end{align*}
We can easily check that
the heterodyne measurement
$\frac{1}{\pi}|\alpha \rangle \langle \alpha |d \alpha$
satisfies the equality of the above inequality.
Using this inequality,
we can evaluate the variance $V_\rho(M)$
of the measuring date in the following way
when the state is given as the density $\rho$
\begin{align*}
V_\rho(M)=
\Tr \rho  \int_\complex
|\alpha|^2 | \alpha \rangle \langle \alpha |
d \alpha
-| \Tr \rho a|^2
\le \Tr \rho a a^\dagger - | \Tr \rho a|^2.
\end{align*}
Hence, if the equality of (\ref{12-1}) holds,
the measurement can be regarded as the optimal measurement for the
annihilation operator $a$.
Since the heterodyne measurement
$\frac{1}{\pi}|\alpha \rangle \langle \alpha |d \alpha$
satisfies its equality,
it can be regarded as the optimal measurement of
$a$.

Moreover, the annihilation operator has the form $a
=\frac{Q + iP}{\sqrt{2}}$,
hence, the heterodyne measurement can be treated as the
optimal joint measurement of the operators
$Q$ and $P$.

In particular, when the state $\rho$ is a Gaussian state
$\rho_{\zeta,\overline{N}}$,
we obtain
\begin{align}
V_{\rho_{\zeta,\overline{N}}}
(
\frac{1}{\pi}|\alpha \rangle \langle \alpha |d \alpha)
=
\overline{N}+1\label{12-28-9}.
\end{align}
Indeed, as is discussed latter,
this POVM is used for the estimation of the unknown
parameter of the Gaussian state.

In order to realize this measurement,
we need to treat quantum correlation between two bosonic systems.
However, the required correlation can be perfumed only by the beam
splitter (Sec 1.5).

\subsection{Formulation of estimation}\label{12-s-3}
In the quantum theory,
in order to obtain any information from the system of interest,
we need to perform a measurement to the system.
Needless to say, the measurement always causes
the demolition of the state of the system.
Hence, the choice of the measurement is crucial.
In this section, in the case of Gaussian state,
we optimize the measurement for the estimation of the prepared unknown
state.

In the estimation theory,
we usually assume that
the true density $\rho$
belongs to a given state family $\cS
= \{\rho_\theta | \theta \in \Theta\}$,
where $\Theta$ is the parameter space.
For simplicity of our analysis,
we often assume that the set $\Theta$
is finite-dimensional.
In this case, the estimator
is given by
a POVM $M(d \hat{\theta})$ taking the measuring data in the parameter space
$\Theta$,
{\it i.e.},
the POVM is continuous-valued.
Hence, when we use the estimator $M$ and
the true parameter is $\theta$,
the estimated value $\hat\theta$ obeys the distribution
$\Tr M(d \hat{\theta})\rho_\theta$.
In order to treat the accuracy of our estimation,
we focus on the risk function
$W(\theta,\hat\theta)$,
which describes the degree of the miss-estimation
when the true parameter is $\theta$ and
our estimated parameter is $\hat\theta$.
Thus, the accuracy of the estimator $M$
is evaluated by
the mean risk:
\begin{align}
\cD_\theta^W(M):=
\int_{\Theta}
W(\theta,\hat\theta)
\Tr  M(d \hat{\theta})\rho_\theta.
\end{align}
In the statistics,
when the parameter space $\Theta$ is a subset of $\real$,
we often adopt the square error $(\theta - \hat\theta)^2$
as the risk function $W(\theta,\hat\theta)$.
In this case, the mean risk:
\begin{align}
{\rm MSE}_\theta (M):=
\int (\hat{\theta}- \theta)^2 \Tr \rho_{\theta} M(d \hat\theta)
\label{12-28-15}
\end{align}
is called the
the mean square error (MSE).
When the parameter space $\Theta$ is $\complex$,
{\it i.e.}, the family has the form
$\{\rho_z|z \in \complex\}$,
the MSE is given by
\begin{align}
{\rm MSE}_z (M):=
\int |\hat{z}- z|^2 \Tr \rho_{z} M(d \hat{z}).
\end{align}
In the multipara-metric case,
we often adopt $\sum_{k=1}^d (\theta^k-\hat\theta^k)^2$
as the risk function $W(\theta,\hat\theta)$.
In the quantum theory, we often use
$1-F^2(\rho_\theta,\rho_{\hat\theta})$ or the square of Bures' distance
$d_b^2(\rho_\theta,\rho_{\hat\theta})$ as the risk function.
Usually, we assume that
$W(\theta,\hat\theta)\ge 0$
and the equality holds if and only if
$\theta=\hat\theta$.

\subsection{Independent and identical condition}\label{12-s-4}
In the statistics,
we often assume that the data are independent, identically
distributed (i.i.d.) with the unknown probability distribution \footnote{%
Currently, many mathematical statisticians treat dependent data,
and they obtained several basic results similar to the i.i.d.\ case.
However, the discussion of the dependent case is so
difficult that we mainly focus on the i.i.d.\ case in this paper.}.
Based on this assumption, we can easily treat estimation theory
in the case where the number $n$ of data is large, {\em i.e.},
in the large sample case.
This is because asymptotic expansions are available in
the large sample case.

In the quantum case,
when each system is prepared in the same state $\rho$,
the state of the $n$-fold system $\cH^{\otimes n}$
is given by the $n$-fold tensor product state $\rho^{\otimes n}$.
This setting is called the (quantum) $n$-i.i.d. case
One may call the state $\rho^{\otimes n}$
$n$ copies of $\rho$.

In the Gaussian case, the $n$-fold tensor product state
$\rho_{\zeta,\overline{N}}^{\otimes n}$
is unitarily equivalent with
$\rho_{\sqrt{n}\zeta,\overline{N}}\otimes
\rho_{0,\overline{N}}^{\otimes n}$.
In order to construct the above unitary $U$,
we generalized the discussion for beam-splitter.
for any $n \times n$ special unitary matrix $M_B$,
there exists a unitary operator $U_B$
on $L^2(\real^n)$
such that
\begin{align*}
U_B ( a_1^\dagger, \ldots, a_n^\dagger)
U_B^\dagger
= ( a_1^\dagger, \ldots, a_n^\dagger) M_B.
\end{align*}
If the special unitary matrix $M_B$ has
the form
\begin{align*}
M_B=
\left(
\begin{array}{cccc}
\frac{1}{\sqrt{n}} & * & \cdots &* \\
\frac{1}{\sqrt{n}} & * & \cdots &* \\
\vdots & \vdots & \ddots & \vdots \\
\frac{1}{\sqrt{n}} & * & \cdots &*
\end{array}
\right),
\end{align*}
the unitary operator $U_B$ satisfies
\begin{align*}
U_B \rho_{\zeta,\overline{N}}^{\otimes n} U_B^{\dagger}
= \rho_{\sqrt{n}\zeta,\overline{N}}\otimes
\rho_{0,\overline{N}}^{\otimes n}.
\end{align*}
Since the state $\rho_{0,\overline{N}}^{\otimes n}$ has no information
concerning the $\zeta$,
the estimation problem with $n$ copies
can be resulted in the estimation with the family
$\{\rho_{\sqrt{n}\zeta,\overline{N}}| \zeta \in \complex\}$.

In the following,
we prove the optimality of the heterodyne measurement
$\frac{1}{\pi}|\alpha\rangle \langle \alpha|
d \alpha$
for the Gaussian family
$\{\rho_{\zeta,\overline{N}}| \zeta \in \complex\}$.
in the respective formulations.

\subsection{Bayesian method}\label{12-s-5}
The minimum value of the mean risk $\cD_\theta^W(M)$ is 0
when the estimated parameter is always $\theta$,
however,
it is impossible to minimize
the mean risk $\cD_\theta^W(M)$ simultanouesly to all $\theta$.
Hence, we have to trade-off $\cD_\theta^W(M)$ with all elements $\theta$ in

the parameter space $\Theta$.
As one method for this trade-off,
Bayesian method is known.
In this approach, we assume that the known parameter $\theta$
is generated the {\it prior} distribution $\mu$ on the parameter space,
and minimize the average of the mean risk
\begin{align*}
\int_{\Theta} \cD_\theta^W(M) \mu(d \theta).
\end{align*}

In the one-parameter case $\{\rho_\theta|
\theta \in \Theta \subset \real\}$,
the Bayes estimator (the optimal estimator of Bayesian approach)
concerning the MSE
is given as follows\cite{Per}.
\begin{align*}
& \int_\real\int_\real
(\hat\theta-\theta)^2 \Tr \rho_\theta M(d\hat\theta) \mu(d \theta)
=
\Tr \int_\real
\left(
\overline{\rho}
\hat\theta^2
-
(L \overline{\rho} + \overline{\rho} L)
\hat\theta
+
S \right)
M(d\hat\theta) \\
=&
\Tr \int_\real
(L-\hat\theta)\overline{\rho} (L-\hat\theta)
M(d \hat\theta)
+S - L \overline{\rho} L,
\end{align*}
where the operator $\overline{\rho}$, $L$, and $S$
are defined by
\begin{align*}
\overline{\rho} :=\int_\real \rho_\theta \mu(d \theta), \quad
S
:=\int_\real \theta^2 \rho_\theta \mu(d \theta), \quad
(L \overline{\rho} + \overline{\rho} L)
=2 \int_\real \theta \rho_\theta \mu(d \theta).
\end{align*}
The equality of the inequality
\begin{align*}
\Tr \int_\real
(L-\hat\theta)\overline{\rho} (L-\hat\theta)
M(d \hat\theta)\ge 0
\end{align*}
holds if and only if
the POVM $M$ is equal to the spectral measure of the operator $L$.
Hence the optimal estimator is given as the spectral measure of the
operator $L$, and its optimal error is equals $\Tr S- L\rho L$.

If we apply the same method to the two-parameter case, we have to consider
a simultaneous spectral measure of two operators, which correspond to the
respective parameter.
Since such a simultaneous spectral measure is impossible generally, this
problem of the two-parameter case is more difficult.
However, its Gaussian case can be solved when the prior is the Gaussian
distribution.
In this case, the Bayes estimator concerning the MSE can be obtained by the
following calculation:
\begin{align*}
&
\frac{1}{\pi \overline{N}_1}
\int_\complex\int_\complex
|\hat\theta-\theta|^2 \Tr \rho_\theta M(d\hat\theta)
e^{-\frac{|\theta|^2}{\overline{N}_1}} d \theta
\\
=&
\Tr \frac{1}{\pi (\overline{N}+\overline{N}_1)}
\int_\complex
\left|\hat\theta
-\frac{\overline{N}_1}{\overline{N}_1+\overline{N}}
\alpha
\right|^2
|\alpha\rangle \langle \alpha|
e^{-\frac{|\theta|^2}{\overline{N}_1+\overline{N}}}
d \alpha
  M(d\hat\theta)
+\overline{N}_1
- \frac{\overline{N}_1^2}{\overline{N}_1+\overline{N}}
\\
=&
\Tr
\Bigl(\rho_{0,\overline{N}_1+\overline{N}}|\theta|^2
-\frac{\overline{N}_1}{\overline{N}_1+\overline{N}+1}
\left(
\theta a^\dagger \rho_{0,\overline{N}_1+\overline{N}}
+
{\theta}^*
\rho_{0,\overline{N}_1+\overline{N}}a
\right) \\
& +
\left(\frac{\overline{N}_1}{\overline{N}_1+\overline{N}}\right)^2
a \rho_{0,\overline{N}_1+\overline{N}}a^\dagger
\Bigr)
M(d\hat\theta)
+\frac{\overline{N}_1 \overline{N}}{\overline{N}_1+\overline{N}}
\\
=&
\Tr
\left(\frac{\overline{N}_1}{\overline{N}_1+\overline{N}+1}
a^\dagger - {\hat\theta}^*\right)
\rho_{0,\overline{N}_1+\overline{N}}
\left(\frac{\overline{N}_1}{\overline{N}_1+\overline{N}+1}
a - \hat\theta\right)
M(d\hat\theta)\\
&
+\frac{\overline{N}_1 \overline{N}}{\overline{N}_1+\overline{N}}
+ \frac{\overline{N}_1^2}
{(\overline{N}_1+\overline{N}+1)(\overline{N}_1+\overline{N})},
\end{align*}
where the second equation follows from (\ref{12-28-3})
and the final equation follows the next equation:
\begin{align*}
&\Tr \left(\frac{\overline{N}_1}{\overline{N}_1+\overline{N}}\right)^2
a \rho_{0,\overline{N}_1+\overline{N}}a^\dagger
-
\left(\frac{\overline{N}_1}{\overline{N}_1+\overline{N}+1}\right)^2
a^\dagger \rho_{0,\overline{N}_1+\overline{N}}a\\
=&\left(\frac{\overline{N}_1}{\overline{N}_1+\overline{N}}\right)^2
(\overline{N}_1+\overline{N})
-\left(\frac{\overline{N}_1}{\overline{N}_1+\overline{N}+1}\right)^2
(\overline{N}_1+\overline{N}+1)\\
=&
\frac{\overline{N}_1^2}
{(\overline{N}_1+\overline{N}+1)(\overline{N}_1+\overline{N})}.
\end{align*}
The equality of the inequality
\begin{align*}
\Tr
\left(\frac{\overline{N}_1}{\overline{N}_1+\overline{N}+1}
a^\dagger - {\hat\theta}^*\right)
\rho_{0,\overline{N}_1+\overline{N}}
\left(\frac{\overline{N}_1}{\overline{N}_1+\overline{N}+1}
a - \hat\theta\right)
M(d\hat\theta)\ge 0
\end{align*}
holds if and only if
\begin{align*}
M(d\hat\theta)=
\left|\frac{\overline{N}_1+\overline{N}+1}{\overline{N}_1}\hat\theta
\right\rangle
\left\langle\frac{\overline{N}_1+\overline{N}+1}{\overline{N}_1}
\hat\theta\right|
d \hat\theta.
\end{align*}
Hence,
the minimum error is equal to
$\frac{\overline{N}_1 \overline{N}}{\overline{N}_1+\overline{N}}
+ \frac{\overline{N}_1^2}
{(\overline{N}_1+\overline{N}+1)(\overline{N}_1+\overline{N})}$,
and the optimal estimator is given by the above POVM, which is realized by
the heterodyne measurement.
This value is smaller than the error of the usual heterodyne measurement
with the following value.
\begin{align*}
 \frac{\overline{N}^2}{\overline{N}_1+\overline{N}}
+ \frac{\overline{N}_1 (2\overline{N}+1)
+\overline{N}^2 +\overline{N}}
{(\overline{N}_1+\overline{N}+1)(\overline{N}_1+\overline{N})}.
\end{align*}
This difference goes to $0$ when $\overline{N}_1$ goes to infinity.

\subsection{Group covariant method}\label{12-s-6}
However, we cannot use the Bayesian method when our prior knowledge is
insufficient to decide the prior distribution.
Hence, several non-Bayesian methods, e.g., the minimax method and minimum
variance unbiased estimation have been proposed.

In the minimax method, we focus on the worst error for an estimator:
\begin{align*}
\max_{\theta \in \Theta}
\cD_\theta^W(M),
\end{align*}
and optimize it.
This method provides an attractive estimator when the family of probability
distribution has a homogenous structure.
In the Gaussian case, the state family $\{\rho_{\zeta,\overline{N}}|\zeta
\in \complex\}$ has the following group symmetry:
\begin{align*}
V(\eta)
\rho_{\zeta,\overline{N}}V(\eta)^{\dagger}
=
\rho_{\zeta+\eta,\overline{N}}, \quad \forall \eta,\zeta\in \complex.
\end{align*}
Hence, if that risk function $W(\theta,\hat\theta)$ is invariant for the
action $U$, {\it i.e.},
$W(\theta,\hat\theta)=W(\theta+\eta,\hat\theta+\eta)$,
the relation
\begin{align}
\min_M \max_{\theta \in \Theta}\cD_\theta^W(M)=
\min_{M:\rm covariant}\cD_\theta^W(M)\label{hunt}
\end{align}
holds, where the POVM $M$ on $\complex$ is called covariant if
\begin{align*}
V(\eta) M(\hat\theta)
V(\eta)^{\dagger}d \hat\theta=
M(\hat \theta + \eta)d \hat\theta,
\end{align*}
which is equivalent with
the following form
\begin{align*}
M(\hat\theta)= \frac{1}{\pi}
V(\hat\theta)^{\dagger}P_0 V(\hat\theta) d \hat\theta.
\end{align*}
The relation (\ref{hunt}) is called Quantum Hunt-Stein's lemma.
(It was obtained by Holevo \cite{HoC} when $G$ is a compact group, and by
Ozawa \cite{Oza-20} and Bogomolov\cite{Bogomolov} when $G$ is a non-compact
group.)

In the Gaussian states family $\{\rho_{\zeta,\overline{N}}|
\zeta \in\complex\}$,
if the risk function $W(\zeta,\hat\zeta)$ can be written
by a monotone increasing function of $|\zeta-\hat\zeta|$,
the optimal estimator is given by the heterodyne measurement:
$\frac{1}{\pi} |\hat\zeta\rangle \langle \hat\zeta| d \hat\zeta$.
That is, the optimal estimator does not depend on the form of the risk
function, and depends only on the group invariance.
This optimality can be proven by treating the following case:
\begin{align*}
W(\zeta,\hat\zeta)
=
\left\{
\begin{array}{cc}
1 & |\zeta-\hat\zeta| \ge R \\
0 & |\zeta-\hat\zeta| < R.
\end{array}
\right.
\end{align*}
In fact, this optimality has been shown by Cerf {\it et
al}.\cite{CKNWW} only when $W$ is $1-$ fidelity and
$\overline{N}=0$.

In the above general case, using the operator
$\hat{W}:= \frac{1}{\pi}\int_{|\zeta|\ge R}
U(\zeta)\rho_{0,\overline{N}}U(\zeta)^\dagger d \zeta$,
we can describe the error
$\cD_\zeta^W(M)$ as
$\cD_\zeta^W(M)=\Tr \hat{W}P_0$.
Hence, it is sufficient to show
$\langle 0 |\hat{W}|0\rangle \le
\langle k |\hat{W}|k\rangle$ for any integer $k$.
For this proof, we focus on the equation
\begin{align*}
\langle k |\hat{W}|k\rangle
=
\int_\complex
\left(\frac{1}{\pi}
\frac{1}{\pi \overline{N}}
\int_{|\zeta|\ge R}
e^{- \frac{|\zeta-\alpha|^2}{\overline{N}}}
d \zeta
\right)
e^{-|\alpha|^2}
\frac{|\alpha|^{2k}}{k\!}
d \alpha.
\end{align*}
Since
$\left(\frac{1}{\pi}
\frac{1}{\pi \overline{N}}
\int_{|\zeta|\ge R}
e^{- \frac{|\zeta-\alpha|^2}{\overline{N}}}
d \zeta
\right)$
and
$\frac{|\alpha|^{2k}}{k\!}$ are monotone increasing function concerning
$|\alpha|$,
the probability distribution
$e^{-|\alpha|^2}d \alpha$
satisfies
\begin{align*}
&
\langle 0 |\hat{W}|0\rangle =
\int_\complex
\left(\frac{1}{\pi}
\frac{1}{\pi \overline{N}}
\int_{|\zeta|\ge R}
e^{- \frac{|\zeta-\alpha|^2}{\overline{N}}}
d \zeta
\right)e^{-|\alpha|^2}d \alpha \\
=&
\int_\complex
\left(\frac{1}{\pi}
\frac{1}{\pi \overline{N}}
\int_{|\zeta|\ge R}
e^{- \frac{|\zeta-\alpha|^2}{\overline{N}}}
d \zeta
\right)e^{-|\alpha|^2}d \alpha
\cdot
\int_\complex
\frac{|\alpha|^{2k}}{k\!}
e^{-|\alpha|^2}d \alpha \\
\le &
\int_\complex
\left(\frac{1}{\pi}
\frac{1}{\pi \overline{N}}
\int_{|\zeta|\ge R}
e^{- \frac{|\zeta-\alpha|^2}{\overline{N}}}
d \zeta
\right)e^{-|\alpha|^2}
\frac{|\alpha|^{2k}}{k\!}
d \alpha
= \langle k |\hat{W}|k\rangle .
\end{align*}
Therefore, we obtain the optimality
of the heterodyne measurement in the group covariant model.

\subsection{Unbiased method}\label{12-s-7}
Finally, we consider
minimum error under the
unbiased estimators.
In the statistics, we often restrict our estimators to
the unbiased estimator,
{\it i.e.},
we assume that the estimator $M$
of the family $\{\rho_\theta| \theta \in \real\}$
satisfies
\begin{align}
{\rm E}_\theta (M):=
\int \hat{\theta} \rho_{\theta} M(d \hat{\theta})= \theta ,
\quad \forall \theta \in \Theta.\label{12-28-7}
\end{align}
And we minimize the mean square error (\ref{12-28-15})
among unbiased estimators.
If the estimator $M$ satisfies the unbiasedness condition,
the MSE is bounded by the inverse of the SLD Fisher information $J_\theta$
(\ref{12-28-8}) as
\begin{align}
{\rm MSE}_\theta (M) \ge (J_\theta)^{-1}.\label{12-28-6}
\end{align}
This inequality is called
SLD Cram\'{e}r-Rao inequality \cite{Helstrom:1967}.
It can be shown by the Schwarz inequality as follows.
Taking the derivative in (\ref{12-28-7}),
we have
\begin{align*}
\Tr ((O(M)-\theta) \circ L_\theta ) \rho
= \Tr (O(M)-\theta) (L_\theta \circ \rho)
= \Tr (O(M) - \theta) \frac{d \rho_\theta}{d\theta}
= 1,
\end{align*}
where $O(M)= \int \hat{\theta} M(d\hat{\theta})$.
the Schwarz inequality implies that
\begin{align*}
& \Tr (O(M)-\theta)^2  \rho
\cdot J_\theta
=
\Tr ((O(M)-\theta) \circ (O(M)-\theta) ) \rho_\theta
\cdot
\Tr (L_\theta  \circ L_\theta ) \rho_\theta \\
\ge &
|\Tr ((O(M)-\theta) \circ L_\theta ) \rho_\theta|^2
=1.
\end{align*}
Since
\begin{align}
&\int (\hat{\theta}-\theta)^2 \Tr \rho M(d \hat{\theta})-
\Tr (O(M)-\theta)^2  \rho \nonumber \\
=&
\Tr
\rho
\int
[(\hat{\theta}-\theta)-(O(M)-\theta)]
M(d \hat{\theta})
[(\hat{\theta}-\theta)-(O(M)-\theta)]\ge 0,\label{12-28-6-3}
\end{align}
we obtain inequality (\ref{12-28-6}).

If the parameter $z$ is a complex number,
we obtain another type of Crame\'{e}r-Rao inequality\cite{YL}:
\begin{align}
{\rm MSE}_\theta (M)
\ge (\overline{J}_\theta)^{-1}.\label{12-28-6-1}
\end{align}
It can be shown similarly.
The operator $O(M)= \int \hat{z} M(d\hat{z})$
satisfies
the Schwarz inequality:
\begin{align*}
& \Tr (O(M)-z)^\dagger(O(M)-z)
\rho_z
\cdot \overline{J}_z
=
\Tr \rho_z(O(M)-z)^\dagger(O(M)-z)
\cdot
\Tr \rho_z \overline{L}_z  \overline{L}_z ^{\dagger} \\
\ge &
|\Tr \rho_z \overline{L}_z (O(M)-z)|^2
=1.
\end{align*}
Hence, using a inequality similar to (\ref{12-28-6-3}),
we obtain inequality (\ref{12-28-6-1}).

If we focus on the weighted MSE $
W_1(\hat{\zeta}_1- \zeta_1)^2
+W_2(\hat{\zeta}_2- \zeta_2)^2 $,
the RLD Fisher information matrix is very useful.
In this case, we focus on the
Covariance matrix:
$
\Cov(M):=
\left(
\int_{\complex}
(\hat{\zeta}_i- \zeta_i)(\hat{\zeta}_j- \zeta_j)
\Tr \rho_{\zeta} M(d \hat\zeta)
\right)$,
and
minimize $\Tr G \Cov (M)$ among unbiased estimators $M$,
where $G$ is a symmetric real matrix.
Applying a discussion similar to
RLD Fisher information,
we can show \cite{HolP}
\begin{align}
\Cov(M)\ge \hat{J}_{\zeta,\overline{N}}^{-1}.
\end{align}
Since the matrix $ \Cov(M)$ is real symmetric,
any unbiased estimator $M$ satisfies\cite{HolP}
\begin{align}
\Tr G \Cov(M)
&\ge
\min_{V:\hbox{ real symmetric matrix}}
\{ \Tr G V|
V \ge \hat{J}_{\zeta,\overline{N}}^{-1}\}\label{1-10-1}\\
&=
(\overline{N}+\frac{1}{2})\Tr G + \sqrt{\det G}.\nonumber
\end{align}
In the last equation, we use equation (\ref{12-28-22}).
This value can be attained by $S= \frac{\sqrt{\det G}}{2}
G^{-1}$ in (\ref{12-28-20})
($V= (\overline{N}+\frac{1}{2})I_2+ \frac{\sqrt{\det G}}{2}G^{-1}$
in (\ref{1-10-1})).
That is, the optimal measurement can be realized by replacing
the coherent state in the ancilla by the squeezed state.
This discussion can be applied to
the asymptotic theory of the state estimation
in the quantum two-level system
because the states in this system can be approximated by
the Gaussian states in the asymptotic sense\cite{HM}.

Now, we apply this inequality to Gaussian states
family $\{\rho_{\zeta,\overline{N}}| \zeta \in\complex\}$.
Then, from (\ref{12-28-10}),
the minimum MSE among unbiased estimators is
$\overline{N}+1$.
From (\ref{12-28-11}),
the heterodyne measurement $\int_{\complex}
\frac{1}{\pi}|\hat\zeta\rangle \langle\hat\zeta| d \hat\zeta$
is an unbiased estimator.
From (\ref{12-28-9}), this value is attained by the heterodyne measurement.
That is, the heterodyne measurement is the optimal estimator even
among unbiased estimators.
Note that this discussion can be applied even in the
pure states case.

Next, we consider the estimation of another parameter $\overline{N}$.
Assume that $\zeta=0$.
Then, the SLD $L_{\overline{N}}$ is equal to
$
\sum_{n=0}^\infty
\left(
\frac{n}{\overline{N}}
-
\frac{n+1}{\overline{N}+1}
\right)
|n \rangle \langle n|
$.
Therefore,
$J_{\overline{N}}
= \frac{1}{\overline{N}(\overline{N}+1)}$.
The MSE of unbiased estimators is
$\overline{N}(\overline{N}+1)$.
This bound is attained by
the number counting $
\{|n \rangle \langle n|\}_{n=0}^{\infty}$.
However,
if we use the heterodyne measurement,
the MSE is $(\overline{N}+1)^2$, {\it i.e.},
this bound cannot be attained.

The advantage of this approach
is that
$n$-copy case can be treated just parallel to one-copy case.
That is,
we can easily construct unbiased estimator
for the $n$-copy family from
an unbiased estimator for the one-copy family.
In this case, the bound $(J_{\theta})^{-1}$
becomes the $\frac{1}{n}$ times of the one of
the one-copy case.
Further, the MSE also behaves similarly.
Therefore,
in the $n$-copy case of the Gaussian states,
the bound of MSE concerning $\zeta$ is
$\frac{\overline{N}+1}{n}$,
and
the bound of MSE concerning $\overline{N}$ is
$\frac{\overline{N}(\overline{N}+1)}{n}$.
For this simultaneous estimation,
we first
perform the state evolution as
$\rho_{\zeta,\overline{N}}^{\otimes n}
\to \rho_{\sqrt{n}\zeta,\overline{N}}\otimes
\rho_{0,\overline{N}}^{\otimes n}$.
Next, we perform the heterodyne measurement for the first state
$\rho_{\sqrt{n}\zeta,\overline{N}}$,
and we perform the number counting for the
remaining states $\rho_{0,\overline{N}}^{\otimes n}$.
In this case,
the MSE concerning $\zeta$ is
$\frac{\overline{N}+1}{n}$,
and
the MSE concerning $\overline{N}$ is
$\frac{\overline{N}(\overline{N}+1)}{n-1}$.
That is,
this method is almost optimal even for the estimation of
$\overline{N}$.

Hence, we can conclude that this measurement realize the simultaneous optimal estimation between the two parameters
$\zeta$ and
$\overline{N}$\cite{Hayashi:2000}.

\subsection{Simple hypothesis testing}
Next, we treat the statistical simple hypothesis
testing. In this problem, it is known that the unknown state is the null hypothesis $\rho_0$ or the alternative hypothesis $\rho_1$.
It is required to reject the null hypothesis $\rho_0$ with a fixed error probability,
and accept the alternative
hypothesis $\rho_1$.
This problem is the simplest case of statistical hypothesis testing, and its mathematical formulation is well established even in the
quantum case.

In this case, if the null hypothesis is rejected despite being correct, it is called the error of the first kind.
Conversely, if the null hypothesis is accepted despite being incorrect, it is
called the error of the second kind.
Then, we make our decision only when we support the alternative hypothesis, and withhold our decision when we support the null one.
Now, we describe our decision by the two-valued POVM $\{T,I-T\}$, where the outcome of $T$ supports $\rho_0$.
This positive semi-definite operator $T$ is called a test.
Then, the first error probability is equal to $\Tr \rho_0 (I-T)$, and the second error probability is equal to $\Tr \rho_1 T$.
In this problem, it is required to keep the first error probability $\Tr \rho_0 (I-T)$ less the fixed error probability.
We minimize the second error probability $\Tr \rho_1 T$ under the above condition for the first error probability $\Tr \rho_0 (I-T)$.
That is, we often focus on the following value:
\begin{align*}
\beta_\epsilon(\rho_0\|\rho_1)
:=
\min_{I \ge T \ge 0} \{ \Tr \rho_1 T|
\Tr \rho_0(I-T)\le \epsilon \}.
\end{align*}
When we prepare the $n$ copies of the unknown state, we treat $\beta_\epsilon^n(\rho_0^{\otimes n}\|\rho_1^{\otimes n})$.
If the error threshold $\epsilon$ is fixed, this value goes to $0$ exponentially.
Hence, we focus on its exponential rate.
In fact, the exponential rate is given as
\begin{align*}
\lim  \frac{-1}{n}
\log \beta_\epsilon^n(\rho\|\sigma)
= D(\rho_0\|\rho_1) ,
\quad 1 \,> \forall \epsilon > 0 ,
\end{align*}
which is called quantum Stein's lemma\cite{HP,Oga-Nag:test}.
In the Gaussian case $\rho_0=\rho_{\zeta_0,\overline{N}}$ and $\rho_1=\rho_{\zeta_1,\overline{N}}$, this bound is equal to $|\zeta_0-\zeta_1|^2
\log \frac{\overline{N}+1}{\overline{N}}$ (\ref{12-29-1}).

Indeed, when $\rho_1$ is not commutative with $\rho_0$, we need quantum correlation in the measuring apparatus for realizing the test $T$ attaining the optimal bound
$D(\rho_0\|\rho_1)$.
That is, if we perform any separable measurement, the bound cannot be attained\cite{H2001}.
In the above Gaussian case, the following test attains the bound.
First, we perform the state evolution as $\rho_{\zeta,\overline{N}}^{\otimes n}
\to \rho_{\sqrt{n} \zeta,\overline{N}}\otimes \rho_{0,\overline{N}}^{\otimes (n-1)}$.
Next, we perform the state evolution as $\rho_{\sqrt{n} \zeta,\overline{N}}\otimes
\rho_{0,\overline{N}}^{\otimes (n-1)}\to \rho_{\sqrt{n} (\zeta-\zeta_1),\overline{N}}\otimes \rho_{0,\overline{N}}^{\otimes (n-1)}$.
Then, we perform the number counting for the first system.
Finally, the detected number $n$ is greater than the threshold, we support $\rho_{\zeta_0,\overline{N}}$.
Otherwise, we support $\rho_{\zeta_1,\overline{N}}$.
For the threshold, see Hayashi \cite{H2001}.
In this test $T_n$, the first error $\Tr \rho_{\zeta_0,\overline{N}}^{\otimes n} (I-T_n)$ goes to $0$ and the second error $\Tr \rho_{\zeta_1,\overline{N}}^{\otimes n} T_n$ goes to $0$ with the exponential rate $|\zeta_0-\zeta_1|^2
\log \frac{\overline{N}+1}{\overline{N}}$\cite{H2001}.
This fact indicates the importance of the quantum correlation using the beam splitter.

\section{Acknowledgement}
This work was started years ago. Before we
complete this work,
 some excellent reviews\cite{GKD01,BL,paris,EP03,EW05} had been presented on the
 similar topics with different emphases. Our work is a complementary
material to all existing related works with our own emphasis. We
hope this review can give a useful overview and/or tutorial material
for researchers such as PhD students, experts in quantum optics who
have interest to QIP,  experts in QIP with two-level systems who
have interest to QIP with Gaussian states and so on.

No one can cover everything in a single review paper. We have chosen
those materials  to which we have special interests.

We thank Prof Imai for support. Wang is partially supported by the
National Fundamental Research Program of China through Grant No.
2007CB807900 and 2007CB807901.

Here is a list of corrections of minors and modifications from the published version. (We use the page number and equation number
of the published version in Physics Reports.)\\
1)After Eq.(15), we consider the $n$ mode Gaussian states. In some places $n$ had been wrongly 
typed by  $m$.\\
2)Page 8, line 7 and line 11, $S\in SP(2n,R)$ is replaced by  $M\in SP(2n,R)$; line 12, $SAS^T$ is replaced
by $MAM^T$.\\
3)The RHS of Eq.(86), the position of matrix and vector should be reversed. See ther right form in Eq.(86) 
of this version.\\
4)Eq.(245) is replaced by $s_\alpha(\mu)=(1+r_\alpha)s_{\alpha}(\mu')$, as used in this version.\\
5)The text after Eq.(246) and before the paragraph prior to Eq.(248) is rewritten.\\
6)Eq.(250) is modified, as Eq.(250) in this version.

%



\begin{thebibliography}{999}
\bibitem{nielsen} M. A. Nielsen and I. L. Chuang,
{\em Quantum Computation and Quantum Information,} (Cambridge University Press, 2000).
\bibitem{shor} P. Shor, Proc. 35th Ann. Symp. on Found. of Computer Science,
(IEEE Comp. Soc. Press, Los Alomitos, CA, 1994) 124.
\bibitem{bene} C.H. Bennett and G. Brassard,
Proc. IEEE Int. Conf. on Computers, Systems, and Signal Processing
(Bangalore, India, IEEE, New York, 1984) 175.
\bibitem{telepor1} C. H. Bennett, C. Crepeau, R. Jozsa, A. Peres, and W. K. Wooters,
Phys. Rev. Lett. 70 (1993) 1895.
\bibitem{glauber} R. J. Glauber, Phys. Rev. 121 (1963) 2766.
\bibitem{KMC65}
J. R. Klauder, J. McKenna, and D. G. Currie, J. Math. Phys. 6 (1965)
734; J. R. K. Klauder and B.-S. Skagerstam, {\em Coherent states: --
applications in physics and mathematical physics} (World Scientific,
Singapore, 1985).
\bibitem{yuenwf} H. P. Yuen, Phys. Rev. A 13 (1976) 2226 .
\bibitem{knight} R. Loudon and P. L. Knight, J. Mod. Opt. 34 (1987) 709.
\bibitem{schu} B. L. Schumaker, Phys. Rep. 135 (1986), 317.
\bibitem{zhangwm} W. M. Zhang and D. H. Feng, Rev. Mod. Phys. 64 (1990) 687.
\bibitem{leonhardt} U. Leonhardt, {\em Measuring the Quantum State of
Light}, Cambridge University Press, 1997.
\bibitem{inns} D. Bouwmeester, J. W. Pan, K. Mattle, M. Eibl, H. Weinfurter and A. Zeilinger,
Nature (London) 390 (1997) 575.
\bibitem{GRTZ02} N. Gisin, G. Ribordy, W. Tittel, and H. Zbinden, Rev. Mod. Phys. 74 (2002) 145.
\bibitem{EPRname} A. Einstein, B. Podolsky, and N. Rosen,
Phys. Rev. 47 (1935) 477.
\bibitem{milburn} D. F. Walls and G. J. Milburn,
{\em Quantum Optics}
(Springer -Verlag, Berlin, 1994).
\bibitem{zoller} G. W. Gardiner and P. Zoller,
{\em Quantum Noise} (Springer-Verlag, Berlin, Heidelberg, 2000).
\bibitem{barnett} S. M. Barnett and P. M. Radmore,
{\it Methods in Theoretical Quantum Optics} (Oxford Science Publication, Oxford, 1997).
\bibitem{orszag} M. Orszag, {\em Quantum Optics}
(Springer-Verlag, Berlin, Heidelberg, 2000).
\bibitem{Pet90}
D. Petz, {\em An Invitation to the Algebra of Canonical Commutation
Relations} (Leuven University Press, Leuven, 1990).
\bibitem{tmsq01} G. J. Milburn, J. Phys. A 17 (1984) 737;
C. M. Caves and B. L. Schumaker, Phys. Rev. D 31 (1989) 3608.
\bibitem{weyl} H. Weyl, Z. Phys. 46 (1927) 1.
\bibitem{wigner} E. Wigner, Phys. Rev. 40 (1932) 749.
\bibitem{wignerr} M. Hillery, R. F. O'Connell, M. O. Scully, and E. P. Wigner,
Phys. Rep. 106 (1984) 121;
H. W. Lee, Phys. Rep. 259 (1995) 147.
\bibitem{quasi8} N. L. Balazs and B. K. Jennings, Phys. Rep. 104 (1984) 347.
\bibitem{moyl} J. E. Moyal, Proc. Cambridge Phil. Soc. 45 (1949) 99.
\bibitem{kimqm} Y. S. Kim  and  M. E. Noz,
{\em Phase space picture of quantum mechanics: a group theoretical approach}
(World Scientific, Singapore, 1991).
\bibitem{gdkthe} G. Giedke,
{\em Quantum Information and Continuous Variable Systems}
Doctoral thesis, University of Innsbruck, 2001.
\bibitem{GKD01}
G. Giedke, B. Kraus, L.-M. Duan, P. Zoller, J. I. Cirac, and M. Lewenstein,
Fortschr. Phys. 49 (2001) 973.
\bibitem{MV68}
J. Manuceau and A. Verbeure, Commun. Math. Phys. 9 (1968) 293.
\bibitem{SSM87}
R. Simon, E. C. G. Sudarshan, and N. Mukunda, Phys. Rev. A 36 (1987)
3868.
\bibitem{Arn89}
V. I. Arnold, {\em Mathematical Methods of Classical Mechanics 2nd.
ed.} (Springer-Verlag, New York, 1989).
\bibitem{bchwang0} X-B. Wang, S. X. Yu, and Y. D. Zhang,
J. Phys. A: Math. Gen. 27 (1994) 6563.
\bibitem{bchwang1} X-B. Wang, C. H. Oh, and L. C. Kwek,
J. Phys. A: Math. Gen. 31 (1998) 4329.
\bibitem{wangE} J. W. Pan, Q. X. Dong, Y. D. Zhang, G. Hou, and X-B. Wang,
Phys. Rev. E 56 (1997) 2553.
\bibitem{bsreview0} B. Yurke, S. L. McCall, and J. R. Klauder, Phys. Rev. A 33 (1986) 4033.
\bibitem{bsreview1} B. Huttner and Y. Ben-Aryeh, Phys. Rev. A 38 (1988) 204;
U. Leonhardt, {\em ibid.} 48 (1993) 3265.
\bibitem{campos} R. A. Campos, B. E. A. Saleh, and M. C. Teich, Phys. Rev. A 40 (1989) 1371;
R. A. Campos, {\em ibid.}, 62 (2000) 013809.
\bibitem{btheorem} X-B. Wang, Phys. Rev. A 66 (2002) 024303; Arvind and N. Mukunda, Phys. Lett. A (1999), 259, 421.
\bibitem{bwang} X-B. Wang, Phys. Rev. A 66 (2002) 064304; M. S. Kim, W. Son, V. Buzek, and P. L. Knight,
Phys. Rev. A (2002), 65, 032323; M. M. Wolf , J. Eisert, and M. B.
Plenio Phys. Rev. Lett., 90(2003), 047904.
\bibitem{wangsingle}X-B. Wang, Phys. Rev. A 68 (2003), 042304.
\bibitem{bscm} C. K. Hong, Z. Y. Ou, and L. Mandel, Phys. Rev. Lett. 59 (1987) 2044;
M. Zukowski {\em et al.}, Phys. Rev. Lett. 71 (1993) 4287;
S. L. Braunstein and A. Mann, Phys. Rev. A 51 (1995) 1727(R), and references therein.
\bibitem{homo1} J. H. Shapiro and H. P. Yuen, IEEE Trans. Inf. Theory. IT-25 (1979) 179;
L. Mandel, Phys. Rev. Lett. 49 (1982) 136.
\bibitem{homo101} B. L. Schumaker, Opt. Lett. 9 (1984) 189.
\bibitem{homo102} B. Yurke, Phys. Rev. A 32 (1985) 311 .
\bibitem{homo2} S. L. Braunstein, Phys. Rev. A 42 (1990) 474;
V. Vogel and J. Grabow, Phys. Rev. A 47 (1992) 4227;
K. Banaszek and K. Wodkiewic, Phys Rev. A 55 (1997) 3117.
\bibitem{homob} B. C. Sanders, Phys. Rev. A 45 (1992) 6811.
\bibitem{wooters} W. K. Wooters and W. H. Zurek, Nature 299 (1982) 802.
\bibitem{swapping} M. Zukowski, A. Zeilinger, M. A. Horne, and A. Ekert,
Phys. Rev. Lett. 71 (1993) 4287.
\bibitem{bdensec} C. H. Bennett and S. J. Wiesner, Phys. Rev. Lett. 69 (1997) 2881.
\bibitem{BL} S. L. Braunstein and P. van Look,
Rev. Mod. Phys. 77 (2005) 427.
\bibitem{cvqt1} L. Vaidman, Phys. Rev. A 49 (1994) 1473.
\bibitem{cvqt2} S. L. Braunstein and H. J. Kimble, Phys. Rev. Lett. 88 (1998) 869.
\bibitem{hofmann} H. F. Hofmann, T. Ide, and T. Kobayashi, Phys. Rev. A 62 (2000) 062304.
\bibitem{takeoka} M. Takeoka, M. Ban, and M. Sasaki, J. Opt. B 4 (2002) 114.
\bibitem{vukics} A. Vukics, J. Janszky, and T. Kobashy, Phys. Rev. A 66 (2002) 023809.
\bibitem{cvswap1} S. M. Tan, Phys. Rev. A 60 (1999) 2752 .
\bibitem{cvswap2} R. E. S. Polkinghorne and T. C. Ralph, Phys. Rev. Lett. 83 (1999) 2095.
\bibitem{cvswap3} P. van Loock and S. L. Braunstein, Phys. Rev. A 61 (2000) 010302(R).
\bibitem{vanpra}van Loock, and S. L. Braunstein, Phys. Rev. A 61 (2000),
010302.
\bibitem{numberphase} G. J. Milburn and S. L. Braunstein, Phys. Rev. A 60 (1999) 937.
\bibitem{numberphase02} J. T. Clausen, T. Opatruy, and D.-G. Welsh,
Phys. Rev. A 62 (2000) 042308.
\bibitem{numberphase03} P. T. Cochrane, G. J. Milburn, and B. Munro,
Phys. Rev. A 62 (2002) 062307; P. T. Cochrane, T. C. Ralph, and G. J. Milburn,
Phys. Rev. A 65 (2002) 062306.
\bibitem{bfk} S. L. Braunstein, C. A. Fuchs, and H. J. Kimble, J. Mod. Opt. 47 (2000) 267.
\bibitem{telee0} A. Furusawa, J. L. Sorensen, S. L. Braunstein, C. A. Fuchs, H. J. Kimble,
and  E. S. Polzik, Science 282 (1998) 706.
\bibitem{barry} T. Rudolph and B. C. Sanders, Phys. Rev. Lett. 87 (2001) 077903.
\bibitem{enk} S. J. van Enk and C. A. Fuchs, Phys. Rev. Lett. 88 (2000) 017901.
\bibitem{telecr2} F. Grosshans and P. Grangier, Phys. Rev. A 64 (2001) 010301(R).
\bibitem{clone3} N. J. Cerf, A. Ipe and X. Rottenberg, Phys. Rev. Lett. 85 (2000) 1754.
\bibitem{telecr3} T. C. Ralph and P. K. Lam, Phys. Rev. Lett. 81 (1998) 5668;
J. Lee, M. S. Kim, and H. Jeong, Phys. Rev. A 62 (2000) 032305;
F.-L. Li, H.-R. Li, J.-X. Zhang, and S.-Y. Zhu, Phys. Rev. A 66
(2002) 024302.
\bibitem{telee2}
W. P. Bowen, N. Treps, B. C. Buchler, R. Schnabel, T. C. Ralph, H.-A. Bachor, T. Symul, and P. K. Lam, Phys. Rev. A 67 (2003) 032302.
\bibitem{telee3} T. C. Zhang, K. W. Goh, C. W. Chou, P. Lodahl, and H. J. Kimble,
Phys. Rev. A 67 (2002) 033802 (2002).
\bibitem{cvdense0} M. Ban, J. Opt. B: Quantum Semiclass. Opt. 1 (1999) L9.
\bibitem{cvdense} S. L. Braunstein and H. J. Kimble, Phys. Rev. A 61 (2002) 042302.
\bibitem{denseee}J. Zhang and K. Peng, Phys. Rev. A 62 (2000) 064302;
J. T. Zhang, J. Zhang, Y. Yan, F. G. Zhao, C. D. Xie, and K. C. Peng,
Phys. Rev. Lett. 90 (2003) 167903.
\bibitem{qecc1} P. W. Shor, Phys. Rev. A 52 (1995) 2493(R).
\bibitem{qecc2} A. M. Steane, Proc. Roy. Soc. London 452 (1996) 2551.
\bibitem{qecc3} A. R. Calderbank and P. W. Shor, Phys. Rev. A 54 (1996) 1098.
\bibitem{qecc4} C. H. Bennett, D. P. DiVincenzo, J. A. Smolin, and W. K. Wooters,
Phys. Rev. A 54 (1996) 3824.
\bibitem{qecc5} E. Knill and R. Laflamme, Phys. Rev. A 55 (1997) 900.
\bibitem{qecc6} A. R. Calderbank, E. M. Rains, P. W. Shor, and N. J. A. Sloane,
Phys. Rev. Lett. 78 (1997) 405.
\bibitem{qecc7} S. L. Braustein, Nature, 394 (1998) 47.
\bibitem{qecc8} S. L. Braustein, Phys. Rev. Lett. 80 (1998) 4084.
\bibitem{qecc9} S. Lloyd and J.-J. E. Slotine, Phys. Rev. Lett. 80 (1998) 4088.
\bibitem{qecc10} D. Gottesman, A. Kitaev, and J. Preskill, Phys. Rev. A 63 (2001) 022309.
\bibitem{clone2} V. Buzek and M. Hilery, Phys. Rev. A 54 (1996) 1844.
\bibitem{clonenm} N. Gisin and S. Massar, Phys. Rev. Lett. 79 (1997) 2153.
\bibitem{clone5} N. J. Cerf and S. Iblisdir, Phys. Rev. A 62 (2000) 040301.
\bibitem{clone6} N. Gisin and S. Massar, Phys. Rev. Lett. 79 (1997) 2153.
\bibitem{clonegisin} V. Scarani, S. Iblisdir, and N. Gisin,
Rev. Mod. Phys. 77 (2005) 1225, and also arXiv:quant-ph/0511088.
\bibitem{clone7} N. J. Cerf and  S. Iblisdir,
In {\em Quantum Communication, Computing, and Measurement 3,} (Kluwer Academic, NewYork, 2001) 11;
S. L. Braustein, N. J. Cerf, S. Iblisdir, P. van Loock, and S. Massar,
Phys. Rev. Lett. 86 (2001) 4938;
J. Fiurasek, Phys. Rev. Lett. 86 (2001) 4942.
\bibitem{paris} A. Ferraro, S. Olivers, M. G. A. Paris,
Phys. Rev. A (2006) 012330.
\bibitem{ppuri} J. W. Pan, S. Gasparoni, R. Ursin, G. Weihs, and A. Zeilinger,
Nature 423 (2003) 8938.
\bibitem{bkcomment} S.I. Braunstein and H.J. Kimble, Nature 394,
840 (1998);  D. Bouwmeester, J. W. Pan, M. Daniell, H. Weinfurter, M. Zukowski, and A. Zeilinger, Nature 394,
840 (1998). 
\bibitem{kok2000}  P. Kok and S.L. Braunstein, Phys. Rev. A 61, 042304 (2000)
\bibitem{wgtele} J. W. Pan, S. Gasparoni, M. Aspelmeyer,
T. Jennewein, and A. Zeilinger, Nature 421 (2003) 721.
\bibitem{swap} X-B. Wang, B.-S. Shi, A. Tomita, and K. Mastumoto,
Phys. Rev. A 69 (2004) 014303.
\bibitem{nonp} X-B. Wang and H. Fan,
Phys. Rev. A 68 (2003) 060302(R).
\bibitem{yamamoto} T. Yamamoto, M. Koashi, S. K. Ozdemir, and N. Imoto,
Nature 421 (2003) 343;
T. Yamamoto, T. Koashi, and N. Imoto, Phys. Rev. A 64 (2001) 012304.
\bibitem{ustc} Z. Zhao, T. Yang, Y. A. Chen, A. N. Zhang and J. W. Pan,
 Phys. Rev. Lett. 90 (2003) 207901.
\bibitem{panature} J.-W. Pan, C. Simon, C. Brukner, and A. Zeilinger,
Nature 410 (2001) 1067.
\bibitem{ins} J.-W. Pan, S. Gasparoni, R. Ursin, G. Weihs, and A. Zeilinger,
Nature 423 (2003) 417.
\bibitem{brownenp} D. E. Browne, J. Eisert, S. Scheel, and M. B. Plenio,
Phys. Rev. A 67 (2003) 062320.
\bibitem{eisertnp} J. Eisert, D. E. Browne, S. Scheel, and M. B. Plenio,
Ann. Phys. 311 (2004) 431.
\bibitem{xywang} X.-Y. Wang, Collision for some hash functions
MD5, MD5 HAVAL-128, primed, in {\em Crypro'04} (2004).
\bibitem {Lutkenhaus00} N. L\"{u}tkenhaus, Phys. Rev. A 61 (2000) 052304.
\bibitem {Hamada03} M. Hamada, arXiv:quant-ph/0308029 (2003).
\bibitem {YT01} A. Yoshizawa and H. Tsuchida, Jpn. J. Appl. Phys. 40 (2001) 200.
\bibitem {BR00} D. S. Bethune and W. P. Risk, IEEE J. Quantum Electron. 36 (2000) 340.
\bibitem {TN02} A. Tomita and K. Nakamura, Opt. Lett. 27 (2002) 1827.
\bibitem {KTNKN03} H. Kosaka, A. Tomita, Y. Nambu, T. Kimura, and K. Nakamura,
Electron. Lett. 39 (2003) 1199.
\bibitem {TRT93} P. D. Townsend, J. G. Rarity, and P. R .Tapster, Electron. Lett. 29 (1993) 634.
\bibitem{ZGGHMT97} H. Zbinden, J. D. Gautier, N. Gisin, B. Huttner, A. Muller, and W. Tittel,
Electron. Lett. 33 (1997) 586.
\bibitem{MHHTZG97} A. Muller, T. Herzog, B. Huttner, W. Tittel, H. Zbinden, and N. Gisin,
Appl. Phys. 70 (1997) 793.
\bibitem{RG3Z98} G. Ribordy, J.D. Gautier, N. Gisin, O. Guinnard, and H. Zbinden,
Electron. Lett. 34 (1998) 2116.
\bibitem {T5N04} A. Tanaka, A. Tomita, A. Tajima, T. Takeuchi, S. Takahashi, and Y. Nambu,
30th European Conference on Optical Communication (ECOC)
(Stockholm, Sweden, Sep. 5-9, 2004) Tu4.5.3.
\bibitem {SGGRZ02} D. Stucki, N. Gisin, O. Guinnard, G. Ribordy, and H. Zbinden,
New J. Phys. 4 (2002) 41.
\bibitem {IDQ} http://www.idquantique.com/.
\bibitem{MAGIQ} http://www.magiqtech.com/.
\bibitem{HNIASM05} T. Hasegawa, T. Nishioka, H. Ishizuka, J. Abe, K. Shimizu, and M. Matsui,
Proc. the 2005 Symposium on Cryptography and Information Security, 2F-3 (in Japanese) (Maiko, Kobe, Japan, Jan. 25-28, 2005).
\bibitem{NEC} http://www.nec.co.jp/press/en/0505/3101.html.
\bibitem{GYS04} C. Gobby, Z. L. Yuan, and A. J. Shields, Appl. Phys. Lett. 84 (2004) 3762.
\bibitem{YS05} Z. L. Yuan and A. J. Shields, Optics Exp. 13 (2005) 660.
\bibitem{MZHGG04} X.-F. Mo, B. Zhu, Z.-F Han, Y.-Z. Gui, and G.-C. Guo,
arXiv:quant-ph/0412023.
\bibitem {NHN04} Y. Nambu, T. Hatanaka, and K. Nakamura, Jpn. J. Appl. Phys. 43 (2004) L1109.
\bibitem {KNHTKN04} T. Kimura, Y. Nambu, T. Hatanaka, A. Tomita, H. Kosaka, and K. Nakamura,
Jpn. J. Appl. Phys. 43 (2004) L1217.
\bibitem{guang}G. Wu et al, Phys. Rev. A 74, 062323(2006)
\bibitem{cvqkd2} B. Huttner, N. Imoto, N. Gisin, and T. Mor, Phys. Rev. A 51 (1995) 1863.
\bibitem{bra} G. Brassard, N. L\"utkenhaus, T. Mor, and B. C. Sanders,
Phys. Rev. Lett. 85 (2000) 1330.
\bibitem{CL04} M. Curty and N. L\"{u}tkenhaus, Phys. Rev. A 69 (2004) 042321.
\bibitem{CL00} M. Curty and N. L\"{u}tkenhaus, Phys. Rev. A 71 (2005) 062301.
\bibitem{FGGNP97} C. A. Fuchs, N. Gisin,  R. B. Griffiths,  C.-S. Niu, and A. Peres,
Phys. Rev. A 56 (1997) 1163.
\bibitem{AGS04} A. Acin, N. Gisin, and V. Scarani, Phys. Rev. A 69 (2004) 012309.
\bibitem{GM97} N. Gisin and S. Massar, Phys. Rev. Lett. 79 (1997) 2153.
\bibitem {BBBSS92} C. H. Bennett, F. Bessate, G. Brassard, L. Salvail, and J. Smolin,
J. Crypt. 5 (1992) 3.
\bibitem{mayer1} D. Mayers, In {\em Advances in Cryptology -- Proc. Crypto'96, Vol. 1109 of Lecture Notes in Computer Science} (Ed. N. Koblitz, Springer-Verlag, New York, 1996) 343;
J. Assoc. Comput. Mach. 48 (2001) 351.
\bibitem {BBCM95} C. H. Bennett, G. Brassard, C. Crepeau, and U. M. Maurer,
IEEE Trans. Inform. Theory 41 (1995) 1915.
\bibitem{LC99} H.-K. Lo and H. F. Chau, Science 283 (1999) 2050.
\bibitem{shor2} P. W. Shor and J. Preskill, Phys. Rev. Lett. 85 (2000) 441, and references therein.
\bibitem{BDSW} C. H. Bennett, D. P. DiVincenzo, J. A. Smolin, and W. K. Wootters,
Phys. Rev. A 54 (1996) 3824.
\bibitem{hwang} W.-Y. Hwang, Phys. Rev. Lett. 91 (2003) 057901.
\bibitem{wang0d} X-B. Wang, Phys. Rev. Lett. 94 (2005) 230503.
\bibitem{lolo} H. K. Lo, X.-F. Ma, and K. Chen, Phys. Rev. Lett. 94 (2005) 230504.
\bibitem{scar} V. Scarani, A. Acin, G. Ribordy, and N. Gisin, Phys. Rev. Lett. 92 (2004) 057901;
A. Acin, N. Gisin, and V. Scarani, Phys. Rev. A 69 (2004) 012309.
\bibitem{srl} C. H. Bennett, Phys. Rev. Lett. 68 (1992) 3121;
M. Koashi, Phys. Rev. Lett. 93 (2004) 120501.
\bibitem{cvgp} D. Gottesman and J. Preskill, Phys. Rev. A 63 (2001) 022309.
\bibitem{decouple} X-B. Wang, arXiv:quant-ph/0409099.
\bibitem{Eke91} A. K. Ekert, Phys. Rev. Lett. 67 (1991) 661.
\bibitem{twogl}D. Gottesman and H.-K. Lo, IEEE Trans. on Info.
Theory 49 (2003), 457.
\bibitem{twochau} H.-F. Chau, Phys. Rev. A66 (2002), 060302.
\bibitem{wangenc1} X-B. Wang, Phys. Rev. Lett. 92 (2004) 077902.
\bibitem{wangenc2} X-B. Wang, Phys. Rev. A 72 (2005) 050304(R);
Y.-K. Jiang, X-B. Wang, B.-S. Shi, and A. Tomita, Opt. Exp. 13 (2005) 9415.
\bibitem{qkdasym} X-B. Wang, Phys. Rev. A 71, 052328(2005).
\bibitem{koashin}M. Koashi, quant-ph/0505108 and quant-ph/0609180.
\bibitem{gllp} H. Inamori, N. L\"{u}tkenhaus, and D. Mayers, arXiv:quant-ph/0107017;
 D. Gottesman, H.K. Lo, N. L\"utkenhaus, and J. Preskill,
Quantum Information and Computation, 4 (2004) 325.
\bibitem{kens1} N. L\"utkenhaus and M. Jahma, New J. Phys. 4 (2002) 44.
\bibitem{lolo1}X.~Ma, B. Qi, Y. Zhao, and H.-K. Lo, Phys. Rev. A {\bf 72},
012326 (2005).
\bibitem{lo4} H.-K. Lo, Proc. 2004 IEEE Int. Symp. on Inf. Theor.
(June 27-July 2, Chicago, 2004) 17.
\bibitem{wang2d} X-B. Wang, Phys. Rev. A 72 (2005) 012322.
\bibitem{harry} J. W. Harrington, J. M. Ettinger, R. J. Hughes, and J. E. Nordholt,
arXiv:quant-ph/0503002.
\bibitem{incontrol1}X.-B. Wang, C.-Z. Peng, J. Zhang and J.-W. Pan,
quant-ph/0612012.
\bibitem{incontrol2}X.-B. Wang, Phys. Rev. A 75, 012301(2007).
\bibitem{incontrol3}X.-B. Wang, C.-Z. Peng and J.-W. Pan, Appl.
Phys. Lett., 90 (2007) 031110.
\bibitem{zeinature}P. Prevedel et al, Nature (London) 445 (2007),
65.
\bibitem{lonothing} H.-K. Lo, Quantum Information and Computation 5,
413(2005).
\bibitem{koashid} M. Koashi, quant-ph/0609180.
\bibitem{ma2w}X. Ma et al, Phys. Rev. A 74 (2006), 032330.
\bibitem{Lo06}
Y.~Zhao, B. Qi, X. Ma, H.-K. Lo and L. Qian, Phys. Rev. Lett. {\bf
96}(2006), 070502 ; Y.~Zhao, B. Qi, X. Ma, H.-K. Lo and L. Qian,
quant-ph/0601168.
\bibitem{peng}  Cheng-Zhi Peng {\em et al.}
Pan  Phys. Rev. Lett. 98 (2007), 010505, D. Rosenberg {em et al.},  Phys. Rev.
Lett. 98 (2007), 010503.
\bibitem{ron}  T. Schmitt-Manderbach {\em et al.}, Phys.
Rev. Lett. 98 (2007), 010504.
\bibitem{yuane} Z.-L. Yuan, A. W. Sharpe, and A. J. Shields,
Appl. Phys. Lett. 90(2007), 011118.
\bibitem{decoypdc}T. Hirikiri and T. Kobayashi, Phys. Rev. A (2006), 73, 032331;
Q. Wang, X.-B. Wang, G.-C. Guo, Phys. Rev. A (2007), 75, 012312; W.
Mauerer and C. Siberborn, e-print quant-ph/0609195; Y. Adachi, T.
Yamamoto, M. Koashi, and N. Imoto, e-print, quant-ph/0610118.
\bibitem{norbertrc} M. Dusek, N. L\"utkenhaus, and  M. Hendrych, quant-ph/0601207.
\bibitem{tkn92} K. Tamaki, M. Koashi, and N. Imoto, Phys. Rev. Lett. 90 (2004) 167904;
K. Tamaki and N. L\"utkenhaus, Phys. Rev. A 69 (2004) 032316.
\bibitem{sarg04} C. Branciard, N. Gisin, B. Kraus, and V. Scarani, Phys. Rev. A 72 (2005) 032301.
\bibitem{sargproof} M. Koashi, arXiv:quant-ph/0507154.
\bibitem{cvqkd1} Y. Mu, J. Seberry, and Y. Zheng, Opt. Commn. 123 (1996) 344.
\bibitem{cvqkd3} T. C. Ralph, Phys. Rev. A 61 (1999) 010303(R);
Phys. Rev. A 62 (2000) 062306.
\bibitem{cvqkd4} M. Hillery, Phys. Rev. A 61 (2000) 022309.
\bibitem{cvqkd5} M. D. Reid, Phys. Rev. A 62 (2000) 062308.
\bibitem{cvqkd6} S. F. Pereria, Z. Y. Oh, and H. J. Kimble, Phys. Rev. A 62 (2000) 042311.
\bibitem{cvqkd7} F. Grossmans and P. Grangier, Phys. Rev. Lett. 88 (2002) 057902.
\bibitem{cvqkd8} Ch. Silberhorn, N. Korolkova, and L. Leuchs, Phys. Rev. Lett. 88 (2002) 167902.
\bibitem{cvqkd9} Ch. Silberhorn, T. C. Ralph, N. L\"ukenhaus, and G. Leuchs,
Phys. Rev. Lett. 89 (2002) 167901.
\bibitem{cvqkd10} R. Namiki and T Hirano, Phys. Rev. Lett. 92 (2004) 117901.
\bibitem{cvqkd11} A. M. Lance, T. Symul, V. Sharma, C. Weedbrook, T. C. Ralph, and P. K. Lam,
Phys. Rev. Lett. 95 (2005) 180503.
\bibitem{cvqkd12} F. Grosshans, G. Van Assche, J. Wenger, R. Brouri, N. J. Cerf, and P. Grangier,
Nature 421 (2003) 238.
\bibitem{mheid} M. Heid and N. L\"ukenhaus, Phys. Rev. A 73, 052316 (2006).
\bibitem{y00} G. A. Barbosa, E. Corndorf, P. Kumar, and H. P. Yuen,
Phys. Rev. Lett. 90 (2003) 227901;
H. P. Yuen, arXiv:quant-ph/0322061.
\bibitem{y001} Z. L. Yuan and A. J. Shields, Phys. Rev. Lett. 84 (2005) 048901;
H. Yuen, E. Corndorf, G. Barbosa, and P. Kumar, {\em ibid.} 94 (2005) 048902.
\bibitem{y002} H. P. Yuen, P. Kumar, E. Corndorf, and R. Nair,
Phys. Lett. A 346 (2005) 1;
T. Nishioka, T. Hasegawa, H. Ishizuka, K. Imafuku, and H. Imai,
Phys. Lett. A 346 (2005) 7.

\bibitem{TWD03} B. M. Terhal, M. M. Wolf, and A. C. Doherty,
Phys. Today 56 (2003) 46.
\bibitem{Wer89} R. F. Werner, Phys. Rev. A 40 (1989) 4277.
\bibitem{LBC00} M. Lewenstein, D. Bru\ss, J. I. Cirac, B. Kraus, M. Ku\'s, J. Samsonowicz, A. Sanpera, and R. Tarrach, J. Mod. Opt. 47 (2000) 2481.
\bibitem{Ter02} B.M. Terhal,
Theoretical Computer Science 287 (2002) 313.
\bibitem{Bru02} D. Bru\ss,
J. Math. Phys. 43 (2002) 4237.
\bibitem{BCH02} D. Bru\ss, J. I. Cirac, P. Horodecki, F. Hulpke, B. Kraus, M. Lewenstein, and A. Sanpera,
J. Mod. Opt. 49 (2002) 1399.
\bibitem{Per96} A. Peres,
Phys. Rev. Lett. 77 (1996) 1413.
\bibitem{HHH96}
M. Horodecki, P. Horodecki, and R. Horodecki,
Phys. Lett. A 223 (1996) 1.
\bibitem{Hor97}
P. Horodecki,
Phys. Lett. A 232 (1997) 333.
\bibitem{Hor01b}
P. Horodecki,
Quantum Information and Computation 1 (2001) 45.
\bibitem{DCL00}
W. D\"ur, J. I. Cirac, M. Lewenstein, and D. Bru\ss,
Phys. Rev. A 61 (200) 062313.
\bibitem{HHH97}
M. Horodecki, P. Horodecki, and R. Horodecki,
Phys. Rev. Lett. 78 (1997) 574.
\bibitem{HHH98}
M. Horodecki, P. Horodecki, and R. Horodecki,
Phys. Rev. Lett. 80 (1998) 5239.
\bibitem{NK01}
M. A. Nielsen and J. Kempe,
Phys. Rev. Lett. 86 (2001) 5184.
\bibitem{VW02a}
K. G. H. Vollbrecht and M. M. Wolf,
J. Math. Phys. 43 (2002) 4299.
\bibitem{Hir03}
T. Hiroshima,
Phys. Rev. Lett. 91 (2003) 057902.
\bibitem{HH99}
M. Horodecki and P. Horodecki,
Phys. Rev. A 59 (1999) 4206.
\bibitem{CAG99}
N. J. Cerf, C. Adami, and R. M. Gingrich,
Phys. Rev. A 60 (1999) 898
\bibitem{Hir001}
T. Hiroshima, Phys. Rev. A 63 (2001) 022305.
\bibitem{WMT01}
X-B. Wang, K. Matsumoto, and A. Tomita,
Phys. Rev. Lett. 87 (2001) 137903.
\bibitem{KGL03}
B. Kraus, G . Giedke, M. Lewenstein, and J. I. Cirac,
Fortschr. Phys. 51 (2003) 305.
\bibitem{Sim00}
R. Simon,
Phys. Rev. Lett. 84 (2000) 2726.
\bibitem{DGC00a}
L.-M. Duan, G. Giedke, J. I. Cirac, and P. Zoller,
Phys. Rev. Lett. 84 (2000) 2722.
\bibitem{WW01}
R. F. Werner and M. M. Wolf,
Phys. Rev. Lett. 86 (2001) 3658.
\bibitem{GKL01a}
G. Giedke, B. Kraus, M. Lewenstein, and P. Zoller,
Phys. Rev. Lett. 87 (2001) 167904.
\bibitem{LB00}
P. van Look and S. L. Braunstein,
Phys. Rev. Lett. 84 (2000) 3482.
\bibitem{LB01}
P. van Look and S. L. Braunstein,
Phys. Rev. A 63 (2001) 022106.
\bibitem{GKL01b}
G. Giedke, B. Kraus, M. Lewenstein, and J. I. Cirac,
Phys. Rev. A 64 (2001) 052303.
\bibitem{GDC01}
G. Giedke, L.-M. Duan, I. Cirac, and P. Zoller,
Quantum Information and Computation 1 (2001) 79.
\bibitem{DGC00b}
L.-M. Duan, G. Giedke, J. I. Cirac, and P. Zoller,
Phys. Rev. Lett. 84 (2000) 4002.
\bibitem{DGC00c}
L.-M. Duan, G. Giedke, J. I. Cirac, and P. Zoller,
Phys. Rev. A 62 (2000) 032304.
\bibitem{FMF03}
J. Fiur$\acute{a}\check{s}$ek, L. Mi$\check{s}$ta Jr., and R. Filip,
Phys. Rev. A 67 (2003) 022304.
\bibitem{GC02}
G. Giedke and J. I. Cirac,
Phys. Rev. A 66 (2002) 032316.
\bibitem{ESP02}
J. Eisert, S. Scheel, and M. Plenio,
Phys. Rev. Lett. 89 (2002) 137903.
\bibitem{Fiu02}
J. Fiur$\acute{a}\check{s}$ek,
Phys. Rev. Lett. 89 (2002) 137904.
\bibitem{BES03}
D. E. Browne, J. Eisert, S. Scheel, and M. Plenio,
Phys. Rev. A 67 (2003) 062320.
\bibitem{Hor01a}
M. Horodecki,
Quantum Information and Computation 1 (2001) 3.
\bibitem{VW02b}
G. Vidal and R. F. Werner,
Phys. Rev. A 65 (2002) 032314.
\bibitem{GWK03}
G. Giedke, M. M. Wolf, O. Kr\"uger, R. F. Werner, and J. I. Cirac,
Phys. Rev. Lett. 91 (2003) 107901.
\bibitem{EP03}
J. Eisert and M. Plenio, Int. J. Quantum Information 1 (2003) 479.
\bibitem{Hay06}
M. Hayashi,
{\em An Introduction to Quantum Information Theory}
(Springer-Verlag, Berlin, 2006).
\bibitem{Hol98a}
A. S. Holevo,
IEEE Trans. Inf. Theory 44 (1998) 269.
\bibitem{SW97}
B. Schumacher and M. D. Westmoreland,
Phys. Rev. A 56 (1997) 131.
\bibitem{AHW00}
G. G. Amosov, A. S. Holevo, and R. F. Werner,
Prob. Inf. Trans. 36 (2000) 305.
\bibitem{BSS02}
C. H. Bennett, P. W. Shor, J. A. Smolin, and A. V. Thapliyal,
IEEE Trans. Inf. Theory 48 (2002) 2637.
\bibitem{Hir01}
T. Hiroshima, J. Phys. A: Math. Gen. 34 (2001) 6907.
\bibitem{Lin00}
G. Lindblad,
J. Phys. A: Math. Gen. 33 (2000) 5059.
\bibitem{HSH00}
A. S. Holevo, M. Sohma, and O. Hirota,
Rep. Math. Phys. 46 (2000) 343.
\bibitem{SEW05}
A. Serafini, J. Eisert, and M. M. Wolf,
Phys. Rev. A 71 (2005) 012320.
\bibitem{HSH99}
A. S. Holevo, M. Sohma, and O. Hirota,
Phys. Rev. A 59 (1999) 1820.
\bibitem{HW02}
A. S. Holevo and R. F. Werner, Phys. Rev. A 63 (2001) 032312.
\bibitem{EW05}
J. Eisert and M. M. Wolf, arXiv:quant-ph/0505151 v1.
\bibitem{GGL04b}
V. Giovannetti, S. Guha, S. Lloyd, L. Maccone, J. H. Shapiro, and H. P. Yuen,
Phys. Rev. Lett. 92 (2004) 027902.
\bibitem{Bek81}
J. D. Bekenstein,
Phys. Rev. D 23 (1981) 287.
\bibitem{Hel81}
C. W. Helstrom,
Opt. Commun. 37 (1981) 175.
\bibitem{OP93}
M. Ohya and D. Petz,
{\em Quantum Entropy and Its Use}
(Springer-Verlag, New York, 1993).
\bibitem{HS04}
A. S. Holevo and M. E. Shirokov,
arXiv:quant-ph/0408176 v1.
\bibitem{Hol98b}
A. S. Holevo,
Russian Math. Surveys 53 (1998) 1295.
\bibitem{Hir05}
T. Hiroshima, Phys. Rev. A 73, 012330 (2007).
\bibitem{WGKWC04}M. M. Wolf, G. Giedke, O. Kr\"{u}ger, R. F. Werner, and J. I. Cirac,
Phys. Rev. A 69 (2004) 052320.
\bibitem{Shor02}
P. W. Shor, J. Math. Phys. 43 (2002) 4334-4340.
\bibitem{King02}
C. King, J. Math. Phys. 43 (2002) 4641-4653.
\bibitem{King03}
C. King, IEEE Trans. Inf. Theory 49 (2003) 221-229.
\bibitem{MY04}
K. Matsumoto and F. Yura, J. Phys. A 37 (2004) L167-L171.
\bibitem{twamley} J. Twamley, J. Phys. A: Math. Gen. 29 (1996) 3723.
\bibitem{scutaru} H. Scutaru, J. Phys. A: Math. Gen. 31 (1998) 3659;
G. S. Paraoanu and H. Scutaru, Phys. Rev. A 61 (2000) 022306.
\bibitem{wang2}  X-B. Wang, C. H. Oh, and L. C. Kwek, Phys. Rev. A 58 (1998) 4186.
\bibitem{multi}  X-B. Wang, L. C. Kwek, and C. H. Oh, J. Phys. A: Math. Gen. 33 (2000) 4925.
\bibitem{Helstrom:1967}
C. W. Helstrom,
{\em Phys. Lett.}, {\bf 25A}(1976) , 101-102 .
\bibitem{HolP} A. S. Holevo,
{\it Probabilistic and Statistical Aspects of Quantum Theory},
(North-Holland, Amsterdam, 1982); Originally published in Russian
(1980).
\bibitem{Na}
S. Amari and H. Nagaoka, {\it Methods of Information Geometry}, (AMS
\& Oxford University Press, 2000).
\bibitem{Petz3}
D. Petz,
{\em Lin. Alg. Appl.}, {\bf 224}(1996) , 81-96 .
\bibitem{Ma-mas}
K. Matsumoto,
Master Thesis, Department of Mathematical Engineering and
Information Physics, Graduate School of Engineering, The University
of Tokyo, Japan (1995). (In Japanese)
\bibitem{Uhlmann:1993}
A. Uhlmann,
{\em Rep. Math. Phys.}, {\bf 33}(1993) , 253--263 .
\bibitem{Hel}
C. W. Helstrom, {\it Quantum Detection and Estimation Theory,}
(Academic Press, New York, 1976).
\bibitem{YL}
H. P. Yuen and M. Lax,
{\em IEEE Trans. Infor. Theory}, {\bf 19} (1973) , 740 .
\bibitem{Per}
S. D. Personick, {\em IEEE Trans. Infor. Theory}, {\bf 17} (1971),
240.
\bibitem{HoC}
A. S. Holevo,
{\em Rep. Math. Phys.}, {\bf 16}(1979) , 385--400 .

\bibitem{Oza-20}
M. Ozawa,
{\em Research Reports on Information Sciences}, {\bf A-74} (1980).

\bibitem{Bogomolov}
N. A. Bogomolov,
{\em Teor. Veroyatnost. i Primenen.}, {\bf 26}, 798--807 (1981)
(English translation: {\em Theory Probab. Appl.}, {\bf 26} (1981) ,
787--795.

\bibitem{CKNWW}N.J. Cerf, O. Krueger, P. Navez, R.F. Werner, M.M. Wolf,
``The optimal cloning of quantum coherent states is non-Gaussian,''
{\em Phys. Rev. Lett.} {\bf 95} (2005), 070501.


\bibitem{HM}
M. Hayashi and K. Matsumoto,
quant-ph/0411073 (2004).


\bibitem{Hayashi:2000}
M. Hayashi,
in {\it Quantum Communication, Computing, and Measurement 2}, edited
by P. Kumar, G. M. D'ariano and O. Hirota, 99--104 (Plenum, New
York, 2000); quant-ph/9809002 (1998). (It is also appeared as
Chapter 14 of {\it Asymptotic Theory of Quantum Statistical
Inference,} M. Hayashi eds.)

\bibitem{HP}
F. Hiai and D. Petz,
{\em Comm. Math. Phys.}, {\bf 143} (1991), 99--114.


\bibitem{Oga-Nag:test}
T.\ Ogawa and H.\ Nagaoka,
{\em IEEE Trans. Infor. Theory},
{\bf 46} (2000), 2428--2433 ;
quant-ph/9906090 (1999).


\bibitem{H2001}
M. Hayashi,
quant-ph/0107004 (2001);
{\em J. Phys. A: Math. and Gen.}, {\bf 35} (2002), 10759-10773 .

\bibitem{H2001n}
M. Hayashi, quant-ph/0702250; J. Hasegawa, M. Hayashi, T Hiroshima, A. Tanaka, and A. Tomita, quant-ph/0705.3081; 
M. Hayashi, quant-ph/0702251.



\end{thebibliography}
\end{document}